# Roadmap on Spin-Wave Computing

A. V. Chumak[1], P. Kabos[2], M. Wu[3], C. Abert[1, 4], C. Adelmann[5], A. Adeyeye[6], J. Åkerman[7], F. G. Aliev[8], A. Anane[9], A. Awad[7], C. H. Back[10], A. Barman[11], G. E. W. Bauer[12,13], M. Becherer[14], E. N. Beginin[15], V. A. S. V. Bittencourt[16], Y. M. Blanter[17], P. Bortolotti[9], I. Boventer[9], D. A. Bozhko[18], S. A. Bunyaev[19], J. J. Carmiggelt[17], R. R. Cheenikundil[20], F. Ciubotaru[5], S. Cotofana[21], G. Csaba[22], O. V. Dobrovolskiy[1], C. Dubs[23], M. Elyasi[12], K. G. Fripp[24], H. Fulara[25], I. A. Golovchanskiy[27, 28], C. Gonzalez-Ballestero[28, 29], P. Graczyk[30], D. Grundler[31], P. Gruszecki[32], G. Gubbiotti[33], K. Guslienko[34, 35], A. Haldar[36], S. Hamdioui[21], R. Hertel[20], B. Hillebrands[37], T. Hioki[12], A. Houshang[7], C.-M. Hu[38], H. Huebl[39], M. Huth[40], E. Iacocca[18], M. B. Jungfleisch[41], G. N. Kakazei[19], A. Khitun[42], R. Khymyn[7], T. Kikkawa[43], M. Kläui[44], O. Klein[45], J. W. Kłos[32], S. Knauer[1], S. Koraltan[1], M. Kostylev[46], M. Krawczyk[32], I. N. Krivorotov[47], V. V. Kruglyak[24], D. Lachance-Quirion[48], S. Ladak[49], R. Lebrun[9], Y. Li[50], M. Lindner[23], R. Macêdo[51], S. Mayr[52,53], G. A. Melkov[54], S. Mieszczak[32], Y. Nakamura[55,56], H. T. Nembach[2,57], A. A. Nikitin[58], S. A. Nikitov[59], V. Novosad[50], J. A. Otalora[60], Y. Otani[61,62], A. Papp[22], B. Pigeau[63], P. Pirro[37], W. Porod[64], F. Porrati[40], H. Qin[65], B. Rana[30], T. Reimann[23], F. Riente[66], O. Romero-Isart[28,29], A. Ross[9], A. V. Sadovnikov[67], A. R. Safin[59], E. Saitoh[12,43,68], G. Schmidt[69,70], H. Schultheiss[71], K. Schultheiss[71], A.A. Serga[37], S. Sharma[16], J. M. Shaw[2], D. Suess[1, 4], O. Surzhenko[23], K. Szulc[32], T. Taniguchi[10], M. Urbánek[72, 73], K. Usami[55], A. B. Ustinov[58], T. van der Sar[17], S. van Dijken[65], V. I. Vasyuchka[37], R. Verba[74], S. Viola Kusminskiy[16, 75], Q. Wang[1], M. Weides[51], M. Weiler[37], S. Wintz[76], S. P. Wolski[55], X. Zhang[77]

[1]Faculty of Physics, University of Vienna, Boltzmanngasse 5, A-1090 Vienna, Austria
[2]National Institute of Standards and Technology, Boulder Colorado 80305, United States
[3]Department of Physics, Colorado State University, Fort Collins, Colorado 80523, United States
[4]Research Platform MMM Mathematics-Magnetism-Materials, University of Vienna, Oskar-Morgenstern-Platz 1, 1090 Vienna, Austria
[5]Imec, 3001 Leuven, Belgium
[6]Department of Physics, Durham University, South Road, Durham, DH1 3LE, United Kingdom
[7]Department of Physics, University of Gothenburg, Gothenburg, 412 96, Sweden
[8]Departamento Física de la Materia Condensada C-III, Instituto Nicolás Cabrera (INC) and Condensed Matter Physics Institute (IFIMAC), Universidad Autónoma de Madrid, Madrid, 28049, Spain
[9]Unité Mixte de Physique CNRS, Thales Université Paris Saclay, Palaiseau 91767 France
[10]Department of Physics, Technical University of Munich, Garching, 85748, Germany
[11]Department of Condensed Matter Physics and Material Sciences, S N Bose National Centre for Basic Sciences, Salt Lake, Kolkata, 700106, India
[12]Advanced Institute for Materials Research, Tohoku University, Sendai, 980-8577, Japan
[13]Zernike Institute for Advanced Materials, University of Groningen, Netherlands
[14]Department of Electrical and Computer Engineering, Technical University of Munich, Munich, 80333, Germany
[15]Laboratory of Magnetic Metamaterials, Saratov State University, Saratov, 410012, Russian Federation
[16]Max Planck Institute for the Science of Light, Erlangen, 91058, Germany
[17]Department of Quantum Nanoscience, Kavli Institute of Nanoscience, Delft University of Technology, 2628 CJ, Delft, The Netherlands
[18]Department of Physics and Energy Science, University of Colorado Colorado Springs, Colorado Springs, 80918 CO, USA
[19]Institute of Physics for Advanced Materials, Nanotechnology and Photonics (IFIMUP)/Departamento de Física e Astronomia, Universidade do Porto, 4169-007 Porto, Portugal
[20]Institut de Physique et Chimie des Matériaux de Strasbourg, CNRS, Université de Strasbourg, F-67000 Strasbourg, France
[21]Department of Quantum and Computer Engineering, Delft University of Technology, 2628 CD Delft, The Netherlands
[22]Faculty for Information Technology and Bionics, Pazmany Peter Catholic University, Prater u. 50/A, Budapest, H-1083, Hungary
[23]INNOVENT e.V. Technologieentwicklung, 07745 Jena, Germany
[24]School of Physics and Astronomy, University of Exeter, Exeter, EX4 4QL, United Kingdom
[25]Department of Physics, Indian Institute of Technology Roorkee, Roorkee 247667, India
[27]Moscow Institute of Physics and Technology, National Research University, Dolgoprudny, 141700, Russian Federation
[28]National University of Science and Technology MISIS, Moscow, 119049, Russian Federation
[28]Institute for Quantum Optics and Quantum Information of the Austrian Academy of Sciences, Innsbruck, A-6020, Austria

[29]Institute for Theoretical Physics, University of Innsbruck, A-6020 Austria

[30]Institute of Molecular Physics, Polish Academy of Sciences, Poznan, 60-179, Poland

[31]Institute of Materials (IMX) and Institute of Electrical and Micro Engineering (IEL), Ecole Polytechnique Federale de Lausanne (EPFL), Lausanne, 1015, Switzerland

[32]ISQI, Faculty of Physics, Adam Mickiewicz University, Poznan, 61-614, Poland

[33]Istituto Officina dei Materiali del CNR (CNR-IOM), c/o Dipartimento di Fisica e Geologia, Universita di Perugia, Perugia, I-06123, Italy

[34]Division de Fisica de Materiales, Depto. Polimeros y Materiales Avanzados: Fisica, Quimica y Tecnologia, Universidad del Pais Vasco, UPV/EHU, 20018 San Sebastian, Spain

[35]IKERBASQUE, the Basque Foundation for Science, 48009 Bilbao, Spain

[36]Department of Physics, Indian Institute of Technology Hyderabad, Kandi, 502284, Telangana, India

[37]Fachbereich Physik and Landesforschungszentrum OPTIMAS, Technische Universität Kaiserslautern, D-67663 Kaiserslautern, Germany

[38]Department of Physics and Astronomy, University of Manitoba, Winnipeg, R3T 2N2, Canada

[39]Walther-Meißner Institut, Bayerische Akademie der Wissenschaften, Garching, 85748, Germany

[40]Physikalisches Institut, Goethe University, 60438 Frankfurt am Main, Germany

[41]Department of Physics and Astronomy, University of Delaware, Newark, 19716, Delaware, United States

[42]Department of Electrical and Computer Engineering, University of California, Riverside, Riverside, 92521, California, United States

[43]Department of Applied Physics, The University of Tokyo, Tokyo, 113-8656, Japan

[44]Institute of Physics, Johannes Gutenberg University Mainz, Mainz, 55099, Germany

[45]Université Grenoble Alpes, CEA, CNRS, Grenoble INP, Spintec, 38054 Grenoble, France

[46]Department of Physics and Astrophysics, University of Western Australia, Perth, 6009, WA, Australia

[47]Department of Physics and Astronomy, University of California, Irvine, 92697, CA, United States

[48]Nord Quantique, Sherbrooke, Québec, J1K 0A5, Canada

[49]School of Physics and Astronomy, Cardiff University, Cardiff, CF24 3AA, United Kingdom

[50]Materials Science Division, Argonne National Laboratory, Argonne, 60439, Illinois, United States

[51]James Watt School of Engineering, Electronics and Nanoscale Engineering Division, University of Glasgow, Glasgow, G12 8QQ, United Kingdom

[52]Paul Scherrer Institut, 5232 Villigen PSI, Switzerland

[53]Laboratory for Mesoscopic Systems, Department of Materials, ETH Zurich, 8093 Zurich, Switzerland

[54]Faculty of Radiophysics, Electronics and Computer Systems, Taras Shevchenko National University of Kyiv, Kyiv, 01601, Ukraine

[55]Research Center for Advanced Science and Technology (RCAST), The University of Tokyo, Meguro-ku, Tokyo, 153-8904, Japan

[56]RIKEN Center for Quantum Computing (RQC), RIKEN, Wako-shi, Saitama, 351-0198, Japan

[57]Department of Physics University of Colorado, Boulder, Colorado 80309 USA

[58]Department of Physical Electronics and Technology, St. Petersburg Electrotechnical University, St. Petersburg, 197376, Russian Federation

[59]Kotel'Nikov Institute of Radio-Engineering and Electronics of RAS, Moscow, 125009, Russian Federation

[60]Departamento de Fí-sica, Universidad Católica del Norte, Av. Angamos 0610, Antofagasta, Chile

[61]Institute for Solid State Physics, University of Tokyo, Kashiwa, 277-8581, Chiba, Japan

[62]RIKEN-CEMS, 2- 1 Hirosawa, Saitama, Wako, 351-0198, Japan

[63]Institut Néel, Université Grenoble Alpes-CNRS:UPR2940, Grenoble, 38042, France

[64]Department of Electrical Engineering, University of Notre Dame, Notre Dame, 46556, IN, United States

[65]Department of Applied Physics, Aalto University School of Science, FI-00076 Aalto, Finland

[66]Department of Electronics and Telecommunications, Politecnico di Torino, Corso Duca degli Abruzzi 24, 10129 Torino, Italy

[67]Laboratory of Magnetic Metamaterials, Saratov State University, Saratov, 410012, Russian Federation

[68]Institute for AI and Beyond, The University of Tokyo, Tokyo 113-8656, Japan

[69]Institut für Physik, Martin-Luther-Universität Halle-Wittenberg, Halle, D-06099, Germany

[70]Interdisziplinäres Zentrum für Materialwissenschaften, Martin-Luther-Universität Halle-Wittenberg, Halle, D-06099, Germany

[71]Helmholtz-Zentrum Dresden-Rossendorf, Institute of Ion Beam Physics and Materials Research, Dresden, 01328, Germany

[72]CEITEC BUT, Brno University of Technology, 612 00 Brno, Czech Republic

[73]Institute of Physical Engineering, Brno University of Technology, 616 69 Brno, Czech Republic

[74]Institute of Magnetism, UKR-03142 Kyiv, Ukraine

[75]Department of Physics, University Erlangen-Nürnberg, Erlangen, 91058, Germany

[76]Max Planck Institute for Intelligent Systems, 70569 Stuttgart, Germany

[77]Center for Nanoscale Materials, Argonne National Laboratory, Lemont, Illinois 60439, United States

Magnonics is a field of science that addresses the physical properties of spin waves and utilizes them for data processing. Scalability down to atomic dimensions, operations in the GHz-to-THz frequency range, utilization of nonlinear and nonreciprocal phenomena, and compatibility with CMOS are just a few of many advantages offered by magnons. Although magnonics is still primarily positioned in the academic domain, the scientific and technological challenges of the field are being extensively investigated, and many proof-of-concept prototypes have already been realized in laboratories. This roadmap is a product of the collective work of many authors that covers versatile spin-wave computing approaches, conceptual building blocks, and underlying physical phenomena. In particular, the roadmap discusses the computation operations with Boolean digital data, unconventional approaches like neuromorphic computing, and the progress towards magnon-based quantum computing. The article is organized as a collection of sub-sections grouped into seven large thematic sections. Each sub-section is prepared by one or a group of authors and concludes with a brief description of the current challenges and the outlook of the further development of the research directions.

*Index Terms— Spin wave (SW), magnon, magnonics, computing, data processing*

OUTLINE

# I. INTRODUCTION

A DISTURBANCE in the local magnetic order can propagate in a magnetic material in the form of a wave. This wave was first predicted by F. Bloch in 1929 and was named a spin wave because it is related to a collective excitation of the electron spin system in magnetically ordered solid bodies [1]. The quanta of spin waves are referred to as magnons. A wide variety of linear and nonlinear spin-wave phenomena attract interests in both fundamental and applied research. The field of science that refers to information transport and processing by spin waves is known as magnonics [2], [3], [4]. Among the key advantages offered by magnons for data processing are the scalability down to atomic dimensions, the compatibility with existing complementary metal–oxide–semiconductor (CMOS) [5] and spintronic technologies [6], the operations in the frequency range from several GHz to hundreds of THz, the possibility to process data in the wide temperature range from ultra-low temperature to room temperature, and the access to pronounced nonlinear phenomena.

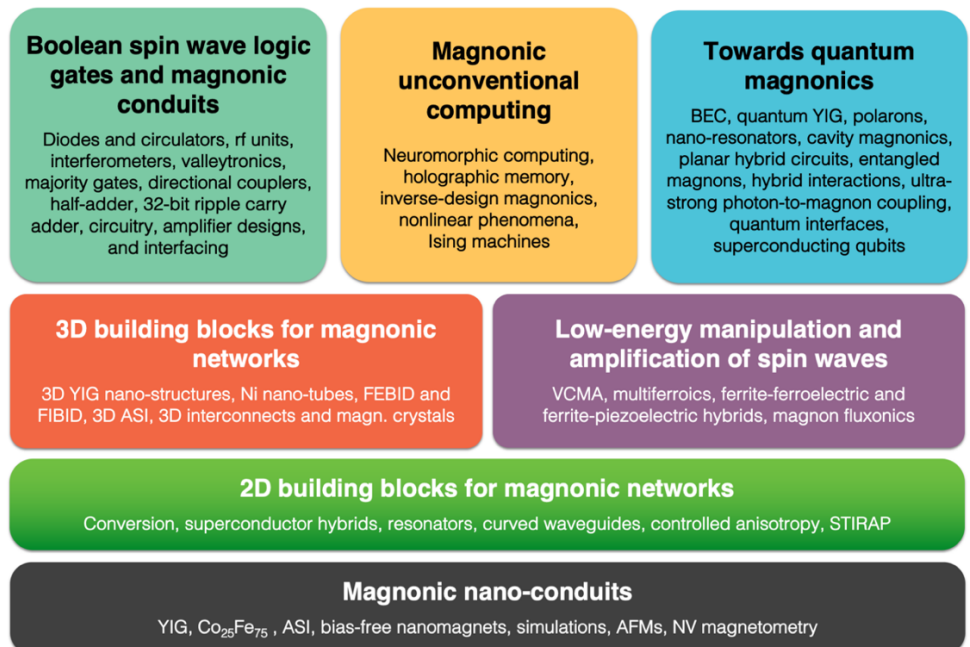

Fig. I.1. Schematic depiction of the roadmap structure, sections, and content. The more advanced and applied building blocks are staying on top of the more fundamental ones. Abbreviations used in the figure: AFM: antiferromagnetic; ASI: artificial spin ice; BEC: Bose-Einstein condensates; FEBID: focused electron beam-induced deposition; FIBID: focused ion beam-induced deposition; NV: nitrogen vacancy; STIRAP: STImulated Raman Adiabatic Passage; VCMA: voltage-controlled magnetic anisotropy; YIG: yttrium iron garnet.

Classically, spin waves were utilized to design passive and active radio-frequency microwave devices [7], [8], [9], [10]. Nowadays, they are attracting considerable attention as data carriers in novel computing devices in addition to (or even instead of) electrons in electronics [6], [11], [12]. Magnon-based computing is a broad field that includes for example the processing of Boolean digital data, unconventional approaches like neuromorphic computing, and quantum magnonics aiming to utilize entangled magnon states to process information, as illustrated in Fig. I.1. The field is very young and is still located primarily in the academic domain, rather than in the engineering or manufacturing domain. The novelty of the field allows for a broad exploitation of the underlying physical phenomena for a broad range of applications. Moreover, magnonics is strongly coupled to other fields of modern physics and technology, such as material science, nanotechnology, spintronics, photonics, semiconductor electronics, physics of superconductors, quantum optics, and electronics. All these synergies explain the dynamic evolution of the field. A simple search in Scopus shows that the number of publications referring to "magnon" per year was relatively constant, around one hundred, from 1970 to 2000, as shown in Fig. I.2(a). However, an exponential growth in the number of papers is observed afterward. In particular, more than 500 papers were published in 2020 (the search on "spin wave" gives about 2500 papers in 2020). If one analyzes the number of patents in which the word "magnon" is mentioned, the dependence is somewhat different. As shown in Fig. I.2(b), there is a growth in the patent number until about the year 2000 and then saturates at about 30 patents per year. The total number of patents provided by Google patent search is 893.

This roadmap gives an overview of magnon-based data processing and discusses future directions and challenges. The structure and thematic sections of the article are shown in Fig. I.1. It covers diverse topics, starting from magnonic nano-conduits and two-dimensional (2D) and three-dimensional (3D) building blocks for magnonic networks, and then proceeding to progress reports towards Boolean, unconventional, and quantum computing. The roadmap is organized in a "self-assembled" manner: The contributors were invited not only by the editors but also by the co-authors themselves. As a result, a kind of "fingerprint" of the current interests in the community was achieved. The roadmap keeps a concise style of presentation.

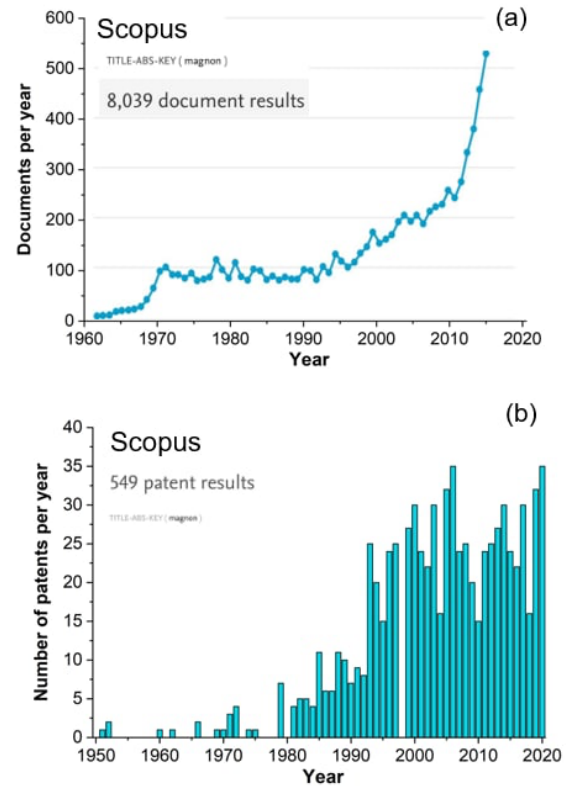

Fig. I.2. Metrics of the field of magnonics. (a) The number of documents per year referring to "magnon" in Scopus. (b) The number of patents per year referring to "magnon" in Scopus. Search performed on 14.10.2021.

For a general overview in the field of magnonics, we would like to attract your attention to already established [2], [3], [4], [6], [13], [14] and very recent [11], [15], [12] reviews that include "Roadmap on Magnonics 2020" by Barman et al. [12]. If you are starting to familiarize yourself with spin-wave physics, we recommend you to read classical books by Gurevich-Melkov [16] and Stancil-Prabhakar [17].

## II. MAGNONIC NANO-CONDUITS

Information can be transferred in a solid magnetic material in the form of a spin wave. The wide variety of physical parameters and phenomena that define spin-wave properties allow for the realization of magnonic conduits with unique (e.g. non-reciprocal or nonlinear) properties or with direct combination of data transport and data processing in reconfigurable systems [18], [19]. The starting aim is rather clear and is defined through the realization of magnonic conduits with the smallest possible cross-sections and largest spin-wave propagation lengths. This aim strongly couples the field of magnonics to modern material science and advanced fabrication technologies. For example, the choice of a magnetic material is of high importance. To ensure a large magnon free path, the spin-wave lifetime, which in the simplest case is inversely proportional to the Gilbert damping parameter of a magnetic material, and the spin-wave group velocity should be maximized [20], [11].

We begin this section with a progress report on the growth of nm-thick yttrium-iron garnet (YIG) films using sputtering and liquid phase epitaxy techniques. These films are the absolute champions with regards to the magnon lifetime. Also, short-wavelength exchange waves in these films allow for data processing with fast magnons. The fabrication of 50-nm-wide waveguides and the investigation of their spin-wave properties in the waveguides are reported in the following section. Further, we discuss the compound $Co_{25}Fe_{75}$ that is very promising due to its CMOS-compatibility and fast dipole-exchange spin waves. The utilization of artificial spin ices (ASIs) as 2D magnonic conduits with reconfigurable properties are discussed afterwards. Section II-F is devoted to reconfigurable and bias-free waveguides in a form of physically separated rhomboid nanomagnets.

Numerical simulations play a pivotal role in the investigations of spin-wave physics and the development of conduits with novel physical properties. We report advances in the computational methods and the combination of purely magnonic simulations with spintronics phenomena, such as spin pumping and spin Hall effects. Next, we discuss spin-wave propagation in nanoscale waveguides with non-uniform magnetization that allow for the control of spin-wave properties. The next two sections discuss the utilization of another essential class of materials for magnonics – antiferromagnetic (AFM) materials. AFM magnonic conduits are robust against magnetic perturbations and also allow for the investigation into the spin dynamics in the hundreds of gigahertz to terahertz regime. We conclude Section II with the discussion of nitrogen-vacancy (NV) magnetometry as a novel technology that enables the highly resolved characterization of spin waves in a diverse set of materials, including monolayer van-der-Waals and complex-oxide magnets.

## A. Sputtering growth of YIG thin films with low damping and perpendicular anisotropy

Magnetic damping in yttrium iron garnet (YIG) $Y_3Fe_5O_{12}$ is lower than in other magnetic materials. As such, YIG materials have been widely used in microwave devices, including phase shifters, isolators, and circulators [21], [22]. As mentioned earlier, yttrium iron garnet is also a material of choice for magnonic device applications. Magnonic applications require YIG thin films with a thickness in the nanometer (nm) range, but it is challenging to grow YIG films that are nanometer thick, and exhibit a damping constant comparable to that of the bulk value.

Recently, Ding and Liu *et al.* demonstrated the growth of nm-thick YIG films via radio-frequency sputtering at room temperature and post-annealing in $O_2$ at high temperature [23]. The optimization of the annealing temperature enabled the realization of YIG films with a Gilbert damping constant of $\alpha \approx 5.2\times10^{-5}$. Two important points should be made here. First, the damping constant here represents the lowest damping reported so far for magnetic films, either metallic or insulating, thinner than 200 nm. Second, previous and ongoing efforts on YIG film growth utilize mainly liquid phase epitaxy, pulsed laser deposition, and sputtering techniques; among them, sputtering is the most industry-friendly approach.

Figure II.A(a) shows representative ferromagnetic resonance (FMR) data obtained on a 75-nm-thick YIG film grown on a $Gd_3Ga_5O_{12}$ substrate. The left and right graphs show the FMR linewidth vs. frequency data measured by shorted rectangular waveguides under an in-plane field and an out-of-plane field, respectively, as indicated. The blue symbols show the data, while the red lines are linear fits. The fitting-yielded two $\alpha$ values which nearly agree with each other. This agreement indicates that two-magnon scattering, if present, is very weak, and the measured $\alpha$ values represent the actual damping in the film. Though not shown in the figure, the film exhibited a surface roughness of only 0.08 nm, a coercivity of only 0.18 Oe, and a saturation induction of 1778 G, which is very close to the bulk value.

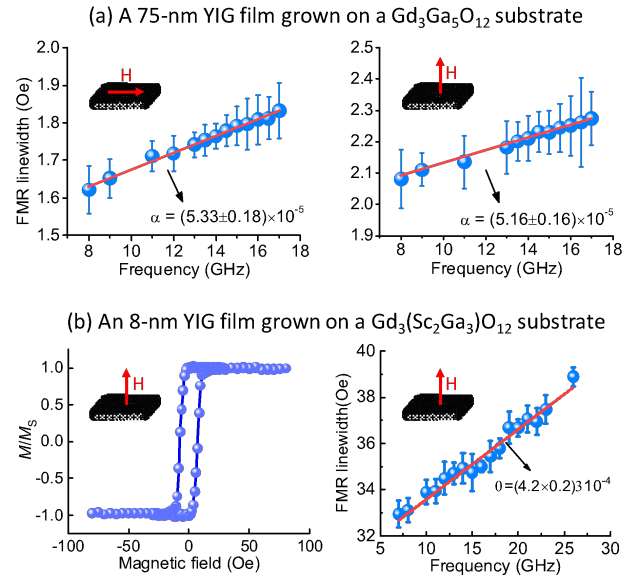

(a) A 75-nm YIG film grown on a $Gd_3Ga_5O_{12}$ substrate

$\alpha = (5.33\pm0.18)\times10^{-5}$

$\alpha = (5.16\pm0.16)\times10^{-5}$

(b) An 8-nm YIG film grown on a $Gd_3(Sc_2Ga_3)O_{12}$ substrate

$\alpha = (4.2\pm0.2)\times10^{-4}$

Fig. II.A. Representative data on YIG thin films grown by sputtering [23], [24]. (a) FMR linewidth vs. frequency data obtained on a 75-nm-thick YIG film grown on a $Gd_3Ga_5O_{12}$ substrate. (b) Magnetic hysteresis loop (left) and FMR linewidth vs. frequency response (right) obtained on a 8-nm-thick YIG film grown on a $Gd_3(Sc_2Ga_3)O_{12}$ substrate. The linewidth data in (a) refer to peak-to-peak FMR linewidth values, while those in (b) are full width at half maximum. The short red arrows indicate the direction of the magnetic field (**H**). The long red lines are linear fits to the data. $M$ and $M_s$ denote the magnetization and saturation magnetization, respectively.

The film described above exhibits weak magnetic anisotropy, but YIG films with perpendicular anisotropy are often preferred in applications. Such films, for example, can have nearly 100% remnant magnetization and thereby do not require an external magnetic field or are self-biased. This self-biasing can significantly facilitate device design and fabrication. Ding, Liu, and Zhang *et al.* recently succeeded in the development of such YIG films [24]. The YIG films in a thickness range of 4-30 nm were deposited by sputtering at room temperature first and then annealed in $O_2$ at high temperature; the same as for the above-described YIG film. The substrates, however, were $Gd_3(Sc_2Ga_3)O_{12}$, not $Gd_3Ga_5O_{12}$. These substrates have a lattice constant slightly larger than that of the YIG material. This lattice mismatching situation resulted in an out-of-plane compressive strain, and the latter induced a perpendicular magneto-elastic anisotropy in the YIG film.

As an example, Fig. II.A(b) presents the data of an 8-nm-thick YIG film grown on a $Gd_3(Sc_2Ga_3)O_{12}$ substrate. The left graph shows a magnetic hysteresis response measured by a vibrating sample magnetometer. The data show a square-like hysteresis loop with a remnant-to-saturation magnetization ratio of 98%, which clearly indicates the presence of perpendicular anisotropy in the film. The right graph presents the FMR linewidth vs. frequency response measured by a vector network analyzer. The blue symbols show the data, while the red line is a linear fit. The fitting yields a damping constant of $\alpha \approx 4.2 \times 10^{-4}$. This value is about eight times larger than the damping constant cited above, but it represents the lowest damping constant measured on nm-thick films with perpendicular anisotropy under out-of-plane fields.

In terms of the applications of the YIG thin films in magnonic computing, the following two future works are of great interest: (1) the optimization of the sputtering process for the realization of YIG films with even lower damping and (2) the patterning of YIG films to nano-conduits. The first one is of particular importance for YIG films with perpendicular anisotropy, because the damping constant obtained so far is still one order of magnitude larger than the intrinsic damping constant in YIG materials. The second could be achieved through photolithography and ion milling processes, as demonstrated recently by Wang *et al.* and Heinz *et al.* [25], [26].

### Acknowledgment

Supported by the U.S. National Science Foundation (EFMA-1641989; ECCS-1915849; DMR-2002980).

### Contributors

This section is authored by Mingzhong Wu (Department of Physics, Colorado State University, USA).

### B. Fabrication of nm-thin YIG films by liquid phase epitaxy

Micrometer-thick, epitaxial iron garnet films, such as yttrium iron garnet ($Y_3Fe_5O_{12}$, YIG), magnetic bubble films, as well as magneto-optical indicator films, are typical representatives of magnetic ferrites that can be grown by liquid phase epitaxy (LPE). Around 1960, bulk magnetic crystals were needed for microwave applications. In 1968, Linares et al. (see e.g. [27]) demonstrated that YIG films can be grown from high temperature solutions. Since then, this class of materials and the LPE technology have been a playground for scientists studying the fundamental aspects of magnetism, microwaves, optics and acoustics (see e.g. [28]). However, modern concepts and forward-looking applications that take advantage of the static and dynamic properties of sophisticated magnetic materials e.g., for spin wave-based data processing (see e.g. [6]) demand microstructurally perfect epitaxial films with nanometer thickness on the largest possible substrate wafer diameter. Therefore, the challenge was to grow nanometer-thin epitaxial YIG films by using classical LPE technology.

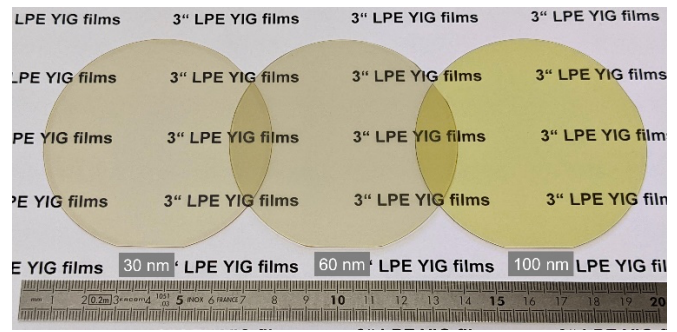

Fig. II.B. 3-inch LPE YIG films with different film thicknesses on (001) (left, middle) and (111) (right) GGG substrate wafers.

During the last five decades, the thicknesses of high-quality LPE-grown garnet films have typically ranged from 1 up to 500 micrometers. It has recently been demonstrated that modern iron garnet LPE films can also be grown on 1-inch $Gd_3Ga_5O_{12}$ (GGG) wafers down to a film thickness of 10 nm [29], [30]. These films are characterized by ideal stoichiometry as well as perfect film lattices and exhibit almost abrupt film/substrate interfaces. Their magnetostatic and dynamic behavior is similar to single crystal bulk YIG, and the Gilbert damping coefficient $\alpha^\parallel$ is independent of the film thickness and is close to $1 \times 10^{-4}$ [30]. The perfection of nm-thin YIG LPE films has already enabled the fabrication of 50 nm wide nano-scaled conduits with spin-wave propagation decay lengths of 8 micrometer [31]. Furthermore, control of intermodal dissipation processes of strongly driven propagating spin waves in nanodevices has been achieved [32] and the first magnon-based data processing devices such as magnonic directional couplers have been realized [33], with all of these devices operating in a single-mode regime.

In addition to the development of nm-thin YIG LPE films on 1-inch substrate wafers, the foreseeable requirement to provide high-performance YIG films on larger GGG wafer diameters, had led us to scale the LPE process to 3-inch film diameters. First, 3-inch YIG films with thicknesses down to 30 nm, were successfully grown on (111) and (001) GGG substrate wafers (see Fig. II.B). The thickness homogeneities ranged from ± 8 % for 30 nm to ± 0.8 % for 100 nm thin films over 85 % of the wafer area. FMR measurements on a 50 nm thin 3-inch film revealed FWHM linewidths @6.5 GHz of 1.7 Oe and a Gilbert damping coefficient $\alpha^{\parallel}$ close to $2\times10^{-4}$.

The main advantages of LPE technology are that up to 3- or 4-inch large films are easy to grow at ambient atmosphere. The perfectly oriented epitaxial films have smooth surfaces with uniform film properties over almost the entire wafer diameter. Due to the straightforwardness of creating complex multicomponent high-temperature solutions, a wide variety of substituted garnet film compositions are accessible and user-defined properties can be set. The LPE setup is not as expensive as vacuum-based deposition techniques, and technological solutions already exist to deposit more than one wafer per run by a multi-wafer technology. If future spin-wave devices for integrated logic circuits e.g., for half- or full-adder architectures, can be microfabricated from only one functional planar layer, and without the necessity of 3D intersections, simplified and cost-efficient device fabrication is realizable.

Therefore, LPE of magnetic garnets plays its part in providing nm-thin, high quality YIG films with large diameters (see e.g. [34]), for the fabrication of nanoscaled spin-wave devices in planar technology to guide dipolar [33] as well as exchange-domitated spin waves by utilizing magnetic arrays of ferromagnetic materials patterned on top of the thin films [35].

If, on the other hand, magnonic circuits require 3D multilayer assembly of nm-thin garnet films, LPE technology comes up against its limits, as the liquid phase equilibrium growth conditions of a second phase could result in a dissolution of the previously grown epitaxial film. Another challenge is the deposition of garnet films on non-lattice matched substrates or semiconductor materials as Si, GaAs, GaN, which have already been successfully demonstrated using PLD and sputtering techniques (see e.g. refs. in [36]). In these cases the lattice misfits could be too large for an epitaxial intergrowth and the solubility of non-oxide substrates in high-temperature solutions of oxides could prevent film growth. Without buffer layers, which are deposited e.g. by gas phase epitaxy, to protect the semiconductor substrates from dissolution, perfect epitaxial films with sharp interfaces cannot be expected. However, the continuous search for new iron garnet materials, e.g. as fully or partially compensated ferrimagnets with perpendicular magnetic anisotropy [37], [38], [39] with the required robust temperature stability and fast domain wall motion [40], or substituted iron garnets with enhanced magneto-optical Faraday and Kerr effect (see e.g. ref. [41]) as swell as enhanced spin Seebeck effect [38]) opens up further opportunities for advanced thin-film LPE technology, which is predestined for the fabrication of multicomponent epitaxial films.

## ACKNOWLEDGMENT

C.D. acknowledges the funding by the Deutsche Forschungsgemeinschaft (DFG, German Research Foundation) – 271741898 and M. L. by the German Bundesministerium für Wirtschaft und Energie (BMWi) - 49MF180119. We also thank B. Wenzel, R. Meyer, M. Reich, P. Lasch and A. Hertzsch for their support.

## CONTRIBUTORS

This section is authored by C. Dubs, O. Surzhenko, M. Lindner, and T. Reimann (INNOVENT e.V. Technologieentwicklung Jena, Germany).

### C. Single-mode sub-100 nm YIG magnonic nano-conduits

Spin-wave conduits are fundamental elements for guiding of spin waves and realization of magnonic circuits. They have been investigated with different widths that range from millimeters [3] down to micrometers [42],[14], with several width modes observed due to the lateral boundary conditions. This multimode character, however, raises the problem of magnon scattering between different width modes after the spin waves pass through bents in the waveguides or impurities [43]. The magnon scatterings change the wavelength of the spin waves and significantly complicate wave-based data processing due to the interference of spin waves with different wavelengths. The most straightforward way to avoid this scattering is to decrease the size of the waveguide to the nanoscale, in which the higher-order width modes are well separated in frequency due to quantized wavevector contribution associated with exchange interactions. Another advantage of reduction of waveguide dimensions to the nanoscale is the waveguide's shape anisotropy, which makes the static magnetization parallel to the long axis of the waveguide even without an external magnetic field [25]. In addition, magnonic nano-conduits can be efficiently coupled with other spintronic nano-elements such as spin-torque and spin-Hall nano-oscillators.

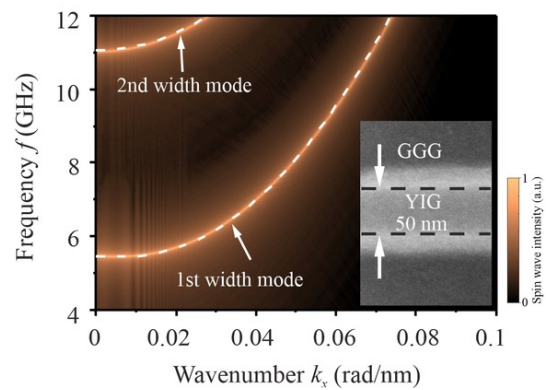

Fig. II.C. Spin-wave dispersion curve of 50 nm wide YIG magnonic waveguide with external field of 108.9 mT along the waveguide. In the frequency range from 5.4 GHz to 11 GHz the operation with only one spin-wave mode is possible.

Recently, magnonic waveguides with widths down to 50 nm have been successfully fabricated, and a new theory has been developed to describe the properties of spin waves in them [25]. Fig. II.C shows the spin-wave dispersion curve in a 50-nm-wide YIG magnonic waveguide in which the first and second width modes have a very large frequency gap of about 6 GHz. This large frequency gap is sufficient for single-mode spin-wave operation. In such a waveguide, propagating spin waves have been measured using micro-focused Brillouin light scattering spectroscopy in both longitudinally and transversally magnetized configurations [26], [31]. A long-range spin-wave propagation has been observed in transversely magnetized waveguides, with a decay length of up to 8 µm and a large spin-wave lifetime of up to 48 ns. Moreover, a frequency non-reciprocity for counter-propagating spin waves is found which is caused by the trapezoidal cross section of the structure. These results pave a path for the development of further nano-scaled, low-energy magnonic devices and their integration into spintronic circuits.

The main challenges in this direction include the further decrease in the dimensions of the waveguides (the fundamental limitation is the lattice constant of YIG materials that is about 1.2 nm), the switch from gadolinium gallium garnet (GGG) substrates to Si substrates without scarifying the YIG film quality, and the switch from dipolar waves to exchange waves (the maximal propagation length in the described 50-nm-wide waveguides is 56 µm for the wave of 63-nm wavelength).

### Acknowledgment

This research has been supported by ERC Starting Grant 678309 MagnonCircuits, and DFG project no. 271741898, the Austrian Science Fund (FWF) through project I 4696-N.

### Contributors

This section is authored by Q. Wang, A. V. Chumak (University of Vienna, Austria), C. Dubs (INNOVENT e.V., Jena, Germany), and P. Pirro (TU Kaiserslautern, Germany).

### D. Ultralow damping metallic ferromagnets for spin-wave waveguides

In order to perform fast and energy efficient computation and information transport, magnonic circuits rely on magnetic materials with low magnetic damping and high spin-wave group velocities. For large-scale integration, these materials should be CMOS-technology compatible polycrystalline metal thin films, that can be grown on top of an arbitrary template without significant thermal processing. Alloys based on 3d transition metals such as $Co_xFe_{1-x}$, $Fe_xV_{1-x}$ and Fe-Al [44], [45], [46] are a particularly relevant material class as are some $Co_2Mn$ based Heusler compounds [47], [48].

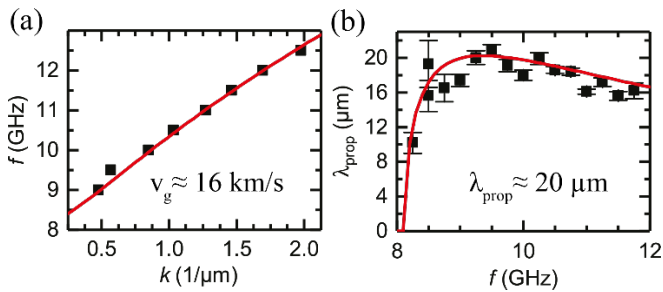

The group velocity of magnetostatic surface spin waves in thin films increases with the total magnetic moment. Therefore, high group velocities are obtained for materials with high saturation moment, leading to long propagation lengths. The $Co_{25}Fe_{75}$ alloy exhibits the lowest magnetic damping of any known polycrystalline metallic ferromagnetic thin film [44] while also featuring a high saturation magnetization $\mu_0 M_s \approx$ 2.4 T. It is worth mentioning that the ultra-low damping $Co_{25}Fe_{75}$ is a result of a sharp minimum in the density of states at the Fermi level at that particular composition [44]. For a similar reason, $Fe_xV_{1-x}$ also has a low intrinsic damping value $\alpha$ < 0.0025 over a wide compositional range, where $\mu_0 M_s$ can be tuned between 0.8 and 2.1 T to suit a particular application [45], [46].

The high $M_s$ of polycrystalline $Co_{25}Fe_{75}$ is particularly useful for spin-wave computation applications since it leads to high dipolar-exchange spin-wave group velocities $v_g \approx$ 16 km/s at a wave vector of $k \approx 0.5$ µm⁻¹ (see the dispersion relation in Fig. II.D.(a)). The data was experimentally obtained by Brillouin Light Scattering in the magnetostatic surface wave configuration in a $Co_{25}Fe_{75}$ magnonic conduit with a width of 5 µm and a thickness of 26 nm for an external magnetic field magnitude of 43 mT. The combination of high group velocity and record low intrinsic damping $\alpha$ < 0.001 [44] results in experimentally observed spin-wave propagation lengths [49] of $\lambda_{prop} \approx$ 20 µm (Fig. II.A (b)) in the same sample. Achieving these high spin-wave propagation lengths requires suppression of two magnon scattering and inhomogeneous relaxation effects [50].

Due to the large moment of $Co_{25}Fe_{75}$, spin waves can be excited and detected by dipolar coupling with high efficiency in magnonic conduits made from this material. Moreover, the metallic system offers high spin-mixing conductance to adjacent heavy metals, enabling spin-orbit torque control of magnetization and large interfacial Dzyaloshinskii-Moriya interaction [51]. The combination of these desirable properties makes $Co_{25}Fe_{75}$ a highly attractive material for magnonic nano-conduits. In a next step, $Co_{25}Fe_{75}$ devices with actual magnonic functionalities will need to be fabricated and tested.

### Acknowledgment

MW acknowledges funding by DFG via projects WE5386/4-1 and WE5386/5-1. HS acknowledges funding by DFG via projects Schu2022/1-3 and Schu2022/4.

### Contributors

This section is authored by M. Weiler (TU Kaiserslautern, Germany), Helmut Schultheiss (Helmholtz-Zentrum Dresden-Rossendorf, Germany), Hans Nembach and Justin Shaw (NIST, Boulder, USA).

Fig. II.D. Spin-wave dispersion (a) and spin-wave propagation length (b) of magnonic conduits made from 30nm thick CoFe (symbols: experimental data, lines: model calculations). Reprinted from [49], with the permission of AIP Publishing.

### E. Harnessing reconfigurability in artificial spin ice for magnonics

2D artificial spin ices (ASIs) are obtained when magnetic thin films are patterned into two-dimensional lattices of strongly coupled elements [52]. This strong coupling leads to frustrated magnetic states that ultimately result in a degenerate energy landscape. Magnons in ASIs are intimately related to the long-range magnetic state [53] and exhibit rich spectra that open opportunities for ASIs as reconfigurable magnonic crystals with on-demand feature toggling [54], [55].

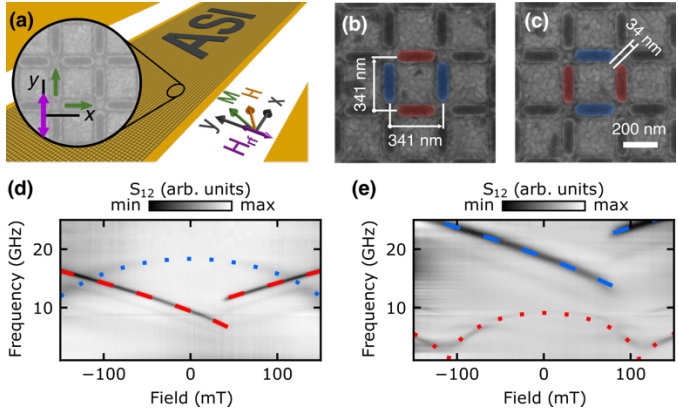

Fig. II.E. Emergent spin dynamics observed in bicomponent ASI made of NiFe and CoFe sublattices. (a) Experimental configuration with ASI patterned on the top of the signal line of a coplanar waveguide. (b) and (c) Scanning electron microscopy images of studied lattices and dimensions. (d) Bicomponent NiFe-CoFe, and (e) bicomponent CoFe-NiFe arrays. The red false color represents nanomagnets made of NiFe, and the blue false color represents nanomagnets made of CoFe. The corresponding absorption spectra at magnetic fields applied parallel to the signal line are shown in (d) and (e). Dashed and dotted lines represent fits to an analytical model. Adapted with permission from [56]. Copyright 2021 American Chemical Society.

To harness reconfigurability, it is fundamental to maintain the strong coupling within the ASI while reliably controlling the long-range magnetic state. For example, Lendinez at al. [56] studied a bicomponent ASI in a square geometry, or square ice, utilizing both NiFe and CoFe elements, such that the combined spectra could be disentangled from each material sublattice. Meanwhile, Gartside et al. [57] presented a scheme where a square ASI is modified by widening a subset of elements relative to the rest of the array enabling the preparation of any microstate via simple global-field protocols and the observation of avoided mode crossings with a coupling strength tunable via microstate selection. An alternative is to control the dynamic energy coupling itself using extended substrates, such as a soft magnetic layer [58]. In addition to the coupling, underlying substrates can enhance the ASI's functionality by the active modification of the magnon's dipolar-exchange character and propagation directions. The main challenge for further progress in utilizing ASIs for magnonics relies on the full 2D control of magnons. Intriguing possibilities include the use of bicomponent ASI to establish topologically protected edge modes or induce exceptional points; or the use of a variety of substrates to unlock novel functionalities relying on surface effects such as spin-orbit coupling when coupled to a heavy metal layer and manipulating superconductivity in an adjacent superconductor.

Another route that has been explored recently relies on connected ASIs, patterned thin films into a mesh. Bhat et al. [59] showed that a connected Kagome lattice exhibits rich spectra with spatially localized modes imaged by Brillouin light scattering. These types of ASIs can further use current-induced spin-transfer torque effects to be reconfigured and, hence, to manipulate the magnetization dynamics. It would be interesting to explore some of the techniques presented before in the context of connected ASIs in conjunction with the unique features that this platform can provide.

#### ACKNOWLEDGMENT

M. B. Jungfleisch acknowledges startup funding from the University of Delaware.

#### CONTRIBUTORS

This section is authored by Ezio Iacocca (University of Colorado Colorado Springs, Colorado Springs, USA) and M. Benjamin Jungfleisch (University of Delaware, USA).

### F. Reconfigurable and bias-free waveguides

One of the building blocks of a magnonic device is a magnetic waveguide that supports the propagation and processing of the spin waves or magnons [60], [61]. Magnetic waveguides are fabricated from low damping magnetic materials such as YIG and Py. Spin waves are broadly characterized by their wavelengths where short and long-wavelength spin waves are known as exchange and dipolar spin waves, respectively. Here, we focus on the dipolar spin waves and their dispersion ($\omega(k)$) which strongly influences the spin-wave propagation and depends on the magnetization orientation resulting in two types of spin waves – known as surface (surface ($M \perp k$) and backward volume ($M \parallel k$) – for in-plane magnetized samples and the third type – known as forward volume ($M \perp k$) spin waves for the out-of-plane magnetized samples. Magnonic devices exploit one of these three types of spin waves and thus an external bias ($\sim 500 - 5000$ Oe) magnetic field is typically used to achieve a particular type of spin-wave dispersion relation. It is a major bottleneck for the on-chip integration of such waveguides.

Here, we discuss one of our recent demonstrations where surface spin-wave propagations are realized without requirement of any external bias magnetic field. In this regard, a chain of dipolar coupled but physically separated rhomboid nanomagnets is used as a waveguide (Fig. II.F (a, b)) that supports surface spin-wave propagation [62]. The shape of the rhomboid nanomagnets allows them to have a unique remanent state where magnetization points along the easy axis of the nanomagnets depending on a simple field initialization either along the short or long axis [63]. Note that one nanomagnet with mirrored orientation to the others has been placed at the gate position as indicated by an arrow in Fig. II.F(b). One of the remanent states of the waveguides is referred to as a ferromagnetically ordered (FO) state where all the nanomagnets point in the same direction when initialized with a field along the long axis of the nanomagnets (Fig. II.F(c)). The gate nanomagnet is switched to the opposite direction to the others (referred to as FO*) when the nanomagnets were initialized along their short axes (Fig. II.F(d)). Spin-wave propagation has been demonstrated by the micro-focused Brillouin light scattering (micro-BLS) technique using two-dimensional (2D) spatial scanning of the laser spot. An uninterrupted flow of spin waves is found for the FO remanent state of the waveguide (Fig. II.F(e)). However, the FO* remanent state shows a discontinuity at the gate position due to the flipped nanomagnet, thereby drastically reducing the spin-wave flow beyond the gate position as indicated by the circle at the output position. Note that such a spin-wave gating operation can also be achieved by using a high-power microwave current using the same input antenna [64]. Another bottleneck in the surface wave geometry using an externally applied field is related to the propagation of the spin waves around a bend as an external global bias magnetic field does not satisfy the required geometry in the bend area. Since the waveguide in Fig. II.F(b) is self-biased, channeling spin waves around a bend is easily possible without any magnetic field. In another demonstration a self-biased magnetic waveguide has been realized for forward volume spin waves [65] which have the advantage of isotropic propagation – a crucial ingredient for complex magnonic circuits.

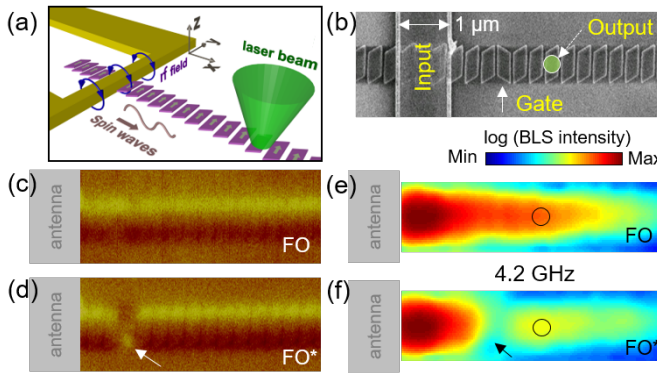

Fig. II.F. (a) Schematic of the self-biased waveguide based on rhomboid nanomagnets, excitation antenna (input) and micro-BLS laser spot (output). (b) Scanning electron micrograph image of the magnonic device. Magnetic force microscopy images for the (c) FO and (d) FO* states. 2D Micro-BLS scan images of spin waves for the (e) FO and (f) FO* states. Reprinted (adapted) with permission from [Haldar2016a]. Copyright 2016 Nature Publishing Group.

The development of functional and reconfigurable self-biased waveguides is essential for the practical implementation of magnonic devices. Now that a few such waveguides are available, it will be of great interest to investigate their performance with electrical input for the spin-wave generation and electrical output for the detection. Therefore, the waveguides can be further downsized with such electrical input and output terminals which generally possess smaller footprints. In addition, the use of emerging phenomena such as spin orbit torque can be exploited for the gating of spin-wave propagation in these waveguides.

## ACKNOWLEDGMENT

A.H. would like to thank the funding under the Ramanujan Fellowship (SB/S2/RJN-118/2016), DST, India. AOA would like to acknowledge the funding from the Royal Society and Wolfson Foundation.

## CONTRIBUTORS

This section is authored by A. Haldar (Indian Institute of Technology Hyderabad, India). and A. O. Adeyeye (Durham University, UK).

### G. Steering of spin waves along domain walls and structure edges

In 1961 Winter predicted the existence of new class of localized quasi-one-dimensional spin waves [66]. These spin waves are nowadays called Winter magnons and could be excited in a wide class of patterned magnetic nanostructures possessing domain walls. Aliev et al. [67] reported on combined FMR and micromagnetic investigations on Winter magnons driven in circular magnetic dots magnetized in a double vortex state, where the magnons travelled along the domain wall connecting the two vortex cores. Garcia-Sanchez et al., [68] proved theoretically and through micromagnetic simulations the non-reciprocal channeling of Winter Magnons in Néel-type domain walls in ultrathin ferromagnetic films with Dzyaloshinskii-Moriya interaction. First experimental evidence by real space imaging of spin waves travelling along domain walls was done by Wagner at al. [69] where domain walls were also suggested as reconfigurable spin-wave nanochannels. Excitation of Winter magnons via coupling to the vortex core was achieved in exchange-coupled ferromagnetic bilayers and has been suggested to be implemented for future spin-wave logic and computational circuits [70].

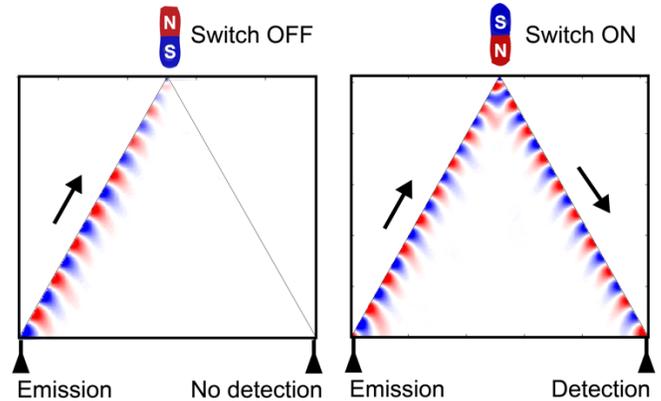

Fig. II.G. A schematic representation of the architecture of spin-wave logic element foreseen as an essential element for phase controlled ultrafast logics. More detailed proposals on using edge spin waves for information processing and transmission have been described in Ref. [71].

The main obstacle is still the limited propagation length of Winter magnons. High spatial damping is caused not only by Gilbert damping losses but also by the instability of certain spin textures. Domain walls can transform, e.g., into cross-tie domain walls, where spin waves are strongly blocked by anti-vortices. The latter challenge can be avoided in a somewhat different approach considering excitation of Winter magnons along device edges [72]. This radically new approach transmits and manipulates data exploiting Winter magnons confined to edge domain walls in ferromagnetic triangles or rectangles [71] – see Fig. II.G. This configuration favors local excitation/detection of SWs via charge-magnon interconversion in magnonic logic gates, enabling their drastic miniaturization. Zang et al. demonstrated tuning of edge-localized spin waves in magnetic microstrips by proximate neighboring magnetic structures [73]. Finally, Gruszecki et al. [74] established with numeric simulations that an edge-localized spin wave can be used to excite plane waves propagating obliquely from the film's edge at a doubled frequency and being over two times shorter in wavelength.

The main obstacles which face edge SWs is control over edge roughness and the presence of a bias field. The second problem however has been proposed to overcome by using ferromagnetic nanostructures exchange biased to antiferromagnets [71].

Future developments in this field will be to actively engineer domain walls or to use domain walls as nanosized, spin torque driven oscillators. Albisetti and coworkers used local heating to imprint nano-sized, non-volatile textures in exchange-bias systems [75] and Sato and coworkers showed that nano-sized domain walls can even be used to create spin-Hall nano-oscillators operating at zero magnetic field [76], where magnons can be locally driven, reprogrammed and moved.

## Acknowledgement

The work in Madrid was supported by Spanish Ministry of Science and Innovation (RTI2018-095303-B-C55, CEX2018-000805-M) and Comunidad de Madrid (NANOMAGCOST-CM P2018/NMT-4321) and "Accion financiada por la Comunidad de Madrid en el marco del convenio plurianual con la Universidad Autonoma de Madrid en Linea 3: Excelencia para el Profesorado Universitario" grants. H.S. acknowledges funding by DFG via projects Schu2022/1-3 and Schu2022/4

## Contributors

This section is authored by F.G. Aliev (Universidad Autonoma de Madrid, Spain) and H. Schultheiss (Helmholtz-Zentrum Dresden-Rossendorf, Germany)

### H. Advanced computational methods for magnonics

Due to the massive increase in computing power over the last decades, numerical simulations have developed into an indispensable tool for the investigation of complex physical phenomena. Theoretical models for the description of magnetic systems range from *ab initio* models that are able to capture quantummechanical effects on an atomistic level up to the macroscopic Maxwell equations where the magnetization is averaged over a multitude of magnetic domains in order to allow for the analytical description of large systems. The theory of choice for the description of magnonics is micromagnetics, a semiclassical theory where the magnetization configuration is represented by a continuous unit-vector field and all energies and interactions are described in terms of nonlinear partial differential equations (PDEs) [77]. The micromagnetic theory accurately resolves the spatial features of domain walls and spin waves without the need to account for individual atomistic moments. However, since analytical solutions are usually not available for realistic systems, modern simulation tools apply numerical methods such as the finite-difference method or the finite-element method to solve the micromagnetic equations [78], [79], [80].

Despite its long history, one of the main challenges in computational micromagnetics remains to be the size of realistic systems. In order to properly resolve spin waves of any wavelength excited in magnonic devices, the spatial discretization usually has to be chosen in the single nm regime leading to tens of millions of degrees of freedom for micron sized systems resulting in very long simulation times. A possible approach to speed up micromagnetic simulations is the use of novel highly parallel computing architectures such as graphics processing units [81], [82], [83]. Alternative strategies include the application of novel algorithms from the rising field of machine learning in order to reduce the complexity of the micromagnetic model [84]. An algorithmic approach that is particularly well suited for magnonics is micromagnetics in the frequency space. By linearization of the Landau-Lifshitz-Gilbert equation, the dynamic modes and the power-spectral density of magnonic systems can be efficiently calculated for any metastable magnetization configuration, yielding tremendous speedups compared to time-domain simulations [85], [86]. While restricted to small angle precessions, this procedure is a powerful tool for the design of magnonic devices. Very recently, this approach has been extended to nonlinear effects [87]. When solving the magnetization dynamics in the normal-mode basis, considerable speedups can be gained by limiting the number of modes according to the desired accuracy making this theory a promising candidate for future high performance magnonic simulations, see Fig. II.H. Besides a strong focus on model-order reduction and tailored algorithms for numerical magnonics, future developments in micromagnetics will certainly also have a strong focus on the use of massively distributed computer systems for the acceleration of simulations. Parallelization on distributed systems itself is already well established for the numerical solution of PDEs. However, the micromagnetic model introduces challenges such as the global demagnetization field which prevent the efficient application of typical off-the-shelf parallelization solutions. Several micromagnetic codes address this difficult matter while still leaving room for improvements [88], [89]. However, the combination of standard distributed parallelization techniques for the spatial discretization with tailored parallel-in-time integration strategies as well as the use of novel highly optimized numerical libraries will pave the way to exascale micromagnetic simulations and allow for the accurate and fast simulation of very large magnonic devices.

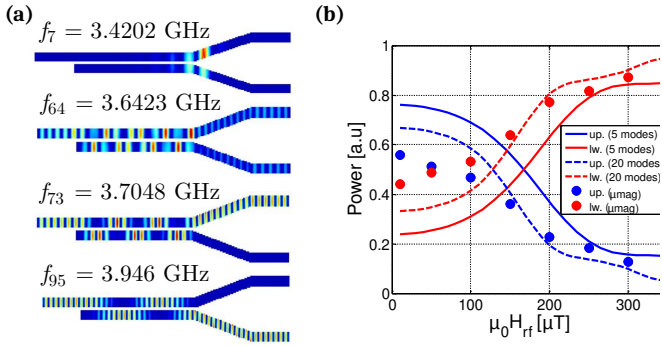

Fig. II.H. Normal mode analysis for a magnonic coupler (a) Spatial profiles of four modes with their corresponding frequencies. (b) Comparison of power transmitted by the magnonic coupler as a function of the microwave field amplitude computed in the normal-mode space (solid lines - 5 modes define the mode space, dashed lines - 20 modes define the mode space) and with micromagnetic simulations (dots) respectively. (red) Power in the upper arm (blue) power in the lower arm.

Device size is not the only challenge in computational magnonics. In addition to the classical magnetic material properties and interactions, magnonic devices often rely on spintronics effects such as spin pumping and the inverse spin-Hall effect in order to translate magnetic signals to electric signals and vice versa. In order to obtain accurate simulation results, these effects have to be considered also in the numerical model. In recent publications the self-consistent coupling of micromagnetics to the spin-diffusion model has been demonstrated, covering a multitude of spintronics effects [90], [80]. However, the integration of spintronics effects that are not yet considered in micromagnetic simulations such as spin pumping will enable more realistic simulations of full magnonic devices and give deep insights into the rich physics of these applications.

Numerical simulations of magnonic devices can help to increase the understanding of experimental findings and predict their behavior for a multitude of external influences such as external fields and currents. Another very important use case for numerical simulations is the design of novel magnonic devices. The possibility to easily change any system parameter such as geometry and material properties allow for the rapid development of novel devices. However, while testing a single parameter set by performing micromagnetic simulations is a fast alternative to experiments, the optimization of complex high dimensional design parameters such as the device shape is a complicated task that requires the application of smart optimization routines in order to ensure feasibility. In the case of relatively small numbers of binary design variables, it has been shown that a simple binary search algorithm in conjunction with a fast forward solver can lead to very good optimization results for magnonic multiplexers [91]. However, with increasing dimensionality of the design space, more sophisticated methods such as gradient based algorithms have to be used. By using an adjoint-variable approach the gradient of arbitrary objective functions subject to PDE constraints is accessible at the cost of a single forward solve, see Sec. VII.C.

## ACKNOWLEDGMENT

We gratefully acknowledge the Austrian Science Fund (FWF) for support through grants P 34671-N and I 4917-N.

## CONTRIBUTORS

This section is authored by C. Abert, S. Koraltan and D. Suess (University of Vienna, Austria).

### I. Spin waves in nanoscale waveguides with nonuniform magnetization

The properties of spin waves in narrow waveguides are strongly affected by quantization and pinning conditions at the edges [92], [25]. At the nanoscale, the shape anisotropy can become comparable to the transverse applied external bias fields (typically 1 to several hundred mT). This situation leads to a nonuniform magnetization distribution inside the conduit with strong repercussions on both dispersion and propagation properties as well as on mode formation. As an example, the magnetization in the common Damon-Eshbach configuration (with a transverse external bias field) becomes tilted towards the direction of the conduit, in particular at its edges [93], [94], [95]. Such effects need to be considered in spin-wave propagation and transmission experiments since they strongly affect the coupling efficiency between, *e.g.*, inductive antenna transducers and different quantized spin-wave width modes [94], [95].

Experimentally, this has been studied in CoFeB waveguides with a width down to 320 nm [95] (see Fig. II.I). At such small widths, which are comparable to their wavelength, spin waves are quantized with each mode having its own dispersion relation. When, in addition, the magnetization is tilted and nonuniform, the dispersion of such quantized modes is further strongly modified. In this case, no analytical solutions for the spin-wave dispersion relation are available, and the magnetization dynamics need to be studied by means of micromagnetic simulations. Experimental results are in good agreement with the simulations and, for low bias fields, indicate backward-volume-like dispersion relations despite the transverse bias field [95]. Depending on the magnitude of the applied field, the nonuniform magnetization modifies the excitation efficiency of the different modes, allowing *e.g.,* for the excitation of even modes that cannot be excited for uniform magnetization. Furthermore, the dispersion relations of the quantized modes can intersect leading to amplitude modulation due to the superposition of different propagating modes, as observed at high magnetic fields.

## ACKNOWLEDGMENT

Contributions from F.C. and C.A. have been supported by imec's industrial affiliate program on beyond-CMOS logic. F.C., C.A., S.H. and S.C. acknowledge funding by the European Union's Horizon 2020 research and innovation program within the FET-OPEN project CHIRON under Grant Agreement No. 801055.

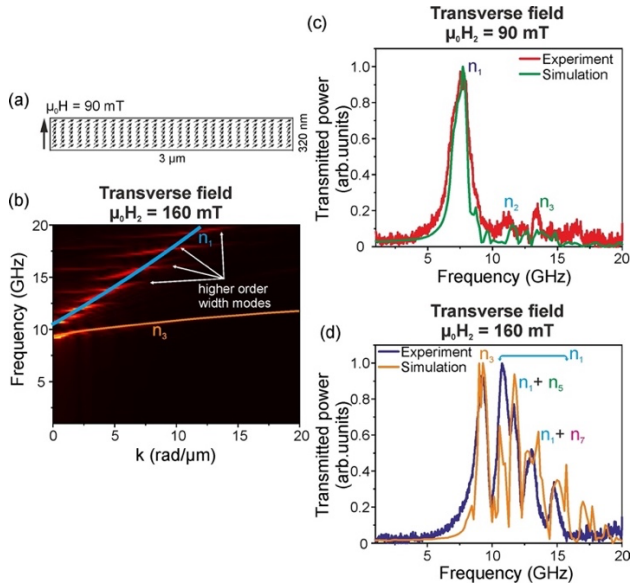

Fig. II.I. (a) Static magnetization distribution in a 320 nm wide CoFeB waveguide for an applied transversal field $\mu_0 H_2 = 90$ mT. (b) Dispersion relations of quantized spin-wave modes extracted from the micromagnetic simulations of an array of 20 magnetic CoFeB waveguides (320 nm width, 180 nm gap) for an applied bias field 160 mT. (c) and (d) Calculated and experimental spin-wave spectra for two magnetic fields as indicated. Reprinted with permission from [95]. Copyright 2021 American Institute of Physics.

## J. Increasing the signal bandwidth - From ferromagnetic to antiferromagnetic magnonic systems

Present spin-wave computing concepts mainly rely on mechanisms such as wave interference and non-linear interaction of waves. With respect to device miniaturization and power consumption, it will be therefore indispensable to utilize ultrashort spin waves with nanoscale wavelengths as signal carrier. Consequently, any technological implementation will require full control over the efficient excitation, low-loss propagation, reliable manipulation, and robust detection of such exchange-governed waves. To that end, significant advances were made in ferromagnetic (FM) and ferrimagnetic (FiM) systems over the past years, e.g., with respect to the generation of coherent, sub-100 nm dipole-exchange waves [96], [97].

Complementary, there is a strong demand for increasing the bandwidth of future magnonic devices, as both the signal transfer rate and processing speed scale with the operation frequency. While decreasing wavelengths typically coincide with increasing frequencies, the fundamental resonances of FMs and FiMs often do not exceed the GHz to 10 GHz range, and only highly anisotropic systems allow to partially circumvent this restriction.

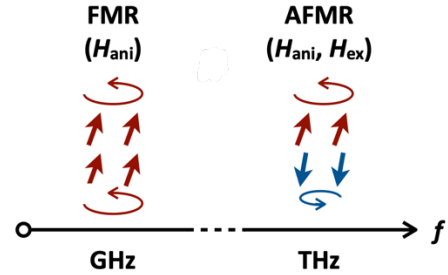

Fig. II.J: Schematics of typical FMR and AFMR frequencies with arrows indicating magnetic moments, $H_{ani}$ and $H_{ex}$ referring to anisotropy and exchange fields, respectively.

Prime candidates to boost the intrinsic frequencies for spin-wave computing are antiferromagnetic (AF) materials with resonances ranging from 10s of GHz up to the THz regime. In contrast to FMs, for which the uniform precession frequency is–apart from external fields–only determined by anisotropy (intrinsic and shape), AFs exhibit dynamic sublattice canting, resulting in an immediate additional exchange contribution and thus very high frequencies (see Fig. II.J). In that sense, resonances in synthetic antiferromagnets (SAFs) [97], [98] and confined exchange modes in thin films [99] can be considered as FM-based intermediates on the way to intrinsic AF systems.

While pioneering investigations of coherent spin dynamics in AFs already go back to the 1950s [100], [101], and seminal advances were reported more recently [102], [103], a clear understanding and direct observations of the real space dynamics of AF spin waves are unavailable to date. One reason for this lack is the difficulty to spatially probe AF systems by reliable means. Here, x-ray magnetic linear dichroism (XMLD) in combination with x-ray microscopy stands out as one of the few techniques that allows for a direct imaging of AFs with nanoscale resolution [104], [105].

In the prospect of magnon computing and for gaining fundamental insights, we therefore propose to employ time-resolved x-ray microscopy for pioneering studies of the real-space spin dynamics in AF systems via direct and phase-resolved imaging at low-alpha synchrotron sources and x-ray free electron lasers. Besides, and in view of application limitations of x-ray techniques, it should be noted that more device-oriented methods for probing AF magnon transport hold great potential as well [106]. In addition to the challenges in the detection of AF spin waves, two other relevant aspects as to AF magnon excitation and propagation need to be addressed.

Being both benefit and drawback, AFs are much less susceptible to external magnetic fields than FMs. This makes AF spin-wave dispersion relations relatively stable towards moderate external magnetic fields. On the other hand, it goes along with a low field-excitation efficiency of spin waves in AFs. More suitable alternatives may exploit spin-transfer torques and in particular spin-orbit torques at heavy-metal interfaces. Other concepts for the excitation of spin waves in AFs may involve FM exchange bias, magnetoelastics, electronically controlled anisotropies and optomagnetics.

Another aspect of spin waves in AFs concerns the transfer of net angular momentum during wave propagation. Depending on crystal orientations, anisotropies (easy axis/easy plane) and external fields, there can be situations where modes are fully linearly polarized, thus not carrying any net angular momentum. Therefore, it needs to be taken care that circularly polarized modes are utilized–at least effectively or locally [107] whenever an operation will require the involvement of net angular momentum.

Besides standard collinear AFs, there can be also more exotic variants considered for additional functionality, such as non-collinear AFs, chiral AFs, and 2D AF systems. Furthermore, FiMs can be designed to exhibit compensation at specific temperatures [108], allowing to combine FM and AF characteristics in a single temperature-controlled device.

In summary, we believe that AFs provide the chance to strongly promote spin-wave computing concepts in terms of signal bandwidth and data processing speed.

### Acknowledgment

S.M. acknowledges funding from the Swiss National Science Foundation (Grant Agreement No. 172517). S.W. acknowledges financial support from the Leibniz association via grant no. K162/2018 (OptiSPIN).

### Contributors

This section is authored by Sina Mayr (Paul Scherrer Institut, Villigen PSI, Switzerland) and Sebastian Wintz (Max Planck Institute for Intelligent Systems, Stuttgart, Germany).

### K. Magnon propagation and spin dynamics in antiferromagnets

The current focus on ferromagnetic materials for spin-wave research gives rise to several key limitations: spin dynamics in the low gigahertz regime, susceptibility to magnetic perturbations and large stray fields resulting in possible unwanted interactions. Furthermore, ferromagnetic materials can only host magnons of a single circular polarization. An alternative class of magnetic material has attracted significant interest in the magnonic and wider spintronic community in recent years, antiferromagnets. With a magnetic state robust against magnetic perturbations and spin dynamics in the hundreds of gigahertz to terahertz regime, this large class of materials is promising for low power, high frequency and robust magnonic applications. Antiferromagnets are in particular capable of hosting both left- and right- circularly, and linearly polarized magnons, thus offering additional distinct advantages for magnonic applications, with the polarization offering an additional degree of freedom. To benefit from the intrinsic advantages of antiferromagnets, a key requirement is to establish the efficient long-distance transport of magnons in antiferromagnetic materials. This was recently achieved in the insulating antiferromagnet hematite ($\alpha$-Fe$_2$O$_3$) by means of electrically exciting circularly polarized magnons transported by the antiferromagnetic Néel vector [106], [109]. Due to its extremely low magnetic damping [110], spin diffusion lengths of up to 9 μm have been shown even in the presence of a complex domain structure which can act as a frequency selective filter (see Fig. II.K.(a)) [109]. However, for hematite the transport based on circularly polarized magnon modes capable of transporting angular momentum requires cryogenic temperatures, as the necessary easy-axis anisotropy is stabilized in pure hematite only at low temperatures. At room temperature, this ubiquitous antiferromagnet adopts an antiferromagnetic (canted) easy-plane anisotropy and the available magnon modes adopt a linear polarization. Nevertheless, despite individual modes not being capable of transporting angular momentum, superpositions of individual magnon modes can lead to efficient magnon transport [107], [110]. The application of a magnetic field modulates this superposition, giving rise to a tunable magnon Hanle effect [111], [112] even in low magnetic symmetry antiferromagnets. This shows that circularly polarized magnon modes are not a pre-requisite for antiferromagnetic magnonic devices, opening up a large range of antiferromagnets as potential candidates for spin-wave devices.

Antiferromagnets are furthermore capable of transporting thermally excited magnons, which present distinct dependences on field and temperature to their electrically excited counterparts [113], [106], [114]. These excited spin currents can also be transported independently by both the Néel vector and emergent net magnetization. Such behaviors and separations are not seen in ferromagnetic materials and offer further opportunities for the tunability of antiferromagnetic magnonic devices to move forward.

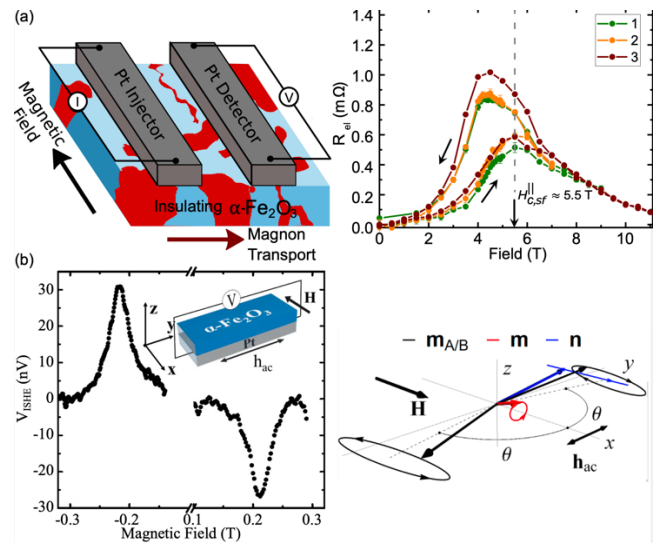

Fig. II.K. (a) Spin transport in antiferromagnets with multi-domain states. Magnon scattering at domain walls filters the propagating spin waves. Magnon transport properties follows the spin-flop states. Adapted with permission from [109]. Copyright 2021 American Chemical Society. (b) Spin—pumping in canted antiferromagnets can arise from the dynamics of the Néel order or from the dynamics of the canted moment. Adapted figure with permission from [115]. Copyright 2021 by the American Physical Society.

With a few exceptions focused on thermally excited magnons at low temperatures, long-distance, antiferromagnetic magnon transport has been confined to $\alpha$-Fe$_2$O$_3$. This material has a sizeable Dzyaloshinskii-Moriya interaction, but it is yet unclear what role this plays for the long distance magnon transport. Other canted orthoferrites share many similarities with α-Fe₂O₃ and may be possible candidates for investigating antiferromagnetic magnon transport in materials with different antiferromagnetic anisotropies. The antiferromagnetic anisotropies can also be tuned through dilute doping, which has already been shown to allow for significant magnon transport even when the magnetic symmetry is low [111]. Finally, there is a need to move from diffusive to coherent spin-transport regimes to make use of the phases of spin waves. A first step could consist in resonantly exciting and detecting propagating spin waves. Moving forward, investigations into antiferromagnets as the host of magnon condensates for non-diffusive, spin-superfluid transport should be a key priority where complex interference effects can be realized for advanced spin-wave based logic [116].

Manipulating and detecting the coherent dynamics of the magnon modes is a pre-requisite to develop magnonic devices that fully benefit from the ultrafast dynamics of antiferromagnetic materials. This requires the ability to detect the *frequency* and wave-vector *resolution* of the magnon modes, which has proven to be challenging with conventional techniques (such as inductive antenna or Brillouin light scattering) due to the lack of magnetic symmetry breaking. Spintronic phenomena (such as spin-pumping and spin-Hall effect) have in the recent years opened new perspectives to solve this key bottleneck. Inverse spin-Hall detection of sub-THz antiferromagnetic magnon modes was recently achieved in collinear antiferromagnets [117], [118]. However, it requires antiferromagnetic materials with a high ratio of anisotropy over exchange field, which generally possess low Néel temperatures. An alternative strategy relies on canted antiferromagnets with a large Dzyaloshinskii-Moryia interaction, for which the dynamical net moment can generate sizeable spin-pumping signals (see Fig. II.K(b)) [115]. Furthermore, canted antiferromagnets such as orthoferrites often exhibit large magneto-optic constants potentially compatible with Kerr and wave-vector selective Brillouin-Light scattering experiments.

## Acknowledgment

We acknowledge funding by the DFG (#423411604 and SFB TRR 173 Spin+X projects A01 and B02 #268565370), and KAUST (Project No. OSR-2019-CRG8-4048.2). This work has received funding from the European Union's Horizon 2020 research and innovation programme under Grant Agreement No. 863155 (s-Nebula) and from the ERC under Grant Agreement No. 856538 (ERC-SyG 3D MAGiC).

## Contributors

This section is authored by A. Ross, R. Lebrun, and I. Boventer (Unité Mixté de Physique CNRS-Thales, France), and M. Kläui (Johannes Gutenberg University Mainz, Germany).

### L. Magnetic imaging of spin waves using spins in diamond

Electron spins associated with nitrogen-vacancy (NV) defects in diamond have emerged as a powerful platform for magnetic imaging of condensed matter systems [119]. NV spins (Fig. II.L(a)) offer long coherence times at room temperature, good sensitivity to magnetic fields, and can be read out at the single-spin level using a table-top photoluminescence microscope. Because NV centers can stably exist at nanometers below the diamond surface, they can be brought into close proximity to a target (Fig. II.L(b, c)). This makes NV magnetometry well suited for imaging spins and currents in materials at the nanoscale.

Probing spin waves in magnetic materials is a new and growing application area of NV magnetometry. In contrast to other techniques, NV spins detect spin waves via their magnetic stray fields. Here, we introduce the technique, discuss its limitations and advantages, and provide an outlook on new experiments it brings within reach.

Three measurement modalities now exist. The first detects changes in the DC magnetic stray field of a sample upon the excitation of a localized spin-wave mode, such as those confined in micromagnets [120] or magnetic films [121]. These DC-field changes cause spatial variations of the NV electron spin resonance frequencies (Fig. II.L(d)) that are detectable via the NV photoluminescence [122] An advantage of this modality is that it is broadband: the spin-wave frequency does not need to match the NV frequency. Combined with a nanotesla sensitivity [122], [123], this approach provides exciting prospects for imaging spin-wave modes in nanosized, atomically thin samples, which will likely play an increasingly important role in spin-wave computing.

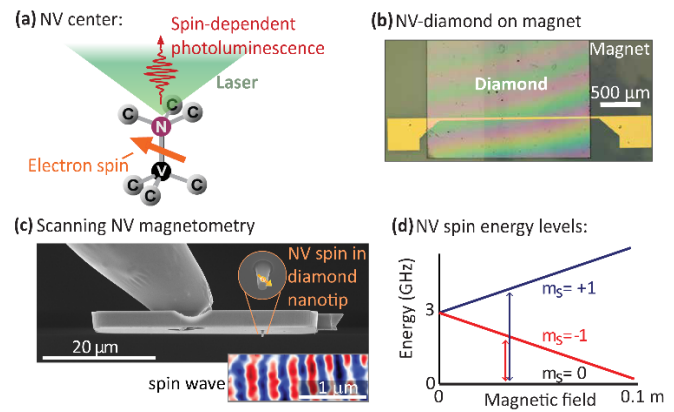

Fig. II.L. Magnetic imaging of spin waves using spins in diamond. (a) The electron spin of the nitrogen-vacancy center can be read out via spin-dependent photoluminescence. (b) Diamond with a dense, shallow NV sensing layer on top of a magnetic film for wide-field spin-wave imaging [124]. (c) Scanning NV tip for high-resolution spin-wave imaging. (d) The Zeeman shift of the NV electron spin energies vs magnetic field.

The second modality images spin waves by their microwave magnetic stray field. By matching the frequency of a spin wave to the NV, the spin-wave field will drive rotations of the NV spin, which can be detected optically [120], [125]. Adding a reference field of the same frequency enables phase-sensitive imaging of the spin-wavefronts (inset Fig. II.L(c)) [124], [126], [127], [128]. Although this modality requires resonance between NV and spin wave, its magnetic nature also provides new opportunities. It enables e.g., imaging spin waves underneath metal electrodes [126], as well as local studies of material properties such as skin depth, permeability and conductivity. Its sensitivity is promising for real-time imaging of spin-wave profiles. The interaction with the NV spins may even enable modulation of spin-wave modes (see Section VIII-K, [129] ) or spin-wave mediated entanglement between spin qubits [130].

Thermal spin fluctuations contain valuable information on a magnet's dissipation, susceptibility and magnonic band occupation. The third measurement modality picks up the stray-field fluctuations generated by thermal spin waves via changes in the NV relaxation rate [120]. A powerful feature is its sensitivity: it can readily detect thermal spin fluctuations in 20 nm thin YIG [131]. This enables studies of the magnonic band occupation under microwave- and spin-Hall-driving [131], Simon2021] and provides exciting prospects for probing magnon condensates such as those in magnon transistors in ultrathin YIG [132].

In the next few years, we expect that NV-probing of spin waves will expand to different materials and phenomena. The detection of spin waves in synthetic antiferromagnets was a recent success in the field [133]. A key challenge is to expand the detection capabilities to spin-wave modes that cannot be tuned into resonance with the NV spins. This may be achieved by new NV-spin-control sequences or by harnessing non-linear spin-wave interactions. The key strengths of the technique are the well-understood coupling between the NV sensor and the spin waves, its table-top use, nanotesla sensitivity, nanometer spatial resolution, and operability over a broad temperature range. This puts the characterization of spin waves in a diverse set of materials, including monolayer van-der-Waals and complex-oxide magnets within reach, and provides opportunities for commercial, high-speed imaging of computational circuits and devices.

### Acknowledgment

This work was supported by the Dutch Research Council (NWO) through the NWO Projectruimte grant 680.91.115 and the Kavli Institute of Nanoscience Delft. The authors thank B.G Simon, S. Kurdi, and I. Bertelli for support and valuable discussions.

### Contributors

This section is authored by J.J. Carmiggelt, Y.M. Blanter, and T. van der Sar (TU Delft, The Netherlands).

### III. 2D building blocks for magnonic networks

Dipolar and dipolar-exchange spin waves are very anisotropic while propagating in in-plane magnetized thin films. The anisotropy opens many additional opportunities but also creates challenges. Thus, even a guiding of a spin wave in a 2D circuit is usually more complex than the guiding of a signal in the form of electric current or light. This section discusses the advantages and the challenges of 2D magnonics networks.

First, we address the opportunity to perform spin-wave wavelength conversion using 2D magnonics structures. Such a conversion and, in particular, the decrease of the wavelength allows for the operation with faster and long-running exchange magnons. The next section addresses the utilization of the physics of superconductors to control spin waves. In particular, the superconductor-ferromagnet-superconductor trilayers are discussed. The following two sections address magnonic Fabry-Pérot and chiral resonators, which might become essential building blocks of future magnonic networks. Controlling the phase and amplitude of waves is fundamental for processing signals, and the transmission of the information should not disturb these characteristics. This question is addressed in the section on spin-wave guiding in curved multimode waveguides (Section III-E). Afterwards, the spin-wave propagation in $Fe_{78}Ni_{22}$ 2D structures with locally-controlled magnetic anisotropy is discussed. Finally, the Magnonic STImulated Raman Adiabatic Passage (M-STIRAP) process is discussed and proposed to control wave propagation in magnonic systems.

### A. Spin-wave wavelength conversion

As one of the most commonly used methods for coherent spin-wave (SW) excitation, a coplanar waveguide (CPW) is attached to a magnetic film and rf Oersted fields are generated by the CPW [134]. In this method, the wavelength of the excited SWs is determined by the size of the CPW. Hence, it is technologically challenging to scale down the SW wavelength, which is a requirement for scaling down magnonic devices. To solve this problem, various concepts for down-conversion of the SW wavelength have been investigated based on using the general CPWs.

The first method is down conversion exploiting Snell's law for SWs. In the dispersion relation of SWs, in the magneto-static regime the wavelength is coupled to the film thickness with via the product $kd$ ($k$ is the wavenumber and $d$ is the thickness) [135]. Hence, once a SW propagates across a thickness step, it changes its wavelength (Fig. III.A(a)) [136], [137], [138]. In the second method a tapered magnetic stripe is used (Fig. III.A(b)) [139], [140]. In the tapered stripe, the internal magnetic field decreases with decreasing the stripe width. As one can find in the dispersion relation, the variation of the internal magnetic field induces wavelength conversion. The third method is based on a conversion of spin-wave modes (Fig. III.A (c, d)). As described in the dispersion relation, the wavenumber can be converted by changing the propagation direction of a SW with respect to the applied magnetic field. In the works reported by T. Brächer *et al.* [141] and A. V. Sadovnikov *et al.* [142], such a conversion has been realized by designing a T-shaped magnonic waveguide.

While the three methods used conversion occurring already in the magnonic waveguide, one can also excite short-wavelength SWs by attaching magnetic objects to a CPW. P. Che *et al.* recently reported the excitation of short-wavelength SWs by designing a magnetic CPW (Fe/Au) (Fig. III.A(e)) [35].

In this work, a dipole field from the magnetic CPW causes a local enhancement of the static internal field and the wavelength is converted to shorter wavelengths outside the CPW region because there the internal field is lower.

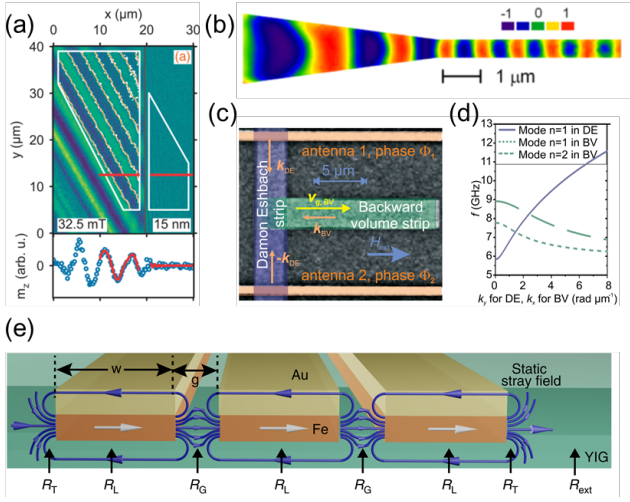

Fig. III.A. (a) SW propagation in Py film from 60 nm to 10 nm [137]. (b) SW propagation in a Py tapered stripe [139]. (c,d) SW conversion from Damon-Eshbach SW to backward volume SW [141]. Reproduced from [139] and [141], with the permission of AIP Publishing. (e) SW conversion using a magnetic CPW [35].

Using the techniques described above, we are reaching a stage to excite coherent SWs in the exchange regime. Since SW conversion especially based on internal magnetic field variation can be utilized even in the exchange regime, it is promising to excite even shorter wavelength SWs by fabricating tinier SW wavelength conversion devices. In order to utilize SW wavelength conversion for magnon computing, it is necessary to have a quantitative comparison of the conversion efficiency between different methods for future.

### Contributors

This section is authored by T. Taniguchi and C. H. Back (TU München, Germany).

### B. Superconducting proximity phenomenon for thin-film magnonic structures

A substantial part of this Roadmap is devoted to 2D magnonic networks, which are based on thin film technologies. Arguably, one of the most important requirements for engineering of magnonic networks is the ability to modify and tune the spin-wave dispersion of different elements of the network. This requirement becomes especially evident in engineering of magnonic crystals with periodic modulation of the spin-wave dispersion. Quite a number of different approaches for fabrication of magnonic crystals can be found in literature, e.g., periodic spatial modulation of external magnetic field, saturation magnetization, exchange properties, magnetic anisotropy, film thickness, mechanical stress, etc.

As an unconventional approach, it was shown that modification of the dispersion in ferromagnetic thin films can be achieved by hybridizing them with superconductors [143]. Also, hybrid magnonic crystals can be formed by hybridizing ferromagnetic films with the ordered superconducting vortex lattice [144] or periodic superconducting microstructure in the Meissner state [145]. Recently a new phenomenon was discovered in hybrid superconductor-ferromagnet-superconductor trilayers [146], which offers new perspectives for engineering of the spin-wave dispersion [147]. It was shown that in presence of both superconducting layers and of superconducting proximity at both superconductor-ferromagnet interfaces a radical increase in ferromagnetic resonance frequency takes place by up to 20 GHz at zero field, depending on the thickness of the ferromagnetic layer. Currently the mechanism of this phenomenon remains an enigma. Yet, its possible application in magnonics seems natural [147]: with superconductor-ferromagnet-superconductor trilayers it is possible to achieve the highest natural ferromagnetic resonance frequency among in-plane magnetized ferromagnetic systems, while spatially periodic trilayered microstructures are expected to form bandgaps with the width in GHz frequency range.

### Acknowledgment

I.A.G. acknowledges support by the Ministry of Science and Higher Education of the Russian Federation in the framework of the State Program (Project No. 0718-2020-0025).

### Contributors

This section is authored by I.A. Golovchanskiy (Moscow Institute of Physics and Technology, and National University of Science and Technology MISIS, Moscow, Russia).

### C. Magnonic Fabry-Pérot resonator

A magnonic Fabry-Pérot resonator can be realized by placing a ferromagnetic stripe near a low-loss YIG film [148]. In this configuration, illustrated in Fig. III.C(a), chiral dipolar coupling between the two magnetic layers produces an asymmetrical spin-wave dispersion relation within the bilayer region (Fig. III.C(b)) [149]. Consequently, the two edges of the bilayer act as interfaces where the wavelength of the spin waves converts upon reflection or transmission. Because the dispersion curve shifts down in frequency inside the bilayer compared to the uncovered YIG film, $\lambda_1 > \lambda_2 > \lambda_3$. Analogous to an optical Fabry–Pérot resonator, destructive interference of the spin waves entering and the spin waves circulating the bilayer region produces gaps at discrete frequencies in the transmission spectrum (Fig. III.C(c)). The destructive interference condition is given by $(|k_2| + |k_3|) \times w + \varphi_0 = (2n + 1) \times \pi$, where $\varphi_0$ accounts for the total phase change caused by the two internal reflections, effects of long-range dipolar fields, and a deviation of the effective resonator width from $w$ [148]. Proper circulation of propagating spin waves within the resonator is aided by low magnetic damping in the YIG film.

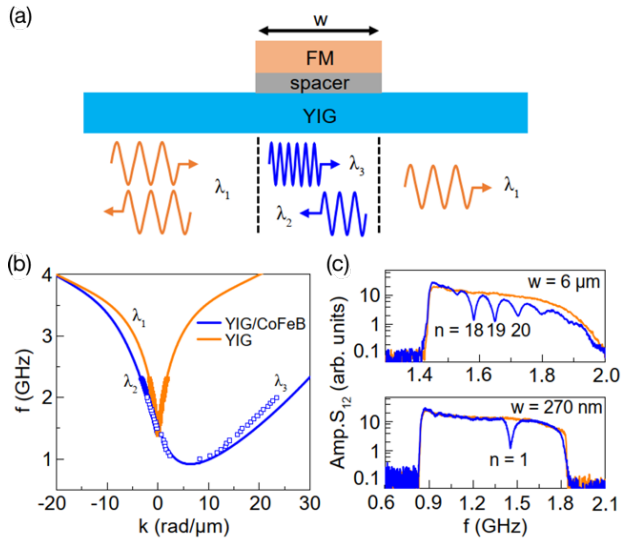

(a)

(b)

(c)

Fig. III.C. (a) Schematic of a magnonic Fabry-Pérot resonator, consisting of a patterned ferromagnetic (FM) stripe, a spacer layer, and a low-loss YIG film. (b) Spin-wave dispersion curve for an uncovered YIG film ($\lambda_1$) and a 100 nm YIG/50 nm CoFeB bilayer with a 5 nm spacer. (c) Spin-wave transmission spectra of the same resonator structure as in (b) for $w = 6$ μm and $w = 270$ nm.

The magnonic Fabry-Pérot resonator could find applications in narrowband microwave filtering, spectroscopy, and new computing concepts. Its transmission losses at allowed frequencies are minimal, as exemplified by the nearly identical transmission characteristics of a YIG film with and without resonator structure (blue and orange curves in Fig. III.C(c)). Moreover, the size of the resonator ($w$) is scalable. The minimal width, which is attained for the $n = 1$ mode, is determined predominantly by the value of $k_3$. In the example shown in the lower panel of Fig. III.C(c), a $n = 1$ transmission gap is measured on a 270 nm wide resonator for an incident spin wave with $\lambda_1 \approx 10$ μm is possible through the engineering of the bilayer dispersion relation. Programmable control of spin-wave transport in the Fabry-Pérot resonator is achieved by magnetic switching, as is done in other magnonic resonators [150], diodes [151], and unidirectional spin-wave emitters [152] made of magnetic double layers (see also Sections III-D and VI-A). Switching between parallel and antiparallel magnetization states changes the bilayer dispersion curve and, thereby, the frequency of the spin-wave transmission gaps [148].

## Acknowledgment

The research leading to these results has received funding from the Academy of Finland (projects 317918, 316857, 321983 and 325480).

## Contributors

This section is authored by H. Qin and S. van Dijken (Aalto University, Finland).

### D. Chiral magnonic resonators

Chiral magnonic resonators (Fig. III.D) – soft magnetic elements chirally coupled, via magneto-dipole interaction, to magnonic media nearby – can deliver all key functionalities required to build realistic nanoscale magnonic devices [150], [153]. Travelling spin waves resonantly couple via their stray magnetic field to magnonic modes of the resonators. This coupling is chiral due to the chirality of the magnetic precession in the resonators, which makes them re-programmable by switching their magnetization. The coupling chirality underpins the resonators' use for unidirectional emission, phase control, transmission (Fig. III.D), and absorption of coherent exchange spin waves, within magnonic sources [150], phase shifters (including inverters) [153], diodes (valves) [153], and detectors (output couplers) [153], respectively. The use of dark modes of the resonators allows one to decouple them from the clock field, without compromise on the control efficiency [154]. Furthermore, they can be used as building blocks of devices ranging from basic logic gates [155] to field programmable gate arrays and artificial neural networks [156], in architectures ranging from 1D to 2D and 3D.

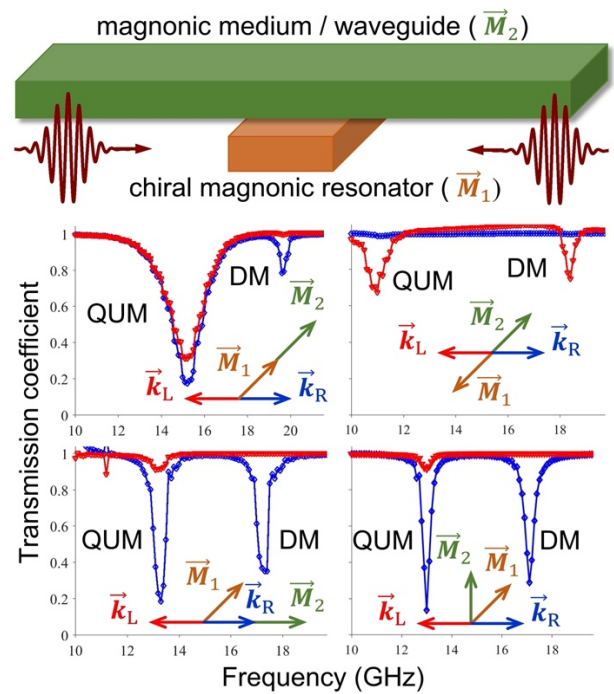

Fig. III.D. The top panel shows schematically a chiral magnonic resonator underneath a thin film magnonic waveguide with packets of right- and left-going spin waves. The bottom panels show spectra of the amplitude transmission coefficient obtained from simulations of permalloy-based structures with sub-100 nm dimensions [154] in different spin-wave geometries, shown as insets. The lower and higher frequency dips in the spectra are due to the incident wave coupling to the quasi-uniform (QUM) and dark (DM) magnon modes of the resonator, respectively.

With the recent advances in fabrication and characterization of magnonic devices with sub-100 nm dimensions (see Section I-C), there does not seem to be fundamental obstacles for building and using similarly sized magnonic devices operating with exchange spin waves. On the practical side, the resonant nature of the devices means that some energy is dissipated in the resonator. The way out could be offered by building the resonators from YIG [157], [158]. Overall, the nearest future will see a greater uptake of the concept by the community and its wider deployment in a variety of device designs and architectures in magnonics, spilling out to spintronics and spin-caloritronics, with neuromorphic and in-memory computing being the primary beneficiaries in longer term.

### Acknowledgment

The research leading to these results has received funding from the Engineering and Physical Sciences Research Council of the United Kingdom under projects EP/L019876/1 and EP/T016574/1.

### Contributors

This section is authored by V. V. Kruglyak and K. G. Fripp (University of Exeter, UK).

### E. Spin-wave guiding in curved multimode waveguides

Controlling the phase and amplitude of waves is fundamental to process the signals. Hence, the transmission of the information (i.e., guiding and routing of the signal in the network of the waveguides) should not disturb these characteristics. However, changing the wave's direction of propagation, which is inevitable in any circuits, introduces scattering and decoherence. This can be overcome by using a single-mode waveguide, where only the fundamental mode is supported. However, to achieve greater bandwidth multi-mode guides have to be also used.

Single-mode magnonic waveguides can be designed in various ways, e.g., by: (i) utilizing domain walls, since they create narrow potential wells for spin waves (SWs) (see Section II-F), (ii) inducing voltage anisotropy in the form of ultra-narrow channels on the film, supporting propagation of SWs only in these regions (see Section V-B), (iii) fabricating sub-100 nm wide waveguides using the state-of-the-art techniques (see Section II-C) or using topologically protected edge states [159]. However, such waveguides are challenging in fabrication (due to the small width), difficult in operation (due to the high frequency of the fundamental mode), and limited bandwidth. Therefore, there is a need to consider multi-mode magnonic waveguides (of larger width and lower operating frequencies) and propose a method of avoiding signal decoherence and scattering at the bends of the waveguide [160].

A solution is to place the graded-index (GRIN) element at [161] the bend that either converts the mode [162] or redirects the signal mode in the desired direction [163]. In the second case, the GRIN element is an inhomogeneous slab (Fig. III.E(b)), which links the waveguide's input and output branches and refracts the SWs strictly into the outgoing section of the waveguide. Knowing how the SW acquires the phase during the propagation through the homogeneous material of modified properties, e.g., reduced magnetization saturation $M_S$, Fig. III.E(a), the GRIN element can be designed, where the gradient of $M_S$ is directly related to the gradient of phase shift and to the anomalous refraction [164].

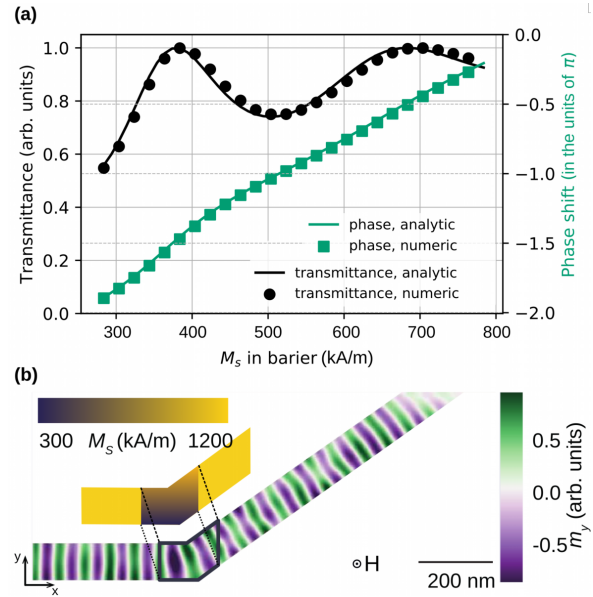

Fig. III.E. Transmittance and phase shift for SWs traveling through the 150 nm-wide and 5 nm thick stripe of homogeneous material with respect to reduction of magnetization saturation $M_S$ at 25 GHz frequency and the external field H = 0.5 T. (b) Propagation of SWs in the waveguide, where the GRIN slab is placed at the bend. The coherent transmission requires a linear change of the phase of the transmitted wave alongside the interface of the input and output branch of the waveguide. The color map represents a dynamical component of the magnetization in the y-direction. A snapshot is taken at the moment when a steady state is reached. Inset presents the distribution of $M_S$ in the waveguide. Waveguide is considered to be made of CoFeB ($M_S$=1200 kA/m and the exchange constant 27 pJ/m). Reprinted (adapted) with permission from [163]. Copyright 2020 American Physical Society.

Proposed device relies on the spatial control of some material parameters, e.g. anisotropy or $M_S$. Thus, the fabrication of such structure is challenging and requires advanced experimental techniques. However, if these obstacles are overcome, the GRIN element in the multi-mode waveguide will ensure the high transmission and preserve the phase, hence it can be the prominent direction in designing the circuits for analog and digital computing based on SWs.

### Acknowledgment

S. M. would like to acknowledge the financial support from the National Science Centre, Poland, project No. UMO-2020/36/T/ST3/00542.

### Contributors

Szymon Mieszczak, Paweł Gruszecki, Jarosław W. Kłos, Maciej Krawczyk (Adam Mickiewicz University in Poznań, Poland).

### F. Spin-wave propagation in materials with locally controlled magnetic anisotropy

One of the elementary premises of complex 2D magnonic circuits is a need for operation in the absence of an external magnetic field. If an external magnetic field is used to stabilize magnetization, even a basic circuit element such as a spin-wave turn exhibits a large dispersion mismatch for regions before and after the turn. Local control of the effective field would stabilize the magnetization of different parts of the magnonic circuit in the desired direction, thus preventing the dispersion mismatch. The required control can be achieved by manipulating the magnetic anisotropy at the local level with nanometer precision. This manipulation can eliminate the need for external magnetic fields and thus presents an important step towards complex magnonic circuits with e.g., waveguides pointing in different directions.

To advance this field, new types of materials possessing additional means of control over their magnetic properties together with good spin-wave propagation need to be developed. Currently, there are two approaches how to control magnetic anisotropy direction and strength at the nanoscale.

The first approach uses manipulation of the crystal structure in order to modify magnetocrystalline anisotropy. The system is based on epitaxially grown metastable face-centered-cubic $Fe_{78}Ni_{22}$ thin films, into which magnetic structures can be directly written by a focused ion beam [165]. There is also a similar system based on $Fe_{60}Al_{40}$ alloy [166]. The focused ion beam transforms the originally nonmagnetic fcc phase into the ferromagnetic bcc phase with additional control over the direction of uniaxial magnetic in-plane anisotropy and saturation magnetization [167]. The direction of the magnetic anisotropy of the transformed areas can be controlled by selecting a proper focused ion beam scanning strategy. This anisotropy is strong enough to stabilize the magnetization in the direction perpendicular to the long axis of narrow waveguides. It has been recently demonstrated that in this system, spin waves can propagate in favorable Damon-Eschbach mode with high group velocities reaching almost 6 km/s in orthogonally magnetised waveguides without the necessity of applying the external magnetic fields [168]. The magnetic anisotropy can also be used to stabilize various spin structures unachievable in common material systems, and these spin structures can be further used to control spin-wave propagation [169].

The second approach uses surface curvature to locally control the magnitude and the direction of the magnetic anisotropy. The surface curvature in films with thicknesses comparable to the amplitude of modulation locally modifies the contributions of dipolar and exchange energies and results in an effective anisotropy term which can be tuned on-demand based on the exact geometry [170]. Advanced nanofabrication techniques such as focused electron beam induced deposition allow the preparation of precisely shaped nanometer-sized structures with defined curvatures. Similarly, to the first system, corrugation-induced magnetic anisotropy can stabilize magnetization in narrow magnonic waveguides perpendicularly to their long axis so that the spin waves can propagate at zero external magnetic fields under the most favorable conditions [171] – see Fig. III.F.

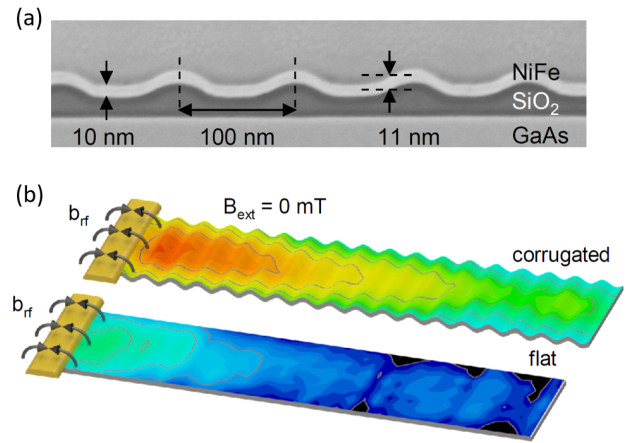

Fig. III.F. (a) Transmission electron microscopy image of the cross-section of the corrugated waveguide. The corrugation induces the effective anisotropy field of 5 mT perpendicular to the corrugation direction. (b) Comparison of zero-field spin-wave propagation in the corrugated and a flat reference waveguide, imaged by micro-BLS. Corrugation-induced anisotropy stabilizes magnetization perpendicularly to the waveguide's long axis. The spin waves then propagate in the most favorable Damon-Eshbach geometry, with high group velocity, over a distance 10× larger than in the flat reference waveguide. Reproduced from [171]; licensed under a Creative Commons Attribution (CC BY) license.

From the two presented material systems, the first one is more suited for rapid prototyping, because the magnonic structures can be written by a focused ion beam at once, in a single fabrication step. The disadvantages are the complicated deposition process of the epitaxial $Fe_{78}Ni_{22}$ thin film in ultra-high vacuum conditions and the need to use expensive single-crystal copper substrates. The latter has been already overcome by using standard silicon substrate and an appropriate buffer layer [172]. The corrugated-waveguide system is more complicated and requires multiple lithography steps, but on the other hand it is potentially more suitable for large scale production via nanoimprint lithography.

Especially in metallic systems with larger damping, it is important to ensure the optimal propagation geometry with a high group velocity in order to be able to propagate spin waves over reasonable distances. The works referenced in this section present the first building blocks – waveguides able to propagate fast spin waves without the presence of the external magnetic field. Now the potential of the systems with locally controlled magnetic anisotropy needs to be further explored. For example, spin-wave turns with dispersion precisely matched along the whole propagation trajectory are yet to be experimentally realized. Other possibilities lie in spin-wave steering and manipulation, either by spatial modification of the magnetization direction (dispersion relation) or via stabilization of arbitrary spin structures in the spin-wave propagation path.

### Contributors

This section is authored by M. Urbánek (CEITEC BUT, Brno University of Technology, Czech Republic).

## G. Magnonic STImulated Raman Adiabatic Passage

The Magnonic STImulated Raman Adiabatic Passage (M-STIRAP) process is an outstanding example from a wide class of quantum-classical analogies which can, in principle, be realized in any wave-based system. M-STIRAP brings the concept of wave computation with dark states to magnonics. Originally, STIRAP [173], [174] describes the population transfer between two atomic states via a third, intermediate state, which is needed since direct transitions between the two states are forbidden. Analogous to the concept in photonics [175], in M-STIRAP the transfer of the spin-wave population takes place between three waveguides that play the role of the three atomic states (see Fig. III. G). In partially curved magnonic waveguides propagating spin waves locally couple to each other via the dynamic dipolar stray fields (see Section VI-F addressing the directional coupler) in well-defined regions of small separation between neighboring waveguides [176].

In the example shown in Fig. III.G, upper panel, spin waves with a frequency of $f = 2.5$ GHz are injected into waveguide 3. Shown is the so-called counter-intuitive case, and displayed are the space-resolved spin-wave intensities in false-color representation. During propagation, i.e., from left to right, the first waveguide 1 comes close to the center waveguide 2, and the next waveguide 3, which, however, carries the injected spin waves. An efficient transfer of intensity from waveguide 3 to waveguide 1 is observed. Note, that the center waveguide 2 is almost not populated during this energy transfer. In contrast, in the intuitive case displayed in Fig. III.G, middle panel, the center waveguide 2 is strongly excited as soon as the spin waves are injected from waveguide 1.

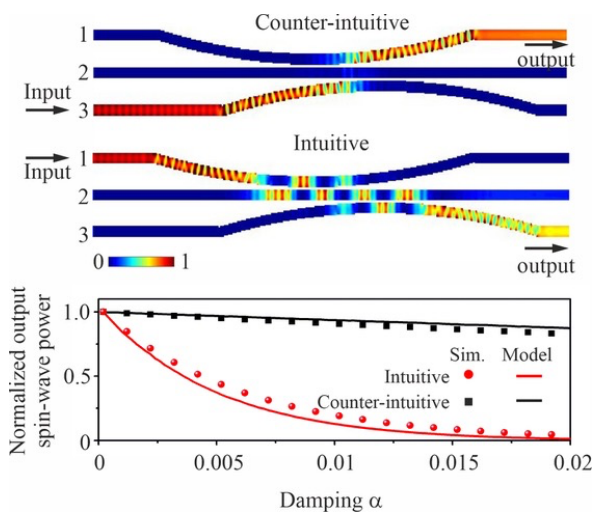

Fig. III.G. Spin-wave intensity distributions for the so-called counter-intuitive coupling scheme (top panel), and the intuitive scheme (middle panel). The output intensity as a function of the Gilbert damping of the center waveguide 2 is shown in the bottom panel. Reprinted (adapted) with permission from [Wang2021]. Copyright 2021 AIP Publishing.

This has important implications if the center waveguide has large losses, since the energy transfer will be strongly damped for the intuitive case. The bottom panel of Fig. III.G shows the output power as a function of the Gilbert damping parameter $\alpha$ of the center waveguide. As expected, it clearly shows that the increase of damping in the center waveguide for the counter-intuitive case is only leading to a rather weak drop of the output power due to the non-adiabatic population of the center waveguide. In contrast, the spin-wave intensity exponentially decreases with $\alpha$ for the intuitive case.

The counter-intuitive scenario demonstrates that propagation across non-populated states (dark states) in a coherent environment is feasible and can be considered in potential applications, such as in magnonic computing. The proposed scheme of M-STIRAP can be experimentally realized using state-of-the-art nanopatterning techniques developed for nanomagnonics [25]. Challenges for further development are the implementation of a tunable control of the decay in the waveguide 2, which can be realized using spintronic effects or parametric spin-wave amplification. Therefore, these results hold great potential for future magnonic device functionalities and designs by bringing together the wealth of quantum-classical analogy phenomena with the wealth of possibilities for controlling wave propagation in magnonic systems.

### Acknowledgment

This research has been supported by the Deutsche Forschungsgemeinschaft (DFG, German Research Foundation) TRR-173 – 268565370 (Collaborative Research Center SFB/TRR-173 'Spin+X', project B01).

### Contributors

This section is authored by Q. Wang (University of Vienna, Austria), P. Pirro and B. Hillebrands (TU Kaiserslautern, Germany).

## IV. 3D BUILDING BLOCKS FOR MAGNONIC NETWORKS

The switch from the two- to three-dimensional architectures attracts special attention probably in most data-processing and storage concepts since it allows for the considerable miniaturization of the devices. In electronics, 3D circuits require an efficient removal of Joule heat from the computing elements that might be a technological challenge. By comparison, magnonics allows for the data transfer with the decreased generation of parasitic heat and spin-wave dynamics in 3D nano-structures is under intensive investigation nowadays.

The section begins with the description of complex free-standing 3D YIG nano-structures. These structures can serve as resonators not only for magnons but also for phonons and photons. Thus, they attract attention also in the view of quantum hybrid systems (see Section VIII). The second section is devoted to vertically standing semiconductor nanowires conformally coated with a 30 nm Ni-shell. Such nanotubes provide an appealing design element for magnonic signal transmission and processing. Section IV-C is devoted to the novel technology of 3D direct writing by focused electron and ion beam-induced deposition (FEBID and FIBID, respectively), which already was successfully used for magnonic nano-architectures. The next section addresses numerical investigations of tunable magnonic excitations in 3D artificial spin ice lattices. The experimental studies of spin-wave dynamics in such structures are presented in the Section IV-E. The section is closed by experimental investigations of magnonic 3D interconnects and 3D magnonic crystals.

### A.  3D YIG nano-structures

Yttrium iron garnet (YIG) holds a unique position in nature for combining the lowest possible magnetic damping, excellent acoustic attenuation (10× better than quartz [177]), and optical transparency. Coupling of magnons with phonons and optical or microwave photons in YIG has already been demonstrated (see for example [178]). However, the YIG resonators employed in these experiments were typically macroscopic spheres of several hundred μm diameter or more. For integrated circuits to be used in quantum sensing or quantum information processing, however, it is imperative to have micron sized YIG resonators. These resonators should ideally be free-standing and have as little magnonic or mechanical coupling to the substrate as possible to avoid energy leakage.

Recently a process was developed that by using a lift-off technique originally developed for 3D metal structures achieves free-standing YIG structures with very low Gilbert damping [157]. The process was enabled by a novel technique for YIG thin film deposition that is based on pulsed laser deposition at room temperature and subsequent annealing [179]. Only for YIG deposition at room temperature is it possible to use resist-based lift-off techniques. By depositing YIG onto a 3D patterned resist, performing lift-off and subsequent annealing free-standing bridges were fabricated that surprisingly exhibit a monocrystalline span with only a single defect that most likely appears during the recrystallization process. The magnonic quality of these structures is excellent. By time resolved magneto optic scanning Kerr microscopy (TR-MOKE) it was possible to visualize standing spin-wave modes in the span of a bridge that exhibit a linewidth (half width at half maximum) of $\mu_0\Delta H\approx140$ μT at a frequency of 8 GHz. The inhomogeneous linewidth was $\mu_0\Delta H_{inh}\approx75$ μT. The Gilbert damping was shown to be as small as $\alpha\approx(2.6\pm0.7)\times10^{-4}$. It should be noted that a calculation of mechanical resonances for a doubly clamped beam of similar size using the material parameters of YIG yields frequencies in the upper MHz or lower GHz regime. The process not only allows for the fabrication of simple bridges but also mor complex structures like rings or drum-like resonators (Fig IV.A).

This process was developed on gallium gadolinium garnet (GGG) substrates that are the only ones on which high quality YIG can be grown up to now. For high frequency application and integration with electronics, however, it is desirable to grow the structures on other substrates or to realize a transfer onto other materials without deteriorating the YIG quality. A first approach was successful in that a large number of bridges were detached from a substrate into a watery suspension [158]. These micron sized YIG slabs could be deposited by drop casting onto various substrates. In a proof of principle, a single YIG slab of $5000\times1500\times100$ nm$^3$ was integrated with a micro coplanar waveguide. For this single YIG structure, ferromagnetic resonance even down to a temperature of 5 K was demonstrated. TR-MOKE investigation showed a quality similar to the one demonstrated in [157].

While the route to micron sized free standing YIG resonators is now open, several challenges and tasks remain. The drop casting has proven that transfer is possible; nevertheless, it does not allow targeting a specific location or orientation. Currently a process for transfer using focused ion beam cutting is under test but for mass production, other solutions need to be found. Ideally, growth on a different substrate material should be developed. Because the crystallization process tends to trap dislocations in the transition region between feet and span, making 3D structures on substrates other than GGG can actually be easier to realize than growth of large area YIG. First tests on RF-suitable materials like MgO or even on silicon show promising results. Now the groundwork for using micron sized YIG resonators has been laid, the different applications in hybrid magnonics need to be explored and numerous experiments for coupling magnons to microwaves, mechanical oscillations or light are currently under way, as well as experiments to determine their mechanical and optical properties. These may finally lead to application for example in the coherent conversion of excitation states between, for example, optical and microwave photons or the readout of Qbits. But also, application in room temperature magnonics seem feasible because YIG might finally get rid of the constraints by the limitations of the GGG substrate.

### Acknowledgment

This work was funded by the DFG in the TRR227 TPB02.

### Contributors

This section is authored by G. Schmidt (Martin-Luther-Universität Halle-Wittenberg, Germany).

### B.  Conformal ferromagnetic coatings for tubular magnon conduits and 3D magnonic networks

During the last decade, research on fundamental aspects and functional properties of spin waves on the nanoscale has significantly grown. Nanotechnology applied to both ferromagnetic metals and the ferrimagnetic insulator YIG enables lateral sizes of spin-wave channels down to sub-100 nm length scale in planar (2D) magnonic crystals [180] and individual magnon waveguides [26]. Microwave-to-magnon transducers based on coplanar waveguides (CPWs) have shown emission and detection of magnons in YIG with wavelengths λ down to 50 nm [181] and promise even shorter λ [35]. It is now key to enhance the integration density, and implement nanotechnologies allowing for the transition from 2D to 3D magnonic devices. Here, the overgrowth of prepatterned substrates is promising. On the one hand, it extends planar thin-film technology to 3D device architectures [182]. On the other hand, overgrowth provides a reliable approach to create curved magnetic nanodevices with particularly the tubular topology. The latter one owns a special potential considering recent theoretical progress that links the curvature of magnetic elements to magnetochiral spin-wave properties [183], [184], [185].

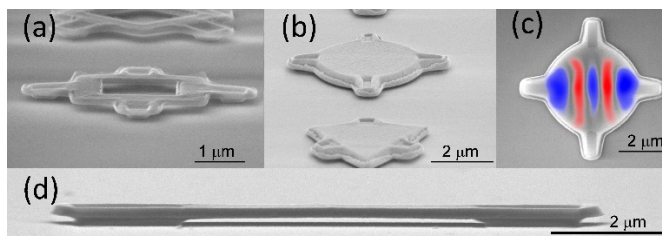

Fig. IV.A: Various free standing YIG structures [157]. c shows a YIG drum superimposed by its measured spin-wave pattern. Reprinted (adapted) with permission from [157].

A main obstacle for the creation of complex 3D networks exhibiting a high structural porosity resides in an adequate deposition technique for low-damping magnetic materials. The largely exploited physical vapor deposition is of restricted use due to shadowing effects. The obstacle has recently been overcome by focused electron beam-induced deposition of e.g., Co-Fe [186] and chemically assisted atomic layer deposition (ALD) [187]. To evidence the versatility of ALD-grown Ni [188], [189] and Ni₈₀Fe₂₀ [190] for large-aspect-ratio 3D magnonic nanodevices, ferromagnetic nanotubes have been prepared as shells around nanotemplates consisting of vertically standing non-magnetic nanowires (Fig. IV.B(a)). On the one hand, such a nanotube represents a prototypical element which serves as a vertical magnonic interconnect in a 3D network. On the other hand, its magnetochiral properties [183], [184], [185] enable spin-wave transport with tailored nonreciprocity in oscillation frequency, amplitude, group velocity and decay length [191]. The mean nanotube curvature renormalizes the dipolar and isotropic exchange interaction in such a way that, depending on the equilibrium magnetic state, the nonreciprocity is controllable on-demand for applications [191].

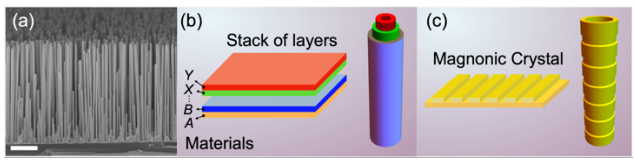

Fig. IV.B. (a) Vertically standing semiconductor nanowires conformally coated with a 30 nm thick Ni shell giving rise to a dense array of ferromagnetic nanotubes. Scale bar: 1 μm. Source: https://doi.org/10.1021/acsami.0c06879. Further permission related to the material excerpted should be directed to the ACS. Sketches of (b) a multi-layered core-shell nanotube consisting of different functional layers labelled, A, B, .., X, Y and (c) a tubular chiral magnonic crystal.

A main obstacle for the theoretical analysis of 3D networks is intrinsic to their geometry and size. They require the well-known description of the spin-wave properties of large-aspect-ratio elements, and analytical techniques which are combined with a fast and versatile numerical solver with special emphasis on curved 3D topology. Such a solver has recently been presented [192] and is ready for implementation. It efficiently calculates spin-wave dispersion relations and spatial mode profiles in arbitrary non-collinear equilibrium magnetization configuration in just few tens of minutes.

The predictions on spin-wave modes and curvature-induced magnetochirality in single-layered ferromagnetic nanotubes are encouraging for 3D magnonics. First sparks of realization already appeared. Recently individual ferromagnetic nanotubes were investigated experimentally and showed curvature-induced asymmetric spin-wave dispersion in the vortex state [193] as well as a large number of confined spin-wave modes in an axially magnetized state [189], [190]. Further experimental efforts should explore selectively the roles of dipolar and isotropic exchange interactions in view of the curvature-induced magnetochiral properties by integrating a single nanotube to CPWs. In Ref. [191] on-demand controllability of nonreciprocities and curvature-induced asymmetries in the spin-wave dispersion at tens of GHz and sub-100 nm wavelength have been predicted. However, the treatment and detailed distinction of spin-wave modes are still challenging when multimode-hybridization takes place. Here, semi-analytical techniques will be needed.

Techniques for conformal coating of patterned 3D nanotemplates are at a promising stage of maturity, such that one can envision the tubular topology as a versatile functional element for 3D magnonic devices. This is realized if so-called core-shell magnetic nanotubes (CSMNs) are formed with incorporate concentric stacks of different thin films (Fig. IV.B(b)). In CSMNs made of ferromagnetic/heavy metal shells the interfacial Dzyaloshinskii-Moriya interaction (DMI) can introduce not only further nonreciprocity, but also nanoscale spin structures such as skyrmions. They offer further engineering of magnon modes at microwave frequencies.

Challenges reside in defining methods and recipes for an optimized structural quality of layers and surfaces such that interfacial interactions across concentric layers are induced in a similar fashion as known from 2D multilayered system (Fig. IV.B(b)). One can thereby conceive a nanotubular platform made of multiple core-shell stacks for exploiting interconversion between magnons, heat and spin-polarized electric charge currents, as well as for devising 3D curvature-induced chiral magnonic crystal (Fig. IV.B(c)). On the one hand, depending on the equilibrium state and the shell composition (Fig. IV.B(b)), phenomena widely explored in 2D systems such as the spin-Hall effect, inverse spin-Hall effect, spin pumping, spin Nernst effect, magnon-drag effect, and spin torque might be enhanced and quantized due to the azimuthal symmetry of nanotubes. For instance, spin-transfer torque effects can be exploited to switch magnetic states and reduce the damping. This might be reached at a lower applied electrical current than in 2D films. The spin-torque effect might be enhanced due to a reduced size of non-collinear spin textures. A sub-50 nm extension of non-collinear spin textures might spontaneously appear in nanotubes with a sub-70 nm diameter, resulting from (i) the dominant exchange interaction and surface anisotropy favoring a radial spin alignment, (ii) the lack of lateral edges, and (iii) the surface-to-volume magnetic charge ratio induced by the curvature. The broken left-right propagation symmetry of spin waves along the long axis due to magnetochirality (translated in a non-zero group velocity along a preferred direction) is expected to induce (i) a preferred propagation direction along which the magnon-drag effect is stronger, and hence (ii) a direction of enhanced flow of electrons and heat. This opens room for modified thermoelectrical transport properties such as both modified Seebeck coefficients and thermal conductivities along the nanotube axis. On the other hand, a periodical modulation of the shell diameter (Fig. IV.B(c)) would lead to a periodical modulation of curvature-induced magnetochiral interaction. As a consequence, a chiral magnonic crystal would be created which features indirect magnonic gaps, flat bands, unidirectional magnon propagation and possibly topological magnons induced by curvature.

We hence expect nanotubes to provide an appealing design element which enhances substantially the signal transmission and information processing functionalities of magnons. Beyond the ALD-grown ferromagnetic metals, optimized ferrimagnetic shells consisting of YIG and α-Fe$_2$O$_3$ will certainly boost 3D magnonics. Fully integrated 3D networks created via conformal coating require specifically engineered nanotemplates. Here, we expect two-photon lithography to play a dominant role to experimentally explore and implement the outlined vision.

### Acknowledgment

D.G. acknowledges funding by SNSF via grant 197360. J.A.O. acknowledges funding by Fondecyt Iniciacion Grant No. 11190184.

### Contributors

This section is authored by D. Grundler (Ecole Polytechnique Fédérale de Lausanne (EPFL), Institute of Materials and Institute of Electrical and Micro Engineering, Switzerland) and J. A. Otalora (Universidad Católica del Norte, Physics Department, Chile).

### C. Direct-write 3D magnonic nano-architectures

Patterned nanomagnets have traditionally been 2D planar structures. Recently, extending nanostructures into the third dimension has become a major research direction in magnetism, superconductivity, and spintronics, because of geometry-, curvature-, and topology-induced phenomena [194]. In magnonics, extension of spin-wave circuits into 3D is required for the reduction of footprints of magnonic logic gates (Section VI) and it allows, e.g., steering of spin-wave beams in graded-index magnonics. In addition, the height of nanomagnets offers an additional degree of freedom in the rapidly developing domain of inverse-design magnonics (Section VII-C). A major challenge is the insufficient suitability of lithographic techniques for demands of 3D magnonics, requiring the development of advanced fabrication techniques.

Nowadays, 3D direct writing by focused electron and ion beam-induced deposition (FEBID and FIBID, respectively) is established as a versatile approach for the fabrication of complex-shaped nano-architectures in magnetism [195], [196], [197], [198], superconductivity [199], and plasmonics [200].

For magnonics of especial interest are direct-write Fe- and Co-based conduits (Section VI-B), exhibiting a spin-wave decay length in the 3 to 6 µm range. Thus, given FEBID's lateral resolution down to about 50 nm (in free-standing magnetic 3D nano-architectures) and its versatility regarding the substrate material, FEBID appears as an interesting nanofabrication technology for 3D magnonics, see Fig IV.C(a) and (b).

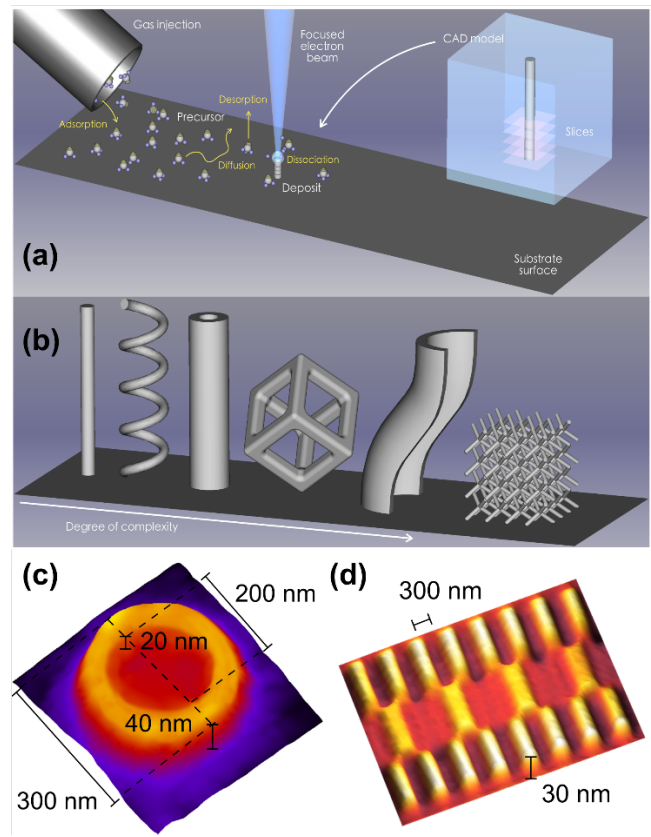

Fig. IV.C. (a) Basics of FEBID process and cartoon of slicing operation needed for shape-true mapping of the 3D CAD target structure (pillar) to the actual nano-deposit. (b) Iconized 3D FEBID structures of increasing complexity in preparation. (c) 3D direct-write nanovolcano, adapted with permission from [186]. (d) Bi-periodic 3D magnonic crystal fabricated by FEBID.

Recently, spin-wave eigenmodes were investigated experimentally in individual direct-write Co-Fe nanovolcanoes (Fig. IV.C(c)) by spin-wave resonance spectroscopy [201]. It was demonstrated that the extension of 2D nanodisks into the third dimension allows the on-demand engineering of their lowest eigenfrequencies by using 3D nanovolcanoes having about 30% smaller footprints. Micromagnetic simulations revealed that the ring encircling the volcano crater leads to an effective confinement of the low-frequency eigenmodes under the volcano crater, because of the strongly non-uniform internal magnetic field, while the higher-frequency eigenmodes are confined in the ring area [186]. The presented 3D nanovolcanoes can be viewed as multi-mode resonators and as 3D building blocks for nanomagnonics. Direct writing by FEBID and FIBID can also be employed for the mask-less fabrication of 3D magnonic crystals, see Fig IV.C(d). Further work should address the dynamics of spin-wave and topological defects (e.g., emerging magnetic monopoles, see also Section IV-E) in hierarchical 3D networks as well as in curvilinear and topologically non-trivial 3D nano-architectures.

### ACKNOWLEDGMENT

O.V.D. acknowledges the Austrian Science Fund (FWF) for support through Grant No. I 4889 (CurviMag). M.H. acknowledges the DFG for support through Grant No. HU 752/16-1. Support through the Frankfurt Center of Electron Microscopy (FCEM) is gratefully acknowledged.

### CONTRIBUTORS

This section is authored by O. Dobrovolskiy (University of Vienna, Austria), F. Porrati and M. Huth (Goethe University Frankfurt am Main, Germany).

### D. Tunable magnonic excitations in 3D artificial spin ice lattices

Three-dimensional artificial spin ice lattice structures in the form of interconnected magnetic nanowire arrays are currently emerging as materials with unexpected potential for magnonic applications. Recent experimental studies have demonstrated coherent spin waves in extended three-dimensional magnetic nanowire arrays and the tunability of their frequency with an external magnetic field. In parallel, simulations predict that the high-frequency spectrum of such artificial spin ice systems depends on their magnetic configuration, which could pave the way towards material with reconfigurable magnonic properties. Given its attractive potential for applications and fundamental research, further developments in this domain can be expected in the near future. We review key aspects of these systems and discuss possible next steps.

Three-dimensional (3D) nanomagnetism [202] has advanced to an essential research topic in recent years. As the technology became available and several articles demonstrated its ability to fabricate magnetic nanostructures with arbitrary shapes, scientists predicted that access to the third dimension would result in new features and functionalities [203]. A long-term goal of this research is to synthesize magnetic nanosystems that could be exploited in future applications. At first, however, it was unclear what specifically these features could be, with one of the best guesses being some symmetry-breaking effects arising from surface curvature [204], as they are treated within the developing specialized niche of so-called curvilinear magnetism. More recently, however, an up-and-coming field of applications for 3D magnetic nanostructures has emerged in an entirely different form. Recent articles point towards 3D *arrays of interconnected magnetic nanowires* as a new type of magnetic material with promising properties in magnonic applications [205], [206].

*Interconnected 3D Artificial spin ice.* These material systems combine features known from various disciplines within research on magnetism, such as artificial spin ice, high-density data storage, neuromorphic computing, and magnonics. The most obvious connection between 3D nanostructures and these other domains concerns artificial spin ice (ASI) systems [52], where 3D interconnected nanowire arrays represent a straightforward extension from two to three dimensions. Provided that the magnetic nanowires connecting the vertex points are sufficiently thin, they display an Ising-type magnetization analogous to single-domain thin-film elements in usual ASI systems. As in traditional ASIs, the vertex configuration plays a decisive role in determining the system's magnetization state. Charged vertices have a magnetic structure characterized by an imbalance in the number of adjacent wires with magnetization pointing towards the vertex and those pointing away from it. The charge of these defect sites arises from the nonzero net divergence of the magnetization at the vertex, which represents a monopole-type source of magnetic fields. More important, in the context of magnonics, the vertices have recently been found to display particular high-frequency magnetic properties [206], [207].

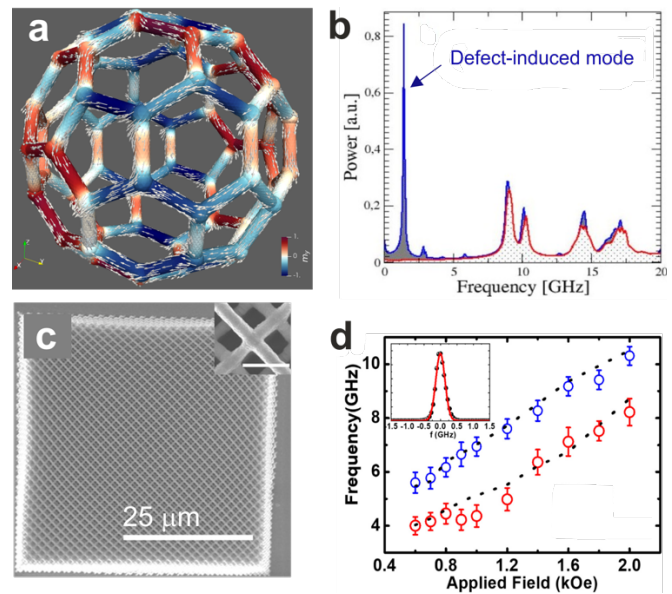

Fig. IV.D. Simulated magnetic structure in a sub-micron sized buckyball nano-architecture (a). The magnonic spectrum (b) changes when the ASI of the buckyball contains defect-type vertices. In an extended 3D ASI (c), the spin-wave mode frequency can be tuned by an external field (d). The sensitivity of high-frequency excitations in 3D ASI to magnetic structure and applied fields makes these systems promising functional materials for magnonic applications. (b) adapted from Ref. [206] with permission from AIP Publishing, (c) and (d) reprinted from Ref. [205] https://doi.org/10.1021/acs.nanolett.1c00650 with permission from ACS. Further permission related to the use of this material should be directed to the ACS.

*High-frequency modes in 3D spin ice.* While the impact of defect sites on the magnonic excitation spectrum was already known from traditional 2D ASI made of disconnected arrays on nanomagnets [208], their role in the high-frequency dynamics of interconnected nanowire structures appears to be even more dominant. In the case of a 3D magnetic nanoarchitecture with the shape of a buckyball, the pivotal role played by defect vertices in the magnonic excitation spectrum was recently demonstrated in a micromagnetic simulation study by Cheenikundil et al. [206], Fig. IV.D(a,b). It was found that these vertices yield a distinct signature in the magnetic excitation spectrum, with a frequency significantly lower than that of typical oscillations of the magnetic material and with unusually strong intensity. The insertion of only two defect states within an array of 60 vertices results in a dominant and sharply defined peak in the magnonic excitation spectrum. So far, the principle has been demonstrated for a quasi-two-dimensional ASI in the form of a buckyball. Despite its three-dimensional spherical shape, the buckyball geometry is essentially restricted to the two-dimensional surface of that sphere.

Nevertheless, the principle can be expected to apply also to extended three-dimensional versions of such artificial magnetic material. The excitation of coherent spin waves in three-dimensional ASI structures has been observed experimentally very recently by Sahoo *et al.* [205]. These authors also showed that the frequency of these spin-wave modes can be tuned with an external magnetic field, cf. Fig. IV.D(c,d). Generalizing the observations for the buckyball-type ASI, one could conclude that a 3D ASI in the form of an interconnected nanowire array represents a reconfigurable magnonic crystal, whose resonance frequency can be modified by introducing or removing magnetic defect sites. Such a switchable control over the high-frequency properties of the magnetic medium would have great potential for magnonic applications. In the case of the buckyball, the insertion and removal of defect sites was demonstrated to be possible through simple quasi-static field variations in the form of minor loops [206]. The controlled insertion of defects into an extended crystal is less straightforward, but should not pose any substantial difficulty. Once a defect is inserted, its controlled propagation through a 3D ASI lattice has recently been demonstrated [209].

*Conclusion and outlook.* Using 3D interconnected nanowire arrays as an effective magnonic medium with reconfigurable properties should be attainable by controlling their magnetic state with ASI characteristics. Given the rapid development in this field, it is likely that scientists will provide a demonstration of such a feature shortly. The unique properties of such artificial magnetic media and their configuration- and field-dependent high-frequency spectrum could open new perspectives in magnonic applications.

For a profound understanding and analysis of the high-frequency properties of such magnetic architectures, theoretical and simulation studies are required in addition to advanced nanofabrication and experimental measurement techniques. Because of these nanowire arrays' extremely low volume occupation and their complex geometry, finite-difference methods are inefficient in simulating such systems. These methods can only be used to simulate single building blocks of such systems, limited to a minimal number of unit cells of the lattice structure. Efficient simulations studies of these interconnected nanowire systems are accessible with advanced finite-element-based micromagnetic software. Moreover, given the large problem size, simulations should exploit GPU acceleration and use efficient data compression methods [210].

ACKNOWLEDGMENT

This work was funded by the LabEx NIE (ANR-11-LABX-0058_NIE) in the framework of the Interdisciplinary Thematic Institute QMat (ANR-17-EURE-0024), as part of the ITI 2021–2028 program supported by the IdEx Unistra (ANR-10-IDEX-0002-002) and SFRI STRATUS (ANR-20-SFRI-0012) through the French Programme d'Investissement d'Avenir.

CONTRIBUTORS

This section is authored by Rajgowrav R. Cheenikundil and Riccardo Hertel (Institut de Physique et Chimie des Matériaux de Strasbourg, CNRS, France).

### E. Magnetic charge transport and spin-wave propagation in 3D nanostructured lattices

The nanostructuring of materials in three-dimensions is an emerging theme in magnetic materials research, providing access to a wide range of novel phenomena including the use of curvature to control spin textures, ultrafast domain wall motion in cylindrical magnetic nanowires, magnetochiral effects and 3D frustrated lattices with tunable ground states [202]. Next generation computing is likely to rely upon complex 3D nanostructured systems, allowing the realization of systems offering ultrahigh density storage, massive connectivity and access to non-von-Neumann architectures. The study of spin-wave propagation in complex 3D magnetic nanostructures is still in its infancy but offers the possibility of systems with tunable transport through intricate control of the spin texture, whilst also exploiting underlying 3D periodicities within the geometry to control band structure. Here, we outline the progress in this field and its future evolution.

*Status of field.* Direct-write technologies such as focused electron beam induced deposition (FEBID) and two-photon lithography (TPL) allow arbitrary 3D geometries to be written. For the case of FEBID, magnetic materials in 3D geometries can be directly written onto a substrate by controlling the interaction between an electron beam and a gaseous precursor [202]. TPL harnesses a femtosecond laser to expose a 3D pattern within a photoresist. Subsequent deposition can then be used to transfer the 3D pattern into a magnetic material [211]. Such a technique was pioneered recently in order to yield complex 3D magnetic nanostructures [212]. Williams et al. utilized TPL with a positive resist in order to produce a series of pores within a tetrapod geometry. Subsequent electrodeposition and resist removal yielded freestanding Co tetrapod structures, which can be considered the most basic building block of a diamond-bond lattice. These novel systems were then subject to optical pump-probe experiments which showed the existence of three spin-wave modes, which had a small variation with magnetic field [213]. Micromagnetic simulations of the tetrapod geometry showed good agreement with the experimental results whilst also elucidating their spatial characteristics. The study demonstrated that time-resolved optical measurements can be performed upon complex 3D magnetic nanostructures. Building upon this, May et al.

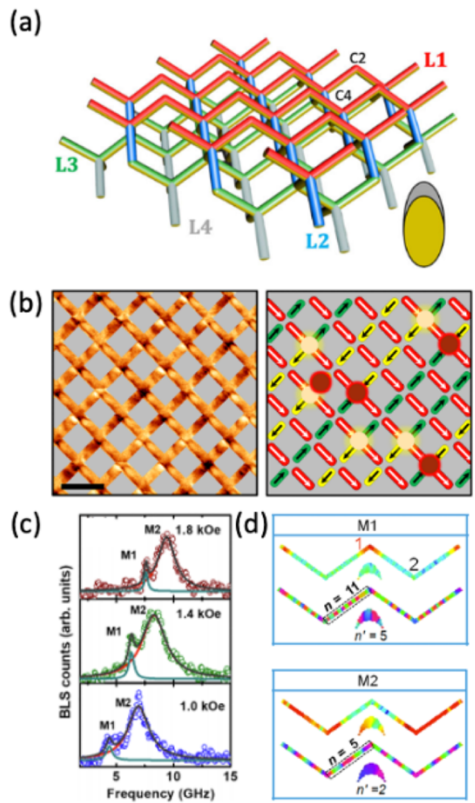

Fig. IV.E. *(a) A schematic of the fabricated 3D artificial spin-ice (3DASI) structure. (b) Left: Magnetic force microscopy image of 3DASI surface. Right: Schematic showing the determined magnetic configuration. Bright red circles indicate positive monopoles and yellow circles indicate negative monopoles. (c) Brillouin light scattering measurements upon the 3DASI for three different field values. Two modes are seen. (d) Spin-wave mode profiles for M1 and M2.*

exploited TPL to fabricate a diamond-bond lattice within a negative resist. Evaporation of NiFe at normal incidence then yields a 3D network of magnetic nanowires within a diamond-bond geometry as shown in Fig. IV.E(a) [214]. Magnetic force microscopy (MFM) demonstrated that the wires were single domain, which was confirmed with optical magnetometry and showed a single abrupt transition in a square hysteresis loop. Further work then showed that these systems behave as 3D artificial spin-ice systems (3DASI) with each of the expected vertex types satisfying the ice-rule as well as the higher energy states that possess finite magnetic charge [209]. Magnetic imaging experiments performed as a function of field, showed very different magnetic charge dynamics, depending upon the direction. Application of a field along the surface termination which consisted of coordination-two vertices, yielded a low frequency of lone monopoles. A field applied transverse to this, showed the creation of many, closely correlated magnetic charge pairs, of opposite sign, which became separated as the field increased. Overall, the study shows that a complex 3D magnetic lattice can be fabricated and its spin configuration imaged (See Fig. IV.E(b)). Such systems are of great interest for future computing applications for several reasons. Firstly, controlled movement of magnetic charge upon a 3D lattice has parallels with racetrack memory type devices. Perhaps the advantage of these systems is the multi-state functionality with 16 possible states upon each four-way junction. When considering this over the entire array, it yields a huge number of possible accessible states within the system which can be exploited for reservoir computing. Subsequently, Sahoo et al. exploited Brillouin light scattering (BLS) spectroscopy to measure spin-waves in this novel 3DASI system [205]. Two spin-wave modes were measured (Fig. IV.E(c)), whose frequencies showed a monotonic variation with magnetic field. Micromagnetic simulations were able to capture these modes and show the mode profile as shown in Fig. IV.E(d). Overall, the power profile suggested that M1 was primarily focused at lattice junctions whereas M2 was extended throughout the entire network. The collective spin-wave modes in the 3DASI structure is very different to that observed in the isolated tetrapod samples, and indicates coherent spin waves can propagate throughout the 3D lattice, suggesting an interesting magnonic band structure.

*Future challenges and evolution of the field.* 3D magnetic lattices are already well-suited for neuromorphic computing applications. They have intrinsic high-connectivity with a random distribution of synaptic weights at junctions, naturally realizing the essential ingredients for reservoir computing as seen in 2D magnetic arrays [215]. Furthermore, the use of coherent spin-wave propagation in such systems allows the exploration of interference-based computing. The challenges in realizing such systems are multifaceted. The fabrication of 3D magnetic lattices poses a number of significant challenges. Current systems, due to the use of line-of-sight deposition are limited to just a unit cell of lattice. Further optimization of other deposition methods which conformally coat the polymer may be a fruitful avenue of investigation. For example, use of a negative resist and electrodepostion can in principle yield wires of single domain character since it avoids dark erosion [216]. Measurement of complex 3D structures also pose a significant challenge. Whilst BLS is able to optically measure the top surface of a 3D lattice, the bulk remains difficult to address. Novel synchrotron techniques such as time-resolved laminography may be the answer and should allow the dynamics to be measured across the entire structure [217].

## Acknowledgment

SL acknowledges funding from the Engineering and Physical Sciences Research Council (EP/R009147/1) and from the Leverhulme Trust (RPG-2021-139). AB acknowledges funding from Department of Science and Technology, Govt. of India (SR/NM/NS-09/2011) and S. N. Bose National Centre for Basic Sciences (SNB/AB/18-19/211).

## Contributors

This section is authored by S. Ladak (Cardiff University, UK) and A. Barman (S. N. Bose National Centre for Basic Sciences, Salt Lake, India).

## F. 3D magnonic crystals and interconnects

In magnonic crystals (MCs), a nanoscale periodic modulation of the magnetic properties allows for tuning the magnon band structure with permitted and prohibited frequency regions, as well as controlling related properties such as the group velocity or magnon interactions. The current state of the art is focused on the understanding of magnonic band formation in planar 1D and 2D MCs which can be fabricated by standard nano-lithographic techniques. [218], [219], [220], [221], [12]

To integrate magnonic micro- and nanocircuits in the existing electronic/spintronic technology and go even beyond that, the development of large-scale three-dimensional (3D) magnetic periodic structures with high precision control of geometry and material composition at the nanometer scale, is an important challenge now. Vertically integrated 3D magnonic structures can increase the density of magnonic and spintronic elements to form scalable and configurable magnonic networks. In addition, these 3D structures can further reduce the total length of interconnects between functional elements and increase the device density, similarly to electronics and semiconductor technologies where vertical connections (vias) are used to create 3D packages and integrated circuits.

First attempts to reach this goal are represented by hybrid heterostructures with different ferromagnetic layers placed in direct contact or separated by a nonmagnetic spacer [152], [223], [224], [33] as well as by 3D nano-objects network [225], [203], [205]. These structures demonstrate novel spin-wave (SW) properties originating from the vertical dynamic coupling between the layers and the increased complexity in magnetic configurations in three dimensions, respectively.

Meander-type ferromagnetic films and multilayers, grown on the top of periodically structured substrates (Fig. IV.F1 (a, c), are candidates of such 3D MCs and represent a wide playground to study the emergent properties of spin waves in 3D systems. [226] Recent studies on 3D meander- shaped films based on metallic (CoFeB, NiFe) [182], [222]] and dielectric (YIG) [227] materials have demonstrated the ability to control the SWs spectra, allowing for SW propagating along the orthogonal out-of-plane film segments.

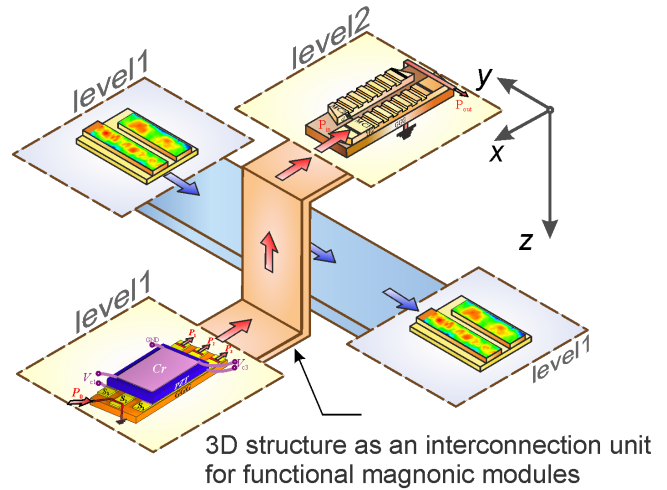

Fig. IV.F2. Schematic drawing of a 3D network of magnonic waveguides connecting functional elements between different levels along the vertical z-coordinate. Signal encoded into SWs are represented by the blue and red arrows.

The magnonic band structures of single CoFeB and bilayer CoFeB/Ta/NiFe meander-shaped films measured by Brillouin light scattering and corroborated by micromagnetic calculations are markedly different, both in observed modes and in their dependences on the wave vector, i. e. whether they are stationary or dispersive in nature. For example, a narrower width of the magnonic band has been observed for the CoFeB/Ta/NiFe structure as compared to the CoFeB sample. This can be related to the interlayer dipolar coupling which changes the SW dispersion relation. The properties of the individual modes have been further characterized by the phase relation (in-phase or out-of-phase) between the magnetization oscillations in the two layers and their localization in the horizontal and vertical segments.

For both structures the mode crossing and the absence of a bandgap were observed at k=$n\pi/a$ ($n$ is an odd number) and explained in terms of gliding-plane symmetry of the sample [182]. An additional feature of the CoFeB/Ta/NiFe system is the presence of the three lowest frequency modes that exhibit a nondispersive character. These modes have a spin precession amplitude concentrated in the topmost NiFe layer with an increasing number of nodes in the horizontal segments.

These results demonstrate that 3D MCs introduce a new degree of freedom to control the excitation of SW modes, or magnon states, with defined dynamic magnetization profiles by the design of the vertical segments which connect horizontal segments.

To assemble functional magnonic units in the multilevel networks, inter-junctions between functional blocks placed on the different layers of the joint magnonic circuitry are required. The concept of 3D magnonics looks promising in this respect, demonstrating the numerous solutions of the realization of vertical interconnects [228]. In particular, the simplest type of interconnects , i. e. a stripe of ferromagnetic material (called the SW bus) could be realized using the part of the unit cell of meander-shaped waveguide.

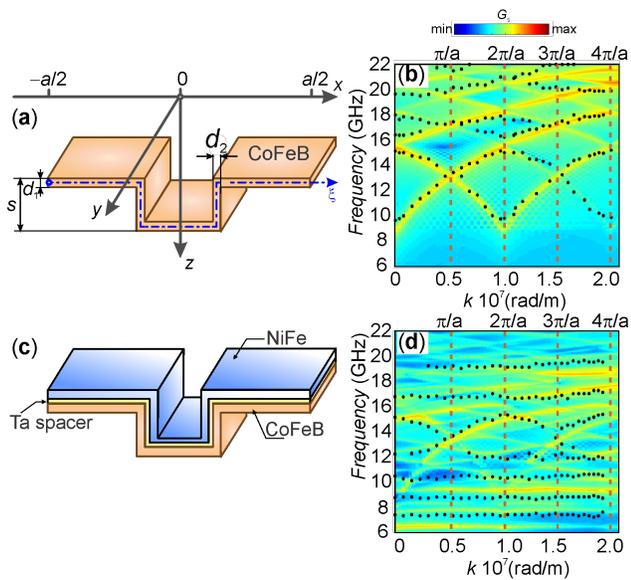

Fig. IV.F1. Cross-section sketch of the meander-shaped (a) CoFeB film and (c) CoFeB/Ta/NiFe bilayer unit cell with lattice constant a=600 nm. Comparison between measured (points) and simulated (color map) dispersion relations for (b) CoFeB and (b) CoFeB/NiFe meander structures. Reprinted (adapted) with permission from [222].

Fig. IV.F2 demonstrates the concept of the 3D meander-shape waveguide as an interconnection between the functional units (frequency selective coupler, MCs, power splitters) located at different levels along the $z$-direction of the whole magnonic network. Moreover, the transmission frequency band of SW propagation in the case of Damon-Eschbach geometry remains the same as for a single magnonic waveguide. Thus, the 3D MCs and buses open new perspectives towards the submicron magnonic circuitry by increasing the density of functional SW units providing flexible interconnection between them.

### Acknowledgment

This work was supported by the Ministry of Science and Higher Education of the Russian Federation (project No. FSRR-2020-0005). Contributions from F.C. and C.A. have been supported by imec's industrial affiliate program on beyond-CMOS logic. F.C. and C.A. acknowledge funding by the European Union's Horizon 2020 research and innovation program within the FET-OPEN project CHIRON under Grant Agreement No. 801055.

### Contributors

This section is authored by G. Gubbiotti (Istituto Officina dei Materiali del CNR, Italy), A.V. Sadovnikov and E. N. Beginin (Saratov State University, Russia), S. A. Nikitov (Kotel'nikov Institute of Radioengineering and Electronics, Russian Academy of Sciences, Russia), C. Adelmann, and F. Ciubotaru (Imec, Leuven, Belgium).

### G. Magnonics in 3D magnetic meta-materials

Magnetic meta-materials (MM) are ordered, periodically patterned systems, where geometrical arrangements, microscopic properties, collective interactions or varied periodicities lead to tailored macroscopic functionalities. Magnons, the quanta of spin waves, show a particular dispersion relation in specially designed artificial magnetic systems, also referred to as magnonic crystals (MS). These crystals are used in numerous applications like magnonic conduits and filters, frequency and time inverters, data buffering systems or components for logic gates [19]. Advances in computational algorithms will give rise to new designs, and thus, new functionalities within the field of magnonics, as discussed in Section II-H.

Making use of magnetic meta materials enlarges the number and the range of applications for the magnonic crystals. While 1D MM are extended periodically along one direction, i.e. nano-stripes, 2D MM make use of two different directions, i.e. 2D artificial spin ices (ASI) [219]. In ferromagnetic MMs the third dimension is introduced in the nonplanarity of the system [229], [230].

Specific applications, e.g., microwave absorption devices [231] or reconfigurable spin-wave channels [208], have been demonstrated in 2D MMs. For the former, arrays of nanodots have been utilized, where different ranges of frequencies are absorbed based on the magnetization of each nanodot, i.e. vortex or uniform states. Here, the total absorption power is restricted by the limited magnetic material in 2D thin films.

A second key feature of MMs is given by its reprogrammability, as demonstrated in tunable magnonic channels in 2D artificial spin ices. ASIs are planar magnetic nano-islands patterned on different 2D lattices [232]. Depending on their magnetic configuration, charge defects are emerging. Spin waves can propagate between these defects, enabling programmable input and output channels.

Transitioning towards 3DMMs, is promising to enable both of these applications, since the increased magnetic material allows for higher power absorption and the third dimension allows for an additional degree for programmability.

Recently, a first disconnected 3D artificial spin ice (3DASI) lattice was theoretically investigated by means of micromagnetic simulations [233], which is very promising to be used as a functional magnonic 3DMM, see Fig. IV.F. Here, rotational ellipsoids are arranged on a lattice with tetrahedral unit cell

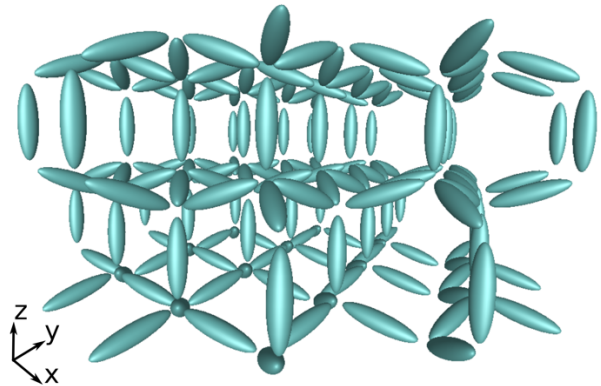

Fig. IV.G. 3D Illustration of the 3DASI lattice, consisting of 3D rotational ellipsoids arranged in tetrahedral unit cell. Reprinted (adapted) with permission from [233].

However, the experimental realization of such a lattice pushes the technological possibilities to their limits. Nanofabrication techniques similar to FEBID or optical lithography are suitable candidates [202]. While the downscaling of single elements is required to create compact devices, the building blocks, i.e., the ellipsoids require non-ferromagnetic connections, such as Nb or Pt. Resolution and thus the purity of the fabrication process is expected to affect the physical properties to certain degree.

With the expected advances in the next years, the 3DASI lattice, or similar constructs, promise to open new ways to explore magnons in 3D MM.

The 3D periodicity allows for formation of bulk crystals, which can enable higher power absorption. Furthermore, the range of allowed or forbidden frequencies can be tuned both by structural changes, i.e., size-distribution of ellipsoids, and variation of magnetic configuration.

The reconfigurability of the inner magnetic configuration gives rise to further potential applications as a reprogrammable MC. Analog to its 2D counterpart, spin waves are expected to propagate between the charge defects. Note that the charge defects are steerable within the 3DASI lattice, and that with more ease then in 2D. Thus, the notion of 3D magnonic cross-transport devices promises future technological advances. By edge ellipsoid engineering charge defects can be injected into the 3DASI, and steered to create an output port. Spin waves excited within a specific range of frequencies, will reach the desired output location, which could cross-transport the magnons to the next waveguide.

With the first experimental realizations, we are awaited by a very promising newly emergent research field. Further theoretical studies and advanced computational methods are required to fully undermine the power of the third dimension, and the novel applications arising from it.

ACKNOWLEDGMENT

S.K., C.A., and D.S. gratefully acknowledge the Austrian Science Fund (FWF) for support through grant No. I 4917 (MagFunc).

CONTRIBUTORS

This section is authored by S. Koraltan, C. Abert and D. Suess (University of Vienna)

## V. LOW-ENERGY MANIPULATION AND AMPLIFICATION OF SPIN WAVES

Efficient interconversion of signals between electric and spin-wave domains and electric control of SW propagation with low energy consumption is required to develop competitive spin-wave computing devices. Conventional methods of spin-wave excitation and manipulation based on the usage of Oersted fields or spin torques from electric currents are prohibitively power-hungry due to significant Ohmic losses. A viable approach to energy-efficient SW devices is the utilization of magneto-electric effects for excitation and control of SWs via electric fields rather than electric currents.

The first two sections describe recent progress obtained in the excitation and manipulation of spin waves through voltage-controlled magnetic anisotropy (VCMA), taking place at the interface between a ferromagnetic (or antiferromagnetic) metal and a nonmagnetic insulator. It is shown that VCMA can even create virtual reconfigurable nanochannels for guiding spin waves. The following four sections address multiferroic (ferrite-ferroelectric or ferrite-piezoelectric) hybrid structures. In strain-coupled systems, total domain correlations between a ferroelectric material and ferromagnetic film can be realized. The final section is devoted to a conceptually different approach for low-energy magnon manipulation by using dissipation-free electric currents in superconductors instead of electric fields.

### A. VCMA-based excitation and amplification of spin waves

Voltage-controlled magnetic anisotropy (VCMA) is a magneto-electric effect observed at an interface between a ferromagnetic metal (FM) and a nonmagnetic insulator (NI) whereby voltage-driven electric charge accumulation at the FM/NI interface modifies perpendicular magnetic anisotropy (PMA) of the FM [234] Given the high conductivity of FMs and low capacitance of nanostructures, VCMA can efficiently modulate PMA at microwave frequencies matching the frequencies of SWs in the FM. Such applied anisotropy modulation can be used for excitation of SWs in the FM both in the linear regime [235], [236], [237]] and via parametric resonance [238], [239]. Furthermore, VCMA can be utilized to control SW propagation [240], [241]. Being essentially free from Ohmic losses in FM/NI junctions with sufficiently thick NI barriers, VCMA SW control appears to be the most promising effect for fast, energy-efficient excitation and manipulation of SWs in thin metallic FMs.

The action of VCMA on SWs depends on the orientation of static magnetization of a FM. Linear excitation of SWs at the microwave voltage frequency is possible only if static magnetization in a FM film is tilted from both the in-plane and normal directions [237], [240].As the application of a global magnetic field in integrated magnonic devices is not desirable, the local tilting of magnetization can be achieved via stray bias fields from nearby nanomagnets.

In the in-plane and normal configurations of static magnetization realized in zero bias magnetic field, VCMA can only excite and amplify SWs parametrically [239], [238] (Fig. V.A). Advanced nonlinear SW operations needed for cascading SW logic circuits, such as SW amplitude normalization and SW phase error correction, can be implemented via VCMA-driven parametric processes, as theoretically shown in [242]. In FM structures with in-plane static magnetization, the parametric coupling of VCMA to SWs differs from the conventional parallel parametric pumping case realized in structures with normal magnetization [239]. Parametric coupling in the in-plane structures is not proportional to SW ellipticity and consequently remains large for short-wavelength exchange-dominated SWs [242]. Therefore, VCMA can address another principal challenge in magnonics – excitation and control of short-wavelength SWs ($\lambda < 100$ nm). Estimates show that total losses in optimally designed VCMA devices are dominated by intrinsic magnetic losses, and thus VCMA SW technology can approach the fundamental limits of magnonic device efficiency.

VCMA-based SW logic can be seamlessly integrated with CMOS electronics because SW signals can be efficiently converted into electric signals. Indeed, material systems exhibiting large VCMA such as Fe/MgO are used in magnetic tunnel junctions (MTJs), which can efficiently convert magnetic oscillations into electric microwave signals via large tunneling magneto-resistance [235], [239]. Alternatively, SW signals can be converted into electric signals via the inverse VCMA effect [243]. Furthermore, the problem of process integration of MTJs with CMOS has been solved in the course of commercial spin transfer torque memory (STT-MRAM) development, which lays a foundation for fabrication of large-scale VCMA SW logic circuits integrated with CMOS electronics.

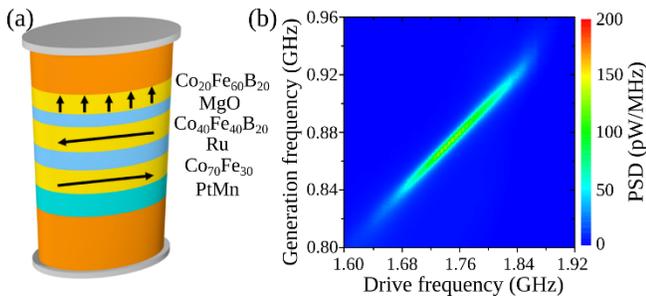

Fig. V.A. VCMA-driven parametric excitation of magnetic tunel junction (MTJ): (a) – a sketch of MTJ, free layer is the 1.58 nm thick CoFeB top layer, driving ac voltage is applied across the MTJ; (b) – generation spectrum showing the parametric resonance – oscillation frequency is exactly the half of the drive frequency, driving ac voltage is 0.185 V. Adapted with permission from [239]. © 2016 American Chemical Society.

To exceed energy efficiency of CMOS logic, VCMA SW logic must be implemented in the form of large integrated circuits where multiple SW logic operations are completed in the magnetic domain before results of the computation are converted into electric signals. This can be implemented in a synchronous computing scheme where a global microwave drive, also serving as a global clock, is used for excitation, amplification and modulation of SWs.

While some elementary operations necessary for VCMA SW logic such as SW excitation and detection have been demonstrated, other key elements such as SW amplification, amplitude modulation, and nonlinear processing remain to be experimentally implemented. The next step will be development of VCMA SW gates such as the majority gate [236] followed by more complex SW logic circuits.

The relatively low SW velocity in thin FM films limits the speed of SW circuits. Implementation of VCMA SW circuits in antiferromagnetic (AFM) materials, predicted recently [244], can solve this problem due to the high group velocities and frequencies of SWs in AFMs. In FM materials, very short-wavelength SWs ($\lambda \sim 10$ nm) must be employed in order to overcome this challenge and achieve high SW group velocities. Development of FM/NI or AFM/NI heterostructures with enhanced VCMA efficiency and low Gilbert damping is another promising direction of research for future VCMA SW technology.

## ACKNOWLEDGMENT

I.N.K. acknowledges support by the National Science Foundation through Awards EFMA-1641989, and ECCS-1708885 and by the Army Research Office through Award W911NF-16-1-0472. R.V. acknowledges support by National Research Foundation of Ukraine, Grant #20020.02/0261.

## CONTRIBUTORS

This section is authored by R. Verba (Institute of Magnetism, Kyiv, Ukraine) and I. N. Krivorotov (University of California, Irvine, USA).

### B. Low-energy manipulation of spin waves by electric fields

The logic operations in a proposed spin-wave (SW) computing device are performed by SW interferences and the interconnections are made by SWs. One of the essential tasks would be manipulating SW properties such as amplitude, frequency, wavevector, band structure, damping, nonreciprocity at the cost of ultralow-power with minimal time delay. The voltage-controlled magnetic anisotropy (VCMA), i.e., electric field ($E$) induced modulation of interfacial magnetic anisotropy (IMA), can fulfill the above requirements (see Fig. V.B).

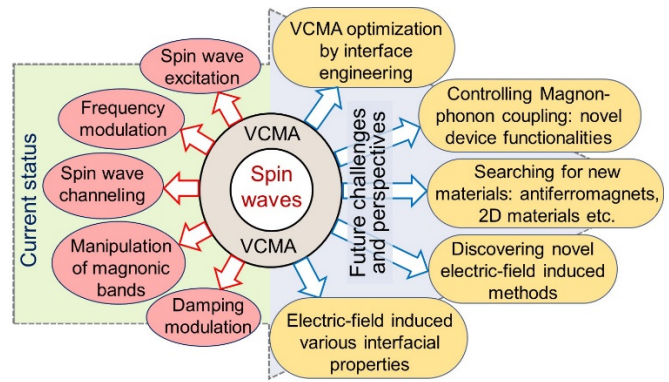

Fig. V.B. A cartoon shows the current status, future challenges and perspectives of electric field induced manipulation of spin waves.

Apart from excitation [235], [237], the VCMA has proven to be an efficient tool for manipulating ferromagnetic resonance [245] and SWs [246]. Rana et al. observed that VCMA induces a significant SW frequency shift in ultrathin ferromagnetic (FM) films [246]. VCMA can even create virtual reconfigurable nanochannels (NCs) for guiding SWs. Interestingly several closely spaced parallel NCs can be formed, and SWs with tunable frequency and wavelength can propagate through a single NC or group of desired NCs. These NCs can be applied to design various SW logic gates [247]. When SWs transmit through parallel NCs, the magnonic band structures and corresponding band gaps are tunable by gate voltage, which may be beneficial for developing SW filters and attenuators [241]. The damping constant of ultrathin FM films can also be tuned by an electric field, just like IMA. Rana et al. observed both linear and nonlinear modulation of damping and explained that the presence of Rashba spin-orbit coupling (RSOC) and the electric field dependence of Rashba strength is the origin behind the observed nonlinear modulation [248]. The choice of the buffer layer and oxide materials solely decides the weight of linear to nonlinear damping behavior. So, by choosing suitable material and through interface engineering, one may achieve a condition to tune damping parameters in a nonlinear fashion to develop low-power SW amplifiers and attenuators.

Even after all these successes, the VCMA induced method faces some critical challenges: the low value of the VCMA coefficient for FM/oxide heterostructures. Although interface engineering, e.g., inserting an ultrathin heavy metallic layer or doping with heavy metals at FM/oxide interfaces, can increase the VCMA coefficient up to a few hundreds of fJ $V^{-1}m^{-1}$, it is still below the minimum criterion of $\sim$ pJ $V^{-1}m^{-1}$. Meanwhile, some alternative ideas such as voltage-controlled redox reactions, charge trapping, electromigration came up, but they are not applicable for developing faster and miniaturized magnonic devices. Unfortunatel, attention has not been paid to investigate the effect of interface engineering on other interfacial properties such as damping, interfacial Dzyaloshinskii-Moriya interaction (iDMI), RSOC, which also affect the SW properties significantly. One of the future directions of research would be to address this issue.

The selection of appropriate materials is the key to improve device performance. For instance, by choosing ferrimagnetic insulators and Heusler alloys, one may achieve a higher VCMA coefficient while reducing the damping constant. Unfortunately, VCMA phenomena have not been observed yet in these materials. Magnetic 2D materials and their heterostructures would be other promising candidates due to their fascinating interfacial properties. Apart from discovering new materials, the study of magnons and various interfacial properties in these materials has started very recently. Antiferromagnetic thin films generate short wavelength coherent SWs with THz frequency, essential for developing ultrafast SW computing devices. Hence, it will be very demanding to investigate and electrically control the physical phenomena at antiferromagnetic interfaces.

Another intriguing research direction would be to find and optimize novel $E$-induced methods to manipulate SWs. An electric field applied perpendicularly to the SW wave vector and magnetization has been shown to induce the Aharanov-Casher effect [249], which can significantly change the SW phase and make SW dispersion anisotropic. Interestingly, the iDMI can also be modulated by an electric field $E$ like IMA [250], which is very important to control nonreciprocity in SW frequency. Novel device functionalities such as nonreciprocal transport, efficient manipulation of magnons may be achieved by coupling SWs with other quasiparticles such as phonons. In principle, the magnon-phonon coupling strength is also tunable by an electric field by modulating iDMI and IMA [251], which has not been demonstrated yet. Consequently, new ideas such as $E$ control of induced in-plane anisotropy at FM/oxide interfaces [252] can also be implemented for manipulating SW properties.

## Acknowledgment

This work was supported by Grants-in-Aid for Scientific Research (S) (No.19H05629) and KAKENHI (B) 20H01865 from the Ministry of Education, Culture, Sports, Science, and Technology of Japan. BR acknowledges RIKEN Incentive Research Project Grant No. FY2019 for financial support.

## Contributors

This section is authored by B. Rana (ISQI, Adam Mickiewicz University in Poznań, Poland), Y. Otani (RIKEN and the University of Tokyo, Japan).

### C. Electromagnonic crystals for spin-wave computing

Multiferroics in a form of waveguide structures composed of ferrite and ferroelectric layers provide a possibility for dual broadband magnetic and fast electric tuning. This unique feature is based on the electrodynamic coupling between the spin and electromagnetic waves (magnon-photon interaction) in the layered ferrite-ferroelectric (multiferroic) heterostructures. These coupled excitations are referred to as spin-electromagnetic waves (SEWs). By analogy to natural multiferroic solids, the quanta of these excitations are considered as electro-active magnons or electromagnons. The multiferroic periodical structures, called electromagnonic crystals (EMCs), demonstrate electrically and magnetically tunable band gaps, where propagation of the electromagnons is forbidden [253], [254] – see Fig. V.C.

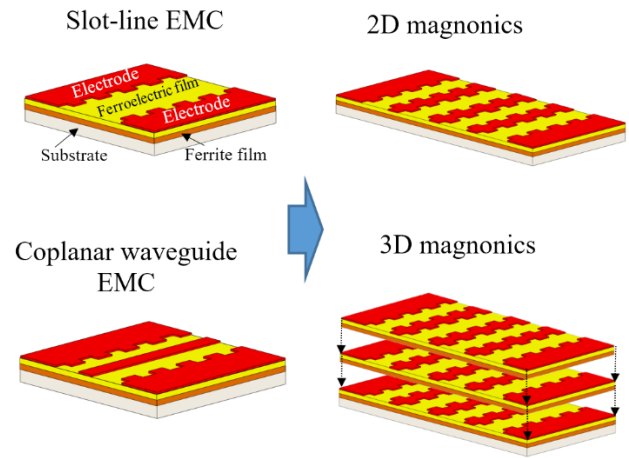

Fig. V.C. From all-thin film one channel electromagnonic crystals (EMC) to 2D and 3D voltage-controlled magnonics.

As it was shown earlier, the effective coupling at microwave frequencies is achieved in the structures fabricated with relatively thick (200-500 μm) ferroelectric layers. Such thicknesses of the ferroelectric layer lead to relatively high control voltages (up to 1000 V) needed for an effective electric tuning of the SEW dispersion. In order to reduce the control voltage, all-thin-film multiferroic heterostructures have been proposed and experimentally tested [255], [256]. Recent advances in this field include the development of theoretical models of thin-film electromagnonic crystals based on a slot transmission line [257] and a coplanar waveguide [258].

Further progress in this direction lies in the experimental verification of the theoretical models, miniaturization of multiferroic heterostructures to the nanoscale, and the development of a technology for thin-film EMCs. The single channel structures can be used as tunable phase shifters and filters. In the general case, multichannel structures can be fabricated (Fig. V.C). They would be promising for controlled magnon switching between channels, and basic elements for spin-wave magnonic logic circuits, reservoir computers and so on.

### ACKNOWLEDGMENT

The work is supported by the Ministry of Science and Higher Education of the Russian Federation (grant number No. FSEE-2020-0005).

### CONTRIBUTORS

This section is authored by Andrey A. Nikitin and Alexey B. Ustinov (St. Petersburg Electrotechnical University, Russia).

### D. Magnon straintronics

Due to the dipolar and exchange interaction the spin-wave quasiparticle (magnon) flows and carries a spin angular momentum in the form of propagating spin waves (SWs). Heat generation and energy consumption in CMOS architecture requires to look for new ways to a data processing paradigm. A spin-wave computing paradigm being based on the interaction of the propagating SWs [259] needs energy-efficient and fast mechanisms of magnon transport control. The tunability by means of both the value and orientation of a bias magnetic field change leads to high energy consumption and is accompanied with a low tunability rate due to unavoidable effects of the electromagnet's inductivity. Providing a reconfigurable landscape of magnetization within the path of the SW via the focusing of laser light on ferromagnetic films is also accompanied by relatively fast heating but slow cooling transition processes. Planar integrated multiferroic structures (ferrite-ferroelectric or ferrite-piezoelectric) [260], [261], [262] are used as a promising alternative in beyond-CMOS computing technology with low-level energy consumption and reduced Joule heating. This way the solution of the problem of the operation speed is limited by the properties of the piezoelectric component of the ferroelectrics. The fabrication of ferromagnetic/semiconductor interfaces leads to reliable transformation of spin-wave spectra accompanied by the nonreciprocal propagation of SWs due to the variation of the conductivity in semiconductor films [263]. The operational speed could be increased due to photo-induced conductivity variation. Local and/or distributed deformation in the magnetic medium can be achieved by static mechanical stress or an electric field acting on one of the phases of the artificial composite magnonic heterostructure that contains the piezoelectric and magnetic phases [264], [265], [246]. The latter could consist of a lateral ensemble of magnonic crystals (MCs) [19] with period $D = 200$ μm, as it is shown in Fig. V.D(a), where MC1 and MC2 refers to adjacent magnonic-crystal waveguides [266], [226]. The piezo-layer (PZT) is strain coupled with MC1 and MC2 in a periodical manner, as it is demonstrated in the inset of Fig. V.D(a). The electrodes on PZT are localized in the area of each MCs. The structure is placed in a bias magnetic field directed along the short axis of each magnonic-crystal waveguide. Lateral coupled MC demonstrate the formation of couple of the band gaps at the spin-wave wavenumber almost equal to the Bragg wavenumber $k_b = \pi/D$ (see Fig. V.D(b)). The variation of the voltage applied between one of the electrodes and the electrode on top of the PZT layer results in the formation of only one tunable band gap at $k_b$: a positive value of the applied voltage leads to a redshift of the gap, while a negative voltage to a blueshift (see Fig. V.D(c)).

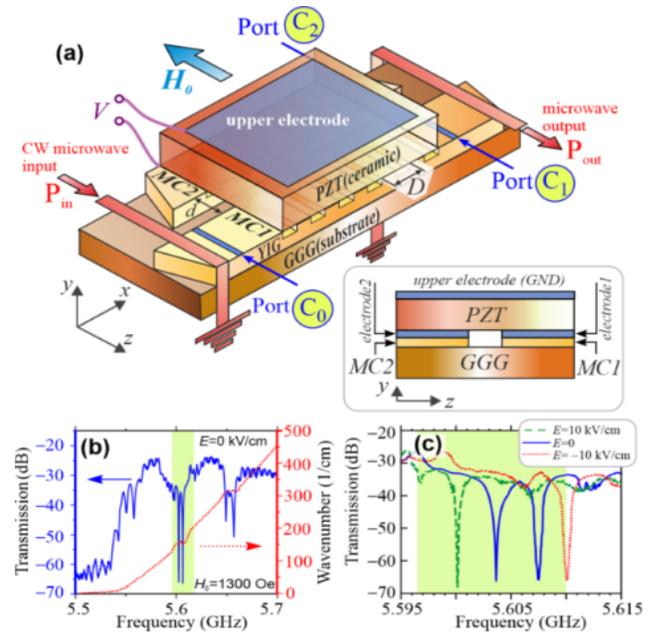

Fig. V.D. (a) Schematic drawing of artificial composite magnonic heterostructure: lateral magnonic crystals and piezoelectric layer with separated electrodes. Cross-section is shown on the inset of panel (a); (b) the transmission response and spin-wave dispersion; (c) tunability of the magnonic band gaps as a result of the variation of applied voltage. The bias magnetic field is $H_0 = 1300$ Oe.

Thus, deformation-induced physical effects in magnonic structures can be used to create energy-efficient complex two- and three-dimensional topologies for magnon devices and heterostructures. The observed effect is associated with the local variation of the internal magnetic field due to inverse magnetostriction (Villari effect) which manifests itself in the area of closest proximity of YIG and PZT phase as in the case of the considered lateral heterostructure MC/PZT [226]. At the same time, the development of technological processes for nanometer scale magnonics [267] and for the manufacture of composite nanometer thick MC/PZT opens the route for next-generation devices for an energy-saving data processing paradigm based on straintronic approaches.

### ACKNOWLEDGMENT

This work was supported by the Russian Science Foundation (project No. 20-79-10191).

CONTRIBUTORS

This section is authored by A.V. Sadovnikov (Saratov State University, Russia), S. A. Nikitov (Kotel'nikov Institute of Radioengineering and Electronics, Russia).

### E. Electric-field control of spin waves in multiferroic heterostructures

Multiferroic heterostructures allow for low-energy manipulation of spin waves through interface coupling of ferromagnetic and ferroelectric parameters. In strain-coupled systems, full domain correlations between a ferroelectric material and ferromagnetic film can be realized (Fig. V.E(a)) [268], [269]. In such composites, the magnetic anisotropy of the ferromagnetic film is modulated regularly, and the magnetic domain walls are pinned onto straight ferroelectric domain boundaries. This pinning effect offers attractive prospects for magnonics, as it allows for the excitation of short-wavelength spin waves [270], [271] and magnetic-field programmable filtering of spin-wave signals [272]. Also, due to interface strain coupling, the magnetization within domains is uniform, which alleviates the need for magnetic bias fields in spin-wave experiments.

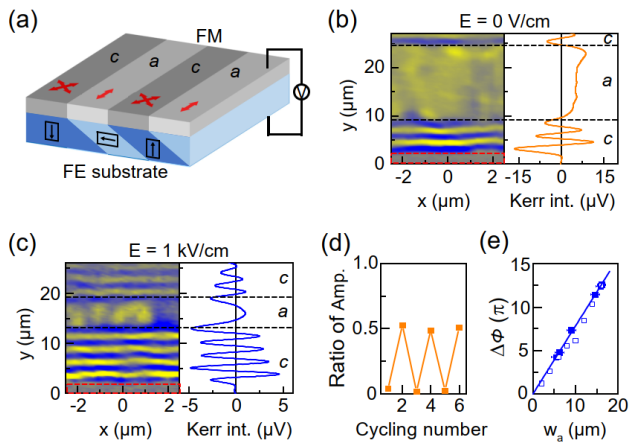

Fig. V.E. (a) Schematic of a multiferroic heterostructure with a fully correlated ferromagnetic (FM)/ferroelectric (FE) domain structure. (b),(c) Kerr microscopy images and line profiles of propagating spin waves in a Fe/BaTiO$_3$ sample before (b) and after (c) the application of an electric field. The excitation frequency is slightly above the FMR of the $a$ domain and well above the FMR of the $c$ domain. (d) Modulation of the spin-wave amplitude. The data compares the amplitude decay in an Fe film over an 8 μm distance when an electric field moves a magnetic domain wall into or away from the transmission area. (e) Modulation of the spin-wave phase as a function of $a$ domain width.

Applying a voltage across the multiferroic heterostructure sketched in Fig. V.E(a), results in the growth of one domain type at the expense of the other via lateral motion of the ferroelectric boundaries and pinned magnetic domain walls [273]. Switching of the voltage polarity reverses the effect. As the two types of domains exhibit different spin-wave dispersion relations, an electric-field induced change of the domain structure alters the transmission of spin waves. Fig. V.E(b, c) illustrates this effect for an epitaxial Fe film grown onto a ferroelectric BaTiO$_3$ substrate [274]. Depending on frequency and measurement configuration, the applied electric field alters the amplitude and phase of the spin waves largely independently. Amplitude modulation is attained at frequencies where one of the magnetic domains allows spin-wave propagation, but the other does not. Moving a single domain wall into or away from the film area between the source and detector then changes the spin-wave amplitude reversibly (Fig. V.E(d)). Phase modulation relies on spin-wave transmission across multiple domains and an electric-field controlled modification of the domain widths. An almost linear variation of the spin-wave phase with domain width has been observed experimentally (Fig. V.E(e)) [274].

Electric-field control of magnetism in multiferroic heterostructures offers a low-power mechanism for spin-wave manipulation at room temperature. To date, most proof-of-principle experiments utilize thick single-crystal ferroelectric substrates, necessitating the use of relatively large voltages. For practical devices, the voltage pulses should be substantially smaller. Epitaxial ferroelectric or multiferroic films could satisfy this requirement (see Section V-F).

ACKNOWLEDGMENT

The research leading to these results has received funding from the Academy of Finland (projects 317918, 316857, 321983 and 325480).

CONTRIBUTORS

This section is authored by H. Qin and S. van Dijken (Aalto University, Finland).

### F. Voltage-controlled, reconfigurable magnonic crystal at the submicron scale

In the past years, there has been tremendous progress towards on-chip magnonics such as the realization of submicron scaled integrated magnonic half adders [Wang2020]. The successful integration of magnonics concepts into CMOS technologies will however require voltage control capabilities to achieve low-energy operations. Logic magnonic circuits will also depend on analog building blocks that will need to be both scalable and reconfigurable at the submicron scale.

Within magnonics, magnonic crystals are particularly interesting for spin wave-based computing due to their ease in controlling and manipulating spin-wave dispersion relation [19]. They are built by introducing a periodic variation of one parameter of the spin-wave free energy. Current approaches either employ static or bulk reconfigurable magnonic crystals. For instance, a static, nano-scaled magnonic crystal based on 20 nm thick width-modulated Yttrium-Iron Garnet waveguides, was recently shown [275] – see Fig. V.F. In parallel, the available reconfigurable magnonic crystals exhibit dimensions at the microscale and require external sources such as lasers [276].

To that respect multiferroics provide a promising road for reconfigurable nanoscale magnonic devices, with built-in coupling between electrical polarization and magnetization degrees of freedom. Their Gilbert damping is however too high to allow for magneto-static spin-wave propagation. Theoretically, it was shown that a control over the ferroelectric domain states modulates the magnetoelectric coupling at the interface and in turn the spin-wave properties could allow for the voltage control of spin-wave propagation by means of such type of magnonic crystal [277]. By fabricating a fully epitaxial heterostructure of the multiferroic BiFeO₃ (BFO) and the ferromagnetic La₂/₃Sr₁/₃O₃ (LSMO) a reconfigurable, voltage-controlled magnonic crystal was recently demonstrated. By means of the magneto-electrical coupling, the reconfigurable control over the spin-wave propagation in LSMO was achieved via modulating the ferroelectric polarization state in BFO. Accordingly, imprinting periodic ferroelectric domains (periodicity of 500 nm) in BFO resulted in an artificial magnonic crystal in LSMO with > 20 dB rejection at the magnonic bandgap [275].

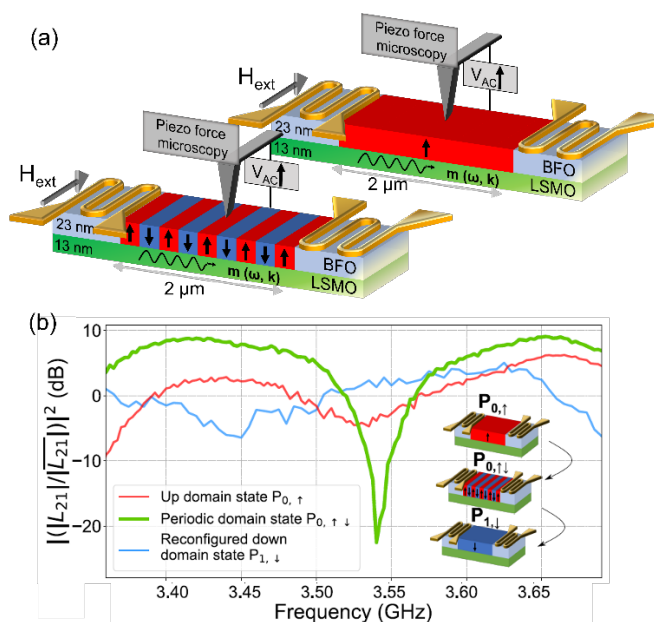

Fig. V.F. Voltage-controlled magnonic crystal (a) Experimental setup with uniform (top) and periodic (bottom) domains in a ferromagnetic (LSMO)/multiferroic (BFO) heterostructure (b) Transmission inductance amplitude from propagative spin-wave spectroscopy. The bandgap with supression >20 dB is clearly visible for the periodic domain configuration, whilst it is vanishing i.e. reconfigured in a subsequent uniform BFO domain polarization. Reprinted (adapted) with permission from [275]. Copyright 2021 American Chemical Society.

This proof-of concept experiment shows that the combination of magnonics and functional oxides and among them multiferroics are highly promising for novel and multifunctional magnonic-oxitronic devices for beyond CMOS spin-wave computing. Despite this progress, the exact physical processes governing the magneto-electric coupling at the interface need to be further investigated. The future control over the spin-wave dispersion via a locally controllable ferroelectric polarization state could also enable new approaches for a spin-wave logic based on reconfigurable conduits, providing important building blocks for magnon based information processing.

ACKNOWLEDGMENT

This work has been supported from the EUROPEAN UNION'S Horizon 2020 research and innovation program within the FET-OPEN project CHIRON under grant agreement No. 801055 and the project OISO under grant agreement ANR-17-CE24-0026 from the Agènce National de la Recherche.

CONTRIBUTORS

This section is authored by I. Boventer, R. Lebrun, P. Bortolotti and A. Anane (Unité Mixte de Physique CNRS/Thales, Palaiseau, France)

### G. Magnon fluxonics

Low-energy manipulation of spin waves is nowadays at the forefront of magnonics research. One way, as previously discussed in this section, implies the manipulation of spin waves by electric fields. Another fundamental approach is the use of magnonic materials in combination with dissipation-free currents in superconductors. In addition to zero resistivity, superconductors behave as ideal diamagnets exhibiting the Meissner effect (see also Sec. III-B) which consists in the expulsion of magnetic flux from their interior. Furthermore, the majority of technologically-relevant nanoscale superconductors are superconductors of type II. In the presence of moderately strong magnetic fields (as those usually used for magnonic devices) they are penetrated by a lattice of magnetic flux quanta (Abrikosov vortices, or fluxons). In this way, the combination of magnonics with fluxonics – the research domain concerned with the confinement and manipulation of magnetic flux quanta in nanoengineered superconductors [278] – makes accessible a rich palet of physical phenomena at the interface between spin-wave and superconductors physics.

Recently, the predicted [279] fluxon-magnon interactions have been realized experimentally in a Nb/Py heterostructure [144], see Fig. V.G(a). The periodic modulation of the local magnetic field emanating from the vortex cores made it possible to observe Bloch-like bandgaps in the magnon frequency spectrum. The reconfigurability of the fluxon-induced magnonic crystal is achieved via tuning the vortex lattice parameter through the biasing magnetic field variation. Further, fluxons were moved by the Lorentz-type force induced by an applied current and, thereby, acting as a moving Bragg grating, leading to Doppler shifts of the magnon bandgap frequencies [144].

In the following investigations [280], the vortex ratchet effect occurring in the asymmetric washboard vortex pinning potential [278] was used to achieve nonreciprocal spin-wave transport. In such a system, the application and polarity reversal of a dc current of 100 μA resulted in bandgap shifts of about 2 GHz (corresponding to 10 GHz/mA tunability). Furthermore, the application of an ac transport current at a power of 2 nW resulted in bandgap frequency shifts by up to ±100 MHz on the 10 ns time scale [280].

Another emerging research direction is related to the regime of ultra-fast vortex motion [281], with the predicted radiation of spin waves via the Cherenkov mechanism [282], when the velocity of a perfectly-ordered vortex lattice is equal to the phase velocity of the spin wave in the same (or adjacent) medium, see Fig. V.G(b). To achieve such a regime, ultra-high vortex velocities (>1 km/s) are required. However, the phenomenon of flux-flow instability [283] prevents its experimental observation. Recently, a Nb-C direct-write superconductor was fabricated by focused ion beam induced deposition (FIBID, see Section IV-C). Featuring high-structural uniformity, short electron relaxation time and close-to-depairing critical current, this system enabled the experimental realization of vortex velocities of up to 15 km/s [281]. In this way, the magnon Cherenkov radiation was discovered experimentally [284]. The observed emission was unidirectional (spin wave propagates in the direction of motion of the vortex lattice), monochromatic (magnon wavelength is equal to the vortex lattice parameter), and coherent (spin-wave phase is self-locked due to quantum interference with eddy currents in the superconductor), with magnon wavelengths down to 36 nm.

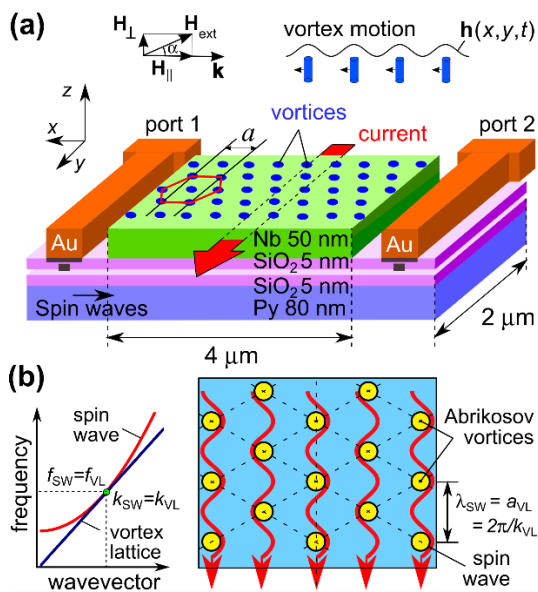

Fig. V.G. (a) Py/Nb hybrid nanostructure for investigations of the magnon-fluxon interactions, adapted with permission from [144]. Spin waves are excited by antenna 1, propagate through the Py waveguide, and are detected by antenna 2. The vortex lattice induces a spatially periodic magnetic field $\mathbf{h}(x,y)$ in Py, which becomes alternating in time when the vortices move under the action of the transport current. (b) Match of the frequency and the wavevector of the spin wave and the fast-moving vortex lattice ($f_{SW} = f_{VL}$, $k_{SW} = k_{VL}$), results in the Cherenkov radiation of spin waves, as demonstrated experimentally in [284].

To summarize, the interplay between superconducting and magnonics physics opens a plethora of new physical phenomena attractive also for future applications in the classical and quantum regimes. The main challenges in this research direction are the need for a perfectly-crystalline fast-moving vortex lattice, stability of fluxons as moving topological objects, and engineering of complex 2D and 3D superconductor-ferromagnet hybrid nanostructures.

## ACKNOWLEDGMENT

Financial support by the German Research Foundation (DFG) through Grant No. 374052683 (DO1511/3-1) and within the ERC Starting Grant no. 678309 MagnonCircuits is gratefully acknowledged. This research was conducted within the COST Action CA16218 (NANOCOHYBRI) of the European Cooperation in Science and Technology.

## CONTRIBUTORS

This section is authored by O. Dobrovolskiy and A. Chumak (University of Vienna, Austria).

## VI. BOOLEAN SPIN-WAVE LOGIC GATES AND MAGNONIC CIRCUITRY

Digital data have thoroughly replaced analog data in modern computers and information technologies. Thus, the focus of magnon-based data processing was placed on the operations with Boolean data. The original idea to use spin waves for such operations was proposed by Hertel et. al [285], and significant progress in this direction took place over the last decade. It is currently the most advanced sub-branch of magnon-based computing, which is already at the edge between physics and engineering.

In the first section, we discuss spin-wave diodes and circulators that are of importance for rf applications and are crucial for complex magnonic networks. Their usage allows for the suppression of parasitic reflections in circuits, the destructive role of which increases with the increase in number of elements. Further, we discuss nano-scaled purpose-engineered magnonic rf units in which information is carried and processed via spin waves. The logic gates based on nonlinear spin waves are addressed in Section VI-C and the spin-wave pseudospin and magnon valleytronics in Section VI-D.

In the following section, spin-wave majority gates are discussed. Such gates attract special attention since, depending on the way of estimation, one majority gate can replace around ten transistors in CMOS. The recently realized inline majority gate allowed for the significant miniaturization of the device and the improvement of its characteristics. Further, a nano-scale directional coupler is presented in Section VI-F. This device is also of interest for rf and unconventional computing and allows for efficient operations with Boolean data. Thus, a magnonic half adder comprising of two directional couplers was demonstrated numerically and benchmarked. It was shown that such an approach in combination with low-energy spin-wave amplification (see Section V-A) allows for the realization of a device with comparable to 7 nm CMOS footprint and ten times smaller energy consumption.

The finalizing three sections in Section VI have engineering character. The first one addresses the model that allows for the realization of complex magnonic circuits based on the magnonic directional couplers. The model is tested on the example of the 32-bit ripple carry adder. Section VI-H addresses the general question of cascading of logic gates and circuits. Finally, the questions of circuitry and amplifier designs for magneto-electric interfaces are discusses. This question is of high importance, considering that magnonic nano-waveguides carry very little power, and interfacing to electronics might be a challenging task.

## A. Spin-wave diodes and circulators

The control of wave propagation (see Section III-E) is an important issue in the computing based on spin waves (Sections VI-E – VI-G). The devices with the property of irreversibility, like diode and circulator, are used to reroute the signal (Section VI-F) or to protect the areas from the unwanted communication. For that reason, they require the nonreciprocal mechanism to be used. The first spin-wave diode model uses the chiral interaction with the resonator magnetized in the direction perpendicular to the waveguide (Section III-D), [150]. The second model takes advantage of domain-wall circuit with the easy-axis surface anisotropy and the asymmetry of the spin-wave propagation in the waveguide with the Dzyaloshinskii-Moriya interaction [286]. This interaction was also used to propose the effect of the unidirectional coupling where the wave can be transmitted between the waveguides only in one direction of propagation and led to design both diode and circulator (Fig. VI.A) [287]. Another circulator was proposed based on the dipolar interaction with the resonator taking advantage of circulating modes [288] and based on the inverse-design method (Section VII-C), [91]. The first experimental realization of the spin-wave diode was made for the double-layer system in which the dispersion modelling was used to slow down a spin wave in one direction of propagation [151]. However, there is still a lack of experimental realization of the spin-wave circulator.

The biggest challenge laying before the design of these devices is to combine all expected properties like size, time of operation, energy loss, etc. The efficiency is higher for a larger device which is connected to more time needed to make the operation on the spin wave. The problem with the increased time required for operation can be also attributed to the devices basing on the resonators as the resonance effect has to be excited properly. The losses of the signal going into the designated direction should be as small as possible simultaneously with the least possible leakage to the undesired directions. Low losses can be achieved using ultralow-damping material like YIG but its properties are hampering the downsizing of the devices which can be done easier with metallic magnetic materials like CoFe, CoFeB, or permalloy, but at the cost of higher damping (Section II-D).

The diodes and circulators are inevitable elements of any complex circuit where the wave-like signals are nonreciprocally guided. Depending on the preferable properties and the particular circuit design, different materials and working principles can be used to fabricate the magnonic diodes and circulators. Nevertheless, the nonreciprocal devices can find many applications, for instance: (i) signal duplexers, where the signal in the bi-directional bus/waveguide is directed to the receiver, and form the transmitter, (ii) signal possessing in one-port devices, where the reflected signal must be separated from the input signal, or (iii) signal isolation.

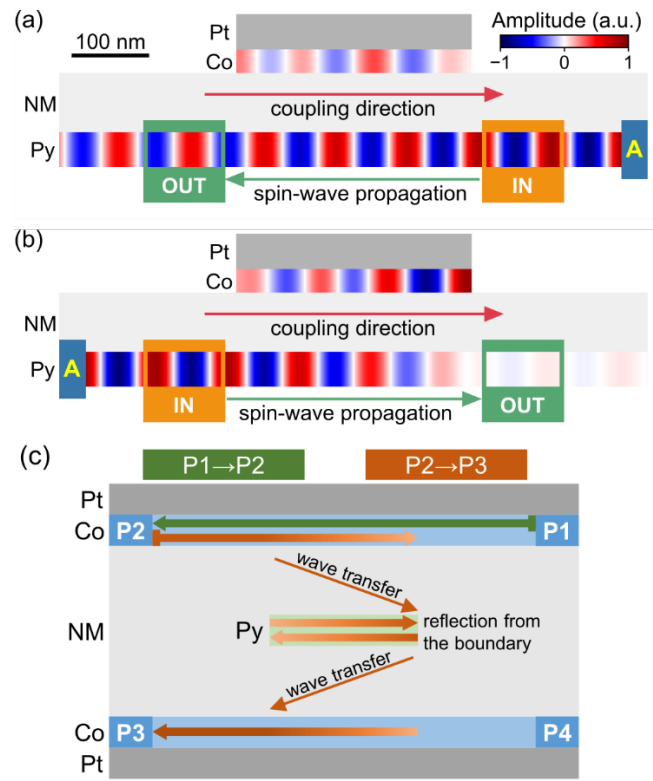

Fig. VI.A. The functionality of the spin-wave diode based on unidirectional coupling in the (a) forward and (b) reverse direction. Reprinted figures (a) and (b) with permission from [287]. Copyright 2020 by the American Physical Society. (c) The mode of operation of the spin-wave circulator based on the unidirectional coupling [287].

### ACKNOWLEDGMENT

The financial support from the National Science Centre, Poland, project No. 2018/30/Q/ST3/00416 is acknowledged.

### CONTRIBUTORS

This section is authored by K. Szulc, J. W. Kłos, M. Krawczyk (Adam Mickiewicz University, Poznań, Poland), and P. Graczyk (Polish Academy of Sciences, Poland).

## B. Nano-scaled purpose-engineered magnonic rf units

Local modification of magnetic properties is a key to design the nanoscale magnonic devices in which information is carried and processed via spin waves. One of the biggest challenges here is to fabricate simple and reconfigurable magnetic elements with broad tunability. While patterning of thin films has been established as a standard technique for the fabrication of "high-magnetic-contrast" magnonic conduits (T-junctions, magnonic crystals, etc.), the combination of additive and subtractive manufacturing by focused ion and electron beams (see Fig. VI.B(a, b)) allows a gradual tailoring of magnetic properties. Thus, milling of nanoholes (antidots) in direct-write

Co microdisks allowed one to realize the magnetization reversal via metastable states with half antivortices [289]. The post-growth irradiation of Co/Pt-based heterostructures with 5 kV electrons resulted in the formation of a hard magnetic CoPt L1$_0$ phase at their interface, with control on the lateral mesoscale [290]. The ion irradiation-induced evolution of the magnetic parameters of thin films and nanostructures is another powerful approach [167].

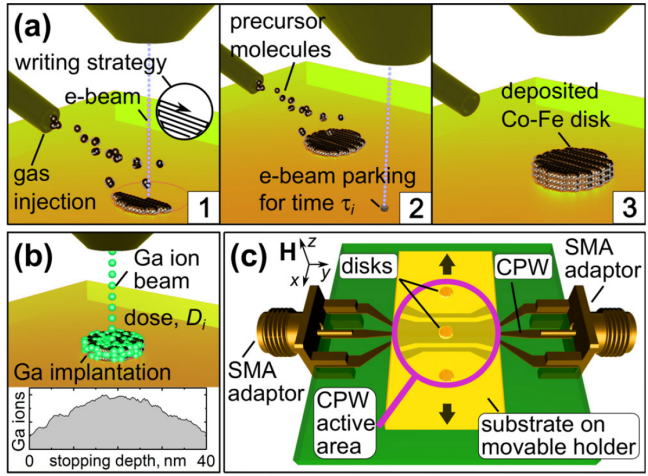

Fig. VI.B. (a) Illustration of the process of focused electron beam induced deposition for the fabrication of magnetic nanodisks with a close-to-bulk saturation magnetization: after each pass over the sample surface (1), the beam is parked outside of the disk for the given time $\tau_i$ (2). The writing process is continued until the desired nanodisk thickness is achieved (3). (b) Irradiation with a Ga ion beam leads to a reduction of magnetization. (c) Experimental geometry for the spatially-resolved microwave spectroscopy of individual magnetic elements (not to scale). A substrate with a series of nanodisks is placed face-to-face to a gold coplanar waveguide for spin-wave excitations in the out-of-plane field geometry. Adapted with permission from [291].

Recently, a miniature highly-tunable spin-wave shifter was experimentally realized upon a single nanogroove milled by a focused ion beam in a Co-Fe microsized magnonic waveguide [292]. The microscopic mechanism of the phase shift in this device is based on the combined action of the nanogroove as a geometrical defect and the decreased saturation magnetization ($M_s$) under the groove because of the incorporation of Ga ions during the milling process. At the same time, the broad tunability of the composition of direct-write nanostructures appealed for the development of methodology to determine the magnetic properties of individual nanoelements. To this end, spatially-resolved microwave spectroscopy was developed (see Fig. VI.B(c)), allowing the precise determination of $M_s$ and the exchange stiffness $A$ of individual circular magnetic elements with volumes down to $10^{-3}$ µm³ [293]. The key component of the setup is a coplanar waveguide whose microsized central part is placed over a movable substrate with well-separated magnetic nanodisks under study. Recently, this technique was employed to experimentally demonstrate a continuous tuning of $M_s$ and $A$ in direct-write Co-Fe disks [291]. The achieved $M_s$ variation in the range from 720 emu/cm³ to 1430 emu/cm³ bridges the gap between the $M_s$ values of widely used magnonic materials such as Permalloy and CoFeB, opening a way toward nanoscale 2D and 3D systems with purpose-engineered and space-varying magnetic properties. Synthesis of novel materials and exploration of static and dynamic magnetic configurations therein represent essential directions of investigations in this domain [194].

### Acknowledgment

S.A.B. and G.N.K acknowledge the Network of Extreme Conditions Laboratories-NECL and the Portuguese Foundation of Science and Technology (FCT) support through Project Nos. NORTE-01-0145-FEDER-022096, PTDC/FIS-MAC/31302/2017, POCI-0145-FEDER-030085 (NOVAMAG), and EXPL/IF/00541/2015. K.G. acknowledges support from IKERBASQUE (the Basque Foundation for Science) and by the Spanish Ministry of Science and Innovation through Grant No. PID2019-108075RB-C33/AEI/10.13039/501100011033.

### Contributors

This section is authored by O. V. Dobrovolskiy (University of Vienna, Austria), K. Yu. Guslienko (Universidad del País Vasco, Spain), S. A. Bunyaev and G. N. Kakazei (University of Porto, Portugal).

## C. Logic gates based on nonlinear spin waves

Development of spin-wave (SW) logic gates is promising for fabrication of the magnonic logic circuits. The first SW logic gate was proposed by Kostylev et. al. [294]. The device had a structure of Mach-Zehnder interferometer controlled by pulses of electric current. After that a considerable amount of theoretical and experimental work was carried out to develop the magnonic logic gates (see e.g., [20]).

Linear and nonlinear spin waves propagating in yttrium iron garnet (YIG) films and magnonic crystals are used as a basis for the magnonic logic development. Special attention is given for realization of all-magnon logic circuits providing the possibility for a cascade connection of logic gates. Such circuits can be fabricated with nonlinear SW components.

To date, nonlinear SW logic gates are realized with two schemes. The first scheme is nonlinear SW interferometer of Mach-Zehnder type [295], [296], [288]. In the work of [296] the interferometer branches comprised magnon transistors. The working effect was the nonlinear magnon-magnon scattering in the transistors. Different phenomenon was utilized in the work of [288]. It was a nonlinear phase shift of spin waves emerging in the interferometer branches with an increase in operating spin-wave power.

The second scheme used for logic gates is a nonlinear SW directional coupler [297], [33]. Principle of operation of this device is based on a power-dependent coupling. It provides a power-dependent switching of spin-wave pulses between two channels. This functionality was realized with structures of a branch-line coupler [297] and a side-coupled waveguides [33]. Note, that the nonlinear spin-wave coupling in the latter structure was investigated only recently [298].

Further progress in this area appears to be in the search for new spin-wave effects and new magnonic circuits for realization of logic functions, as well as a miniaturization of the components. An example of successful logic gate miniaturization is reported recently in [33].

### ACKNOWLEDGMENT

The work is supported by the RFBR, project # 21-52-50006.

### CONTRIBUTORS

This section is authored by A.B. Ustinov (St. Petersburg Electrotechnical University, Russia).

### D. Spin-wave pseudospin and magnon valleytronics

Traditionally, spin-wave amplitude and phase have been considered as the main resources to transmit and manipulate information in magnonics. However, there is another way to encode the Boolean data into spin waves – that are based on spin-wave pseudospin [299]. This method uses the peculiar double-valley shape of the spin-wave dispersion relation in the backward volume geometry, which leads to existence of two degenerate spin-wave modes at frequencies below that of the ferromagnetic resonance (FMR). Each of the two modes resides in its own valley and may therefore be assigned a pseudospin. This pseudospin can be switched by scattering the spin wave from decreases of the bias magnetic field, while it is immune to scattering from field increases [299]. Chiral magnonic resonators (discussed in Section III-D) can be used not only to create spin waves with a pseudospin defined by their magnetization [150] but also to read it out [153], [154].

The combination of our abilities to create, manipulate and then read out the spin-wave pseudospin could enable us to develop a spin-wave version of valleytronics – "magnon valleytronics". As with any emerging technology, the first step is to prove experimentally the theoretical predictions from [299] and then to optimize the device performance. The recent results obtained in the studies of magnonic Bose-Einstein condensates [300] prove experimental feasibility of magnon valleytronics. Fabrication of downscaled samples with relatively steep gradients of the magnonic index [301] is likely to be the main challenge for this concept. Theoretically, it needs to be extended to 2D problems, e.g., oblique incidence, and graded-index profiles, e.g., lenses [302]. Overall, it remains to see whether this idea will transform into a topic of practical value.

### ACKNOWLEDGMENT

The research leading to these results has received funding from the Engineering and Physical Sciences Research Council of the United Kingdom under project EP/T016574/1.

### CONTRIBUTORS

This section is authored by V. V. Kruglyak and K. G. Fripp (University of Exeter, UK).

### E. Spin-wave majority gates

Spin-wave logic has particularly focused on the design and realization of compact spin-wave majority gates. Majority gates [303] and inverters form a universal logic gate set that allow for the design of any logic circuit while potentially resulting in significant circuit complexity and area reductions with respect to conventional CMOS Boolean gate-based counterparts [304].

The most promising approach for spin-wave majority gate design relies of spin-wave phase information encoding and makes use of spin-wave interference for computation. This is because the phase of the output wave generated by the interference of an odd number of input waves with phases 0 or $\pi$ corresponds to the majority of the phases of the input waves. In such an approach, an inverter corresponds simply to a delay line with a length of half of the spin-wavelength.

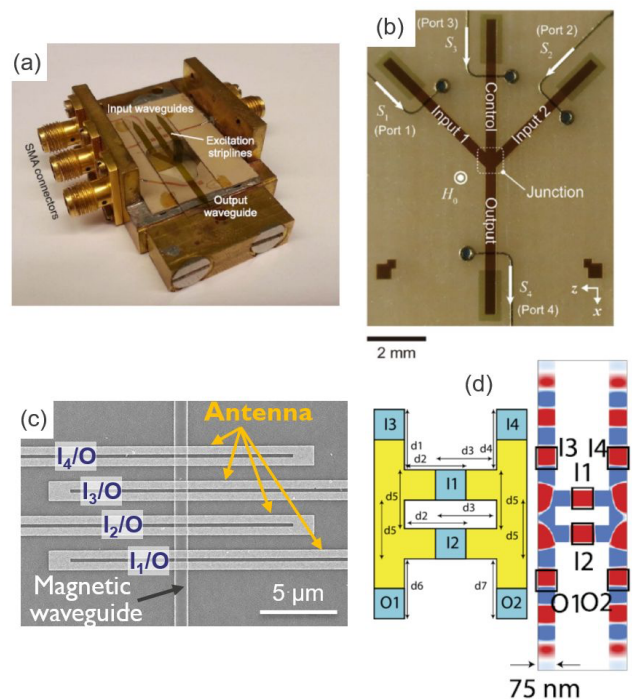

Fig. VI.E. Designs of spin wave majority gates based (a) and (b) on mm-scaled YIG waveguides, (c) sub-µm wide CoFeB waveguide, and (d) nano-scaled CoFeB waveguides. (a) Reprinted with permission from [305]. Copyright 2017 American Institute of Physics. (b) Reprinted with permission from [306]. Copyright 2017 Nature. (c) Reprinted with permission from [267]. Copyright 2020 American Association for the Advancement of Science. (d) Reprinted with permission from [307]. Copyright 2020 American Institute of Physics.

Several designs of spin-wave majority gates have been proposed, such as trident [308], [309], [305], [306], inline [267], or ladder-shaped [307] designs (see *e.g.*, Fig. VI.E). A first proof of concept was realized by mm-scale trident designs using YIG as the magnetic conduit material and inductive antennas as transducers between the spin-wave and the microwave domains [305]. The operation of these devices allowed for the demonstration of the full majority function truth table including the possibility to distinguish between strong and weak majority. However, the trident design is difficult to scale to submicron dimensions due to strong reflection of spin waves at the bends of the structure, although the operation in scaled devices has been shown by micromagnetic simulations for certain material parameters and dimensions in the μm range.

Smaller sizes can be reached by inline or ladder-shaped majority gate designs. Recently, the operation of a sub-μm inline spin-wave majority gate based on a CoFeB conduit with a width of 850 nm has been demonstrated [267]. The behavior of these devices has been imaged using time-resolved x-ray microscopy and studied in more detail by electrical spectroscopy. As their much larger YIG counterparts, these devices allowed for the demonstration full majority truth table including the distinction between strong and weak majority within a frequency band of approximately 300 MHz. In addition, the devices allowed for the realization of frequency-division multiplexing as well as the possibility to use different transducers as flexible inputs or outputs. Micromagnetic simulations indicated potential gate delay of less than one ns.

## Acknowledgment

Contributions from F.C. and C.A. have been supported by imec's industrial affiliate program on beyond-CMOS logic. F.C., C.A., S.H. and S.C. acknowledge funding by the European Union's Horizon 2020 research and innovation program within the FET-OPEN project CHIRON under Grant Agreement No. 801055.

## Contributors

These sub-sections are authored by F. Ciubotaru and C. Adelmann (imec, Belgium), S. Hamdioui and S. Cotofana (TU Delft, The Netherlands).

### F. Spin-wave directional coupler for magnonic half-adder

A directional coupler is a standard device that has been widely used in the field of microwave engineering and photonics to guide or separate signals in circuits. A macroscopic magnonic directional coupler has been investigated experimentally using Brillouin light scattering spectroscopy by Sadovnikov et at. [310], [298]. It has been shown that SW coupling efficiency between the arms of the coupler depends on both the geometry of magnonic waveguides and the characteristics of the interacting SW modes. Following this idea, a directional coupler was investigated numerically at the nanoscale [311] and, recently, it has been realized and investigated experimentally – see Fig. VI.F [33].

The physical mechanism of the spin-wave direction coupler is relatively straightforward. The spin-wave dispersion curve splits into antisymmetric and symmetric modes due to the dipolar coupling between two closely placed waveguides [310], [311]. The interference between these modes in each waveguide results in an energy oscillation between the waveguides. The distance which a spin wave propagates before a complete energy transfer from one waveguide to another has taken place is defined as the coupling length and strongly depends on the geometrical sizes of the waveguides, the distance between them, the relative magnetization orientation, and the spin-wave wavelength. The control of these parameters allows realizing different functionalities. For example, the directional coupler can be used as power splitter, filter, delay line or frequency multiplexer in the linear regime due to the frequency-dependent coupling length [311], [33]. Moreover, the coupling length is nonreciprocal in special designed structures, which can be used for realization of spin-wave diodes and circulators [287] (see also Section VI-A).

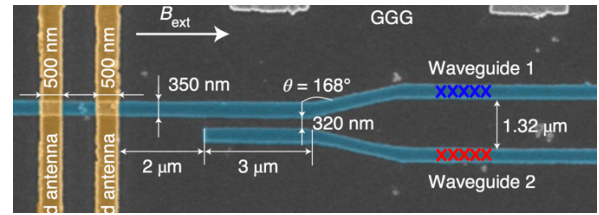

Fig. VI.F. (a) Scanning electron microscopy (SEM) image of the directional coupler (shaded in blue) with the U-shaped antenna for the spin-wave excitation. An external magnetic field of 56 mT is applied along the YIG conduits, and spin waves are detected using micro-focused BLS. Reprinted (adapted) with permission from [Wang2020]. Copyright 2021 Springer Nature.

The main challenge in the directional couplers' further evolution is the need to switch from dipolar to exchange spin waves with higher velocities and smaller wavelengths. The dipolar coupling between the exchange waves vanishes even in closely placed waveguides, so it should be replaced by the exchange coupling or any other mechanism in specialized hybrid structures.

One of the key advantages offered by spin waves is their pronounced natural nonlinearities compared to the relatively low intrinsic nonlinearity of light. The operation of a directional coupler in a nonlinear regime was originally investigated in [298]. Magnonic crystals (see Section IV-F) were embedded in the arms of the directional coupler, and it was demonstrated that the nonlinear spin-wave phase shift and nonlinear transformation of spin-wave stray field profile [160] leads to a nonlinear switching regime at the frequencies near the forbidden magnonic gap. In a weaker nonlinear regime, the coupling length depends on the spin-wave amplitude, and the device acts as a signal switch when spin waves of different amplitude are guided to different outputs of the directional coupler [33]. This nonlinear switch can be used for rf applications (e.g., as power limiters or signal-to-noise enhancers), for future neuromorphic computing (see Section VII-D), or, as it has been shown in [33], for the processing of Boolean data.

A half-adder suitable for cascading with other elements is an essential building element of a computational circuit. It was demonstrated numerically that a magnonic half adder can be constructed using two spin-wave directional couplers [33]. One operates in the linear regime as a power splitter, and another as the described above nonlinear switch to perform AND logic operation. The benchmarking of the half adder was performed, and it was found that such a device consisting of three planar nanowires replaces around 14 transistors in CMOS implementations, the total footprint of the 30 nm-based YIG magnonic structure is comparable to the 7 nm CMOS, but the magnonic half adder is around two to three orders of magnitude slower. A still missing element of such a half adder is an amplifier which should be installed at the "S" output to amplify the signal 4 times. The analysis shows that only the parametric amplifier based on the voltage-controlled magnetic anisotropy (VCMA) is suitable for the required low-energy amplification [33]. The energy consumption of the half adder with accounting for the theoretically-estimated energy needed for the amplification, is about 5 aJ and seven times smaller compared to the 7 nm CMOS. The first steps in the experimental realization of such a spin-wave amplifier are taken (see Section V-A), but the fully functioning amplifier remains a challenge.

## ACKNOWLEDGMENT

This research has been supported by ERC Starting Grant 678309 MagnonCircuits, FET-OPEN project CHIRON (contract no. 801055).

## CONTRIBUTORS

This section is authored by Q. Wang and A. V. Chumak (University of Vienna, Austria), P. Pirro (TU Kaiserslautern, Germany), A.V. Sadovnikov (Saratov State University, Russia), S. A. Nikitov (Kotel'nikov Institute of Radioengineering and Electronics, Russia).

### G. Magnonic ripple carry adders

The aim of magnonics is to deliver high-performance circuits taking advantage of the high non-linearity, the low power consumption and the scalability of the technology. Boolean logic requires a functional complete set of logic gates to implement any logic function. Recent work from [33] showed the half adder structure as a potential building block for all-magnonic digital circuits. Two technology nodes have been proposed, the 100nm and the 30nm, where the node corresponds to the width of the waveguide. The magnonic half adder has the peculiarity to integrate both the XOR and the AND logic functions. It consists of a combination of directional couplers [298], one operating in the linear regime (DC1) and the second one operating in the non-linear regime (DC2). By the tuning the gap between the couplers [311] and their physical geometries (length and thickness) it is possible to obtain a logic behavior, which is encoded in the presence of the excited spin wave (logic 1) and its absence (logic 0) [33]. However, the lack of tools and simplified computational models makes it difficult for researchers the understanding of drawbacks/advantages of proposed devices.

In all computer architectures, implementing the summation is one of the most important operation. Moreover, the full adder and the ripple carry adder are among the circuits considered when benchmarking beyond CMOS technologies, see the methodology proposed in [312]. The core element developed in [33] can be used to implement more complex circuits. The major challenge of exploring larger all-magnonic circuits is to have methodologies able to simulate such complex designs with good accuracy in evaluating the physical behavior of the devices. Recently, a compact model to study magnonic circuits has been presented in [313], starting from works of [311], [314], [315], [316] and it is openly available on Zenodo [317]. The proposed model makes it possible to study larger circuits, like the ripple carry adder depicted in Fig. VI.G(a). It enables the extraction of metrics like circuit area, power consumption, and propagation delay useful to make comparison with the CMOS counterpart. The model takes into account the nonlinear behavior the of the directional couplers [318].

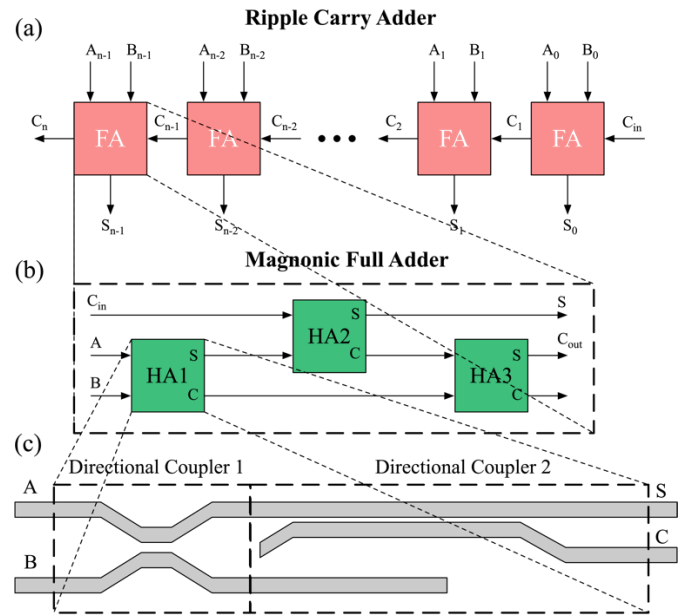

Fig. VI.G. (a) Schematic representation of the magnonic ripple carry adder. (b) Internal implementation of every full adder. (c) Internal structure of the half adders composed of two directional couplers.

The design presented in [313] shows a possible implementation of an adder. In particular, for a 32bit ripple carry adder, built hierarchically from half adders (Fig. VI.G(b)) and then directional couplers (Fig. VI.G(c)), the area occupation is about 4x larger in the case of the 30nm magnonic circuit when compared with a 15nm CMOS technology. On the contrary, analyzing the energy dissipated, it is possible to observe that it is more than 14x lower in the 30nm magnonic circuits. This study shows good potential for all-magnonic logic circuits, especially for low power electronic devices. Moreover, it worth noticing that the design proposed is only a proof of concept and it is not optimized. However, the device concept and the methodology proposed has great potential for future development and optimization of all-magnonic logic circuits.

## CONTRIBUTORS

This section is authored by F. Riente (Politecnico di Torino, Italy).

## H. Cascaded logic gates and circuits

Compared to the work on spin-wave logic devices, little effort has been devoted to design spin-wave circuits [308], [319], [320], [321], [322], [323], [324]. To create circuits based on spin-wave logic gates, the gates must be cascaded [11]. This means that the output signal of a gate can be used as input signal of the subsequent gate(s). Moreover, this also requires that the spin-wave signal is not degraded during propagation in the circuit and that signal reflections that can interfere with the signals in previous gates are kept to a minimum. If the signal degradation cannot be avoided (e.g., due to strong spin-wave attenuation), the intensity loss must be compensated by spin-wave amplifiers [11], [325], [326], albeit at an additional energy cost.

For spin-wave majority gates, cascading is not straightforward since the intensity of the output wave is different for weak and strong majority. Therefore, an additional device with nonlinear output characteristics is needed to "renormalize" the amplitude of the output spin wave of a spin-wave majority gate [11]. It has been demonstrated by micromagnetic simulations that directional couplers [311], [33] in the nonlinear regime can act as such devices, enabling the cascading of spin-wave majority gates based on wave interference [327]. We note that delay line inverters do not exhibit input data dependent intensity output spin waves, thus, they can be directly cascaded. However, also in their case potential signal attenuation may occur and it should be compensated by means of amplifiers or repeaters based on the phase-sensitive switching of a nanomagnet [328]. Using a set of basic elements comprising of majority gates, inverters, and directional couplers, micromagnetic simulations have demonstrated the ability to cascade logic devices in complex spin-wave circuits [327], [329]. An experimental realization of such cascaded logic gates is however still pending.

A second key property of logic gates that is required to design efficient circuits is the possibility to achieve fan-out, i.e., the ability of a logic gate to drive the input of *several* subsequent logic gates. In general, this can be achieved by spin-wave splitters [330]. However, maintaining the amplitude of the signals remains crucial and may require spin-wave amplifiers to ensure correct circuit functionality. Alternatively, some designs of spin-wave majority gates can provide intrinsic capabilities for fan-out. As an example, inline majority gates can provide a fan-out to two gates [267]. To further increase fan-out capabilities, H-shaped majority gates have been proposed that allow for a fan-out of up to 4 gates [307], [331]. Spin-wave majority gates also allow for additional features, such as reconfigurability [267], [331], pipelining [329], or frequency-division multiplexing, which enables multibit parallel computation [267], [307]. Moreover, the gates can also comprise control gates, which render the gates reprogrammable [331].

These functionalities can be exploited in spin-wave circuits to reduce circuit footprint and increase the computational throughput. However, despite the introduction of these concepts, spin-wave circuits beyond individual logic gates have not been realized so far. In particular, the cascading of two spin-wave gates has not been reported, although the realization and operation of a directional coupler has been demonstrated [33]. A major current hurdle is to achieve signal integrity during propagation, which requires spin-wave amplitude renormalization and amplification to compensate for attenuation. The combination of these effects effectively limits the maximum size and complexity of potentially realizable spin-wave circuits.

Beyond spin-wave combinational circuits, no concept presently exists for the realization of a "spin-wave computer", i.e., a computing system that entirely relies on spin waves and includes spin-wave memory as well as a spin-wave based interconnect. Therefore, it has been argued that spin-wave circuits are best integrated in hybrid systems that also comprise CMOS circuitry [11]. In such hybrid schemes, the transducers between the CMOS and spin-wave domains are critical elements with a strong impact on the performance of the system [11], [323]. Typical magnonic experiments and device realizations employ inductive antenna transducers, which are not well scalable and possess low energy efficiency at nanoscale dimensions. SOT-enhanced antennas have been proposed [332] but their efficiency still needs to be enhanced for practical utilization. Magnetoelectric transducers have been proposed [308], [11], [333], [334] but both, the thorough device level understanding as well as experimental realization of scaled transducers, are still lacking.

Finally, hybrid systems require peripheral CMOS circuits that should efficiently link with magnonic circuits. So far, only little attention has been devoted to such circuits and solutions are only emerging [335]. The design of the CMOS periphery needs to progress in parallel with the development of efficient spin-wave transducers. In this field, future breakthroughs are needed to obtain an efficient CMOS periphery, in which magnonic circuits can then be embedded. Advances in this research topic are essential and enabling factors for making the step forward from spin-wave logic gates to circuits, and ultimately for the realization of hybrid CMOS-spin-wave chips that can be integrated in commercial ultra-low power applications.

### ACKNOWLEDGMENT

Contributions from F.C. and C.A. have been supported by imec's industrial affiliate program on beyond-CMOS logic. F.C., C.A., S.H. and S.C. acknowledge funding by the European Union's Horizon 2020 research and innovation program within the FET-OPEN project CHIRON under Grant Agreement No. 801055.

### CONTRIBUTORS

This section is authored by F. Ciubotaru and C. Adelmann (imec, Belgium), S. Hamdioui and S. Cotofana (TU Delft, The Netherlands).

*I. Waveguide and amplifier designs for magneto-electric interfaces*

Sub-micrometer-scale magnonic waveguides carry very little power (about nanowatts for a micrometer-wide structure), which may lead to fundamental inefficiencies when have to be matched to electronic circuits that operate at higher (microwatt or milliwatts) power levels.

If there is no significant power mismatch, then waveguide-based input devices can achieve remarkably high transduction efficiency [336]. The key to achieve such efficiencies is a well-designed matching network or meandering waveguides - both of them effectively preventing power waste on resistive loads of the waveguide. In the circuit model of the spin-wave transducer, the spin-wave medium appears as a lossy inductor – therefore in order to design an optimal waveguide, one needs to accurately model the inductive load that magnetic medium represents toward the waveguide. No matter which geometry (meander, fan, simple coplanar, grating) is used, there will be a tradeoff between the bandwidth and the efficiency. Low bandwidth results in low data throughput and slow computing. In [336] a meander transducer design on YIG is presented generating $\lambda = 500$ nm spin waves with an efficiency of -4.45 dB and a 3 dB-bandwidth of 134 MHz – a surprisingly high efficiency with a practically usable bandwidth.

The construction of the output amplifier very much depends on the targeted application. A full-blown amplifier / mixer circuitry (similar to a standard RF receiver but optimized for spin-wave readout) is presented in [337]. – The chip consumes few-tens milliwatts of power, which is several orders of magnitude higher than the power dissipated in the magnetic domain and at input waveguides. More efficient designs may be possible, such as a power detector that does not require mixers or using large antennas on large-area magnonic outputs. It is likely, though, that the output circuitry remains the biggest obstacle in the way of a truly efficient magnonic computing system.

It seems that for competitively energy-efficient magnonic systems, there are two routes. One is to place magnonic devices at RF telecommunication system front-ends, close to the antenna as a pre-processor system where only low power levels have to be handled. A power-hungry RF amplifier would be required anyway in this device, so picking up the magnonic signals does not come at an additional cost. The second route, is relevant if nanoscale computing is the targeted application. In this case the number of outputs have to be minimized and as a complex as possible computation must be achieved in the magnetic domain – this will amortize the huge energy overhead of output circuitry.

CONTRIBUTORS

This section is authored by A. Papp and G. Csaba (Pazmany Peter Catholic University, Budapest, Hungary), W. Porod (University of Notre Dame, USA), M. Becherer (Technische Universität München, Germany).

## VII. MAGNONIC UNCONVENTIONAL COMPUTING

The previous section was devoted to the processing of Boolean data with spin waves. For this, the data was coded either in spin-wave amplitude or in its phase. All the rest of the carried information by the wave (frequency, wavelength, amplitude, or phase) was neglected, limiting the strength of magnon-based computing. Therefore, the versatility of unconventional computing approaches is expected to be more efficient and should enable the development of novel data-processing units to solve specific tasks. These units will be compatible with the existing semiconductor CMOS technology.

We begin this section with the discussion about nonlinear spin-wave phenomena in metallic ferromagnets, utilized for unconventional computing. Section VII-A discusses the magnonic holographic memory that emerges as promising technology for the realization of memory and logic devices. Furthermore, we report on the very new direction of inverse-design magnonics. In this approach, any functionality can be specified first, and a feedback-based computational algorithm is used to obtain the device design. Section VII is a collective work of many Authors and addresses magnonic neuromorphic computing inspired by the functioning of the (human) brain and its neuronal structures. The last section is dedicated to voltage-controlled spin Hall nano-oscillator-based computing and Ising machines. Compared to existing commercial solutions, they can lead to orders of magnitude smaller, faster, and less power-hungry solvers of combinatorial optimization problems.

### A. Nonlinear spin-wave phenomena in metallic ferromagnets

The intrinsic nonlinearity of the Landau-Lifshitz equation gives rise to an abundance of nonlinear phenomena. In metallic systems, however, magnon-magnon scattering was considered to be a parasitic effect increasing the line-width of magnetic resonances and reducing the magnetization after a pulsed excitation due to generation of secondary magnons [338]. In contrast to low damping ferrimagnets, in metallic ferromagnets the direct observation of magnons originating from nonlinear scattering processes was hampered for a long time due to the typically much larger damping. Even though the larger magnetization of metallic ferromagnets lowers the critical amplitude for higher-order magnon-magnon scattering processes compared to insulating ferrimagnets, discretization of the spin-wave spectrum through spatial confinement allows for direct detection of the scattered magnons [339].

Recently, 3-magnon splitting in micron-sized magnetic vortices was utilized to convert k=0 magnons into whispering-gallery magnons (see Fig. VII.A) with unprecedented large wave vectors [340]. Active control of splitting processes in a magnetic vortex by magnons propagating in a nearby waveguide [341] opened new perspectives for magnon-based computing, where nonlinear magnon-magnon scattering is at the core of the neuromorphic engine. The development of new metallic materials with ultralow damping is further accelerating the use-cases of nonlinear magnon interactions. While 4-magnon scattering showed to be the limiting mechanism for magnon transport in micron sized $Co_{25}Fe_{75}$ waveguides above certain magnon amplitudes [342], it could recently be shown in the same material and waveguide geometry that 4-magnon scattering can be used to generate a magnon frequency comb (see Fig. VII.A) [343].

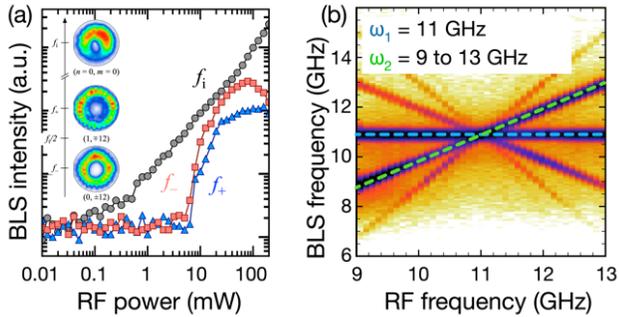

Fig. VII.A. Nonlinear magnon-magnon scattering. (a) 3-magnon-scattering in a 5 μm $Ni_{80}Fe_{20}$-disc Reprinted (adapted) with permission from [340]. (b) Spin-wave frequency comb generated by stimulated 4-magnon-scattering in a 30 nm thick and 2μm wide $Co_{25}Fe_{75}$ waveguide Reprinted (adapted) with permission from [343].

Metallic ferromagnets with high moment and low damping allow us to reach the realm of nonlinear phenomena in samples with thicknesses orders of magnitude smaller compared to insulator-based magnonic devices. As such interfacial spin-orbit torque phenomena can be utilized for amplification of magnons. In combination with the compatibility to standard CMOS processes this allows for direct integration in the fabrication nodes aiming at ultrahigh integration density as already in operation for MRAM storage devices.

### Acknowledgment

KS. acknowledges funding within the Helmholtz Postdoc Programme. HS acknowledges funding by DFG via projects Schu2022/1-3 and Schu2022/4, MW acknowledges funding by DFG via projects WE5386/4-1 and WE5386/5-1.

### Contributors

This section is authored by K. Schultheiss and H. Schultheiss (Helmholtz-Zentrum Dresden-Rossendorf, Germany), H. Nembach (University of Colorado and NIST, Boulder, USA), and J. Shaw (NIST, Boulder, USA), and M. Weiler (TU Kaiserslautern, Germany).

## B. Magnonic Holographic Memory

There are several physical properties inherent to spin waves that appear very promising for building memory and logic devices. In the Table below, we have briefly outlined physical properties and possible practical applications.

Magnonic Holographic Memory (MHM) is one of the possible applications to benefit unique spin-wave characteristics [344]. The essence of this approach is to utilize spin waves for parallel magnetic bit read-out somewhat similar to optic holographic memory. The first working prototype demonstrated two magnetic bit read-out [345]. The magnetic state of each magnet was recognized by the inductive voltage produced by the interfering spin waves [345]. The ability to use spin-wave interference opens a new horizon for magnetic database search. For example, MHM device was utilized for pattern recognition [346]. In Fig. VII.B, there are shown the schematics of MHM, a photo of the eight-terminal prototype along with the experimental data demonstrating the correlation between the input phases of the spin waves and the output inductive voltage. Coding input information into the spin-wave phases (e.g., logic 0 corresponds to phase 0 and logic 1 to phase π) makes it possible to check a number of possible input combinations in parallel.

| Physical Property | Practical Application |
|---|---|
| Dispersion depends on magnetic field | Magnetic bit read-out, built-in memory |
| Inductive voltage produced by spin waves | Compatibility with electronic devices |
| Exchange spin waves | Sub-micrometer scale devices |
| Relatively long coherent length at room temperature | Phase in addition to amplitude for data processing |

MHM devices can be also utilized for special task data processing such as parallel database search and prime factorization [346], [347]. The utilizing of spin-wave superposition makes it possible to apply some of the algorithms developed for quantum computers to speed up the database search. The detailed description of the search procedure and experimental data can be found in Ref. [348].

It is expected to see more MHM prototypes to be demonstrated in the coming five years. Parallel database search through magnetic database using spin-wave superposition is the most appealing/promising application for MHM. It is also expected that MHM devices with built-in memory will be utilized for special task data processing (e.g., prime factorization). The lack of convenient (i.e., non-optical) devices for sub-micrometer wavelength spin-wave generation is the main technical obstacle to be overcome.

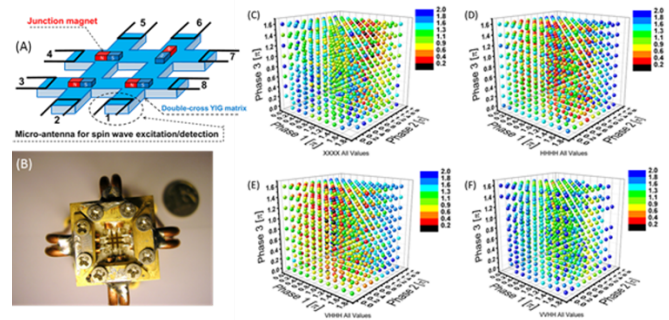

Fig. VII.B. a) Schematics of the eight-terminal MHM. (b) Photo of the Prototype. (c)-(f) Collection of the experimental data showing the correlation between the input spin-wave phases and the output inductive voltage. Reprinted (adapted) with permission from [Khivintsev2016].

ACKNOWLEDGMENT

The current work on MHM is supported in part by the National Science Foundation (NSF) under Award # 2006290 and by INTEL CORPORATION (Award #008635).

CONTRIBUTORS

This section is authored by A. Khitun (University of California, Riverside, USA).

### C. Inverse-design magnonics

Many magnonic devices were demonstrated recently [12], but the development of each of them requires specialized investigations and, usually, one device design is suitable for one function only. In contrast, electronic circuit design is much more convenient since it uses automatic design software, especially for large-scale integrated CMOS circuits with billions of transistors. Moreover, the so-called inverse-design method was introduced into photonics to design and optimize photonic circuits [349]. In this approach, any functionality can be specified first, and a feedback-based computational algorithm is used to obtain the device design. Recently, inverse-design was also utilized and validated numerically in the field of magnonics [91], [350].

A three-port proof-of-concept prototype is based on a rectangular ferromagnetic area that can be patterned using square-shaped voids (see. Fig. VII.C(a)) [91]. To demonstrate the universality of this approach, linear, nonlinear and nonreciprocal magnonic functionalities were explored and the same algorithm was used to create a magnonic (de-multiplexer, a nonlinear switch and a circulator. In Ref. [350], the authors took a more complex functionality and designed a neural network by inverse-design (see. Fig. VII.C(b)). It is shown that all neuromorphic computing functions, including signal routing a nonlinear activation, can be performed by spin-wave propagation and interference. In particular, vowel recognition is used for the demonstration of the inverse-design magnonic neural network.

From a computing perspective, a particular advantage of machine learning methods is that they enable the design of spin-wave devices in the weakly nonlinear regime. By tuning the excitation amplitude used for launching the waves, one can tune their nonlinearity as well. Linear spin-wave interference is limited in its computing capabilities, while strong nonlinearities and modulation instabilities prevent the design of robust devices. Weakly nonlinear spin waves greatly outperform linear waves in neuromorphic operations (such as classification), and machine learning methods appear to work very well in this regime [350].

A challenge of machine learning methods for device design is the required huge computational workload of learning. Backpropagation in time is computationally expensive, and state-of-the-art GPUs can design magnonic devices smaller than a few ten times the spin-wave wavelength in each dimension. The complexity of inverse-designed devices is likely limited by the available GPU memory and not the computing capacity of the magnonic substrate.

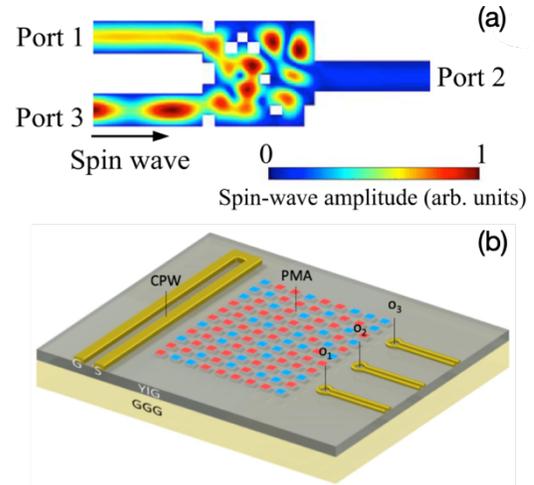

Fig. VII.C. (a) The inverse-design magnonic Y-circulator [Wang2021]. Reprinted (adapted) with permission from [Wang2021]. Copyright 2021 Springer Nature. The design region is divided into 10×10 elements each with a size of 100×100 nm². White regions represent voids. The color map of the simulated spin-wave amplitude shows the particular 3→1 operation of the circulator. (b) The schematics of the envisioned computing device [350]. The input signal is applied on the coplanar waveguide (CPW) on the left, and the magnetic state (up/down) of programming nano-magnets on top of the YIG film defines the weights.

From a numerical perspective, inverse magnonics can be formulated as an optimization problem constrained by the system of partial differential equations (PDE) defined by the micromagnetic model. One of the main challenges in PDE constrained optimization is the computation of the gradient that is required for the efficient minimization of the objective function. PDE constrained optimization problems usually have a continuous design space which translates to a very high number of degrees of freedom after discretization. A finite-difference discretization of the gradient is prohibitively expensive in this case since this method requires the solution of the complete forward problem for every degree of freedom. However, the gradient of such a problem can be efficiently computed by solving the adjoint problem to the constraining PDE [351]. Solving the adjoint problem usually has the same computational complexity as solving the forward problem. Hence, the gradient of an arbitrary objective function can be accessed at the additional cost of a single forward solve independently of the number of degrees of freedom.

In the context of micromagnetics, this procedure has already been successfully applied to the stray field problem in order to perform inverse field computations and topology optimization [352], [353]. This method could already be used in magnonics to optimize the excitation efficiency of antennas. Extending this approach to time-dependent micromagnetics will significantly extend the novel field of inverse magnonics and allow for time-dependent optimization of various continuous design variables such as the device topology or spatially dependent material parameters.

First steps in this direction were already taken in [350] where the authors use the back-propagation capabilities of a machine-learning framework in order to compute the gradient of a time-dependent objective. The combination of this technique with highly optimized micromagnetic algorithms will give rise to a very powerful toolkit for magnonic device design.

## Acknowledgment

This research has been supported by the Austrian Science Fund (FWF) through project I 4917-N and by the DFG TRR-173 – 268565370 (Collaborative Research Center SFB/TRR-173 'Spin+X', project B01).

## Contributors

This section is authored by Q. Wang, A. V. Chumak, C. Abert, D. Süss (University of Vienna, Austria), A. Papp and G. Csaba (Pazmany Peter Catholic University, Budapest, Hungary), W. Porod (University of Notre Dame, USA), G. A. Melkov (Taras Schevchenko University of Kyiv, Ukraine), P. Pirro (TU Kaiserslautern, Germany).

### D. Magnonic neuromorphic computing

Many unconventional ways of computing were originally inspired by the functioning of the (human) brain and its neuronal structures. These so-called neuromorphic or "brain-inspired" approaches attempt to match the unrivaled efficiency of the biological brain for many types of tasks in terms of performance and energy requirements. Applications for this subarea of artificial intelligence include, e.g., pattern recognition, classification, and prediction.

Neuromorphic computing uses Artificial Neural Networks (ANN) consisting of nonlinear signal processing elements, the so-called artificial neurons that are highly interconnected by artificial synapses. Nowadays, the most powerful ANNs are based on software models which simulate the network using CMOS-based, binary hardware. A direct implementation of the artificial neurons and their synaptic connections as physical elements [354] is generally envisaged to lead to higher efficiencies and better scalability. This architecture, if realized in a CMOS-based circuit, faces a lot of challenges, since it needs at least dozens of transistors to represent a single neuron. The electric connections between neurons show a poor scalability, and its physical separation of memory and processing units limits efficiency. Also, the interconnectivity of the brain ($10^4$ connections per neuron) cannot be matched by planar IC technologies. In the following, we discuss why the special properties of magnonics make it a promising candidate for the hardware-based realization of ANNs.

Magnonics fulfills the general criteria for a powerful physical realization of an ANN like room-temperature operation and scalability to the nanoscale (see previous sections). Its wave-based nature allows for a convenient use of continuous variables common to ANNs. Typically, an artificial neuron collects input data streams, which it receives from other neurons via the synapses. This can be achieved, e.g., using magnonic resonators with active damping compensation [355]. Just like in photonic neural networks [356], spin waves allow all-to-all interconnections between the neural building blocks. The available spin-wave waveguides, multiplexers and directional couplers can be used to construct these magnonic synapses. Magnonic frequency-multiplexing techniques [357] similar to those demonstrated in photonic ANNs, and time multiplexing shown for both photonic and magnonic reservoir networks [358], [359], [360] can increase the interconnectivity drastically without the need to create large amounts of individual waveguides. The intrinsic nonlinearity of spin waves evident in the intensity-depended shifts of frequency and wave vector, or by spin-wave instabilities represent one more key property. Artificial neurons respond to input signals via a so-called "activation function" which needs to be highly nonlinear. A common example is a threshold function that lets the neuron "fire" an output signal only if a certain input threshold is exceeded. This functionality can be implemented using, e.g., nonlinear magnonic resonators [361] or directional couplers [297], [33]. Furthermore, spin waves possess an important property of short-term memory due to a significant delay time associated with their propagation in thin magnetic films.

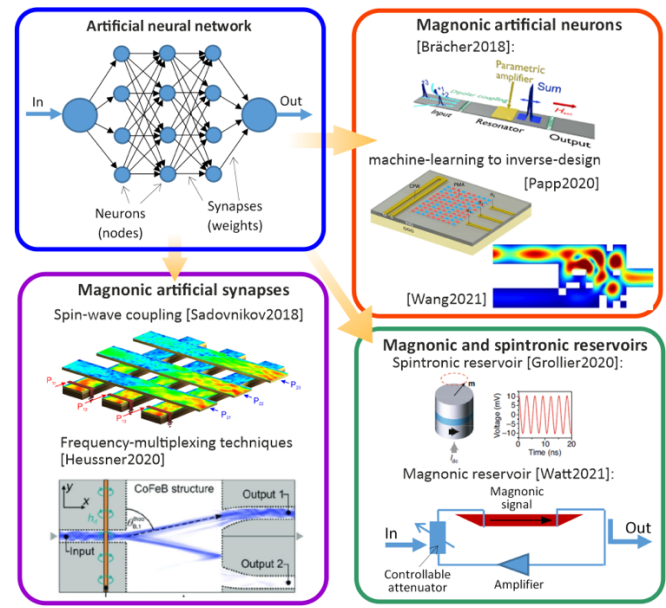

Fig. VII.D. Magnonic neuromorphic computing: from the general concept of the artificial neural network to its magnonic implementations. Reprinted (adapted) with permission from [355], [350], [91], [362], [357], [363], [360].

ANNs are trained to perform certain tasks by adjusting the synaptic weights, i.e., the strength of the connections between the individual neurons. In magnonic ANNs, these weights can be realized using parametric [364] or spintronic amplifiers as well as reconfigurable magnonic ground states [365] (domains and domain walls, reconfigurable nanomagnets). So-called "supervised learning" schemes, where the weights are adjusted by an external circuit, are comparatively straightforward to construct. "Unsupervised learning" however, where magnonic signals can change weights directly, to create, e.g., spike-time-dependent plasticity, are harder to realize due to the low energy associated with a magnon flux. Potentially, the displacement of a domain wall by spin waves [366] in combination with nonlinear effects can constitute a realization of such a weight which can be reconfigured by magnons themselves.

Another strategy to adjust the weights of an ANN is to use machine-learning to inverse-design the scattering parameters, a strategy that was explored in [91], [350]. The ANN device is a YIG film which is structured by a magnetic field distribution on its top. The non-uniform field distribution scatters the waves to define a desired computing function. Crucially, the scatterer does not merely act as a linear filter, and the computing function is greatly enhanced if the spin-wave intensity is sufficiently strong to display non-linear effects. Weak nonlinearity is best for neuromorphic computing: linear spin waves have limited computing power, while strongly nonlinear waves display seemingly chaotic and uncontrollable behaviors.

In a nonlinear, magnonic scatterer, neurons and their interconnections are not spatially separated – this solves the interconnection problem, but makes designing the weights very challenging. The challenge is analogous to the inverse-design problem in optics, where a complex scattering geometry has to be designed in such a way that desired input-output relations are fulfilled – and machine learning methods developed for optics work well for spin waves, where nonlinearities are significantly stronger. A big advantage of the machine-learning-based design is that the same learning mechanism works regardless of the details of the physics – for example, if parametric pumping is added to the scatterer to compensate for damping, then the scattering geometry could be designed with the same method.

A way to circumvent the adjustment of many weights is by using the reservoir computing approach, which does not require the layered structure and dedicated memories for weights as typical for ANNs. The combination of the significant short-term-memory capacity and strong intrinsic nonlinearity of dynamics makes spin waves a promising candidate for reservoir computing. The concept of treating a reservoir as a non-linear black box greatly simplifies the design of magnonic networks and removes one major roadblock in magnonics, which is the negative effect of magnetic damping on information transport based on magnons. The quality (internal complexity) of a reservoir can be quantitatively characterized by Kernel Rank and Generalization Rank metrics – and spin-wave scatterers fare well in these metrics [367]. An alternative approach to using the scatterers is spin-wave reservoirs based on travelling-spin-wave ring resonators [359], [360]. Compensating the spin-wave decay by adding amplification gain to the ring enables creation, storage and efficient coupling of multiple virtual neurons in the resonator due to a large delay time and strong nonlinearity available with travelling spin waves. This approach of time multiplexing ultimately leads to efficient nonlinear mapping of input reservoir data to a higher-dimensional space that is essential for processing time-dependent data.

Furthermore, recent studies [368], [369], [370], [371], [363] have revealed that reservoir computing can also be performed based on mutually coupled magnonic oscillators using synchronization effects. Synchronization is the fundamental nonlinear effect [372] that can drive the magnonic oscillator network to a collective state, where the phases and frequencies of oscillators are equal or may differ by a constant.

A high-level oscillator-based operational process can be outlined as follows. The input, which might represent an image for recognition, is a pattern in the phase or frequency mapped to the initial states of the individual oscillators. Due to the strong coupling between oscillators, phases or frequencies evolve after removing the inputs. Synchronization moves the network to the energy minimum. Finally, the result of the computation can be read out by extracting relative phases of frequencies from the network.

One of the main characteristics of oscillator-based RC is that externally injected oscillations can tune mutual couplings or bring noninteracting oscillators into the coupling or synchronization. The geometry of the magnonic oscillator-based networks may strongly limit the realization of couplings, e.g., magnons propagate only a few hundred nanometers in NiFe [15] resulting in strongest coupling between nearby oscillators [373], [362]. Thus, it is better to use spin-Hall nano oscillators mutually coupled by magnonic low-damping YIG-based waveguides [181], where magnons travel several tens of microns. However, it is a technological challenge to integrate such a high-level network of oscillators on such a magnetic film [374]. The presented oscillatory RC concept can be implemented not only in the FM case at GHz frequencies, but also in the AFM case at THz frequencies [375], [376], [377], which can significantly speed up the signal and image processing devices.

Overall, magnonic neuromorphic devices have the potential to combine the best of two worlds: the possibility of high interconnections from optics, with the tunable nonlinearities available in electronic devices.

## Acknowledgment

This work was supported by the Russian Science Foundation (project No.20-79-10191, 21-79-10396), Government of the Russian Federation for state support of scientific research conducted under the guidance of leading scientists in Russian higher education institutions, research institutions, and state research centers of the Russian Federation (project number 075-15-2019-1874), the Ministry of Science and Higher Education of the Russian Federation under "Megagrant" (Agreement # 075-15-2021-609), and Deutsche Forschungsgemeinschaft (DFG, German Research Foundation) - TRR 173 - 268565370, Project B01 and the Defense Advanced Research Projects Agency (DARPA) grant 'Nature as Computer'.

## Contributors

This section is authored by M. Kostylev (University of Western Australia, Perth, Australia), G. Csaba (Pazmany Peter Catholic University, Budapest, Hungary), H. Schultheiss and K. Schultheiss (Helmholtz-Zentrum Dresden-Rossendorf, Germany), A. Sadovnikov and A. R. Safin (Saratov State University, Russia), A. Ustinov (St. Petersburg Electrotechnical University, Russia), and P. Pirro (TU Kaiserslautern, Germany).

### E. Voltage controlled spin Hall nano-oscillator based computing and Ising machines

Nano-constriction based spin Hall nano-oscillators (SHNOs) [378], [379] are a novel class of spin current driven, CMOS compatible [380], microwave nano-oscillators, which can be fabricated as small as 20 nm [381], [382], [383], and can be mutually synchronized in both one [384] and two dimensions [385]. Complete mutual synchronization of 64 such SHNOs in an 8x8 square array was recently shown to improve the quality factor an order of magnitude (170,000) compared to the best literature values for any spin current driven nano-oscillator [385].

While the nano-constriction geometry will typically create a spin-wave energy landscape where localized auto-oscillation modes are favored [386], this restriction can be lifted with strong out-of-plane magnetic fields or perpendicular magnetic anisotropy materials, such as ultrathin CoFeB, where also propagating spin waves can be generated [387]. Combining the voltage controlled magnetic anisotropy of electric field gated W/CoFeB/MgO stacks with a nano-constriction SHNO, where the auto-oscillations are tuned to be on the border between localized and propagating, can then lead to a giant voltage modulation of the effective SHNO damping as the auto-oscillation mode can be tuned between localized (low dissipation) and propagating (high dissipation) [387]. It is hence possible to implement low-power voltage control of both the frequency and the threshold current of individual nano-constrictions in large SHNO arrays, which is required for programming such arrays for oscillator-based computing.

Using memristive gates, it is furthermore possible to both add non-volatile memory storage to each nano-constriction [388] and operate the SHNO gate in non-volatile tunable low-resistance state, such that individual gates can add additional drive current to the chain at different locations, which can be used for rapid synchronization-based pattern recognition.

Two-dimensional arrays of SHNOs can also be phase-binarized and then used as oscillator based Ising Machines [389], [390], where they can lead to orders of magnitude smaller, faster, and less power-hungry solvers of combinatorial optimization problems compared to existing commercial solutions.

These recent results highlight the potential of voltage controlled SHNO arrays for both microwave signal generation and emerging unconventional computing schemes. Remaining challenges include finding material combinations that can further reduce the auto-oscillation threshold currents, use ferrimagnetic layers to increase the operating frequencies and hence the computational speed, add magnetic tunnel junctions for improved read-out, improving the robust memristor functionality of the voltage-controlled gates, and scaling up to larger arrays. More fundamental questions also remain, such as a better understanding of the intrinsic synchronization dynamics, e.g., to design optimal annealing schemes, and the impact of changes to the SHNO array layout, connectivity, and topology.

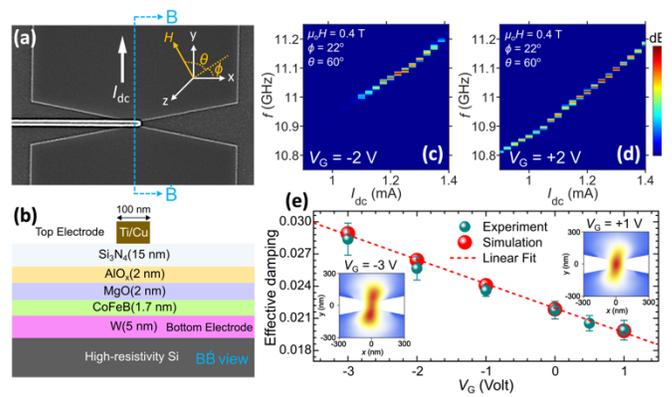

Fig. VII.E. (a) SEM image of a voltage controlled nano-constriction spin Hall nano-oscillator (SHNO). (b) Sideview schematic of the material stack of the device in (a). (c) Power spectral density (PSD) vs. SHNO current with -2V applied to the gate. (d) PSD for the same SHNO when the gate voltage is switched to +2V. (e) Effective damping vs. gate voltage of the SHNO in its linear (sub-threshold) regime as measured by spin torque ferromagnetic resonance spectroscopy together with a micro magnetic simulation. Insets show how the spatial extent of the spin-wave mode varies with applied voltage, favoring a more extended mode when the voltage increases the perpendicular magnetic anisotropy (PMA) at negative voltage and a more localized mode from a reduced PMA at positive voltage. The larger mode volume increases the dissipation, which then leads to the large change in threshold current seen in (c) and (d). Figures adapted from [Fulara2020] with permission under a Creative Commons Attribution 4.0 International Licensethe, http://creativecommons.org/licenses/by/4.0/.

### ACKNOWLEDGMENT

This work was supported by the Swedish Research Council (VR Grant No. 2016-05980) and the Horizon 2020 research and innovation program (ERC Advanced Grant No.~835068 "TOPSPIN").

### CONTRIBUTORS

This section is authored by Afshin Houshang, Ahmad Awad, Roman Khymyn, and Johan Åkerman (University of Gothenburg) and Himanshu Fulara (IIT Roorkee).

## VIII. TOWARDS QUANTUM MAGNONICS

All previous sections discussed the application of classical spin-wave properties for computing. To fully harness the power of spin waves for computation, the quantum nature of magnons and the underlying physics of entangled magnonic or hybrid quantum states should be exploited. The quantum mechanical analogue to a classical bit is a quantum bit (qubit), which is the superposition of the coherent wave functions of two system states. The field of quantum magnonics is just being established. But as shown by many contributions to this section, it attracts widespread attention. Recently, the group of Nakamura [391] reached a milestone and demonstrated a single magnon state experimentally. The ground-breaking nature of such experiments and the extensive research effort in the field, promise a very bright future for quantum magnonics.

The first section discusses the use of macroscopic quantum states in form of Bose-Einstein condensates (BECs) for computation. The different aspects of YIG for quantum magnonics at ultra-low temperatures are discussed in Section VIII-B. Nano-mechanical magnetic resonators for magnon-phonon quantum hybrid systems are presented in the following section. When magnon-phonon interaction surpasses the relaxation rates of magnons and phonons, the two excitations exhibit coherently coupled dynamics, leading to the formation of magnon-phonon polaron covered by Section VIII-D. Further, magnons are able to couple to photons at microwave and optical frequencies. These coupling mechanisms are covered by the modern research direction of cavity magnonics, which is discussed in the following two sections. The utilization of planar resonator-based hybrid magnonic circuits for similar purposes are described in Section VIII-G. The following section explains the preconditions for the experimental realization of entangled magnonic states. Section VIII-I is focused on the manipulation of hybrid magnonic interactions. The realization of ultra-strong photon-to-magnon coupling with superconducting hybrids is discussed in the further section. The Section VIII-K is devoted to the quantum interfaces between magnons and paramagnetic spins, and the Section VIII-L addresses the question of the distillation of entanglement in quantum magnonics.

The integrated generation, manipulation, and detection of non-classical magnonic states are envisioned as some of the ultimate goals, which still remain to be realized. By employing techniques, originally developed in circuit quantum electrodynamics (circuit QED) to manipulate microwave photons with superconducting qubits, we are now at the verge of realizing these goals. Magnon circuit QED and magnon quantum sensing aims to solve some of these challenges. Section VIII-M of this Roadmap discusses quantum sensing of magnons with superconducting qubits and the final Section VIII-N focuses on two-tone cavity-magnon polariton spectroscopy, and controlled on-chip coupling.

## A. Computing with magnonic macroscopic quantum states

Future computing technologies are expected to use quantum computing, in which the basic unit of computation is a quantum bit, or a qubit for short. A qubit is the superposition of the coherent wave functions of two system states. In quantum computing, these are the quantum mechanical wave functions, and the quantum mechanical principles determine the population of the superposition state. In the classical domain, the superposition of two quasiparticle wave states with different energies or different quasi-momenta can be used for qubit computing.

Although entanglement is necessary for many quantum computing algorithms, there are examples of problems that can be solved using a qubit calculus without entanglement and still be more efficient than the best possible classical algorithms. For instance, the BB84 quantum key distribution protocol [392] and the quantum Fourier transform [393], [394] do not employ entanglement and can be implemented by the classical qubit functionality.

Magnonics offers a very convenient and functional implementation of a classical qubit. Macroscopic magnonic quantum states such as Bose–Einstein condensates (BECs) [395] are excellent candidates for this aim [396], [397], [398], particularly due to their inherent coherency and zero group velocity [399]. Using numerical simulation methods, we have recently demonstrated the first realization of a magnon qubit with the two BEC condensates in an in-plane magnetized yttrium iron garnet film [400]. Protocols are developed to allow both for single- and multi-qubit calculus.

The proposed qubit consists of the superposition of the two magnon BEC wave functions, which exist at the minimum magnon spectrum frequency at $\pm q$ wavevectors: $BEC^+$ and $BEC^-$ (see Fig. VIII.A(a)). These wave functions form a standing wave pattern [401], and they are non-propagating because a BEC has zero group velocity. Therefore, they are ideal for representing a classical qubit. In a Bloch sphere representation (see Fig. VIII.A(b)), the amplitude ratio of the BECs defines the polar angle, and the phase difference between the two BECs corresponds to the azimuthal angle. It has been shown that qubit initialization can be performed with different regimes of parametric pumping. Coherent manipulation protocols using Rabi-like oscillations and manipulation of the relative BEC phase are also possible, as confirmed by realistic numerical simulations close to experimental conditions [400]. Inter-qubit coupling can be implemented and controlled using magnon supercurrents and Josephson oscillations [402], [403], [404].

Although qubit calculus with classical wave functions is superior to conventional Boolean logic, it does not compete with quantum computing technologies. Nevertheless, it works under ambient conditions, and the properties of scalability and stability are advantageous. Moreover, the magnon qubits can be used in solid-state devices with standard microwave interface technology [400].

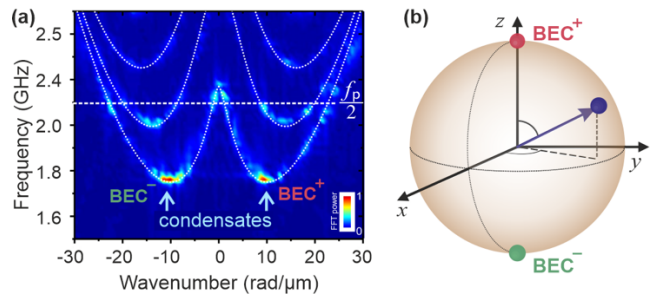

Fig. VIII.A. (a) Numerical simulation of the condensation process in a parametrically populated magnon gas (adapted under the terms of the Creative Commons Attribution 4.0 licence from [405]) leading to a qubit state lying on the equator of the Bloch sphere shown in panel (b). (b) Representation of a magnon BEC qubit (blue dot) on the surface of a Bloch sphere. The $BEC^+$ (North Pole) and $BEC^-$ (South Pole) states are indicated by the red and green dots.

Linking the magnonic macroscopic quantum states with the quantum computing functions opens a new perspective for magnonics. We foresee a growth of experimental and theoretical research on computing with magnonic macroscopic quantum states, which can help cope with long-term challenges, such as building and controlling magnon multi-qubit on-chip circuits. We envision the future development of this direction in the experimental implementation of qubit logic gates, the realization of interqubit coupling, and the study of non-separable magnon BEC states.

### Acknowledgment

We acknowledge funding by the European Research Council within the Advanced Grant No. 694709 SuperMagnonics and by the DFG within the Transregional Collaborative Research Center–TRR 173–268565370 "Spin+X" (project B04).

### Contributors

This section is authored by A. A. Serga, V. I. Vasyuchka, B. Hillebrands (TU Kaiserslautern, Germany).

## B. YIG-based quantum magnonics

*Introduction.* Although superconducting quantum circuits are assumed to be well scalable, constructing a large network of coupled qubits is still a major obstacle to the creation of an ultimate quantum computer. Up to now, only a 12 entangled superconducting qubits state was reported [406]. In other quantum systems, like optical photons and ultra-cold atoms, it is already possible to realize entanglement of up to 20 qubits [407]. Thus, it is of great importance to continue investigating other physical systems that might become companions—or even replacements—for current superconducting computing elements. Another challenge faced by quantum computing is how to connect these systems to conventional superconducting circuits. For example, optical photons have frequencies in the THz range and the typical transition frequencies of ultra-cold atoms systems are below 1 GHz, making interfacing challenging, sometimes even impossible. One of the possible ways to solve connectivity as well as scaling problems is to take advantage of magnetic systems and quasiparticles associated with the elementary disturbance of magnetic order – magnons.

*Yttrium Iron Garnet in modern quantum magnonics.* One of the main interests in magnetic materials for quantum devices is, perhaps, the well-known resonance frequency range; in particular, ferromagnetic resonances typically appear in the low GHz region—the frequency range commonly used in superconducting quantum circuits (4-16 GHz). While this is true for several ferromagnetic structures, most of the interest so far has centered around one material, as this has been the only one to approach reasonable practical requirements on the data carrier lifetime to be competitive. This material is Yttrium Iron Garnet ($Y_3Fe_5O_{12}$, YIG) [410]. The lifetime of magnons in YIG can reach microseconds, especially at cryogenic temperatures. Large tunability [411], intrinsic nonlinearities [412], excellent ability to connect to other systems (like optical and microwave photons, or acoustic phonons) [413], [409], [408], and easy creation of macroscopic quantum states like magnon Bose-Einstein Condensate (BEC) [395], [414], [132], [415], are some of the other main features of YIG which are already available for the advancement of quantum computing.

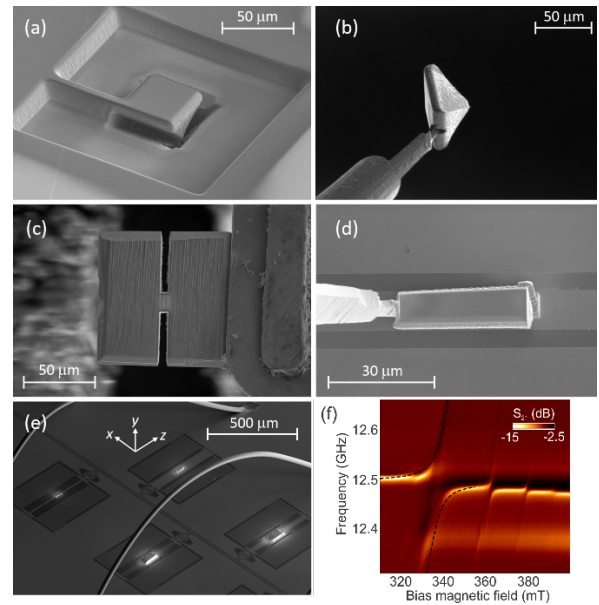

Fig. VIII-B1. (a)-(e) FIB fabrication steps of YIG-on-superconductor samples [408]. (f) Ultra-strong coupling evidence obtained at 80 mK for the structure shown in (e) together with theoretical calculation based on [409] (Reproduced from [408], with the permission of AIP Publishing).

The first steps for employing magnon states in YIG as quantum information carriers have already been taken by studying the possibility to strongly couple bulk YIG crystal samples (spheres) to superconducting circuits [391], [416]. However, utilization of bulk YIG is not scalable and it is not possible to integrate it into on-chip technology. This is because to make planar YIG circuits, it is necessary to pattern it. This task nowadays is performed routinely [26] on high-quality, thin, YIG films, but these are only grown epitaxially on Gadolinium Gallium Garnet (GGG) single-crystal substrates. This substrate is very fine for room temperature applications since its magnetization is negligible. However, at low temperatures this paramagnetic substrate introduces huge losses [417], [418]. Alternative substrates, like Silicon, have been investigated, but those were reported to result in damping of orders of magnitude higher than that of YIG grown on GGG [419]; such loses make them not suitable for applications in quantum computing. More recently, the method developed in [408] has not only allowed the engineering of planar YIG components, but this has also made planar YIG suitable for implementation into superconducting circuits. This method is based on the Xe-Plasma Focused Ion Beam extraction of the high-quality top layer of YIG from the GGG substrate (see Fig. VIII.B1(a-e)). This method also allows for producing long YIG conduits in the order of only tens of micrometers to achieve strong coupling with superconducting resonators (Fig. VIII.B1(f)).

In addition to integration and scaling, further development in this field will also rely on deeper understanding, as well as on-demand manipulation, of magnon-photon hybridization. Works on this are ongoing, but so far it has already been demonstrated that there can exist various regimes of hybridization, including level repulsion, attraction, and merging, which can be achieved through different physical mechanisms [420], [421], [422], [423]. Moreover, the temporal behavior of magnon-photon systems will also be of key for any quantum applications. This has also been investigated, and the time-dependent control of the coupling has been achieved in [424]. Finally, theoretical understanding of such coupling, developed in [409], has allowed predicting the strength of coupling using a perturbation approach. This method works even for ultra-strong coupling regimes, micro-structured planar samples, and can be employed and both high and low temperature regimes using simple analytical expressions.

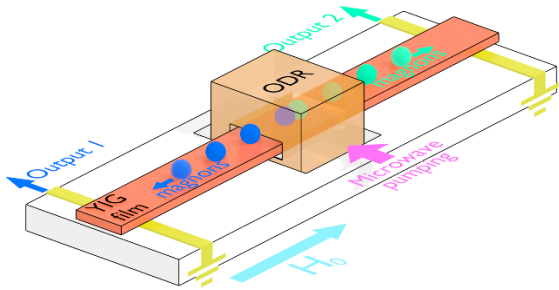

Fig. VIII.B2. Heralded parametric single magnon source concept. Magnon pairs are generated in the pumping area by means of parametric pumping. Magnons propagate in opposite directions. One of the outputs is used to determine (herald) the number of magnons at the other output.

*Outlook.* The great potential of YIG as the key magnetic material for quantum magnonics is yet to be realized. For instance, there are only a few experiments done at room temperature to demonstrate the possibility of magnon state squeezing [425], magnon BEC supercurrents, and Bogoliubov waves [402], [403], [426]. The only attempt to study magnon BEC at cryogenic temperatures was reported in [427] but a thorough investigation in the quantum limit is still needed.

Realization of many of these—and other ideas—require quite strong bias magnetic fields to operate. So far, such fields were considered a disturbing factor for superconducting systems, especially for aluminum-based superconducting circuits. In modern magnetism, this issue could—and should—be resolved using intrinsic or shape-induced anisotropies [311], localized stray fields from neighboring magnetic structures, synthetic magnets/metamaterials [428], etc.

In addition to coupling to superconducting circuits, further attention should be paid to development of YIG-based quantum computing elements, such as qubits. For this purpose, as magnons are bosons, a single magnon source is needed; and it could be realized without use of superconducting elements. The concept of such device is shown in Fig. VIII.B2. If one considers a uniform rf pumping field applied parallel to the magnetization direction of a ferromagnet—the so-called parallel pumping process—it could lead to the creation of a pair of magnons at half of the pumping frequency [429]. In the quantum limit, this process is analogous to the spontaneous parametric down-conversion process in quantum optics, but it offers many more possibilities for control over the generated frequency of magnons as well as their momentum. In a practical experiment, one can use an open dielectric resonator (ODR) with a resonant frequency of about 13-14 GHz to enhance the pumping efficiency. The width of the pumping area should be large compared to the wavelength of the generated magnons, resulting in their propagation in opposite directions towards two detection antennas. In the quantum limit, such a device can act as a single magnon source if one of the outputs is used for heralding the generated magnon number. This device could be used to create a variety of other devices and circuits, such as all-magnon qubits, quantum gates, and spin-based quantum memory. The possible designs and experimental realizations of such elements are in active development.

## Acknowledgment

D. A. Bozhko acknowledges support from UCCS Committee on Research and Creative Works.

## Contributors

This section is authored by D. A. Bozhko (University of Colorado Colorado Springs, USA), R. Macêdo, and M. Weides (University of Glasgow, UK).

### C. Nano-mechanical magnetic resonators for magnon-phonon quantum hybrid systems

The efficient inter-conversion of coherent signals between different types of elementary excitations, that are operating in the microwave energy range, is an important operation in quantum sciences, as these functions represents the fundamental building block for memory, transport or sensing applications. It requires a tailored coupling rate between the subsystems which must be carefully balanced with their intrinsic relaxation or loss rates. The regime of strong coupling, which manifests in form of a level repulsion, when tuned to degeneracy, is of particular interest as the properties of subsystems hybridize. There has been recently a renewed interest in exploiting acoustic phonons in materials with low ultrasonic attenuation as potential candidates to establish strong coupling with collective spin modes in a magnet. For this particular combination the strength of the magneto-elastic interaction needs to be greater than the relaxation time of the magnon and phonon modes in the sample. In this respect, yttrium iron garnet (YIG) holds a unique position in nature for combining the lowest reported magnetic damping [430] combined with excellent acoustic attenuation (10× better than quartz). Several experiments have demonstrated that indeed one could achieve strong coupling between magnons and phonons [431], [432], [433]. But so far, most of the work has concentrated on performing experiments with macroscopic, millimeter sized YIG crystals, usually in the form of a perfectly polished sphere. Indeed, the difficulty that has so far hampered the development of garnet thin films for integrated solutions is that their epitaxial growth could only be achieved on gadolinium gallium garnet (GGG) substrates. However, for both photons and phonons, GGG substrates must be considered for all practical purposes as matched medium preventing in consequence the confinement of their oscillating energy within the YIG layer. To overcome this problem, recently, a new process developed by the group of G. Schmidt in Halle has demonstrated the fabrication of freestanding YIG microstructures with high magnon life time [157]. These new garnet objects, which benefit from greatly reduced phononic and photonic energy leakage through the substrate, have the potential to become game-changing coherent microwave transducers exploiting their ultra-high finesse in integrated components.

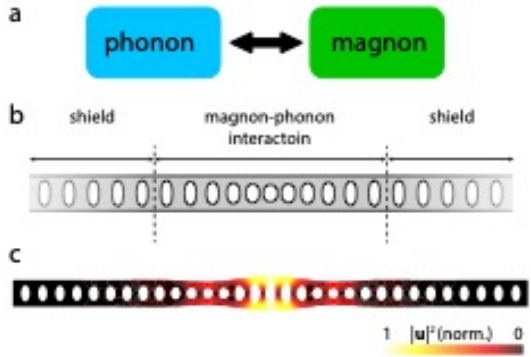

Fig. VIII.C. Tailored magnon phonon interaction in suspended nanostructures. The formation of defects in magnonic and phononic crystals allow to co-localize spatially selected eigen-modes. Due to the comparable velocity of phonons an magnons, this results in an optimal mode overlap while realizing a resonant mode coupling. Panel b shows a schematic of a top view of a structure supporting this interaction, c) represents modeling for the confinement of the phononic mode at the defect location.

The options offered by this new approach are numerous. Within a bounded range of variation, suspended structures of different sizes and shapes can be produced to allow for different spatio-temporal mechanical patterns. In single clamped cantilever type structures, the fundamental vibration mode occurs at a few tens of MHz. In doubly clamped bridges, one can shift the eigen-frequency of the fundamental vibration mode to several GHz. Additional engineering can be obtained by chosing the polarization (transverse vs longitudinal) and propagation direction. Meanwhile, it is even possible to create more complex shapes that allow post-patterning by focused ion beam techniques to create structures as in Fig. VIII.C or for transfer onto other substrates and devices.

The weak mechanical coupling between the magnetic system and the substrate inherent to suspended structure holds the promise of high quality factors as dissipation channels originating from the substrate (traditionally GGG) can be suppressed. Here low temperature measurements hold the promise to improve the signal to noise ratio by reducing the parasite effect of thermal fluctuations.

A future vision is to build on the technological advances in the nano- and microfabrication of suspended YIG structures and combine this with the established concepts developed in the field of optomechanics [434]. Fig. VIII.C shows a conceptual device layout highlighting the potential of this ansatz. In detail, the suspended nanostructure is tailored to host a phononic mode, which controllably interacts with the magnonic mode co-localized at the center region. It is expected that this engineered mode overlap combined with the intrinsic magnon-phonon interaction allows to realize the strong coupling required for the formation of hybridized magnon-phonon states. For conversion applications, the concept presented here can be extended to host a photonic mode enabling tripartite systems for quantum signal conversion applications [435].

Our vision marries the fields of magnonics and phononics inside integrated components, profiting from the best of both worlds, viz. the high tunability of magnets and superior quality of mechanical systems to realize strong coupling in hybrid systems uniting quantum information and quantum communication. We believe that phonons in garnet single crystals represent an excellent way to coherently couple distant spins in solid-state devices at microwave frequencies, while keeping the layout compact.

We expect that the future steps will merge structures similar to the ones presented in Fig. VIII.C with microwave cavities, again aiming for strong coupling and hence hybridization with microwave photons [436]. This together with the (magneto)optical properties will open up the possibilities for quantum science applications even further. While most of the experiments are so far performed at room temperature, we envisage larger impact when the quantum information exchange and distant entanglement of magnons, phonons, and microwave photons is performed at low temperatures, where the number of thermal excitations present at GHz frequencies are reduced to zero and quantum fluctuations dominate [437]. We envisage extension of the present experiments by studying lateral configurations and anticipate novel quantum effects when driving the magnets into the non-linear regime at low temperatures.

### Acknowledgment

This work was funded by the DFG via the TRR227 TPB02 and Germany's Excellence Strategy EXC-2111-390814868. This work was supported in part by the Grants No.18-CE24-0021 from the ANR of France.

### Contributors

This section is authored by Hans Huebl (Walther-Meissner-Institut, Garching, Germany), Georg Schmidt (Martin-Luther-Universität Halle-Wittenberg, Germany), Olivier Klein (SPINTEC, France), and Benjamin Pigeau (Institut Néel, France).

### D. Magnon-phonon polaron

When magnon-phonon interaction surpasses the relaxation rates of magnons and phonons, a situation called strong coupling, the two excitations exhibit coherently coupled dynamics, leading to the formation of magnon phonon polaron. Such hybrid systems play a crucial role in quantum and classical information processing since it enables coherent transfer of information from one physical system to another. Magnon phonon polaron has both the characteristics of magnons and phonons, i.e. spin information transfer, long lifetime, and mode degrees of freedom are realized altogether, advantageous for information transfer and processing via magnons.

The longevity of magnon-phonon polaron originates from its long-lived phononic constituent, which makes magnon-phonon polarons an efficient spin information carrier than pure magnons. In fact, recent experiments revealed that the magnon-phonon polarons sharply enhance the spin-Seebeck effect (SSE) voltage at certain conditions (see Fig. VIII.D(a) for the $H$ dependence of SSE for a YIG/Pt bilayer [438]). Here, the peaks show up at the field values $H_{TA}$ and $H_{LA}$ at which the (TA or LA) phonon dispersion curves become tangential to the magnon dispersion curve [Fig. VIII.D(b)], i.e., when the magnon and phonon frequencies $\omega$, wave numbers $k$, and group velocities become the same. Under these "touching" conditions, the magnon and phonon modes can be coupled over the largest volume in the $k$ space (see also Fig. 2 in Ref. [439]), such that the phase space portion over which the lifetimes of magnonic spin current are enhanced is maximal, leading to the enhancement of SSE [439]. The magnon-phonon polaron induced anomalies in the SSE have been observed also for other magnetic insulators such as antiferromagnetic $Cr_2O_3$ [437] and (partially) compensated ferrimagnetic $Lu_2BiFe_4GaO_{12}$ [38] and $Gd_3Ga_5O_{12}$ [440]. In 2020, Yahiro et al. reported that the anomalies show up also in the spin Peltier effect, the reciprocal of the SSE [441]. The formation of magnon-phonon polarons is predicted to affect magnonic spin conductivity and thermal conductivity [439], which await experimental discovery.

Another feature that makes the spin information transfer by magnon-phonon polarons unique is the mode degrees of freedom of phonons. Recent optical imaging experiments demonstrate the control of propagation orientation of magnon-phonon polarons by using the mode degrees of freedom (see Fig. VIII.D(c) for magneto-optical image of bi-reflection of magnon- phonon polarons [442]). Here the reflected beam of magnon-phonon polarons splits into the $k_1$ and $k_2$ orientations at which frequency and wavenumber along the edge is conserved. The condition is met for TA and LA phonons at two different wavevectors, leading to splitting of reflected beams. The propagation dynamics of magnon-phonon polarons has also been investigated under microwave excitation, revealing anisotropic propagation dynamics owing to non-linear multi-magnon phonon interactions [443], [444].

Propagation dynamics of magnon-phonon polarons may enable the encoding to networks of magnonic computing elements, such as a magnetic parametron. A magnetic parametron is the non-linear magnetic oscillator characterized by magnetization precession phases discretized into 0 or $\pi$ [445]. The function of parametron is demonstrated by a YIG/Pt thin disk under parallel parametric pumping. As the microwave excitation power increases, the precession phase exhibits Ising-like fluctuation between 0 and $\pi$ phases (Fig. VIII.D(d)). Therefore, the network of parametrons would enable reservoir and probabilistic computing or annealing in which the interconnection strength defines a problem to solve. The solution can be read out via the patterns of the phases on the network. Control of the spin information transfer by magnon-phonon polarons may play crucial role in controlling the interconnection among the parametrons, which may open a new device concept for magnonic computing.

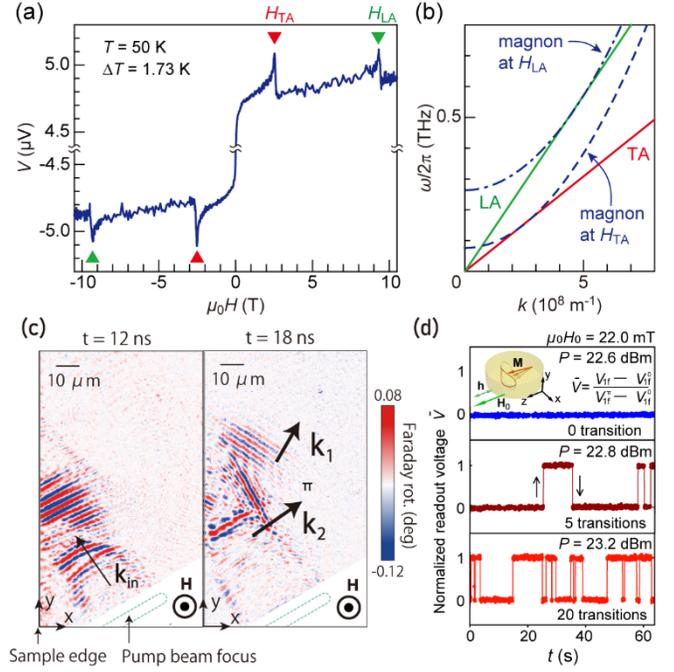

Fig. VIII.D. (a) H dependence of SSE voltage $V$ for a Pt/YIG-film system at 50 K under the temperature difference $\Delta T$ of 1.73 K [438]. Reprinted (adapted) with permission from [438]. (b) Magnon, TA-, and LA-phonon dispersion relations of YIG at the touching field $H_{TA}$ (= 2.6 T) and $H_{LA}$ (= 9.3 T). (c) Time-dependence of magneto-optical imaging of magnon phonon polaron [442]. Reprinted (adapted) with permission from [442]. (d) Microwave power dependence of temporal evolutions of the readout voltage of a magnetic parametron. A voltage value 1(0) corresponds to the precession with the initial phase of $\pi$(0). The inset shows schematic of a magnetic parametron [445] Reprinted (adapted) with permission from [445].

### ACKNOWLEDGMENT

This work was supported by JST CREST (JPMJCR20C1 and JPMJCR20T2), JSPS KAKENHI (JP19H05600, JP20H02599, and JP21H04643).

### CONTRIBUTORS

This section is authored by E. Saitoh and T. Kikkawa (University of Tokyo, Japan), T. Hioki (Tohoku University, Japan).

## E. Cavity Magnonics

Magnons can couple to photons at both microwave (MW) and optical frequencies, albeit via different mechanisms [446]. At MW frequencies, the spins couple to the magnetic field via Zeeman splitting, resulting, after a rotating wave approximation, in a beam-splitter type interaction

$$H_{mw} = \hbar g_{mw}(m^\dagger a_{mw} + a_{mw}^\dagger m),$$

where $m^{(\dagger)}$ and $a_{mw}^{(\dagger)}$ are the magnon and MW-photon annihilation (creation) operators and $\hbar$ the reduced Planck constant. A MW cavity loaded with a magnetic material of high spin density, like Yttrium Iron Garnet (YIG), can be used to boost the coupling strength $g_{mw}$, which can reach a significant fraction of the MW frequency. The frequency of the magnons can be tuned by an external magnetic field and brought into resonance with the MW cavity field. In the strong coupling regime, the magnon and photon modes hybridize, signaled by an avoided crossing at resonance. The coupling of magnons to optical photons (typically in the near-infrared) occurs instead via Brillouin scattering, a second order process in the electric field where light is scattered by absorbing or creating a magnon. The corresponding interaction Hamiltonian is parametric,

$$H_{\text{opt}} = \hbar g_{\text{opt}}\big(a_2^\dagger a_1 m^\dagger + a_1^\dagger a_2 m\big),$$

where $a_{1,2}^{(\dagger)}$ are the annihilation (creation) photon operators. For a dielectric magnetic material like YIG, the photonic cavity can be realized by the material itself, by total internal reflection. For optimal mode matching, the coupling $g_{\text{opt}} \sim 1/\sqrt{V}$ and could reach the MHz range for a diffraction limited cavity volume $V \sim 1 \mu m^3$ [447]. For a laser-driven cavity, $g_{\text{opt}}$ can be enhanced as $\tilde{g}_{\text{opt}} = \sqrt{n_c} g_{\text{opt}}$, where $n_c$ is the steady state number of photons trapped in the cavity. Under driving, $H_{\text{opt}}$ can be turned into a beam-splitter or a parametric amplifier Hamiltonian by choosing the laser detuning [448]. This corresponds to a linearization procedure, $a^\dagger a \rightarrow \sqrt{n_c}(a^\dagger + a)$, that conceals the intrinsic non-linear character of $H_{\text{opt}}$.

Whereas experiments with MWs can reach the strong coupling regime routinely, realizations with *opto*magnonic cavities are still in the weak coupling regime, in which the coupling is smaller than the typical photon ($\kappa$) and magnon ($\Gamma$) decay rates. A useful figure of merit in this context is the *cooperativity* $C = g^2/\kappa\Gamma$, where $g$ is the coupling strength. $C > 1$ is required e.g. for quantum state transfer.

The strong coupling regime can enable applications at the quantum level, such as the generation and manipulation of nonclassical states of magnons. A protocol to optically write and read single magnon Fock states was proposed in [448] by using $H_{\text{opt}}$ together with a combination of driving and single photon detection (heralding). In turn, a heralding protocol in the MW regime was put forward for generating magnetic cat states, where the magnetization is in a superposition of two semiclassical states pointing in different directions [449]. The protocol requires a squeezed quantum ground state of the magnetization (a state with anisotropic zero-point fluctuations), which can be easily obtained by using an anisotropic, elongated magnet. Cats with components separated by $\sim 5\hbar$ and involving $> 10^8$ spins are predicted to be feasible in YIG for mK temperatures.

Systems incorporating other degrees of freedom are also being explored. It has been proposed that the parametric coupling of magnons to kHz mechanical vibrations in MW cavity magnonics can be harnessed for quantum thermometry, with the possibility of realizing a primary thermometer at sub-Kelvin temperatures [450]. Dynamical backaction in these systems has been demonstrated, including the magnon spring effect on the mechanics, paving the way e.g. for entanglement, squeezing, or magnon-mediated ground-state cooling of massive objects [451]. Levitated micromagnets interacting with light have been moreover proposed to optically probe and control the coupled spin-mechanics [452].

Antiferromagnetic insulators have been put forward as an alternative platform for cavity optomagnonics with potential for on-demand tunable coupling by an external magnetic field, light-induced interaction between the two magnon branches, and THz transduction [453]. Two-magnon/two-photon Raman scattering processes at optical frequencies have been predicted to generate chiral magnon currents in two dimensional antiferromagnetic samples [454] and could be explored for stimulated magnon-pair generation in a cavity setup.

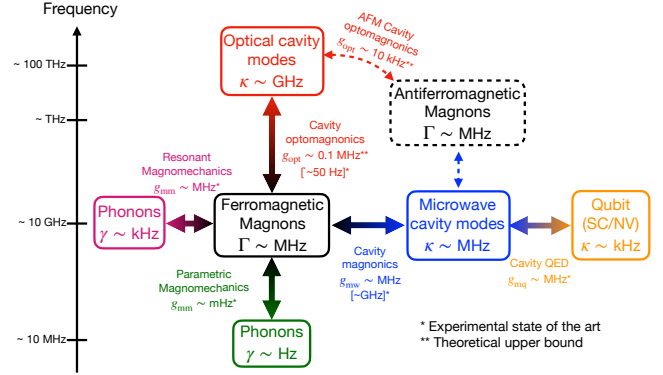

Fig. VIII.E. Hybrid cavity magnonics: characteristic frequencies and state-of-the-art decay rates for each degree of freedom. A comparison between the experimental state of the art (when available) and the upper bound for the coupling rates is given.

The different hybrid platforms for quantum magnonic systems are schematically depicted in Fig. VIII.E.

*Outlook & Challenges.* Cavity magnonics is a promising approach for quantum magnonics. Incorporating magnons into hybrid quantum systems can bring new functionalities into play, such as easy MW integration and nonreciprocal tunable quantum transduction. MW cavity magnonics is ripe for exploring the quantum regime, incorporating e.g. superconducting qubits as a nonlinear element [446]. Truly nonclassical magnon states, such as Fock or cat states, have not yet been demonstrated in these platforms, but it is to be expected that these can be realized in the near future.

One of the main challenges for quantum applications in the optical regime is to boost the state-of-the-art coupling rates and cooperativities. Although the predicted upper bound for $g_{\text{opt}}$ is large, theoretical proposals for concrete platforms still fall short from predicting $C > 1$. These include interfacing optics with magnetic textures [455] and the design of optomagnonic crystals at the microscale [435]. The main challenge resides in achieving optimal mode matching between the magnon and photon modes. Optomagnonic crystals could also be designed to realize strong resonant magnon-phonon coupling in the GHz regime. Interfacing efficiently also with MWs and phonons requires optimization and mode matching at all levels, which is extremely challenging. This however could open the door for quantum state transfer between microwave and optics or between microwaves and phonons.

The proposed protocols for generating magnonic nonclassical states rely on projective measurements [448], [449] and are probabilistic by nature. It is challenging to design deterministic protocols instead. A possibility is to systematically add magnons one-by-one via their coupling to a qubit, e.g., a superconducting qubit or an NV center. The intrinsic properties of magnetic systems (such as nonreciprocity and nonlinearities) are also attractive resources for the generation of nonclassical states to be explored.

Magnetic materials beyond YIG, and in particular antiferromagnetic materials in the context of cavity magnonics, remain still mostly to be explored, as are other frequency regimes for the photons. This is challenging both from a theoretical and experimental point of view, but could bring new exciting phenomena at the quantum level.

We acknowledge funding from the Max Planck Society through a Max Planck Research Group and from the Deutsche Forschungsgemeinschaft (DFG, German Research Foundation) through Project-ID 429529648–TRR 306 QuCoLiMa ("Quantum Cooperativity of Light and Matter").

### CONTRIBUTORS

This section is authored by V. A. S. V. Bittencourt, S. Sharma (Max Planck Institute for the Science of Light, Erlangen, Germany), and S. Viola Kusminskiy (Max Planck Institute for the Science of Light, Department of Physics, University of Erlangen-Nürnberg, Germany).

### F. Dissipative cavity magnonics

Cavity magnonics set initially its foot on the coherent magnon-photon coupling (as highlighted in the previous section). The physics of coherent coupling, as shown in the example of spring-coupled pendulums, has broad relevance. Yet, coherent coupling is only one of the two typical forms of coupling that enables hybridization. Dissipative coupling, the other form, has been less studied in the labs but is equally ubiquitous in nature. Examples include the synchronization of pendulum clocks mounted on a common base, the tidal locking of the Earth and the Moon, and the interacting heat currents in the Sun.

In 2018, dissipative magnon-photon coupling was discovered by setting a YIG sphere in a Fabry-Pérot-like cavity that consists of both standing and travelling waves [456]. Other experiments, performed soon after, observed similar effects by setting YIG spheres in different types of cavities [457], [421], [458], [459]. In contrast to coherent magnon-photon coupling that is characterized by level repulsion in the dispersion and beating pattern in the time-domain response, dissipative magnon-photon coupling is characterized by level attraction [456] and synchronization [460]. It is an indirect magnon-photon coupling effect mediated through a common reservoir such as the travelling wave, which provides a common source of dissipation that removes energy from the open system [461].

Since then, two distinct capabilities have been demonstrated in developing dissipative cavity magnonics: (i) Utilizing the directional interference between coherent and dissipative couplings, time-reversal symmetry is broken in cavity magnonics, which enables nonreciprocal microwave transmission [462]. (ii) Utilizing purely dissipative coupling, anti-parity-time symmetric cavity magnonics is demonstrated. In such a system, two types of singularities have been found (see Fig. VIII.F): the exceptional points that are square-root singularities appearing in non-Hermitian systems, and bound states in the continuum that simultaneously exhibit maximal coherent superposition and slow light capability [463].

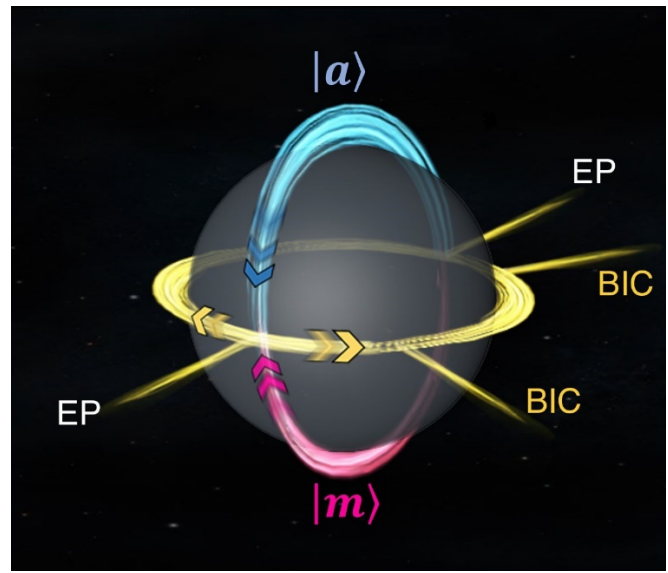

Fig. VIII.F. Bloch sphere representation for dissipative cavity magnonics showing the evolution of eigenmodes hybridized by the magnon (m) and photon (a). Two types of singularities emerge in such a system: the exceptional points (EP) that are square-root singularities appearing in non-Hermitian systems, and the bound state in the continuum (BIC) that exhibits slow light capability.

Utilizing the dissipative coupling, dissipation is no longer a nuisance. On the contrary, it becomes a resourceful ingredient of open systems. So far, all experiments in this direction have been performed in the classical regime. Expending the research to the quantum regime is of significant interest. However, it is still an open question whether the travelling wave is feasible to mediate a strong dissipative coupling between a single magnon and a single cavity photon. In contrast, coherent magnon-photon coupling in such a single quanta regime has already been demonstrated 7 years ago (see Section VIII-M). Achieving both coherent and dissipative couplings in the quantum regime will open new perspectives, such as enabling dissipation engineering in cavity magnonics devices, developing quantum sensors based on non-Hermitian physics, and creating resourceful magnon-photon states mediated by travelling photons, etc. All these may inspire cross applications that combine diverse capabilities such as eigenmodes control, quantum states manipulation, singularity-based sensing, and microwave confinement.

### ACKNOWLEDGMENT

Financial support of NSERC Discovery Grants and NSERC Discovery Accelerator Supplements is acknowledged.

### CONTRIBUTORS

This section is authored by Can-Ming Hu (University of Manitoba, Canada).

### G. Planar resonator-based hybrid magnonic circuits

The emergent behaviors of hybrid quantum systems promise to advance quantum information science with versatile capabilities by combining different physical platforms. A new breed of hybrid quantum systems is based on dynamic magnetic excitations, or magnons, which are collective excitations of magnetically ordered materials [178], [464]. The field of magnonics is dedicated to exploring magnons for spin-wave-based computing applications. The highly tunable dispersion of magnons makes them ideal candidates for coherent information processing and transduction, potentially down to the single-quantum level.

Planar microwave resonators are well suited for coupling magnons with microwave excitations and building hybrid magnonic networks and circuits. They offer great freedom in terms of circuit design and are compatible with lithographic fabrication processes and the modern complementary metal-oxide-semiconductor (CMOS) platform. Planar resonators have much smaller effective volumes compared with 3D cavities and therefore allow for better coupling efficiency with magnetic dipoles. For example, Li et al. [465] and Hou et al. [466] demonstrated strong magnon-photon coupling between a lithographically patterned permalloy (NiFe) thin-film device and a superconducting coplanar resonator. To overcome the challenge of a small coupling volume in magnetic nanostructures, one could replace the widely used metallic NiFe by materials with a larger spin density such as yttrium iron garnet (YIG). Indeed, recent progress showed that nanofabrication of YIG is feasible [467]. Moreover, this would enable the fabrication of arrays of nano-magnets [56] and magnonic crystals on planar resonators offering novel functionalities to tune the magnon resonance and propagation characteristics.

As a future perspective, one promising architecture is the use of split-ring resonators for studying magnon-photon polaritons. By employing optical techniques on this open geometry design, one can gain information about the localization of the hybrid excitations. Furthermore, it simplifies up- and down-conversion of electromagnetic excitations from the microwave to the optical domain. Kaffash et al. recently studied magnon-photon interaction between a split-ring resonator (SRR) and a 2.46 μm-thin YIG film sample in the strong coupling regime using Brillouin light scattering spectroscopy [54]. Complementary inductive and BLS measurements are shown in Fig. VIII.G.(a,b). Strong coupling between magnons and photons with a coupling strength of $g_{eff}/2\pi$=132 MHz was observed for a split-ring resonator with inner dimensions of $1.5 \times 1.5$ mm², outer dimensions of $4.5 \times 4.5$ mm² and a 0.2 mm gap. This demonstrates that a larger coupling strength for a much smaller system can be achieved compared to a 3D cavity since the effective volume of a planar SRR is smaller. Comparing the magnon-photon polariton results obtained by the two techniques reveals that BLS senses both magnonic and photonic characteristics (Fig. VIII.G.(b)), while microwave absorption mainly probes the photonic character of the hybrid excitation (Fig. VIII.G.(a)). Additional modes observed by BLS are attributed to spin-wave modes [468].

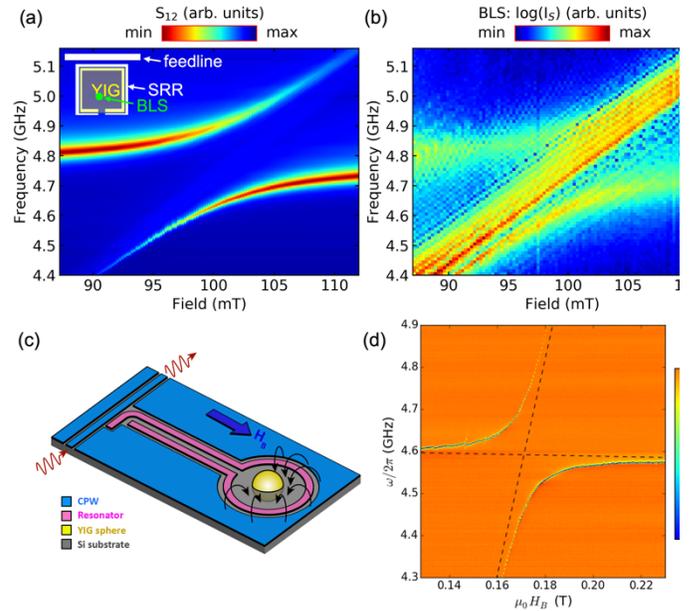

Fig. VIII.G. Demonstration of microwave-to-optical upconversion using strong magnon-photon coupling in a split-ring resonator/YIG thin film hybrid circuit. (a) Microwave transmission parameter $S_{12}$ as a function of magnetic field and microwave frequency. (b) Corresponding Brillouin light scattering (BLS) intensity versus magnetic field and frequency, effectively representing an upconversion. Avoided level crossing and uncoupled Kittel mode are detected by BLS. Adapted with permission from [469]. (c-d) Coherent coupling between a chip-embedded YIG sphere and a coplanar superconducting resonator. (c) Schematic of the coplanar superconducting resonator design. (d) Avoided crossing between the magnon mode and the resonator mode. Adapted with permission from [470].

Another promising planar architecture is embedding single-crystal YIG spheres onto coplanar superconducting circuits. Substrate-free YIG crystals enables state-of-the-art magnon coherence time at cryogenic temperatures for real quantum applications. Li et al. recently demonstrated hybrid circuits by mounting a YIG sphere into a deeply etched hole that is lithographically defined on a Si substrate [470]. The design of the superconducting resonator resembles a λ/4 resonator with the shorted end forming a circular antenna to couple with the YIG sphere (Fig. VIII.G.(c)). As shown in Fig. VIII.G.(d), a clear avoided crossing is observed between the resonator mode and the magnon mode, both of which show ultra-narrow linewidths down to 1 MHz for the entire frequency and field range. A magnon-photon coupling strength of 130 MHz and a cooperativity of 13,000 have been achieved for a 250-μm-diameter YIG sphere. Negligible parasite magnon modes are excited owing to the uniformity of the microwave field from the circular antenna. This is important for eliminating magnon decoherence processes in the dispersive regime. The YIG-sphere-embedded planar geometry can be further applied in coupling two remote YIG spheres, with the magnon-magnon coupling quantitatively controllable by the magnon-resonator frequency detuning [470].

We expect planar hybrid magnonic circuits to act as a leading contender for future development of quantum magnonics.

Steady advances in optical techniques at ultralow temperatures may unlock the full potential of quantum magnonics by studying magnon propagation in the quantum limit. The capability of integrating multiple magnon and resonator components in planar circuit systems enables on-chip magnon-based quantum networks and the realization of advanced magnon computing with quantum-tolerant excitation coherence.

### Acknowledgment

M.B.J. acknowledges support by the U.S. Department of Energy, Office of Basic Energy Sciences, Division of Materials Sciences and Engineering under Award DE-SC0020308. Y.L. and V.N. acknowledge support by the U.S. Department of Energy, Office of Science, Basic Energy Sciences, Materials Sciences and Engineering Division for preparation of this contribution to the "Roadmap on spin-wave computing concepts". The authors also acknowledge support from the University of Delaware for collaboration with Argonne National Laboratory.

### Contributors

This section is authored by M. B. Jungfleisch (University of Delaware, USA), Y. Li and V. Novosad (Argonne National Laboratory, USA).

### H. Towards entangled magnons in hybrid quantum systems

The field of magnonics usually operates with coherent magnons at room temperature. The density of the so-called thermal magnons that are in equilibrium with a phononic bath of a solid body is around $10^{18}$ cm$^{-3}$. Thus, the operations with single magnons are not possible without using cryogenic techniques. The most straightforward estimation using Bose-Einstein distribution shows that the thermal magnon population at 10 GHz and 100 mT is around 0.01, and nowadays mT temperatures are readily accessible with commercial dilution refrigerators. The field of quantum magnonics already took its first but essential steps, as our colleagues in this section describe it. Compared to the area of quantum optics, which is already a well-established field of modern physics, magnonics offers a set of unique features inherent also to a usual room-temperature magnonics [20], [12]: scalability down to the atomic lattice scale, frequency range from GHz up to hundreds of THz, straightforward control of magnons by electric currents and fields, pronounced natural nonlinearity, and a manifold of nonreciprocal phenomena.

Moreover, hybrid quantum magnonics shows a promising path to bridge the gap between different quantum technologies. It offers the unique combination of spin-, phonon-, and photon- (including microwave) quantum systems and ultimately highly integrable hybrid quantum systems. Having achieved longer coherence times and lengths, integrated quantum magnonics will allow the coupling to other quantum systems, promising efficient interaction between the sub-systems and their mutual interactive characterization and examination [277]. Using machine learning algorithms [471], one may then exploit the coupling to find true quantum models of the integrated quantum magnonics circuits, typically obtained from empirical measurements and analytical models.

One of the major challenges facing quantum magnonics, in our opinion, is the transition from a single magnon in the form of a uniform precession described in Section VIII-L, to propagating single magnons. The comparatively large spin-wave loss is is a challenge which will limit the number of magnons reaching a detector region. Nevertheless, the recently-developed 50 nm-wide single-mode spin-wave waveguides offer relatively large propagation lengths of about 10 μm [31] which should be sufficient for quantum operations at the nanoscale. Thus, besides a set of technical questions, the main challenges will involve the realization of efficient single-magnon sources and detectors. The superconducting qubit approach described in Section VIII-L seems to be the most viable solution. Superconducting systems on the basis of nonlinear Josephson circuits with discrete energy spectra allow for the detection of microwave photons and magnons with single-quantum sensitivity. These systems are analogs of a single atom described with a discrete Hamiltonian and allow for efficient electromagnetic coupling between constituents in hybrid systems [472].

A further fundamental challenge in the field of quantum magnonics is the generation of entangled magnon states. Fortunately, there are already many well-studied physical mechanisms that generate such states. For example, Brillouin Light Scattering (BLS) generates two magnons that are entangled between themselves and with a photon, parallel parametric pumping couples two magnons with a microwave photon. Three-magnon splitting [16] generates two entangled magnons as well. The required powers and the efficiencies of all these phenomena will possibly be an eminent challenge. The realization of the entangled magnons will enable a variety of experimental investigations, such as quantum teleportation, quantum simulation, quantum computation and quantum sensing [473].

### Contributors

This section is authored by A. V. Chumak, S. Knauer, O. V. Dobrovolskiy (University of Vienna).

### I. Manipulating hybrid magnonic interactions

*Status.* Tuning the interaction between a magnonic device and another resonator is important for spin-wave computing. In its various forms of implementation, hybrid magnonics has exhibited great high flexibility and versatility, with a number of parameters can be used for manipulating the system dynamics. As shown by the simplified mode in Fig. VIII.H (a), where a magnon mode ($m$) is coupled with a resonance mode ($a$), the available tuning parameters include: the frequency detuning ($\Delta f$), coupling strength ($g$), and dissipation rates ($\kappa_a$ and $\kappa_m$).

The detuning frequency is the mostly widely used tuning parameter and is used in most hybrid magnonic experiments. This also represent the greatest advantage of hybrid magnonics – the highly tunable magnon frequency. Simply changing the strength of bias magnetic field for magnonic device will tune the magnon frequency over a broad range, making it extremely convenient to match the frequency of the other mode.

The coupling strength is another easily tunable parameter that strongly affects the interaction dynamics. It can be controlled by changing the system configuration [474], such as the relative position and size of the two coupled resonators, the geometry of the resonator design, the type of magnetic materials, the direction of the bias magnetic field, etc.

The dissipation rate also controls hybrid magnonic systems, associating hybrid magnonic interactions with different regimes such as strong coupling, Purcell effect, or magnetically induced transparency [474], and may lead to novel phenomena such as level attraction [475].

In addition, magnon states can be directly manipulated by microwave [424] or optical signals [476]. Moreover, specific properties of the resonator also affect the system dynamics. For instance, in hybrid magnon-microwave photon systems, the microwave polarization is a key parameter to be tuned to produce nonreciprocal behaviors [477].

*Current and future challenges.* With all the possible approaches for tuning the hybrid magnonic interactions, one major challenge is how to obtain real-time tuning. As of today, most experimentally demonstrated tunings in hybrid magnonics are static or semi-static. The tuning operations, such as changing the bias magnetic field (direction or strength) or modifying the device configuration, are usually on a time scale that is much longer than the lifetime of the magnon mode, which is no more than a few microseconds even in yttrium iron garnet (YIG), the material with lowest magnon damping.

In practical magnon-based computing applications, real-time tuning is desired, which requires a tuning operation to be finished within the finite magnon lifetime. A pulse signal that temporally changes the detuning frequency, coupling strength or dissipation rates will change the interaction dynamics and accordingly the final states of the system, as shown in Fig. VIII.G(b-d). However, in order to apply the available tuning mechanism to such transient response, a series of challenges will need to be addressed.

(1) Approaches for applying rapid magnetic pulses need to be developed, which is the most straightforward way of tuning the magnon frequency. It requires both large tuning range and a fast tuning speed, and therefore is highly challenging. Such principle has been recently demonstrated using an ultrasmall coil [478], but significant further improvements are required for practical applications. This approach can also be applied to resonance mode $a$ if magnetic tunability is available on the resonator [391].

(2) Another approach for detuning tuning is to utilize frequency-tunable resonators to couple with magnons. For instance, the resonator can be integrated with high-speed mechanical components [479] such as micro- or nano-electromechanical systems (MEMS or NEMS), tunable kirigami or origami structures, etc.

(3) Novel physics need to be explored to obtain tunable coupling strengths because the beam-splitter type interaction is fixed for a given device configuration. For example, Floquet engineering has been proven as a promising approach [480]. Moreover, utilizing the nonlinearity of the mode coupled with magnons, tunable parametric coupling can be obtained [481].

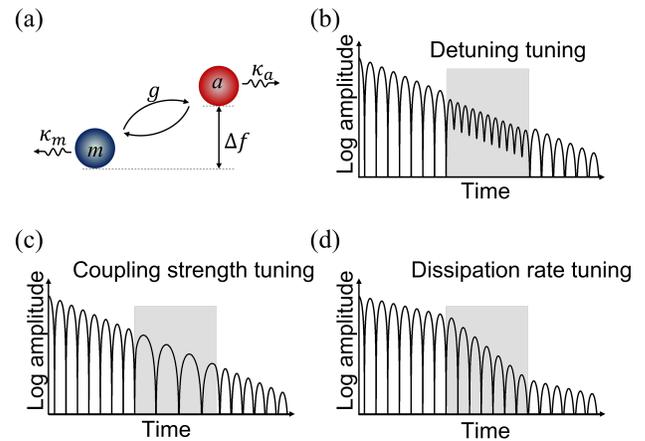

Fig. VIII.I. (a) List of available parameters for manipulating the interaction dynamics in hybrid magnonic systems. (b)-(d) Dynamics of magnon amplitude in a hybrid magnonic system where the detuning, coupling strength, and dissipation rate are temporarily tuned (shaded regions), respectively.

(4) New materials with novel tuning mechanisms for magnon (frequency, dissipation, etc) are desired. For instance, multiferroic, optomagnetic, or magnetoelastic materials can introduce novel electrical/optical/mechanical approaches for real-time tuning of the magnon dynamics, if novel low-loss materials with strong magnonic responses can be discovered.

*Concluding remarks.* Hybrid magnonics intrinsically possesses rich tunability thanks to its great versatility. Incorporating real-time dynamical controls will revolutionize hybrid magnonics research which is restricted to static or semi-static situations, allowing complicated gating operations that are at the core of modern signal processing. This points to a new pathway towards a broad range of promising spin-wave computing applications such as low-power computing and quantum signal processing.

### Acknowledgment

This part of the work was performed at the Center for Nanoscale Materials, a U.S. Department of Energy Office of Science User Facility, and was supported by the U.S. Department of Energy, Office of Science, under Contract DE-AC02-06CH11357.

### Contributors

This section is authored by X. Zhang (Argonne National Laboratory, USA).

### J. Realization of ultrastrong photon-to-magnon coupling with superconducting hybrids

Quantum magnonics is one of emerging subfields of hybrid quantum technologies [482]. By considering hybridization of photons with magnons quantum magnonics offers various novel approaches, including hybrid quantum platforms [391], magnon memory [483], and quantum transducers [484]. Developments in quantum magnonics are partially obstructed by a weak coupling strength between photons and magnons. In fact, the strong and the ultra-strong photon-to-magnon coupling regimes were originally reached in macroscopic cavity-based systems by utilizing the Dicke coupling relation, i.e., by introduction of a large number spins into the system with poor single-spin coupling strength (typically below 1 Hz). For practical realization on-chip quantum systems are highly desired where high single-spin coupling strength is required.

Recently realization of on-chip ultrastrong photon-to-magnon coupling was demonstrated using superconducting hybrid structure [485], [486]. The structure is based on superconductor-insulator-superconductor trilayer resonator where the photon phase velocity is suppressed by several orders of magnitude following the Swihart relation, and on the ferromagnetic layer placed between superconducting layers. By a corresponding reduction in dimensions of the resonator the electromagnetic volume becomes comparable with the ferromagnetic volume, thus ensuring efficient photon-to-magnon coupling. For instance, the following parameters of interaction have been reached: the coupling strength above 6 GHz, the single-spin coupling strength about 350 Hz, the coupling ratio about 0.6 [486]. Interestingly, with achieved coupling ratio the system approaches the so-called deep-strong coupling regime and allows to verify the interaction model. As it appears, the superconducting hybrid structure obeys the Hopfield light-matter interaction model instead of the commonly accepted Dicke model, which indicates plasmonic contribution into the energy of the hybrid system. The demonstrated technology offers new perspectives in quantum magnonics.

### Acknowledgment

I.A.G. acknowledges support by the Ministry of Science and Higher Education of the Russian Federation in the framework of the State Program (Project No. 0718-2020-0025).

### Contributors

This section is authored by I.A. Golovchanskiy (Moscow Institute of Physics and Technology, and National University of Science and Technology MISIS, Moscow, Russia).

### K. Quantum Interfaces between magnons and paramagnetic spins

Magnons are promising components for quantum technology due to their unique properties [178], such as tunable dispersion relations and large nonlinearities. Since these useful properties are absent in other platforms, harnessing them could unlock novel, flexible, magnon-based approaches to areas such as quantum information processing. In analogy with its photon-based counterpart, there are two potential approaches toward magnon-based quantum computing. First, magnon buses coupled to several information processing nodes, commonly qubits, can passively mediate the coupling and information exchange between them. Alternatively, magnons can act as flying qubits holding the quantum information. This information is processed on-the-fly via local operations enabled by controlling local strong non-linear elements, usually also qubits.

Both approaches require strong coherent coupling between propagating magnons and qubits. Solid-state paramagnetic spin impurities (PS), e.g. nitrogen-vacancy centres in diamond, are promising qubit candidates due to their long coherence times. They can be coherently manipulated and optically initialized at room temperature, thus being able to operate above cryogenic temperatures. Moreover, the high densities reachable by PS ensembles allow to concentrate many PS within a single magnon mode volume ($\sim\lambda^3$ with $\lambda$ the magnon wavelength), thereby collectively enhancing magnon-qubit coupling.

Motivated by applications in quantum technologies, recent works have focused on devising magnon-PS interfaces. Experiments have demonstrated the coupling between classical spin waves and single PS [120], high-density PS ensembles [124], and low-density ensembles with individually addressable PS [487]. Coupling to a PS ensemble has enabled spatially-resolved spin-wave detection by tracking the spin wave-induced change of the PS fluorescence. Moreover, the coupling between spin waves and PS has been shown to be coherent [125], paving the way toward the quantum regime. The development of a quantum theory describing magnon-PS interfaces [129] has enabled to explore this regime and provide several theoretical predictions. These include magnon-mediated entanglement generation between two PS [488], [130] and the use of high-density PS ensembles both to mold the propagation of quantum magnonic states and to detect magnonic quantum vacuum fluctuations [129].

The above results reveal promising prospects for magnon-PS interfaces in the context of quantum information processing. First, the flexibility of these interfaces could be an asset in networks where PS nodes (qubits) interact via mediating magnons. Such a „magnon reservoir" can be engineered with unprecedented flexibility using useful properties intrinsic to magnons, such as band minima at finite momentum or spectra tunable through external fields. This could enable novel, more efficient protocols for quantum computing and simulation, devised specifically for magnonic quantum networks. Moreover, dense PS ensembles can be used to control spin-wave propagation and thus have great potential to route and perform quantum operations on flying magnon qubits. Indeed, the quantum state of PS ensembles, and thus their linear and non-linear interaction with magnons, can be time-modulated and spatially patterned with sub-wavelength resolution (e.g. via spatially-resolved optical pumping). This will allow to externally imprint and erase the desired functionalities, from magnonic amplifiers to magnonic nonlinear gates, directly on the spin ensemble. Such functionalities can be optimized for the quantum regime using nanophotonic-inspired engineering. Devising coherent quantum repeaters and two-qubit gates would enable efficient and universal magnonic quantum computing. Exploration of these research lines could provide a magnonic analog – and much more flexible alternative – to photonic quantum computing.

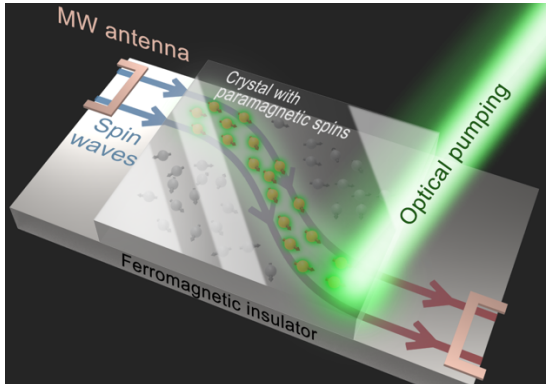

Fig. VIII.J. The coupling between spin waves and solid-state paramagnetic spin ensembles can be tuned in time domain and with sub-wavelength spatial resolution (e.g. via optical pumping), enabling flexible quantum interfaces. These allow, among others, to route and process quantum information stored in the magnons or in the paramagnetic spins.

The development of magnon-PS-based quantum information platforms faces several challenges. The main are the high losses (decoherence) of magnons as compared to e.g. photons. Low-temperature operation can mitigate decoherence but will hardly suffice to devise scalable devices. The design of error correction protocols adapted for magnonic platforms, such as bang-bang protocols for flying qubits or coherent spin control of PS, will be necessary. Whether these protocols can efficiently be implemented in magnon-PS interfaces remains an open question, whose answer requires a deep theoretical understanding of the sources of decoherence.

Even if decoherence is decreased by several orders of magnitude, quantum magnonic platforms would still be too lossy to scale up into fully magnonic quantum networks. Future platforms will likely exploit low-loss superconducting microwave buses to distribute quantum information between small, highly flexible, high-performance processing nodes based on magnon-PS interfaces. To realize such hybrid networks, it is key to devise methods to transduce quantum states between a propagating microwave photon and a propagating magnon with high fidelity. This quantum engineering challenge might require recruiting other degrees of freedom, e.g. phonons [489], [490]. Ultimately, the potential of magnon-PS interfaces for quantum information processing will only be attested by experimental observation of quantum coherence in a magnon-PS system. This will require to develop novel tools in magnonics such as reproducible methods to generate quantum magnonic states or to certify their quantum nature (e.g. quantum state tomography).

In summary, magnon-PS interfaces could provide a flexible quantum information processing platform, incorporating novel and powerful designs for devices and protocols. Realizing these platforms requires input from nanophotonics, quantum optics, and material science. The increasing research on magnon-PS interfaces shows the promising potential of a nascent field: hybrid quantum spin-magnonics.

## CONTRIBUTORS

This section is authored by C. Gonzalez-Ballestero and O. Romero-Isart (University of Innsbruck and IQOQI).

### L. How to distill entanglement in quantum magnonics

In a spin-singlet state the projection of one of the two spins dictates the projection of the other spin, even when they are spatially separated and non-interacting. This spooky action at a distance is an example of quantum entanglement of the 'discrete variable' type, i.e. between two systems with finite Hilbert spaces. 'Continuous variable' entanglement exists in infinite Hilbert spaces such as position and momentum of harmonic oscillators (HO). The vanishing uncertainty of the relative position and total momentum of two HO indicates 'continuous variable' quantum entanglement. Entanglement is necessary for quantum communication, teleportation, and quantum encryption. The advantages of continuous over discrete variable entanglement are deterministic rather than probabilistic and heralded observation, as well as the relative simplicity to observe electric/magnetic field amplitudes instead of photon "clicks", especially at microwave frequencies.

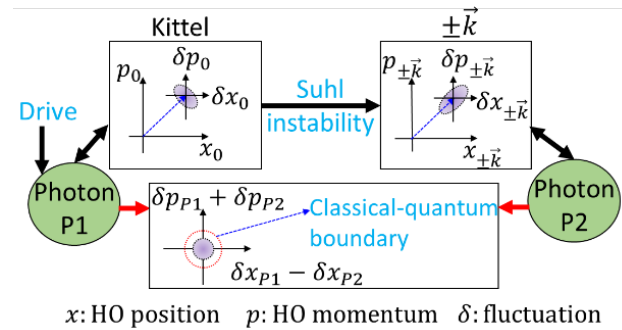

Fig. VIII.L. Continuous variable entanglement in the phase space of Kittel mode and a $\pm\vec{k}$ magnon pair, accessed by two photon modes. The shaded ellipse in the bottom panel indicates the uncertainty that vanishes in fully entangled states.

Lachance-Quirion et al. [391] demonstrated discrete variable entanglement of the lowest number (Fock) states of Kittel mode magnons with a qubit mediated by photons in a high quality cavity. Magnons are superior to other bosonic quasiparticles by their anisotropic dispersion and various types of nonlinearities. The dissipative dynamics of driven magnon systems leads to phenomena such as Bose condensation and an entanglement that does not require auxiliary systems such as a superconducting qubit [491]. When driving a Kittel mode beyond the threshold of a Suhl instability, 4-magnon scattering excites a degenerate $\pm\vec{k}$ magnon pair. We theoretically showed that in a steady state in which Kittel and $\pm\vec{k}$ modes are fixed points of the dynamics in phase space, interactions "squeeze" the fluctuations and generate continuous variable entanglement between them at low (but not ultralow) temperatures (see Fig. VIII.L). The partial entanglement can be transformed into a pure one such as in a spin singlet state, so it is "distillable" and therefore useful for quantum information purposes. In order to measure, distill, and implement the entanglement, we need two distinct microwaves sources, i.e. $\vec{k} \rightarrow 0$ cavity photons that drive and probe the Kittel mode, and spatially modulated ones that couple directly to the $\pm\vec{k}$ pair (Fig. VIII.L). Distillation protocols based on local operations and classical communications can then transform the partial entanglement of the output to perfect entanglement bits for quantum information by, e.g., lattices of interacting magnets.

The bistability of the "magnon parametron", a parametrically driven magnetic disk, can be used as an Ising spin memory and stochastic "p-bit" for probabilistic computations [445]. The entanglement principle sketched above for resonantly excited magnets works as well for the magnon parametron [492] at even lesser drive powers. The magnetoelastic interaction helps to generate [493], transfer, and manipulate magnon-magnon entanglement via phonons. Analogous to atoms in Bose condensate and photonic solitons, magnons can be entangled in real rather than reciprocal-space in interacting magnon condensates or soliton wave functions whose fluctuations are squeezed by four magnon interactions [425].

The developments sketched above provide a clear roadmap to realize magnon entanglement in all-magnetic circuits and devices that execute quantum encryption and other protocols.

### Acknowledgment

We acknowledge support by JSPS KAKENHI Grants with Nos. 19H00645, and 21K13847.

### Contributors

This section is authored by M. Elyasi (Tohoku University, Japan), Y. M. Blanter (TU Delft, The Netherlands), and G. E. W. Bauer (Tohoku University, Japan).

### M. Quantum sensing of magnons with superconducting qubits

Magnons, the quanta of spin-wave excitations, are the mainstay concept of magnonics, yet, their genuine quantum natures have rarely been revealed. The generation, manipulation, and detection of non-classical magnon states are envisioned as some of the ultimate feats in magnonics that remain to be realized. Employing techniques developed in circuit quantum electrodynamics (circuit QED) for manipulating microwave photons with superconducting qubits [494], [495], however, we are now on the verge of realizing these feats. In the quantum sensing frontier, major progress has already been achieved exploiting the magnonic analogue of the circuit QED architecture.

The schematic representation of the architecture [496], [178] is shown in Fig. VIII.M. Here the uniform magnetostatic mode, or the Kittel mode, of a single-crystal spherical YIG sample and a transmon-type superconducting qubit of frequencies $\omega_m$ and $\omega_q$ are coupled to a common microwave cavity mode of frequency $\omega_c$ through magnetic and electric dipole interactions of coupling strengths $g_{m\text{-}c}$ and $g_{q\text{-}c}$, respectively. This leads to a second-order effective coupling between the magnetostatic mode and the superconducting qubit of coupling strength $g_{q\text{-}m}$. The strong coupling regime is reached when $g_{q\text{-}m}$ is larger than both magnon and qubit linewidths $\gamma_m$ and $\gamma_q$, respectively, and has been experimentally realized in this system [481]. When the qubit and the Kittel mode are detuned, the coupling between them becomes off-resonant, suppressing direct energy exchange, yet produces a state-dependent frequency shift in either the qubit or Kittel mode, quantified by the dispersive shift $2\chi_{q\text{-}m}$. Experimentally, the strong dispersive regime has been achieved, demonstrating the qubit spectrum splitting according to the magnon Fock states in the Kittel mode [497].

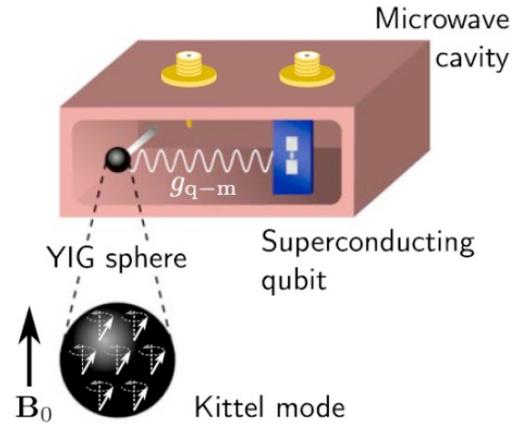

Fig. VIII.M. Schematic representation of the architecture to demonstrate the strong coupling between the uniform magnetostatic mode, or Kittel mode, of a YIG sphere and a transmon-type superconducting qubit. The frequency of the magnons in the Kittel mode is tunable through the amplitude of the external magnetic field B_0. Both systems are placed inside a multimode three-dimensional microwave cavity, leading to an effective qubit--magnon coupling. Reprinted (adapted) from [178]. Copyright (2019) The Japan Society of Applied Physics.

Through the strong dispersive interaction, the qubit can be excited conditionally on the number of magnons in the Kittel mode. Leveraging this capability together with a high-fidelity qubit readout technique, the single-shot detection of a single magnon was demonstrated [391]. The quantum efficiency of up to 0.71 was reached, primarily limited by the qubit decoherence. The dark count probability was 0.24, and was further reduced to 0.12 with an alternative protocol. The strong dispersive interaction can also be exploited for monitoring how many magnons are excited by a continuous process. Employing the so-called $T_2^*$ relaxometry with the Ramsey interrogation scheme for a qubit [498], a quantum sensing of a steady-state magnon population was performed [499]. The demonstrated magnon detection sensitivity was $1.55\times10^{-3}$ magnons/$\sqrt{(\text{Hz})}$, which is equivalent to a magnetic field sensitivity of $1.35\times10^{-15}$ T/$\sqrt{(\text{Hz})}$. Here, the noise primarily stemmed from the quantum projection noise in measuring the qubit state, which is either in the ground state or the excited state.

Among other things, the qubit $T_2^*$, primarily limited by conductive losses in the copper microwave cavity, constitutes both a dominant source of error in the single-shot single magnon detection and a bottleneck of the sensitivity for the $T_2^*$ relaxometry. Improvements in this vein must reconcile with the need to apply a strong external magnetic field $B_0$ (typically 0.3 T) to the YIG sphere from outside the cavity, preventing a simple replacement of the cavity material with a superconductor such as aluminium. To resolve this, use of a cavity made of a low-loss dielectric material such as rutile would be an option. With a large dielectric constant, a rutile cavity can also boost the magnetic field confinement within the cavity and therefore the coupling between the microwave cavity mode and the Kittel mode. With a microwave cavity of higher quality factor and a qubit with an improved $T_2^*$, non-classical magnon states such as Fock states and Schrödinger's cat states can be generated and controlled.

Finally, to be a little bit more ambitious, let us ask ourselves whether the quantum magnonics architecture can contribute to quantum information technologies. More specifically, can magnons be used as a quantum information carrier and play a similar role to microwave photons [495] in this context? The answer hinges on how much we can increase the magnon lifetime $\tau_m = 1/\gamma_m$. In order to compete with microwave photons in a high-quality superconducting cavity, the magnon lifetime has to be improved by at least two orders of magnitude from the current best value in YIG spheres, which is of the order of 100 ns. Even if it is hard to compete with microwave photons directly, magnons can potentially offer advantages in auxiliary roles that supplement photonic elements of the quantum information pipeline, such as microwave-to-optical transduction [484]. In any case, there is certainly a demand for new ideas!

### ACKNOWLEDGMENT

This work is supported by JSPS KAKENHI (18F18015), and JST ERATO (JPMJER1601).

### CONTRIBUTORS

This section is authored by K. Usami and S. P. Wolski (University of Tokyo, Japan), D. Lachance-Quirion (Nord Quantique, Canada), Y. Nakamura (University of Tokyo and RIKEN, Japan).

### N. Towards tunable, dissipative quantum magnonics: Two-tone cavity-magnon polariton spectroscopy and controlled on-chip coupling

As part of cavity magnonics, quantum magnonics comprises different hybridized physical systems which includes the manipulation and control over single magnons, i.e., the quanta of spin waves, and the coherent coupling (direct or indirect) to linear and nonlinear physical systems such as microwave resonators and qubits. The associated polaritons emerging from the strong coupling describe a coherent exchange of information between all involved systems. Its hallmark is the observation of avoided level crossings. Coherent coupling is at the heart of all applications of quantum magnonics for quantum information processing. Experimentally, cavity magnonics only emerged in the past decade, and the observation of the quantum regime has been realised only within bulk three-dimensional (3D) systems. Among others, pivotal experiments included the observation of a coherent but fixed coupling of cavity photons to magnons in the quantum regime at millikelvin and in the classical regime at room temperatures [416], [474], and the coupling of qubits based on superconducting circuits to magnons via the cavity photons [481]. However, in terms of decoherence, these qubits are very sensitive to a direct coupling to the environment and magnetic fields [500]. In turn, magnons require both static and time-dependent magnetic fields to be excited and tuned. Thus, magnons and qubits (or other nonlinear elements) are usually physically separated and indirectly coupled via cavity resonator photons. However, for a robust, large-scale integration benchmarking against CMOS-based technologies, at least two key challenges remain. *First*, downscaling from the predominantly bulk, quantum magnonic architectures is inevitable for a full on-chip realization of this form of quantum-based information processing. *Second*, a dynamic control over the coherent information exchange is necessary for tunable and nonreciprocal realizations for enhanced sensing applications or computing capabilities. To date, the physical mechanisms, and implementations of individual magnon-photon and qubit-photon couplings are well understood [461]. However, in most experiments the coherent coupling strength remains mostly static. Further, despite experimental progress [475], [408], the implementation of quantum magnonics in 2D remains elusive as well. In parallel, the emergence of dissipative cavity magnonics has opened new paths towards the active control of the coupling strength but it has not been applied to quantum magnonics yet. Among other dissipative approaches [408], the dynamical control of the coupling strength using microwave tones is most appealing as it does not require any internal modification of the system. In the two-tone approach, the split microwave source signal steers individually the magnon and the cavity photons. A variable phase shifter and attenuator in one path are used to modify the net cavity photon-magnon coupling and hence modify the effective magnon-qubit coupling via the cavity photon coupling. Experimentally, the key ingredients for overcoming both challenges simultaneously are already at hand but need to be combined. The control of the coherent interaction via the above two-tone approach has successfully been demonstrated by using a bulk magnonic system [422] and strongly coupled cavity photon-magnon states could be heavily scaled-down towards thin film, 2D architectures as well [464], [408]. Thus, as a first step, the static coherent coupling in 2D needs now to be controlled by the above two-tone approach which means harnessing dissipative cavity magnonics in micro- or nanostructured circuits. Such on-chip hybrid magnonic systems could be based e.g. on microstructured magnets precision-placed on microwave circuits [408] or multilayered microstructures of superconducting, insulating, and ferromagnetic layers and harnessing modified photon phase velocities and magnon eigenfrequencies [485], and integrated local field drives. Their enhanced coupling strength is provided by the radically reduced photon-mode volume relative to bulk microwave cavities.

The ubiquity of coherent and dissipative interactions in nature based on light and matter (quasi) particles could allow adding various types of nonlinear systems such as skyrmions, 2D electron quantum gases or quantum dots for new types of quantum magnonics beyond the utilisation in superconducting circuits. In addition, the introduction of dissipative coupling will establish pathways to open quantum systems and associated non-hermiticity in quantum magnonics. Such non-Hermitian quantum hybrid systems could, for instance, enable the realization of controlled entangled hybrid quantum systems, and lead to completely new, exciting avenues for quantum information communication, sensing and processing with magnonic circuits.

## Acknowledgements

This work has been supported from the European Union's Horizon 2020 research and innovation program within the FET-OPEN project CHIRON under grant agreement No. 801055. Financial support of the European Research Council within the Consolidator Grant 648011 'QuantumMagnonics' is gratefully acknowledged.

## Contributors

This section is authored by I. Boventer, R. Lebrun, P. Bortolotti and A. Anane (Unité Mixte de Physique CNRS/Thales, Palaiseau, France), and M. Weides (University of Glasgow, UK).

## IX. Conclusion

This roadmap is constructed in a self-assembled way and reflects the essential current and future challenges in magnonics. In the 60 sub-sections, the authors report recent achievements on the path towards magnon-based computing and share their vision on the future evolution of the field. The pronounced interest of the community in this roadmap, which is reflected by a large number of contributors, and the extensive scope of the scientific and technological problems covered by the article, clearly illustrate a very positive dynamic in the development of the field. An overview of this roadmap suggests that the field of magnonics has reached the state in which the amount of the physical phenomena discovered and the material and device fabrication techniques developed is sufficiently large to switch the gear from the fundamental research to engineering and applied research in the near future.

## References

[1] F. Bloch, "Zur Theorie des Ferromagnetismus," *Zeitschrift für Phys.*, vol. 61, no. 3–4, 1930, doi: 10.1007/BF01339661.

[2] V. V Kruglyak, S. O. Demokritov, and D. Grundler, "Magnonics," *J. Phys. D. Appl. Phys.*, vol. 43, no. 26, p. 264001, 2010, doi: 10.1088/0022-3727/43/26/264001.

[3] A. A. Serga, A V Chumak, and B Hillebrands, "YIG magnonics," *J. Phys. D Appl. Phys*, vol. 43, no. 26, p. 264002, 2010, doi: 10.1088/0022-3727/43/26/264002.

[4] B. Lenk, H. Ulrichs, F. Garbs, and M. Münzenberg, "The building blocks of magnonics," *Phys. Rep*, vol. 507, p. 107136, 2011.

[5] B. Dieny *et al.*, "Opportunities and challenges for spintronics in the microelectronics industry," *Nature Electronics*, vol. 3, no. 8. Nature Research, pp. 446–459, Aug. 2020, doi: 10.1038/s41928-020-0461-5.

[6] A. V Chumak, V. I. Vasyuchka, A. A. Serga, and B. Hillebrands, "Magnon spintronics," *Nat. Phys.*, vol. 11, no. 6, pp. 453–461, 2015, doi: 10.1038/nphys3347.

[7] J. M. Owens, J. H. Collins, and R. L. Carter, "System applications of magnetostatic wave devices," *Circ., Syst. Sign. Proc*, vol. 4, p. 317, 1985.

[8] J. D. Adam, "Analog signal processing with microwave magnetics," *Proc. IEEE*, vol. 76, p. 159, 1988.

[9] I. V Zavislyak and M. A. Popov, "Microwave properties and applications of yttrium iron garnet (YIG) films: current state of art and perspectives," in *Yttrium: Compounds, Production and Applications*, B. D. Volkerts, Ed. 2009, pp. 87–125.

[10] V. G. Harris, "Modern microwave ferrites," *IEEE Trans. Magn.*, vol. 48, no. 3, 2012, doi: 10.1109/TMAG.2011.2180732.

[11] A. Mahmoud *et al.*, "An Introduction to Spin Wave Computing," *J. Appl. Phys.*, vol. 128, no. 16, p. 161101, 2020, doi: 10.1063/5.0019328.

[12] A. Barman *et al.*, "The 2021 Magnonics Roadmap," *J. Phys. Condens. Matter*, vol. 33, p. 413001, 2021.

[13] V. E. Demidov and S. O. Demokritov, "Magnonic waveguides studied by microfocus brillouin light scattering," *IEEE Trans. Magn.*, vol. 51, no. 4, 2015, doi: 10.1109/TMAG.2014.2388196.

[14] T. Sebastian, K. Schultheiss, B. Obry, B. Hillebrands, and H. Schultheiss, "Micro-focused Brillouin light scattering: imaging spin waves at the nanoscale (Review paper)," *Front. Phys*, vol. 3, p. 35, 2015.

[15] P. Pirro, V. Vasychka, A. Serga, and B. Hillebrands, "Advances in coherent magnonics," *Nat. Rev. Mat*, 2021.

[16] A. G. Gurevich and G. A. Melkov, *Magnetization oscillations and waves*. Boca Raton: CRC Press, 1996.

[17] D. D. Stancil and A. Prabhakar, *Spin Waves: Theory and applications*. Springer, 2009.

[18] M. Krawczyk and D. Grundler, "Review and prospects of magnonic crystals and devices with reprogrammable band structure," *J. Phys. Condens. Matter*, vol. 26, no. 12, 2014, doi: 10.1088/0953-8984/26/12/123202.

[19] A. V. Chumak, A. A. Serga, and B. Hillebrands, "Magnonic crystals for data processing," *J. Phys. D. Appl. Phys.*, vol. 50, no. 24, p. 244001, 2017, doi: 10.1088/1361-6463/aa6a65.

[20] A. V Chumak, "Magnon spintronics: Fundamentals of magnon-based computing," in *Spintronics Handbook: Spin Transport and Magnetism*, 2nd ed., CRC Press, 2019, pp. 247–302.

[21] M. Pardavi-Horvath, "Microwave applications of soft ferrites," *J. Magn. Magn. Mater*, vol. 215–216, pp. 171–183, 2000.

[22] V. G. Harris *et al.*, "Recent advances in processing and applications of microwave ferrites," *J. Magn. Magn. Mater*, vol. 321, pp. 2035–2047, 2009.

[23] J. Ding, T. Liu, H. Chang, and M. Wu, "Sputtering growth of low-damping yttrium iron garnet thin films," *IEEE Magn. Lett*, vol. 11, p. 5502305, 2020.

[24] J. Ding *et al.*, "Nanometer-thick yttrium iron garnet films with perpendicular anisotropy and low damping," *Phys. Rev. Appl*, vol. 14, p. 14017, 2020.

[25] Q. Wang *et al.*, "Spin Pinning and Spin-Wave Dispersion in Nanoscopic Ferromagnetic Waveguides," *Phys. Rev. Lett.*, vol. 122, no. 24, p. 247202, 2019, doi: 10.1103/PhysRevLett.122.247202.

[26] B. Heinz *et al.*, "Propagation of Spin-Wave Packets in Individual Nanosized Yttrium Iron Garnet Magnonic Conduits," *Nano Lett.*, vol. 20, no. 6, p. 4220, 2020, doi: 10.1021/acs.nanolett.0c00657.

[27] E. A. Giess, "Liquid phase epitaxy of magnetic garnets," *J. Cryst. Growth*, vol. 31, pp. 358–365, 1975.

[28] G. Winkler, "Magnetic Garnets," in *Vieweg tracts in pure and applied physics*, vol. 5, Braunschweig/Wiesbaden, 1981.

[29] N. Beaulieu *et al.*, "Temperature dependence of magnetic properties of a ultrathin yttrium-iron garnet film grown by liquid phase epitaxy: Effect of a Pt overlayer," *IEEE Magn. Lett*, vol. 9, p. 3706005, 2018.

[30] C. Dubs *et al.*, "Low damping and microstructural perfection of sub-40nm-thin yttrium iron garnet films grown by liquid phase epitaxy," *Phys. Rev. Mater.*, vol. 4, no. 2, 2020, doi: 10.1103/PhysRevMaterials.4.024416.

[31] B. Heinz *et al.*, "Long-range spin-wave propagation in transversely magnetized nano-scaled conduits," *Appl. Phys. Lett*, vol. 118, p. 132406, 2021.

[32] M. Mohseni *et al.*, "Controlling the Nonlinear Relaxation of Quantized Propagating Magnons in Nanodevices," *Phys. Rev. Lett*, vol. 126, p. 97202, 2020.

[33] Q. Wang *et al.*, "A magnonic directional coupler for integrated magnonic half-adders," *Nat. Electron.*, 2020, doi: 10.1038/s41928-020-00485-6.

[34] S. Maendl, J. Stasinopoulos, and D. Grundler, "Spin waves with large decay length and few 100 nm wavelengths in thin yttrium iron garnet grown at the wafer scale," *Appl. Phys. Lett*, vol. 111, p. 12403, 2017.

[35] P. Che, K. Baumgaertl, A. Kúkol'ová, C. Dubs, and D. Grundler, "Efficient wavelength conversion of exchange magnons below 100 nm by magnetic coplanar waveguides," *Nat. Commun.*, vol. 11, no. 1, 2020, doi: 10.1038/s41467-020-15265-1.

[36] L. Sheng, J. Chen, H. Wang, and H. Yu, "Magnonics Based on Thin-Film Iron Garnets," *J. Phys. Soc. Jap*, vol. 90, p. 81005, 2021.

[37] S. Geprägs *et al.*, "Origin of the spin Seebeck effect in compensated ferrimagnets," *Nat. Commun*, vol. 7, no. 10452, 2016.

[38] E. Ramos, R. and Hioki, T. and Hashimoto, Y. and Kikkawa, T. and Frey, P. and Kreil, A.J.E. and Vasyuchka, V.I. and Serga, A.A. and Hillebrands, B. and Saitoh, "Room temperature and low-field resonant enhancement of spin Seebeck effect in partially compensated magnets," *Nat. Commun.*, vol. 10, no. 1, p. 5162, 2019.

[39] J. J. Carmiggelt, O. C. Dreijer, C. Dubs, O. Surzhenko, and T. van der Sar, "Electrical spectroscopy of the spin-wave dispersion and bistability in gallium-doped yttrium iron garnet," *arXiv:2109.05045*, 2021.

[40] J. Finley and L. Liu, "Spintronics with compensated ferrimagnets," *Appl. Phys. Lett.*, vol. 116, p. 11050, 2020, doi: 10.1063/1.5144076.

[41] J. Xu *et al.*, "Quantum Spin-Wave Materials, Interface Effects and Functional Devices for Information Applications," *Front. Mater*, vol. 7, no. 594386, 2020.

[42] S. O. Demokritov and V. E. Demidov, "Micro-brillouin light scattering spectroscopy of magnetic nanostructures," *IEEE Trans. Magn.*, vol. 44, no. 1, pp. 6–12, 2008, doi: 10.1109/TMAG.2007.910227.

[43] P. Clausen *et al.*, "Mode conversion by symmetry breaking of propagation spin waves," *Appl. Phys. Lett*, vol. 99, p. 162505, 2011.

[44] M. A. W. Schoen *et al.*, "Ultra-low magnetic damping of a metallic ferromagnet," *Nat. Phys*, vol. 12, p. 839, 2016.

[45] D. A. Smith *et al.*, "Magnetic Damping in Epitaxial Fe Alloyed with Vanadium and Aluminum," *Phys Rev. Appl*, vol. 14, p. 34042, 2020.

[46] M. Arora *et al.*, "Magnetic Damping in Polycrystalline Thin-Film Fe-V Alloys," *Phys. Rev. Appl*, vol. 15, p. 54031, 2021.

[47] C. Guillemard *et al.*, "Ultralow Magnetic Damping in Co2Mn Based Heusler Compounds: Promising Materials for Spintronics," *Phys. Rev. Appl*, vol. 11, p. 64009, 2019.

[48] J. M. Shaw *et al.*, "Magnetic damping in sputter-deposited Co2MnGe Heusler compounds with A2, B2, and L21 orders: Experiment and theory," *Phys. Rev. B*, vol. 97, p. 94420, 2018.

[49] L. Flacke *et al.*, "High spin-wave propagation length consistent with low damping in a metallic ferromagnet," *Appl. Phys. Lett*, vol. 115, p. 122402, 2019.

[50] H. S. Körner *et al.*, "Magnetic damping in poly-crystalline Co25Fe75: Ferromagnetic resonance vs. spin wave propagation experiments," *Appl. Phys. Lett*, vol. 111, p. 132406, 2017.

[51] L. Flacke, "Robust formation of nanoscale magnetic skyrmions in easy-plane thin film multilayers with low damping," *arXiv:2102.11117*, 2021.

[52] S. H. Skjærvø, C. H. Marrows, R. L. Stamps, and L. J. Heyderman, "Advances in artificial spin ice," *Nat. Rev. Phys*, vol. 2, pp. 13–2, 2020.

[53] M. B. Jungfleisch *et al.*, "Dynamic response of an artificial square spin ice," *Phys. Rev. B*, vol. 93, p. 100401, 2016.

[54] M. T. Kaffash, S. Lendinez, and M. B. Jungfleisch, "Nanomagnonics with artificial spin ice," *Phys. Lett. A*, vol. 402, p. 127364, 2021.

[55] S. Gliga, E. Iacocca, and O. G. Heinonen, "Dynamics of reconfigurable artificial spin ice: toward magnonic functional materials," *APL Mater.*, vol. 8, p. 040911, 2020.

[56] S. Lendinez, M. T. Kaffash, and M. B. Jungfleisch, "Emergent Spin Dynamics Enabled by Lattice Interactions in a Bicomponent Artificial Spin Ice," *Nano Lett*, vol. 21, pp. 1921–1927, 2021.

[57] J. C. Gartside *et al.*, "Reconfigurable magnonic mode-hybridisation

[58] E. Iacocca, S. Gliga, and O. G. Heinonen, "Tailoring Spin-Wave Channels in a Reconfigurable Artificial Spin Ice," *Phys. Rev. Appl*, vol. 13, p. 044047, 2020.

[59] V. S. Bhat *et al.*, "Magnon Modes of Microstates and Microwave-Induced Avalanche in Kagome Artificial Spin Ice with Topological Defects," *Phys. Rev. Lett*, vol. 125, p. 11720, 2020.

[60] A. Haldar and A. O. Adeyeye, "Functional magnetic waveguides for magnonics," *Appl. Phys. Lett*, vol. 119, no. 6, p. 60501, 2021.

[61] H. A. Haldar and A. O. Adeyeye, "Reconfigurable and self-biased magnonic metamaterial," *J. Appl. Phys*, vol. 128, no. 24, p. 240902, 2020.

[62] H. A. Haldar, D. Kumar, and A. O. Adeyeye, "A reconfigurable waveguide for energy-efficient transmission and local manipulation of information in a nanomagnetic device," *Nat. Nanotechnol*, vol. 11, no. 5, pp. 437–443, 2016.

[63] H. A. Haldar and A. O. Adeyeye, "Deterministic control of magnetization dynamics in reconfigurable nanomagnetic networks for logic applications," *ACS Nano*, vol. 10, no. 1, pp. 1690–1698, 2016.

[64] H. A. Haldar and A. O. Adeyeye, "Microwave assisted gating of spin wave propagation," *Appl. Phys. Lett*, vol. 116, no. 16, p. 162403, 2020.

[65] A. Haldar, C. Tian, and A. O. Adeyeye, "Isotropic transmission of magnon spin information without a magnetic field," *Sci. Adv*, vol. 3, no. 7, p. 1700638, 2017.

[66] J. M. Winter, "Bloch Wall Excitation. Application to Nuclear Resonance in a Bloch Wall," *Phys. Rev*, vol. 124, p. 452, 1961.

[67] F. G. Aliev, A. A. Awad, D. Dieleman, A. Lara, V. Metlushko, and K. Y. Guslienko, "Localized domain-wall excitations in patterned magnetic dots probed by broadband ferromagnetic resonance," *Phys. Rev. B*, vol. 84, p. 144406, 2011.

[68] F. Garcia-Sanchez, P. Borys, R. Soucaille, J.-P. Adam, R. L. Stamps, and J.-V. Kim, "Narrow magnonic waveguides based on domain walls," *Phys. Rev. Lett*, vol. 114, p. 247206, 2015.

[69] K. Wagner, A. Kákay, K. Schultheiss, A. Henschke, T. Sebastian, and H. Schultheiss, "Magnetic domain walls as reconfigurable spin-wave nanochannels," *Nat. Nanotech*, vol. 11, p. 432, 2016.

[70] V. Sluka *et al.*, "Emission and propagation of 1D and 2D spin waves with nanoscale wavelengths in anisotropic spin textures," *Nat. Nanotechnol.*, vol. 14, no. 4, pp. 328–333, 2019, doi: 10.1038/s41565-019-0383-4.

[71] A. Lara, "Information processing in patterned magnetic nanostructures with edge spin waves," *Sci. Rep.*, vol. 7, p. 5597, 2017.

[72] A. Lara, "Observation of propagating edge spin waves modes," *J. Appl. Phys*, vol. 114, p. 213905, 2013.

[73] Z. Zhang, M. Vogel, M. B. Jungfleisch, A. Hoffmann, Y. Nie, and V. Novosad, "Tuning edge-localized spin waves in magnetic microstripes by proximate magnetic structures," *Phys. Rev. B*, vol. 100, p. 174434, 2019.

[74] P. Gruszecki, I. L. Lyubchanskii, K. Y. Guslienko, and M. Krawczyk, "Local non-linear excitation of sub-100 nm bulk-type spin waves by edge-localized spin waves in magnetic films," *Appl. Phys. Lett*, vol. 118, p. 62408, 2021.

[75] E. Albisetti *et al.*, "Nanopatterning reconfigurable magnetic landscapes via thermally assisted scanning probe lithography," *Nat. Nanotech*, vol. 11, p. 545, 2016.

[76] N. Sato *et al.*, "Domain Wall Based Spin-Hall Nano-Oscillators," *Phys. Rev. Lett*, vol. 123, p. 57204, 2019.

[77] W. F. Brown, *Micromagnetics*. New York, London: J. Wiley, 1963.

[78] M. J. Donahue, *OOMMF user's guide: version 1.0., US Department of Commerce*. National Institute of Standards and Technology, 1999.

[79] D. Suess *et al.*, "Time resolved micromagnetics using a preconditioned time integration method," *J. Mag. Mag. Mat*, vol. 248, pp. 298–311, 2002.

[80] C. Abert, "Micromagnetics and spintronics: Models and numerical methods," *Eur. Phys. J. B*, vol. 92, no. 6, p. 1–45, 2019.

[81] A. Kakay, E. Westphal, and R. Hertel, "Speedup of fem micromagnetic simulations with graphical processing units," *IEEE Trans. Magn.*, vol. 46, no. 6, pp. 2303–2306, 2010.

[82] R. Chang, S. Li, M. Lubarda, B. Livshitz, and V. Lomakin, "Fastmag: Fast micromagnetic simulator for complex magnetic

structures," *J. Appl. Phys*, vol. 109, no. 7, p. 07 358, 2011.

[83] A. Vansteenkiste, J. Leliaert, M. Dvornik, M. Helsen, F. Garcia-Sanchez, and B. V. Waeyenberge, "The design and verification of mumax3," *AIP Adv.*, vol. 4, no. 10, pp. 107–133, 2014.

[84] A. Kovács *et al.*, "Defects in paramagnetic Co-doped ZnO films studied by transmission electron microscopy," *J. Appl. Phys.*, vol. 114, no. 24, 2013, doi: 10.1063/1.4851015.

[85] M. D'Aquino, C. Serpico, G. Miano, and C. Forestiere, "A novel formulation for the numerical computation of magnetization modes in complex micromagnetic systems," *J. Comput. Phys.*, vol. 228, 17, pp. 130–6149, 2009.

[86] F. Bruckner, M. D'Aquino, C. Serpico, C. Abert, C. Vogler, and D. Suess, "Large scale finite-element simulation of micromagnetic thermal noise," *J. Magn. Magn. Mater*, vol. 475, pp. 408–414, 2016.

[87] S. Perna, F. Bruckner, C. Serpico, D. Suess, and M. D'Aquino, "Computational micromagnetics based on normal modes: Bridging the gap between macrospin and full spatial discretization," *arXiv:2105.08829*, 2021.

[88] W. Scholz, J. Fidler, T. Schrefl, D. Suess, H. Forster, and V. Tsiantos, "Scalable parallel micromagnetic solvers for magnetic nanostructures," *Comput. Mater. Sci*, vol. 28, no. 2, pp. 366–383, 2003.

[89] T. Fischbacher, M. Franchin, G. Bordignon, and H. Fangohr, "A systematic approach to multiphysics extensions of finite-element-based micromagnetic simulations: Nmag," *IEEE Trans. Magn*, vol. 43, no. 6, pp. 2896–2898, 2007.

[90] C. Abert *et al.*, "A self-consistent spin-diffusion model for micromagnetics," *Sci. Rep*, vol. 6, no. 1, pp. 1–7, 2016.

[91] Q. Wang, A. V Chumak, and P. Pirro, "Inverse-design magnonic devices," *Nat. Commun.*, vol. 12, no. 1, p. 2636, 2021, doi: 10.1038/s41467-021-22897-4.

[92] M. Kostylev, V. E. Demidov, U.-H. Hansen, and S. O. Demokritov, "Nonlinear mode conversion in monodomain magnetic squares," *Phys. Rev. B - Condens. Matter Mater. Phys.*, vol. 76, no. 22, 2007, doi: 10.1103/PhysRevB.76.224414.

[93] M. Bailleul, D. Olligs, and C. Fermon, "Micromagnetic phase transitions and spin wave excitations in a ferromagnetic stripe," *Phys. Rev. Lett*, vol. 91, p. 137204, 2003.

[94] F. Vanderveken, M. Heyns, B. Sorée, C. Adelmann, and F. Ciubotaru, "Excitation and propagation of spin waves in non-uniformly magnetized waveguides," *J. Phys. D Appl. Phys*, vol. 53, p. 495006, 2019.

[95] G. Talmelli *et al.*, "Electrical spin-wave spectroscopy in nanoscale waveguides with nonuniform magnetization," *Appl. Phys. Lett*, vol. 118, p. 152410, 2021.

[96] H. Yu *et al.*, "Approaching soft X-ray wavelengths in nanomagnet-based microwave technology," *Nat. Commun.*, vol. 7, 2016, doi: 10.1038/ncomms11255.

[97] S. Wintz *et al.*, "Magnetic vortex cores as tunable spin wave emitters," *Nat. Nanotechnol*, vol. 11, p. 984, 2016.

[98] E. Albisetti *et al.*, "Optically inspired nanomagnonics with non-reciprocal spin waves in synthetic antiferromagnets," *Adv. Mater*, vol. 32, p. 1906439, 2020.

[99] G. Dieterle *et al.*, "Coherent excitation of heterosymmetric spin waves with ultrashort wavelengths," *Phys. Rev. Lett*, vol. 122, p. 117202, 2019.

[100] C. Kittel, "Theory of antiferromagnetic resonance"," *Phys. Rev.*, vol. 82, p. 565, 1951.

[101] F. M. Johnson and A. N. Nethercot Jr, "Antiferromagnetic resonance in MnF2," *Phys. Rev*, vol. 114, p. 705, 1959.

[102] T. Kampfrath *et al.*, "Coherent terahertz control of antiferromagnetic spin waves," *Nat. Photonics*, vol. 5, p. 31, 2011.

[103] J. R. Hortensius *et al.*, "Coherent spin-wave transport in an antiferromagnet," *Nat. Phys*, vol. 17, pp. 1001–1006, 2021.

[104] F. Nolting *et al.*, "Direct observation of the alignment of ferromagnetic spins by antiferromagnetic spins," *Nature*, vol. 405, p. 767, 2000.

[105] H. Jani *et al.*, "Antiferromagnetic half-skyrmions and bimerons at room temperature," *Nature*, vol. 590, pp. 74–79, 2021.

[106] R. Lebrun *et al.*, "Tunable long-distance spin transport in a crystalline antiferromagnetic iron oxide," *Nature*, vol. 561, no. 7722, pp. 222–225, 2018, doi: 10.1038/s41586-018-0490-7.

[107] J. Han *et al.*, "Birefringence-like spin transport via linearly polarized antiferromagnetic magnons," *Nat. Nanotech*, vol. 15, no. ue 7, pp. 563–568, 2020.

[108] C. Kim *et al.*, "Distinct handedness of spin wave across the compensation temperatures of ferrimagnets"," *Nat. Mater*, vol. 19, pp. 980–985, 2020.

[109] R. A. Ross *et al.*, "Propagation Length of Antiferromagnetic Magnons Governed by Domain Configurations," *Nano Lett*, vol. 20, pp. 306–313, 2019.

[110] R. Lebrun *et al.*, "Long-distance spin-transport across the Morin phase transition up to room temperature in ultra-low damping single crystals of the antiferromagnet $a$-Fe2O3," *Nat. Commun*, vol. 11, p. 6332, 2020.

[111] R. A. Ross *et al.*, "An insulating doped antiferromagnet with low magnetic symmetry as a room temperature spin conduit," *Appl. Phys. Lett*, vol. 117, no. 24, p. 242405, 2020.

[112] T. Wimmer *et al.*, "Observation of Antiferromagnetic Magnon Pseudospin Dynamics and the Hanle Effect," *Phys. Rev. Lett*, vol. 125, no. 24, p. 247204, 2020.

[113] G. Hoogeboom and J. B. van Wees, "Van Wees "Non-local spin Seebeck effect in the bulk easy-plane antiferromagnet NiO," *Phys. Rev. B*, vol. 102, p. 214415, 2020.

[114] A. Ross *et al.*, "Exceptional sign changes of the nonlocal spin Seebeck effect in antiferromagnetic hematite," *Phys. Rev. B*, vol. 103, no. no 22, p. 224433, 2021.

[115] I. Boventer, H. T. Simensen, A. Anane, M. Kläui, A. Brataas, and R. Lebrun, "Room-Temperature Antiferromagnetic Resonance and Inverse Spin-Hall Voltage in Canted Antiferromagnets," *Phys. Rev. Lett*, vol. 126, no. 18, p. 187201, 2021.

[116] Y. Tserkovnyak and M. Kläui, "Exploiting Coherence in Nonlinear Spin-Superfluid Transport," *Phys. Rev. Lett*, vol. 119, p. 187705, 2017.

[117] P. Vaidya *et al.*, "Subterahertz spin pumping from an insulating antiferromagnet," *Science (80-. ).*, vol. 368, no. ue 6487, pp. 160–165, 2020.

[118] J. Li *et al.*, "Spin current from sub-terahertz-generated antiferromagnetic magnons," *Nature*, vol. 578, pp. 70–74, 2020.

[119] F. Casola, T. Sar, and A. Yacoby, "Probing condensed matter physics with magnetometry based on nitrogen-vacancy centers in diamond," *Nat. Rev. Mat*, vol. 3, p. 17088, 2018.

[120] T. van der Sar, F. Casola, R. Walsworth, and A. Yacoby, "Nanoscale probing of spin waves using single electron spins," *Nat. Commun*, vol. 6, p. 7886, 2015.

[121] G. Wu, Y. Cheng, S. Guo, F. Yang, D. V Pelekhov, and P. C. Hammel, "Nanoscale imaging of Gilbert damping using signal amplitude mapping," *Appl. Phys. Lett*, vol. 118, p. 42403, 2021.

[122] L. Rondin, J.-P. Tetienne, T. Hingant, J.-F. Roch, P. Maletinsky, and V. Jacques, "Magnetometry with NitrogenVacancy Defects in Diamond, Rep," *Prog. Phys*, vol. 77, p. 56503.

[123] R. Santagati *et al.*, "Magnetic-Field Learning Using a Single Electronic Spin in Diamond with One-Photon Readout at Room Temperature," *Phys. Rev. X*, vol. 9, p. 21019.

[124] I. Bertelli *et al.*, "Magnetic resonance imaging of spin-wave transport and interference in a magnetic insulator," *Sci. Adv.*, vol. 6, p. 3556, 2020.

[125] P. Andrich *et al.*, "Long-range spin wave mediated control of defect qubits in nanodiamonds," *npj Quantum Inf*, vol. 3, p. 28, 2017.

[126] I. Bertelli *et al.*, "Imaging spin-wave damping using electron spins in diamond." .

[127] B. G. Simon *et al.*, "Directional excitation of a high-density magnon gas using coherently driven spin waves," *Nano Lett*, vol. 21, pp. 8213–8219, 2021.

[128] T. X. Zhou *et al.*, "A magnon scattering platform," in *Proceedings of the National Academies of Sciences*, 2021, vol. 118, p. 2019473118.

[129] C. Gonzalez-Ballestero, T. Sar, and O. Romero-Isart, "Towards a quantum interface between spin waves and paramagnetic spin baths," *arXiv:2012.00540*, 2020.

[130] M. Fukami, D. R. Candido, D. D. Awschalom, and M. E. Flatté, "Opportunities for long-range magnon-mediated entanglement of spin qubits via on- and off-resonant coupling," *arXiv: 2101.09220*, 2021.

[131] C. Du *et al.*, "Control and local measurement of the spin chemical potential in a magnetic insulator," *Science (80-. ).*, vol. 357, 2017.

[132] M. Schneider *et al.*, "Bose–Einstein condensation of quasiparticles by rapid cooling," *Nat. Nanotechnol.*, vol. 15, no. 6, p. 457, 2020, doi: 10.1038/s41565-020-0671-z.

[133] A. Finco *et al.*, "Imaging non-collinear antiferromagnetic textures

via single spin relaxometry," *Nat. Commun*, vol. 12, p. 767, 2021.

[134] V. Vlaminck and M. Bailleul, "Spin-wave transduction at the submicrometer scale: Experiment and modeling," *Phys. Rev. B*, vol. 81, p. 14425, 2010.

[135] B. A. Kalinikos and A. N. Slavin, "Theory of dipole-exchange spin wave spectrum for ferromagnetic films with mixed exchange boundary conditions," *J. Phys. C Solid State Phys.*, vol. 19, no. 35, pp. 7013–7033, 1986, doi: 10.1088/0022-3719/19/35/014.

[136] J. Stigloher *et al.*, "Snell's Law for Spin Waves," *Phys. Rev. Lett*, vol. 117, p. 37204, 2016.

[137] J. Stigloher *et al.*, "Spin-wave wavelength down-conversion at thickness steps," *Appl. Phys. Express*, vol. 11, p. 53002, 2018.

[138] T. Li, T. Taniguchi, Y. Shiota, T. Moriyama, and T. Ono, "Chromatic aberration effect in refraction of spin waves," *J. Magn. Soc. Jpn*, vol. 44, p. 133, 2020.

[139] V. E. Demidov, M. P. Kostylev, K. Rott, J. Mnchenberger, G. Reiss, and S. O. Demokritov, "Excitation of short-wavelength spin waves in magnonic waveguides," *Appl. Phys. Lett.*, vol. 99, no. 8, 2011, doi: 10.1063/1.3631756.

[140] D. V Kalyabin, A. V Sadovnikov, E. N. Beginin, and S. A. Nikitov, "Surface spin waves propagation in tapered magnetic stripe," *J. Appl. Phys*, vol. 126, p. 173907, 2019.

[141] T. Brächer *et al.*, "Generation of propagating backward volume spin waves by phase-sensitive mode conversion in two-dimensional microstructures," *Appl. Phys. Lett*, vol. 102, p. 132411, 2013.

[142] A. V Sadovnikov *et al.*, "Magnonic beam splitter: The building block of parallel magnonic circuitry," *Appl. Phys. Lett*, vol. 106, p. 192406, 2015.

[143] I. A. Golovchanskiy *et al.*, "Ferromagnet/Superconductor Hybridization for Magnonic Applications," *Adv. Func. Mater.*, vol. 28, no. 33, p. 1802375, 2018, doi: 10.1002/adfm.201802375.

[144] O. V Dobrovolskiy *et al.*, "Magnon–fluxon interaction in a ferromagnet/superconductor heterostructure," *Nat. Phys.*, vol. 15, no. 5, pp. 477–482, 2019, doi: 10.1038/s41567-019-0428-5.

[145] I. A. Golovchanskiy *et al.*, "Ferromagnet/Superconductor Hybrid Magnonic Metamaterials," *Adv. Sci*, vol. 6, p. 1900435, 2019.

[146] L. Yi, Y. Zhao, X. Zhang, and Y. Sun, "Possible Evidence for Spin-Transfer Torque Induced by Spin-Triplet Supercurrents," *Chin. Phys. Lett*, vol. 35, p. 77401, 2018.

[147] I. A. Golovchanskiy *et al.*, "Magnetization Dynamics in Proximity-Coupled Superconductor-Ferromagnet-Superconductor Multilayers," *Phys. Rev. Appl*, vol. 14, p. 24086, 2020.

[148] H. Qin *et al.*, "Nanoscale magnonic Fabry-Pérot resonator for low-loss spin-wave manipulations," *Nat. Commun*, vol. 12, p. 2293, 2021.

[149] P. Grünberg, "Magnetostatic spin-wave modes of a heterogeneous ferromagnetic double layer," *J. Appl. Phys*, vol. 52, pp. 6824–6829, 1981.

[150] Y. Au, E. Ahmad, O. Dmytriiev, M. Dvornik, T. Davison, and V. V Kruglyak, "Resonant microwave-to-spin-wave transducer," *Appl. Phys. Lett*, vol. 100, p. 182404, 2012.

[151] M. Grassi *et al.*, "Slow-wave-based nanomagnonic diode," *Phys. Rev. Appl*, vol. 14, p. 24047, 2020.

[152] J. Chen *et al.*, "Excitation of unidirectional exchange spin waves by a nanoscale magnetic grating," *Phys. Rev. B*, vol. 100, p. 104427, 2019.

[153] Y. Au, M. Dvornik, O. Dmytriiev, and V. V Kruglyak, "Nanoscale spin wave valve and phase shifter," *Appl. Phys. Lett*, vol. 100, p. 172408, 2012.

[154] F. K. G. Fripp, A. V Shytov, and V. V Kruglyak, "Spin-wave control using dark modes in chiral magnonic resonators," *Phys. Rev. B*, vol. 104, p. 54437, 2021.

[155] V. V Kruglyak *et al.*, "Graded Magnonic Index and Spin Wave Fano Resonances in Magnetic Structures: Excite, Direct, Capture," in *Spin wave confinement. Propagating Waves*, S. O. Demokritov, Ed. New York: Jenny Stanford Publishing, 2017, pp. 11–46.

[156] V. V Kruglyak, "Chiral Magnonic Resonators: Rediscovering the Basic Magnetic Chirality in Magnonics." .

[157] F. Heyroth *et al.*, "Monocrystalline Freestanding Three-Dimensional Yttrium-Iron-Garnet Magnon Nanoresonators," *Phys. Rev. Appl*, vol. 12, p. 54031, 2019.

[158] P. Trempler, R. Dreyer, P. Geyer, C. Hauser, G. Woltersdorf, and G. Schmidt, "Integration and characterization of micron-sized YIG structures with very low Gilbert damping on arbitrary substrates," *Appl. Phys. Lett*, vol. 117, no. 23, p. 232401, 2020.

[159] Y.-M. Li, J. Xiao, and K. Chang, "Topological magnon modes in patterned ferromagnetic insulator thin films," *Nano Lett*, vol. 18, pp. 3032–3037, 2018.

[160] A. V Sadovnikov, S. A. Odintsov, E. N. Beginin, S. E. Sheshukova, Y. P. Sharaevskii, and S. A. Nikitov, "Toward nonlinear magnonics: Intensity-dependent spin-wave switching in insulating side-coupled magnetic stripes," *Phys. Rev. B*, vol. 96, p. 144428, 2017.

[161] D. C. S. Davies *et al.*, "Towards graded-index magnonics: Steering spin waves in magnonic networks," *Phys. Rev. B*, vol. 92, p. 20408, 2015.

[162] M. Vogel, R. Aßmann, P. Pirro, A. V Chumak, B. Hillebrands, and G. von Freymann, "Control of Spin-Wave Propagation using Magnetisation Gradients," *Sci. Rep.*, vol. 8, no. 1, p. 11099, 2018, doi: 10.1038/s41598-018-29191-2.

[163] S. Mieszczak, O. Busel, A. N. Gruszecki Paweland Kuchko, J. W. Kłos, and M. Krawczyk, "Anomalous refraction of spin waves as a way to guide signals in curved magnonic multimode waveguides," *Phys. Rev. Appl*, vol. 13, p. 54038, 2020.

[164] N. Yu *et al.*, "Efficient classical simulation of the approximate quantum Fourier transform," *Science (80- ).*, vol. 334, pp. 333–337.

[165] J. Gloss *et al.*, "Ion-beam-induced magnetic and structural phase transformation of Ni-stabilized face-centered-cubic Fe films on Cu(1009," *Appl. Phys. Lett*, vol. 103, no. 26, p. 262405, 2013.

[166] M. Nord *et al.*, "Strain Anisotropy and Magnetic Domains in Embedded Nanomagnets," *Small*, vol. 15, no. 52, p. 1970287, 2019.

[167] M. Urbánek *et al.*, "Research Update: Focused ion beam direct writing of magnetic patterns with controlled structural and magnetic properties," *APL Mater.*, vol. 6, no. 6, 2018, doi: 10.1063/1.5029367.

[168] L. Flajšman *et al.*, "Zero-field propagation of spin waves in waveguides prepared by focused ion beam direct writing," *Phys. Rev. B*, vol. 101, no. 1, p. 14436, 2020.

[169] O. Wojewoda *et al.*, "Propagation of spin waves through a Néel domain wall," *Appl. Phys. Lett*, vol. 117, no. 2, p. 22405, 2020.

[170] O. A. Tretiakov, M. Morini, S. Vasylkevych, and V. Slastikov, "Engineering Curvature-Induced Anisotropy in Thin Ferromagnetic Films," *Phys. Rev. Lett*, vol. 119, no. 7, p. 77203, 2017.

[171] I. Turčan, L. Flajšman, O. Wojewoda, V. Roučka, O. Man, and M. Urbánek, "Spin wave propagation in corrugated waveguides," *Appl. Phys. Lett*, vol. 118, no. 9, p. 92405, 2021.

[172] J. Gloss *et al.*, "The growth of metastable fcc Fe78Ni22 thin films on H-Si(1 0 0) substrates suitable for focused ion beam magnetic patterning," *Appl. Surf. Sci*, vol. 469, pp. 747–752, 2019.

[173] K. Bergmann, H. Theuer, and B. W. Shore, "Coherent population transfer among quantum states of atoms and molecules," *Rev. Mod. Phys*, vol. 70, pp. 1003–1025, 1998.

[174] K. Bergmann *et al.*, "Roadmap on STIRAP applications," *J. Phys. B At. Mol. Opt. Phys*, vol. 52, p. 20202001, 2019.

[175] S. Longhi, G. D. Valle, M. Ornigotti, and P. Laporta, "Coherent tunneling by adiabatic passage in an optical waveguide system," *Phys. Rev. B*, vol. 76, p. 201101, 2007.

[176] Q. Wang, T. Brächer, M. Fleischhauer, B. Hillebrands, and P. Pirro, "Stimulated-Raman-adiabatic-passage mechanism in a magnonic environment," *Appl. Phys. Lett*, vol. 118, p. 182404, 2021.

[177] R. C. LeCraw and R. Comstock, "Piezoelectric and Piezomagnetic Materials and Their Function in Transducers," in *Physical Acoustics*, vol. 3, W. P. Mason, Ed. Amsterdam: Elsevier, 1964, pp. 127–199.

[178] D. Lachance-Quirion, Y. Tabuchi, A. Gloppe, K. Usami, and Y. Nakamura, "Hybrid quantum systems based on magnonics," *Appl. Phys. Expr*, vol. 12, p. 70101, 2019.

[179] C. Hauser *et al.*, "Yttrium Iron Garnet Thin Films with Very Low Damping Obtained by Recrystallization of Amorphous Material," *Sci. Rep*, vol. 6, p. 20827, 2016.

[180] S. Watanabe, V. S. Bhat, K. Baumgaertl, and D. Grundler, "Direct Observation of Worm-Like Nanochannels and Emergent Magnon Motifs in Artificial Ferromagnetic Quasicrystals," *Adv. Funct. Mater*, vol. 30, p. 2001388, 2020.

[181] C. Liu *et al.*, "Long-distance propagation of short-wavelength spin waves," *Nat. Commun.*, vol. 9, no. 1, 2018, doi: 10.1038/s41467-018-03199-8.

[182] G. Gubbiotti *et al.*, "Magnonic Band Structure in Vertical Meander-Shaped Co40Fe40B20 Thin Films," *Phys. Rev. Appl*, vol. 15, p. 14061, 2021.

[183] J. A. Otálora, M. Yan, H. Schultheiss, R. Hertel, and A. Kákay,

"Curvature-induced asymmetric spin-wave dispersion'," *Phys. Rev. Lett*, vol. 117, p. 227203, 2016.

[184] J. A. Otálora, M. Yan, H. Schultheiss, R. Hertel, and A. Kákay, "Asymmetric spin-wave dispersion in ferromagnetic nanotubes induced by surface curvature," *Phys. Rev. B*, vol. 95, p. 184415, 2017.

[185] J. A. Otálora *et al.*, "Frequency linewidth and decay length of spin waves in curved magnetic membranes," *Phys. Rev. B*, vol. 98, p. 14403, 2018.

[186] O. V Dobrovolskiy *et al.*, "Spin-wave eigenmodes in direct-write 3D nanovolcanoes," *Appl. Phys. Lett.*, vol. 118, no. 13, 2021, doi: 10.1063/5.0044325.

[187] J. Bachmann *et al.*, "Ordered iron oxide nanotube arrays of controlled geometry and tunable magnetism by atomic layer deposition," *JACS*, vol. 129, pp. 9554–9555, 2007.

[188] D. Rüffer *et al.*, "Magnetic states of an individual Ni nanotube probed by anisotropic magnetoresistance," *Nanoscale*, vol. 4, pp. 4989–4995, 2012.

[189] M. C. Giordano *et al.*, "Plasma-enhanced atomic layer deposition of nickel nanotubes with low resistivity and coherent magnetization dynamics for 3D spintronics," *ACS Appl. Mater. Interfaces*, vol. 12, pp. 40443–40452, 2020.

[190] M. C. Giordano, S. E. Steinvall, S. Watanabe, A. F. Morral, and D. Grundler, "Ni80Fe20 Nanotubes with Optimized Spintronic Functionalities Prepared by Atomic Layer Deposition," *Nanoscale*, vol. 13, pp. 13451–13462, 2021.

[191] M. M. Salazar-Cardona *et al.*, "Nonreciprocity of spin waves in magnetic nanotubes with helical equilibrium magnetization," *Appl. Phys. Lett*, vol. 118, p. 262411, 2021.

[192] L. Körber, G. Qaserbarth, A. Otto, and A. Kákay, "Finite-element dynamic-matrix approach for spin-wave dispersions in magnonic waveguides with arbitrary cross section," *arXiv: 2104.06943v1*, 2021.

[193] L. Körber *et al.*, "Experimental observation of the curvature-induced asymmetric spin-wave dispersion in hexagonal nanotubes," *arXiv:2009.02238*, 2020.

[194] D. Makarov, O. M. Volkov, A. Kakay, O. P. Pylypovskyi, B. Budinska, and O. V Dobrovolskiy, "New Dimension in Magnetism and Superconductivity: 3D and curvilinear geometry," *Adv. Mater*, vol. 33, p. 2101758, 2021.

[195] L. Keller *et al.*, "Direct-write of free-form building blocks for artificial magnetic 3D lattice," *Sci. Rep*, vol. 8, p. 6160, 2018.

[196] M. Al Mamoori, C. Schröder, L. Keller, M. Huth, and J. Müller, "First-order reversal curves (FORCs) of nano-engineered 3D Co-Fe structures," *AIP Adv.*, vol. 10, p. 15319, 2020.

[197] D. Sanz-Hernández *et al.*, "Artificial double-helix for geometrical control of magnetic chirality," *ACS Nano*, vol. 14, pp. 8084–8092, 2020.

[198] A. Fernandez-Pacheco *et al.*, "Writing 3D nanomagnets using focused electron beams," *Mater.*, vol. 13, no. 17, p. 3774, 2020, doi: 10.3390/MA13173774.

[199] F. Porrati *et al.*, "Crystalline Niobium Carbide Superconducting Nanowires Prepared by Focused Ion Beam Direct Writing," *ACS Nano*, vol. 13, no. 6, pp. 6287–6296, 2019, doi: 10.1021/acsnano.9b00059.

[200] R. Winkler, J. D. Fowlkes, P. D. Rack, and H. Plank, "3D nanoprinting via focused electron beams," *J. Appl. Phys*, vol. 125, p. 210901, 2019.

[201] O. V Dobrovolskiy *et al.*, "Kakazei Spin-wave eigenmodes in direct-write 3D nanovolcanoes," *Appl. Phys. Lett*, vol. 118, p. 132405, 2021.

[202] A. Fernández-Pacheco, R. Streubel, O. Fruchart, R. Hertel, P. Fischer, and R. P. Cowburn, "Three-dimensional nanomagnetism," *Nat. Commun*, vol. 8, p. 15756, 2017.

[203] P. Fischer, D. Sanz-Hernandez, R. Streubel, and A. Fernandez-Pacheco, "Launching a new dimension with 3D magnetic nanostructures," *APL Mater.*, vol. 8, p. 10701, 2020.

[204] R. Hertel, "Curvature-Induced Magnetochirality," *SPIN*, vol. 3, no. 3, p. 1340009, 2013.

[205] S. Sahoo, A. May, A. Den Berg, A. K. Mondal, S. Ladak, and A. Barman, "Observation of coherent spin waves in three-dimensional artificial spin ice structures," *Nano Lett*, vol. 21, p. 4629 4635, 2021.

[206] R. Cheenikundil and R. Hertel, "Switchable magnetic frustration in buckyball nanoarchitectures," *Appl. Phys. Lett*, vol. 118, no. 21, p.

212403, 2021.

[207] A. Frotanpour, J. Woods, B. Farmer, A. P. Kaphle, and L. E. D. Long, "Vertex dependent dynamic response of a connected Kagome artificial spin ice," *Appl. Phys. Lett*, vol. 118, no. 4, p. 42410, 2021.

[208] S. Gliga, A. Kákay, R. Hertel, and O. G. Heinonen, "Spectral Analysis of Topological Defects in an Artificial Spin-Ice Lattice," *Phys. Rev. Lett*, vol. 110, no. 11, p. 117205, 2013.

[209] A. May, M. Saccone, A. Berg, J. Askey, M. Hunt, and S. Ladak, "Magnetic charge propagation upon a 3D artificial spin-ice," *Nat. Commun*, vol. 12, no. 1, art. no. 1, 2021.

[210] R. Hertel, S. Christophersen, and S. Börm, "Large-scale magnetostatic field calculation in finite element micromagnetics with H2-matrices," *J. Mag. Mag. Mat*, vol. 477, pp. 118–123, 2018.

[211] M. Hunt *et al.*, "Harnessing multi-photon absorption to produce three-dimensional magnetic structures at the nanoscale," *Materials (Basel).*, vol. 13, p. 761, 2020.

[212] G. Williams *et al.*, "Two-photon lithography for 3D magnetic nanostructure fabrication," *Nano Res*, vol. 11, p. 845, 2018.

[213] S. Sahoo, S. Mondal, G. Williams, A. May, S. Ladak, and A. Barman, "Ultrafast magnetization dynamics in a nanoscale three-dimensional cobalt tetrapod structure," *Nanoscale*, vol. 10, p. 9981, 2018.

[214] A. May, M. Hunt, A. Den Berg, A. Hejazi, and S. Ladak, "Realisation of a frustrated 3D magnetic nanowire lattice," *Commun. Phys.*, vol. 2, p. 13, 2018.

[215] R. W. Dawidek, "Dynamically Driven Emergence in a Nanomagnetic System," *Adv. Func. Mat*, vol. 31, p. 2008389, 2021.

[216] J. Askey, M. Hunt, W. Langbein, and S. Ladak, "Use of two-photon lithography with a negative resist and processing to realise cylindrical magnetic nanowires," *Nanomaterials*, vol. 10, p. 429, 2020.

[217] C. Donnelly, "Time-resolved imaging of three-dimensional nanoscale magnetization dynamics," *Nat. Nano*, vol. 15, p. 356, 2020.

[218] B. Obry *et al.*, "A micro-structured ion-implanted magnonic crystal," *Appl. Phys. Lett.*, vol. 102, no. 20, p. 202403, 2013, doi: 10.1063/1.4807721.

[219] M. Krawczyk and D. Grundler, "Review and perspectives of magnonic crystals and devices withreprogrammable band structures'," *J. Phys. Condens. Matter*, vol. 26, p. 123202, 2014.

[220] S. Tacchi, G. Gubbiotti, M. Madami, and G. Carlotti, "Brillouin light scattering of 2D magnonic crystals," *J. Phys.-Cond. Matt*, vol. 9, p. 73001, 2017.

[221] K. Zakeri, "Magnonic crystals: towards terahertz frequencies," *J. Phys.-Cond. Matt*, vol. 32, p. 363001, 2020.

[222] G. Gubbiotti, "Magnonic band structure in CoFeB/Ta/NiFe meander-shaped magnetic bilayers," *Appl. Phys. Lett*, vol. 118, p. 162405, 2021.

[223] H. Yu *et al.*, "Omnidirectional spin-wave nanograting coupler," *Nat. Comm*, vol. 4, p. 2702, 2013.

[224] P. Graczyk, M. Krawczyk, S. Dhuey, W.-G. Yang, H. Schmidt, and G. Gubbiotti, "Magnonic band gap and mode hybridization in continuous permalloy films induced by vertical coupling with an array of permalloy ellipses," *Phys. Rev. B*, vol. 98, p. 174420, 2018.

[225] G. Gubbiotti, *Three-Dimensional Magnonics: Layered, Micro- and Nanostructures*. Singapore: Jenny Stanford Publishing, 2019.

[226] A. V Sadovnikov and S. A. Nikitov, "Using Mandelstam–Brillouin Spectroscopy to Study Energy-Efficient Devices for Processing Information Signals on the Basis of Magnon Straintronics," *Bull. Russ. Acad. Sci. Phys*, vol. 85, pp. 595–598, 2021.

[227] V. K. Sakharov *et al.*, "Spin waves in meander shaped YIG film: Toward 3D magnonics," *Appl. Phys. Lett*, vol. 117, p. 22403, 2020.

[228] A. A. Martyshkin, E. N. Beginin, A. I. Stognij, S. A. Nikitov, and A. V Sadovnikov, "Vertical spin-wave transport in magnonic waveguides with broken translation symmetry," *IEEE Trans. Magn*, vol. 10, p. 5511105, 2019.

[229] P. A. Popov *et al.*, "Spin wave propagation in three-dimensional magnonic crystals and coupled structures'," *J. Magn. Magn. Mater*, vol. 476, pp. 423–427, 2019.

[230] D. Van Qudenbosch, G. Hukic-Markosian, S. Ott, C. Abert, and M. H. Bartl, "An Experiment-Based Numerical Treatment of Spin Wave Modes in Periodically Porous Materials'," *Phys. Status Solidi B*, vol. 257, p. 1900296, 2020.

[231] K. Y. Guslienko, G. N. Kakazei, Y. V Kobljanskyj, G. A. Melkov, V. Novosad, and A N Slavin, "Microwave absorption properties of

permalloy nanodots in the vortex and quasi-uniform magnetization states," *New J. Phys*, vol. 16, p. 63044, 2014.

[232] C. Nisoli, R. Moessner, and P. Schiffer, "Colloquium: Artificial spin ice: Designing and imaging magnetic frustration," *Rev. Mod. Phys.*, vol. 85, pp. 1473–1490, 2013.

[233] S. Koraltan, F. Slanovc, and F. Bruckner, "Tension-free Dirac strings and steered magnetic charges in 3D artificial spin ice," *npj Comput Mater*, vol. 7, p. 125, 2021.

[234] S. Miwa *et al.*, "Perpendicular magnetic anisotropy and its electric-field-induced change at metal-dielectric interfaces," *J. Phys. D Appl. Phys*, vol. 52, no. 6, p. 63001, 2019.

[235] J. Zhu *et al.*, "Voltage-Induced Ferromagnetic Resonance in Magnetic Tunnel Junctions," *Phys. Rev. Lett*, vol. 108, no. 19, p. 197203, 2012.

[236] J. G. Alzate, "Spin wave nanofabric update," in *Proceedings of the 2012 IEEE/ACM International Symposium on Nanoscale Architectures*, 2012, pp. 196–202,.

[237] B. Rana, Y. Fukuma, K. Miura, H. Takahashi, and Y. Otani, "Excitation of coherent propagating spin waves in ultrathin CoFeB film by voltage-controlled magnetic anisotropy," *Appl. Phys. Lett*, vol. 111, no. 5, p. 52404, 2017.

[238] R. Verba, V. Tiberkevich, I. Krivorotov, and A. Slavin, "Parametric Excitation of Spin Waves by Voltage-Controlled Magnetic Anisotropy," *Phys. Rev. Appl.*, vol. 1, no. 4, p. 044006, 2014.

[239] Y.-J. Chen *et al.*, "Parametric Resonance of Magnetization Excited by Electric Field," *Nano Lett*, vol. 17, no. 1, pp. 572–577, 2016.

[240] R. Verba, M. Carpentieri, G. Finocchio, V. Tiberkevich, and A. Slavin, "Parametric Excitation and Amplification of Spin Waves in Ultrathin Ferromagnetic Nanowires by Microwave Electric Field," in *Spin Wave Confinement*, 2nd ed., Singapore: Jenny Stanford Publishing, 2017, pp. 385–425.

[241] S. Choudhury *et al.*, "Voltage controlled on-demand magnonic nanochannels," *Sci. Adv.*, vol. 6, no. 40, p. eaba5457, 2020.

[242] R. Verba, M. Carpentieri, G. Finocchio, V. Tiberkevich, and A. Slavin, "Excitation of Spin Waves in an In-Plane-Magnetized Ferromagnetic Nanowire Using Voltage-Controlled Magnetic Anisotropy," *Phys. Rev. Appl.*, vol. 7, no. 6, p. 064023, 2017.

[243] S. A. Shanker *et al.*, "Generation of charge current from magnetization oscillation via the inverse of voltage-controlled magnetic anisotropy effect," *Sci. Adv.*, vol. 6, no. 32, p. eabc2618, 2020.

[244] P.-H. Chang, W. Fang, T. Ozaki, and K. D. Belashchenko, "Voltage-controlled magnetic anisotropy in antiferromagnetic MgO-capped MnPt films," *Phys. Rev. Mater.*, vol. 5, no. 5, p. 054406, 2020.

[245] S. Yuta, "All-optical detection of magnetization precession in tunnel junctions under applied voltage," *Appl. Phys. Express*, vol. 10, no. 2, p. 23002, 2017.

[246] B. Rana, S. Choudhury, K. Miura, H. Takahashi, A. Barman, and Y. Otani, "Electric field control of spin waves in ultrathin CoFeB films," *Phys. Rev. B*, vol. 100, no. 22, p. 224412, 2019.

[247] B. Rana and Y. Otani, "Voltage-controlled reconfigurable spin-wave nanochannels and logic devices," *Phys. Rev. Appl*, vol. 9, no. 1, p. 14033, 2018.

[248] B. Rana, C. A. Akosa, K. Miura, H. Takahashi, G. Tatara, and Y. Otani, "Nonlinear control of damping constant by electric field in ultrathin ferromagnetic films," *Phys. Rev. Appl*, vol. 14, no. 1, p. 14037, 2020.

[249] X. Zhang, T. Liu, M. E. Flatté, and H. X. Tang, "Electric-field coupling to spin waves in a centrosymmetric ferrite," *Phys. Rev. Lett*, vol. 113, no. 3, p. 37202, 2014.

[250] T. Koyama, Y. Nakatani, J. Ieda, and D. Chiba, "Electric field control of magnetic domain wall motion via modulation of the Dzyaloshinskii-Moriya interaction," *Sci. Adv*, vol. 4, no. 12, p. 265, 2018.

[251] M. Xu, "Nonreciprocal surface acoustic wave propagation via magneto-rotation coupling," *Sci. Adv*, vol. 6, no. 32, p. 1724, 2020.

[252] A. Deka *et al.*, "Electric-field control of interfacial in-plane magnetic anisotropy in CoFeB/MgO junctions," *Phys. Rev. B*, vol. 101, no. 17, p. 174405, 2020.

[253] A. B. Ustinov *et al.*, "Dynamic electromagnonic crystal based on artificial multiferroic heterostructure," *Commun. Phys*, vol. 2, p. 137, 2019.

[254] A. A. Nikitin, A. A. Nikitin, I. L. Mylnikov, A. B. Ustinov, and B. A. Kalinikos, "Electromagnonic crystals based on ferrite-ferroelectric-ferrite thin-film multilayers," *IET Microw. Antennas Propag*, vol. 14, pp. 1304–1309, 2020.

[255] A. A. Nikitin, A. B. Ustinov, A. A. Semenov, B. A. Kalinikos, and E. Lähderanta, "All-thin-film multilayered multiferroic structures with a slot-line for spin-electromagnetic wave devices," *Appl. Phys. Lett*, vol. 104, p. 93513, 2014.

[256] A. A. Nikitin *et al.*, "Dispersion characteristics of spin-electromagnetic waves in planar multiferroic structures," *J. Appl. Phys*, vol. 118, p. 183901, 2015.

[257] A. A. Nikitin, A. A. Nikitin, A. V Kondrashov, A. B. Ustinov, B. A. Kalinikos, and E. Lähderanta, "Theory of dual-tunable thin-film multiferroic magnonic crystal," *J. Appl. Phys*, vol. 122, p. 153903, 2017.

[258] A. A. Nikitin, A. A. Nikitin, A. B. Ustinov, E. Lähderanta, and B. A. Kalinikos, "Theory of spin-electromagnetic waves in planar thin-film multiferroic heterostructures based on a coplanar transmission line and its application for electromagnonic crystals," *IEEE Trans. Mag*, vol. 54, no. 11, pp. 1–5, 2018.

[259] T. Schneider, A. A. Serga, B. Leven, and B. Hillebrands, "Realization of spin-wave logic gates," *Appl. Phys. Lett*, vol. 92, p. 22505, 2008.

[260] Y. K. Fetisov and G. Srinivasan, "Nonlinear electric field tuning characteristics of yttrium iron garnet–lead zirconate titanate microwave resonators," *Appl. Phys. Lett*, vol. 93, p. 33508, 2008.

[261] A. B. Ustinov, Y. K. Fetisov, S. V Lebedev, and G. Srinivasan, "Electrics witching in bistable ferrite-piezoelectric microwave resonator," *Tech. Phys. Lett*, vol. 36, pp. 166–169, 2010.

[262] A. A. Bukharaev, A. K. Zvezdin, A. P. Pyatakov, and Y. K. Fetisov, "Strain-tronics: a new trend in micro- and nanoelectronics and materials science," *Physics-Uspekhi*, vol. no. 61, pp. 1175–1212, 2018.

[263] A. V Sadovnikov *et al.*, "Route toward semiconductor magnonics: Light-induced spin-wave nonreciprocity in a YIG/GaAs structure," *Phys. Rev. B*, vol. 99, p. 54424, 2019.

[264] A. V Sadovnikov *et al.*, "Magnon straintronics: Reconfigurable spin-wave routing in strain-controlled bilateral magnetic stripes," *Phys. Rev. Lett*, vol. 120, p. 257203, 2018.

[265] B. Rana and Y. Otani, "Towards magnonic devices based on voltage-controlled magnetic anisotropy," *Commun. Phys*, vol. 2, 2019.

[266] A. A. Grachev *et al.*, "Strain-mediated tunability of spin-wave spectra in the adjacent magnonic crystal stripes with piezoelectric layer," *Appl. Phys. Lett*, vol. 118, p. 26, 2021.

[267] G. Talmelli *et al.*, "Reconfigurable nanoscale spin wave majority gate with frequency-division multiplexing," *Sci. Adv*, vol. 6, p. 4042, 2019.

[268] T. H. E. Lahtinen, J. O. Tuomi, and S. van Dijken, "Pattern transfer and electric-field-induced magnetic domain formation in multiferroic heterostructures," *Adv. Mater*, vol. 23, pp. 3187–3191, 2011.

[269] T. H. E. Lahtinen, K. J. A. Franke, and S. van Dijken, "Electric-field control of magnetic domain wall motion and local magnetization reversal," *Sci. Rep*, vol. 2, p. 258, 2012.

[270] B. Van de Wiele, S. J. Hämäläinen, P. Baláž, F. Montoncello, and S. van Dijken, "Tunable short-wavelength spin wave excitation from pinned magnetic domain walls," *Sci. Rep.*, vol. 6, p. 21330, 2016.

[271] S. J. Hämäläinen, F. Brandl, K. J. A. Franke, D. Grundler, and S. van Dijken, "Tunable short-wavelength spin-wave emission and confinement in anisotropy-modulated multiferroic heterostructures," *Phys. Rev. Appl*, vol. 8, p. 14020, 2017.

[272] S. J. Hämäläinen, M. Madami, H. Qin, G. Gubbiotti, and S. van Dijken, "Control of spin-wave transmission by a programmable domain wall," *Nat. Commun*, vol. 9, p. 4853, 2018.

[273] K. J. A. Franke, B. Van de Wiele, Y. Shirahata, S. J. Hämäläinen, T. Taniyama, and S. van Dijken, "Reversible electric-field-driven magnetic domain-wall motion," *Phys. Rev. X*, vol. 5, p. 11010, 2015.

[274] H. Qin, R. Dreyer, G. Woltersdorf, T. Taniyama, and S. van Dijken, "Electric-field control of propagating spin waves by ferroelectric domain-wall motion in a multiferroic heterostructure," *Adv. Mater*, vol. 33, p. 2100646, 2021.

[275] H. Merbouche, I. Boventer, and V. Haspot, "Voltage-Controlled Reconfigurable Magnonic Crystal at the Submicron Scale," *ACS Nano*, vol. 15, pp. 9775–9781, 2021.

[276] M. Vogel *et al.*, "Optically reconfigurable magnetic materials," *Nat.

*Phys.*, vol. 11, no. 6, p. 487, 2015, doi: 10.1038/nphys3325.

[277]  Q. Wang, A. V Chumak, L. Jin, H. Zhang, B. Hillebrands, and Z. Zhong, "Voltage-controlled nanoscale reconfigurable magnonic crystal," *Phys. Rev. B*, vol. 95, no. 13, p. 134433, 2017, doi: 10.1103/PhysRevB.95.134433.

[278]  O. V Dobrovolskiy, "Abrikosov fluxonics in washboard nanolandscapes," *Phys. C*, vol. 533, pp. 80–90, 2017, doi: 10.1016/j.physc.2016.07.008.

[279]  S. E. Barnes, J. L. Cohn, and F. Zuo, "The possibility of flux flow spectroscopy," *Phys. Rev. Lett*, vol. 77, pp. 3252–3255, 1996.

[280]  O. V Dobrovolskiy and A. V Chumak, "Nonreciprocal magnon fluxonics upon ferromagnet/superconductor hybrids," *J. Magn. Magnet. Mater*, vol. 543, p. 168633, 2021, doi: 10.1016/j.jmmm.2021.168633.

[281]  O. V Dobrovolskiy *et al.*, "Ultra-fast vortex motion in a direct-write Nb-C superconductor," *Nat. Commun.*, vol. 11, no. 1, p. 3291, 2020, doi: 10.1038/s41467-020-16987-y.

[282]  A. A. Bespalov, A. S. Mel'nikov, and A. I, "Buzdin 'Magnon radiation by moving Abrikosov vortices in ferromagnetic superconductors and superconductor-ferromagnet multilayers,'" *Phys. Rev. B*, vol. 89, p. 054516, 2014.

[283]  O. V Dobrovolskiy, V. M. Bevz, E. Begun, R. Sachser, R. V Vovk, and M. Huth, "Fast Dynamics of Guided Magnetic Flux Quanta," *Phys. Rev. Appl.*, vol. 11, no. 5, p. 054064, 2019.

[284]  O. V Dobrovolskiy *et al.*, "Cherenkov radiation of spin waves by ultra-fast moving magnetic flux quanta," *arXiv:2103.10156*, 2021.

[285]  R. Hertel, W. Wulfhekel, and J. Kirschner, "Domain-wall induced phase shifts in spin waves," *Phys. Rev. Lett*, vol. 93, p. 257202, 2004.

[286]  J. Lan, W. Yu, R. Wu, and J. Xiao, "Spin-wave diode," *Phys. Rev*, vol. X, 5, p. 41049, 2015.

[287]  K. Szulc, P. Graczyk, M. Mruczkiewicz, G. Gubbiotti, and M. Krawczyk, "Spin-wave diode and circulator based on unidirectional coupling," *Phys. Rev. Appl*, vol. 14, p. 140363, 2020.

[288]  J. W. K. P. Roberjot K. Szulc, M. Krawczyk, A. B. Ustinov, E. Lähderanta, M. Inoue, and B. A. Kalinikos, "Nonlinear spin-wave logic gates," *IEEE Magn. Lett*, vol. 10, p. 5508204, 2019.

[289]  A. Lara, O. V Dobrovolskiy, J. L. Prieto, M. Huth, and F. G. Aliev, "Magnetization reversal assisted by half antivortex states in nanostructured circular cobalt disks," *Appl. Phys. Lett.*, vol. 105, no. 18, p. 182402, 2014, doi: 10.1063/1.4900789.

[290]  O. V Dobrovolskiy *et al.*, "Tunable magnetism on the lateral mesoscale by post-processing of Co/Pt heterostructures," *Beilstein J. Nanotechnol.*, vol. 6, no. 1, pp. 1082–1090, 2015, doi: 10.3762/bjnano.6.109.

[291]  S. A. Bunyaev *et al.*, "Engineered magnetization and exchange stiffness in direct-write Co-Fe nanoelements," *Appl. Phys. Lett*, vol. 118, p. 22408, 2021.

[292]  O. V Dobrovolskiy *et al.*, "Spin-Wave Phase Inverter upon a Single Nanodefect," *ACS Appl. Mater. Interfaces*, vol. 11, no. 19, pp. 17654–17662, 2019, doi: 10.1021/acsami.9b02717.

[293]  O. V Dobrovolskiy *et al.*, "Spin-wave spectroscopy of individual ferromagnetic nanodisks," *Nanoscale*, vol. 12, p. 21207, 2020.

[294]  M. P. Kostylev, A. A. Serga, T. Schneider, B. Leven, and B. Hillebrands, "Spin-wave logical gates," *Appl. Phys. Lett*, vol. 87, p. 153501, 2005.

[295]  A. B. Ustinov and B. A. Kalinikos, "Nonlinear microwave spin wave interferometer," *Tech. Phys. Lett*, vol. 27, pp. 403–405, 2000.

[296]  A. V Chumak, A. A. Serga, and B. Hillebrands, "Magnon transistor for all-magnon data processing," *Nat. Commun.*, vol. 5, p. 4700, 2014, doi: 10.1038/ncomms5700.

[297]  A. B. Ustinov and B. A. Kalinikos, "The power-dependent switching of microwave signals in a ferrite-film nonlinear directional coupler," *Appl. Phys. Lett*, vol. 89, p. 172511, 2006.

[298]  A. V Sadovnikov *et al.*, "Nikitov "Nonlinear spin wave coupling in adjacent magnonic crystals," *Appl. Phys. Lett*, vol. 109, p. 42407, 2016.

[299]  K. G. Fripp and V. V Kruglyak, "Spin-wave wells revisited: From wavelength conversion and Möbius modes to magnon valleytronics"," *Phys. Rev. B*, vol. 103, p. 184403, 2021.

[300]  I. V Borisenko, V. E. Demidov, V. L. Pokrovsky, and S. O. Demokritov, "Spatial separation of degenerate components of magnon Bose–Einstein condensate by using a local acceleration potential," *Sci. Rep.*, vol. 10, no. 1, 2020, doi: 10.1038/s41598-020-71525-6.

[301]  C. S. Davies and V. V Kruglyak, "Graded-index magnonics"," *Low Temp. Phys*, vol. 41, p. 760, 2015.

[302]  N. J. Whitehead, S. A. R. Horsley, T. G. Philbin, and V. V Kruglyak, "Graded index lenses for spin wave steering," *Phys. Rev. B*, vol. 100, p. 94404, 2019.

[303]  L. Amarú, P. E. Gaillardon, S. Mitra, S., and G. D. Micheli, "New Logic Synthesis as Nanotechnology Enabler," in *Proceedings of the IEEE*, 2015, vol. 103, pp. 2168–2195.

[304]  P. Radu *et al.*, "Spintronic majority gates," in *Proceedings of the IEEE International Electron Devices Meeting (IEDM)*, 2015, pp. 32 5 1----32 5 4,.

[305]  T. Fischer *et al.*, "Experimental prototype of a spin-wave majority gate," *Appl. Phys. Lett.*, vol. 110, no. 15, p. 152401, 2017, doi: 10.1063/1.4979840.

[306]  N. Kanazawa *et al.*, "The role of Snell's law for a magnonic majority gate," *Sci. Rep*, vol. 7, p. 7898, 2017.

[307]  A. Mahmoud, F. Vanderveken, C. Adelmann, F. Ciubotaru, S. Hamdioui, and S. Cotofana, "Fan-out enabled spin wave majority gate," *AIP Adv*, vol. 10, p. 35119, 2020.

[308]  A. Khitun and K. L. Wang, "Non-volatile magnonic logic circuits engineering," *J. Appl. Phys*, vol. 110, p. 34306, 2011.

[309]  S. Klingler, P. Pirro, T. Brächer, B. Leven, B. Hillebrands, and A. V Chumak, "Design of a spin-wave majority gate employing mode selection," *Appl. Phys. Lett.*, vol. 105, no. 15, p. 152410, 2014, doi: 10.1063/1.4898042.

[310]  A. V Sadovnikov, E. N. Beginin, S. E. Sheshukova, D. V Romanenko, Y. P. Sharaevskii, and S. A. Nikitov, "Directional multimode coupler for planar magnonics: Side-coupled magnetic stripes," *Appl. Phys. Lett*, vol. 107, p. 202405, 2015.

[311]  Q. Wang, P. Pirro, R. Verba, A. Slavin, B. Hillebrands, and A. V Chumak, "Reconfigurable nanoscale spin-wave directional coupler," *Sci. Adv.*, vol. 4, no. 1, p. e1701517, 2018, doi: 10.1126/sciadv.1701517.

[312]  D. Nikonov and I. Young, "Benchmarking of Beyond-CMOS Exploratory Devices for Logic Integrated Circuits," *IEEE J. Explor. Solid-State Comput. Devices Circuits*, vol. 1, pp. 3–11, 2015.

[313]  U. Garlando, Q. Wang, O. V Dobrovolskiy, A. V Chumak, and F. Riente, "Numerical model for 32-bit magnonic ripple carry adder." .

[314]  K. Y. Guslienko and A. N. Slavin, "Boundary conditions for magnetization in magnetic nanoelements," *Phys. Rev. B*, vol. 72, p. 1446, 2005.

[315]  R. Verba, G. Melkov, V. Tiberkevich, and A. N. Slavin, "Collective spin-wave excitations in a two-dimensional array of coupled magnetic nanodots," *Phys. Rev. B*, vol. 85, p. 014427, 2012.

[316]  R. Verba, M. Carpentieri, G. Finocchio, V. Tiberkevich, and A. Slavin, "Excitation of propagating spin waves in ferromagnetic nanowires by microwave voltage-controlled magnetic anisotropy," *Sci. Rep*, vol. 6, p. 25018, 2016.

[317]  F. Riente and U. Garlando, "f-riente/spinwaves-model: Spinwave computational model for all-magnon circuits," *(v1.0.0). 10.5281/zenodo.5520990*, 2021.

[318]  M. Beleggia, S. Tandon, Y. Zhu, and M. De Graef, "On the magnetostatic interactions between nanoparticles of arbitrary shape," *J. Magn. Magn. Mat*, vol. 278, pp. 1–2, 2014.

[319]  A. Khitun, D. E. Nikonov, M. Bao, K. Galatsis, and K. L. Wang, "Feasibility study of logic circuits with a spin wave bus," *Nanotechnology*, vol. 18, p. 465202, 2007.

[320]  A. Khitun, M. Bao, and K. L. Wang, "Spin Wave Magnetic NanoFabric: A New Approach to Spin-Based Logic Circuitry," *IEEE Trans. Magn*, vol. 44, pp. 2141–2152, 2008.

[321]  A. Khitun, M. Bao, and K. L. Wang, "Magnonic logic circuits," *J. Phys. D. Appl. Phys.*, vol. 43, no. 26, 2010, doi: 10.1088/0022-3727/43/26/264005.

[322]  O. Zografos *et al.*, "System-level assessment and area evaluation of Spin Wave logic circuits," in *2014 IEEE/ACM International Symposium on Nanoscale Architectures (NANOARCH)*, 2014, pp. 25–30.

[323]  O. Zografos *et al.*, "Design and benchmarking of hybrid CMOS-Spin Wave Device Circuits compared to 10nm CMOS," in *2015 IEEE 15th International Conference on Nanotechnology (IEEE-NANO)*, 2015, pp. 686–689.

[324]  E. Albisetti *et al.*, "Nanoscale spin-wave circuits based on engineered reconfigurable spin-textures," *Commun. Phys*, vol. 1, p. 56, 2018.

[325]  A. Khitun, D. E. Nikonov, and K. L. Wang, "Magnetoelectric spin

wave amplifier for spin wave logic circuits," *J. Appl. Phys*, vol. 106, p. 123909, 2009.

[326] T. Brächer *et al.*, "Time- and power-dependent operation of a parametric spin-wave amplifier," *Appl. Phys. Lett*, vol. 105, p. 232409, 2014.

[327] A. Mahmoud, F. Vanderveken, C. Adelmann, F. Ciubotaru, S. Cotofana, and S. Hamdioui, "Spin Wave Normalization Toward All Magnonic Circuits," *IEEE Trans. Circuits Syst. I Regul. Pap*, vol. 68, pp. 536–549, 2020.

[328] S. Dutta *et al.*, "Non-volatile Clocked Spin Wave Interconnect for Beyond-CMOS Nanomagnet Pipelines," *Sci. Rep*, vol. 5, p. 9861, 2015.

[329] A. Mahmoud, F. Vanderveken, C. Adelmann, F. Ciubotaru, S. Hamdioui, and S. Cotofana, "Achieving Wave Pipelining in Spin Wave Technology," in *2021 22nd International Symposium on Quality Electronic Design (ISQED)*, 2021, pp. 54–59.

[330] F. Heussner, A. A. Serga, T. Brächer, B. Hillebrands, and P. Pirro, "A switchable spin-wave signal splitter for magnonic networks," *Appl. Phys. Lett*, vol. 111, p. 122401, 2017.

[331] A. Mahmoud, F. Vanderveken, C. Adelmann, F. Ciubotaru, S. Hamdioui, and S. Cotofana, "4-output Programmable Spin Wave Logic Gate," in *2020 IEEE 38th International Conference on Computer Design (ICCD)*, 2020, pp. 332–335.

[332] G. Talmelli *et al.*, "Spin wave emission by spin-orbit torque antennas," *Phys. Rev. Appl*, vol. 10, p. 44060, 2018.

[333] S. Cherepov *et al.*, "Electric-field-induced spin wave generation using multiferroic magnetoelectric cells," *Appl. Phys. Lett*, vol. 104, p. 82403, 2014.

[334] D. Tierno, F. Ciubotaru, R. Duflou, M. Heyns, I. P. Radu, and C. Adelmann, "Strain coupling optimization in magnetoelectric transducers," *Microelectron. Engin*, vol. 187–188, pp. 144–147, 2018.

[335] E. Egel, C. Meier, G. Csaba, and S. B. Gamm, "Design of a CMOS integrated on-chip oscilloscope for spin wave characterization," *AIP Adv*, vol. 7, p. 56016, 2017.

[336] D. A. Connelly *et al.*, "Efficient Electromagnetic Transducers for Spin-Wave Devices," *Sci. Rep.*, vol. 11, no. 1, pp. 1–14, 2021.

[337] S. B. Gamm *et al.*, "Design of on-chip readout circuitry for spin-wave devices," *IEEE Magn. Lett.*, vol. 8, pp. 1–4, 2016.

[338] H. T. Nembach, "Magneto-optical observation of four-wave scattering in a 15-nm Ni81Fe19 film during large-angle magnetization precession," *Phys. Rev. B*, vol. 84, p. 184413, 2011.

[339] H. Schultheiss, "Direct observation of nonlinear four-magnon scattering in spin-wave microconduits," *Phys. Rev. B*, vol. 86, p. 54414, 2012.

[340] K. Schultheiss, "Excitation of Whispering Gallery Magnons in a Magnetic Vortex," *Phys. Rev. Lett*, vol. 122, p. 97202, 2019.

[341] L. Körber, "Nonlocal Stimulation of Three-Magnon Splitting in a Magnetic Vortex," *Phys. Rev. Lett*, vol. 125, p. 207203, 2020.

[342] T. Hula, "Nonlinear losses in magnon transport due to four-magnon scattering," *Appl. Phys. Lett*, vol. 117, p. 42404, 2020.

[343] T. Hula, "Spin-wave frequency combs," *arXiv:2104.11491*, 2021.

[344] A. Khitun, "Magnonic holographic devices for special type data processing," *J. Appl. Phys*, vol. 113, 2013.

[345] F. Gertz, A. Kozhevnikov, Y. Filimonov, and A. Khitun, "Magnonic Holographic Memory," *Ieee Trans. Magn.*, vol. 51, 2014.

[346] Y. Khivintsev *et al.*, "Prime factorization using magnonic holographic devices," *J. Appl. Phys*, vol. 120, 2016.

[347] A. Khitun, "Parallel database search and prime factorization with magnonic holographic memory devices," *J. Appl. Phys*, vol. 118, 2015.

[348] M. Baliskiy, C. H., D. Gutierrez, A. V Kozhevnikov, F. Y., and K. A, "Quantum Computing without Quantum Computers: Database Search and Data Processing Using Classical Wave Superposition," *arXiv:2012.08401*, 2020.

[349] S. Molesky, Z. Lin, A. Y. Piggott, W. Jin, J. Vucković, and A. W. Rodriguez, "Inverse design in nanophotonics," *Nat. Phot.*, vol. 12, pp. 659–670, 2018.

[350] A. Papp, W. Porod, and G. Csaba, "Nanoscale neural network using non-linear spin-wave interference," *arXiv:2012.04594*, 2020.

[351] M. Hinze, R. Pinnau, M. Ulbrich, and S. Ulbrich, *Optimization with PDE constraints*, vol. 23. Berlin/Heidelberg: Springer Science & Business Media, 2009.

[352] F. Bruckner *et al.*, "Solving large-scale inverse magnetostatic problems using the adjoint method," *Sci. Rep*, vol. 7, 2017.

[353] C. Abert, C. Huber, F. Bruckner, C. Vogler, G. Wautischer, and D. Suess, "A fast finite-difference algorithm for topology optimization of permanent magnets," *J. Appl. Phys*, vol. 122, no. 11, p. 113 904, 2017.

[354] D. Marković, A. Mizrahi, D. Querlioz, and J. Grollier, "Physics for neuromorphic computing," *Nat. Rev. Phys*, vol. 2, p. 499, 2020.

[355] T. Brächer and P. Pirro, "An analog magnon adder for all-magnonic neurons," *J. Appl. Phys*, vol. 124, no. 15, p. 152119, 2018.

[356] J. Feldmann, N. Youngblood, C. D. Wright, H. Bhaskaran, and W. H. P. Pernice, "All-optical spiking neurosynaptic networks with self-learning capabilities," *Nature*, vol. 569, no. 7755, pp. 208–214, 2019.

[357] F. Heussner *et al.*, "Experimental Realization of a Passive Gigahertz Frequency-Division Demultiplexer for Magnonic Logic Circuits," *Phys. Status Solidi RRL*, vol. 14, no. 4, p. 1900695, 2020.

[358] L. Appeltant *et al.*, "Information processing using a single dynamical node as complex system," *Nat. Commun*, vol. 2, pp. 466–468, 2011.

[359] M. K. S.Watt, A. B. Ustinov, and B. A. Kalinikos, "Implementing a Magnonic Reservoir Computer Model Based on Time-Delay Multiplexing," *Phys. Rev. Appl*, vol. 5, p. 64060, 2021.

[360] S. Watt, M. Kostylev, and A. B. Ustinov, "Enhancing computational performance of a spin-wave reservoir computer with input synchronization," *J. Appl. Phys*, vol. 129, p. 44902, 2021.

[361] Q. Wang *et al.*, "A nonlinear magnonic nano-ring resonator," *npj Comput. Mater*, vol. 6, no. 1, p. 453, 2020.

[362] A. V Sadovnikov *et al.*, "Neuromorphic Calculations Using Lateral Arrays of Magnetic Microstructures with Broken Translational Symmetry," *JETP Lett.*, vol. 108, pp. 312–317, 2018.

[363] J. Grollier, D. Querlioz, K. Y. Camsari, K. Everschor-Sitte, S. Fukami, and M. D. Stiles, "Neuromorphic spintronics," *Nat. Electron*, vol. 3, pp. 360–370, 2020.

[364] T. Brächer, P. Pirro, and B. Hillebrands, "Parallel pumping for magnon spintronics: Amplification and manipulation of magnon spin currents on the micron-scale," *Phys. Rep*, vol. 699, pp. 1–34, 2017.

[365] H. Yu, J. Xiao, and H. Schultheiss, "Magnetic texture based magnonics," *Phys. Rep*, vol. 905, pp. 1–59, 2021.

[366] J. Han, P. Zhang, J. T. Hou, S. A. Siddiqui, and L. Liu, "Mutual control of coherent spin waves and magnetic domain walls in a magnonic device," *Science (80-. ).*, vol. 366, no. 6469, pp. 1121–1125, 2019.

[367] A. Papp, G. Csaba, and W. Porod, "Characterization of nonlinear spin-wave interference by reservoir-computing metrics," *Appl. Phys. Lett*, vol. 119, no. 11, p. 112403, 2021.

[368] J. Torrejon *et al.*, "Neuromorphic computing with nanoscale spintronic oscillators," *Nature*, vol. 547, pp. 428–431, 2017.

[369] T. Tsunegi *et al.*, "Physical reservoir computing based on spin torque oscillator with forced synchronization," *Appl. Phys. Lett*, vol. 114, p. 164101, 2019.

[370] D. Markovic *et al.*, "Reservoir computing with the frequency, phase, and amplitude of spin-torque nano-oscillators," *Appl. Phys. Lett*, vol. 114, p. 12409, 2009.

[371] T. Kanao, H. Suto, K. Mizushima, H. Goto, T. Tanamoto, and T. Nagasawa, "Reservoir Computing on Spin-Torque Oscillator Array," *Phys. Rev. Appl*, vol. 12, p. 24052, 2019.

[372] A. Pikovsky, M. Rosenblum, and J. Kurths, *Synchronization: A Universal Concept in Nonlinear Sciences*. Cambridge: Cambridge University Press, 2001.

[373] M. Dvornik and V. V Kruglyak, "Dispersion of collective magnonic modes in stacks of nanoscale magnetic elements'," *Phys. Rev. B*, vol. 84, p. 140405, 2011.

[374] G. Csaba and W. Porod, "Coupled oscillators for computing: A review and perspective," *App. Phys. Rev*, vol. 7, p. 11302, 2020.

[375] R. Khymyn, I. Lisenkov, V. Tiberkevich, B. Ivanov, and A. Slavin, "Antiferromagnetic THz-frequency Josephson-like Oscillator Driven by Spin Current," *Sci. Rep.*, vol. 7, p. 43705, 2017.

[376] R. Khymyn *et al.*, "Ultra-fast artifcial neuron: generation of picosecond duration spikes in a current-driven antiferromagnetic auto-oscillator," *Sci. Rep.*, vol. 8, p. 15727, 2018.

[377] O. Sulymenko *et al.*, "Ultra-fast logic devices using artificial 'neurons' based on antiferromagnetic pulse generators," *J. Appl. Phys*, vol. 124, p. 152115, 2018.

[378] V. E. Demidov, S. Urazhdin, A. Zholud, A. V Sadovnikov, and S. O. Demokritov, "Nanoconstriction-based spin-Hall nano-oscillator,"

*Appl. Phys. Lett.*, vol. 105, no. 17, 2014, doi: 10.1063/1.4901027.

[379]   V. E. Demidov, S. Urazhdin, A. Zholud, A. Sadovnikov, and S. O. Demokritov, "Nanoconstriction-based spin-Hall oscillators," 2015, doi: 10.1109/INTMAG.2015.7157174.

[380]   M. Zahedinejad *et al.*, "CMOS compatible W/CoFeB/MgO spin Hall nano-oscillators with wide frequency tunability," *Appl. Phys. Lett*, vol. 112, p. 132404, 2018.

[381]   P. Dürrenfeld, A. A. Awad, A. Houshang, R. K. Dumas, and J. Åkerman, "A 20 nm spin Hall nano-oscillator," *Nanoscale*, vol. 9, pp. 1285–1291, 2017.

[382]   A. A. Awad, A. Houshang, M. Zahedinejad, R. Khymyn, and J. Åkerman, "Width dependent auto-oscillating properties of constriction based spin Hall nano-oscillators," *Appl. Phys. Lett*, vol. 116, p. 232401, 2020.

[383]   M. Haidar, A. A. Awad, M. Dvornik, R. Khymyn, A. Houshang, and J. Åkerman, "A single layer spin-orbit torque nano-oscillator," *Nat Commun*, vol. 10, p. 2362, 2019, doi: 10.1038/s41467-019-10120-4.

[384]   A. A. Awad *et al.*, "Long-range mutual synchronization of spin Hall nano-oscillators," *Nat. Phys*, vol. 13, pp. 292–299, 2017.

[385]   M. Zahedinejad *et al.*, "Two-dimensional mutually synchronized spin Hall nano-oscillator arrays for neuromorphic computing," *Nat. Nanotech*, vol. 15, pp. 47–52, 2020.

[386]   A. A. A. M. Dvornik and J. Åkerman, "Origin of magnetization auto-oscillations in constriction-based spin Hall nano-oscillators," *Phys. Rev. Appl*, vol. 9, p. 14017, 2018.

[387]   H. Fulara *et al.*, "Spin-orbit torque-driven propagating spin waves," *Sci. Adv.*, vol. 5, p. 8467, 2019.

[388]   M. Zahedinejad *et al.*, "Memristive conrol of mutual SHNO synchronization for neuromorphic computing," *arXiv:2009.06594*, 2020.

[389]   A. Houshang, M. Zahedinejad, S. Muralidhar, J. Checinski, A. A. Awad, and J. Åkerman, "A Spin Hall Ising Machine," *arXiv:2006.02336*, 2020.

[390]   D. I. Albertsson, M. Zahedinejad, A. Houshang, R. Khymyn, J. Åkerman, and A. Rusu, "Ultrafast Ising Machines using spin torque nano-oscillators," *Appl. Phys. Lett*, vol. 118, p. 112404, 2021.

[391]   D. Lachance-Quirion, S. P. Wolski, Y. Tabuchi, S. Kono, K. Usami, and Y. Nakamura, "Entanglement-based single-shot detection of a single magnon with a superconducting qubit," *Science (80-. )*., vol. 367, no. 6476, pp. 425–428, 2020, doi: 10.1126/science.aaz9236.

[392]   E. Rieffel and W. Polak, *Quantum Computing. A Gentle Introduction*. Cambridge: MIT Press, 2011.

[393]   D. E. Browne, "Efficient classical simulation of the quantum Fourier transform," *New J. Phys*, vol. 9, p. 146, 2007.

[394]   N. Yoran and A. J. Short, "Efficient classical simulation of the approximate quantum Fourier transform," *Phys. Rev. A*, vol. 76, p. 42321, 2007.

[395]   S. O. Demokritov *et al.*, "Bose-Einstein condensation of quasi-equilibrium magnons at room temperature under pumping," *Nature*, vol. 443, no. 7110, pp. 430–433, 2006, doi: 10.1038/nature05117.

[396]   T. Byrnes, K. Wen, and Y. Yamamoto, "Macroscopic quantum computation using Bose-Einstein condensates," *Phys. Rev. A*, vol. 85, p. 40306, 2012.

[397]   T. Byrnes *et al.*, "Macroscopic quantum information processing using spin coherent states," *Opt. Commun*, vol. 337, pp. 102–109, 2015.

[398]   S. N. Andrianov and S. A. Moiseev, "Magnon qubit and quantum computing on magnon Bose-Einstein condensates," *Phys. Rev. A*, vol. 90, p. 42303, 2014.

[399]   T. B. Noack, V. I. Vasyuchka, A. Pomyalov, V. S. L'vov, A. A. Serga, and B. Hillebrands, "Evolution of room-temperature magnon gas: Toward a coherent Bose–Einstein condensate," *Phys. Rev. B*, vol. 104, p. 100410, 2021.

[400]   A. A. S. M. Mohseni, V. I. Vasyuchka, V. S. L'vov and B. Hillebrands, "Magnon Bose–Einstein-condensate-based qubit calculus," *Prep.*

[401]   P. Nowik-Boltyk, O. Dzyapko, V. E. Demidov, N. G. Berloff, and S. O. Demokritov, "Spatially non-uniform ground state and quantized vortices in a two-component Bose-Einstein condensate of magnons," *Sci. Rep.*, vol. 2, p. 482, 2012, doi: 10.1038/srep00482.

[402]   D. A. Bozhko *et al.*, "Supercurrent in a room-temperature Bose–Einstein magnon condensate," *Nat. Phys*, vol. 12, no. 11, pp. 1057–1062, 2016.

[403]   D. A. Bozhko *et al.*, "Bogoliubov waves and distant transport of

[404]   A. J. E. Kreil *et al.*, "Experimental observation of Josephson oscillations in a room-temperature Bose-Einstein magnon condensate," *Phys. Rev. B*, vol. 104, p. 144414, 2021.

[405]   M. Mohseni, A. Qaiumzadeh, A. A. Serga, B. Brataas, B. Hillebrands, and P. Pirro, "Bose–Einstein condensation of nonequilibrium magnons in confined systems," *New J. Phys*, vol. 22, p. 83080, 2020.

[406]   M. Gong *et al.*, "Genuine 12-Qubit Entanglement on a Superconducting Quantum Processor," *Phys. Rev. Lett*, vol. 122, no. 11, p. 110501, 2018.

[407]   A. Omran *et al.*, "Generation and manipulation of Schrödinger cat states in Rydberg atom arrays," *Science (80-. )*., vol. 365, no. 6453, pp. 570–574, 2019.

[408]   P. G. Baity *et al.*, "Strong magnon–photon coupling with chip-integrated YIG in the zero-temperature limit," *Appl. Phys. Lett*, vol. 119, no. 3, p. 33502, 2021.

[409]   R. Macêdo *et al.*, "Electromagnetic Approach to Cavity Spintronics," *Phys. Rev. Appl*, vol. 15, no. 2, p. 024065, 2021.

[410]   V. Cherepanov, I. Kolokolov, and V. L'vov, "The saga of YIG: Spectra, thermodynamics, interaction and relaxation of magnons in a complex magnet," *Phys. Rep*, vol. 229, no. 3, pp. 81–144, 1993.

[411]   O. V Prokopenko *et al.*, "Recent trends in microwave magnetism and superconductivity," *Ukr. J. Phys.*, vol. 64, no. 10, p. 888, 2019, doi: 10.15407/ujpe64.10.888.

[412]   A. A. Serga, A. Andre, S. O. Demokritov, B. Hillebrands, and A. N. Slavin, "Black soliton formation from phase-adjusted spin wave packets," *J. Appl. Phys.*, vol. 95, no. 11 II, pp. 6607–6609, 2004, doi: 10.1063/1.1669292.

[413]   D. D. Awschalom *et al.*, "Quantum Engineering With Hybrid Magnonics Systems and Materials," *IEEE Trans. Quantum Eng*, p. 1, 2021.

[414]   A. A. Serga *et al.*, "Bose-Einstein condensation in an ultra-hot gas of pumped magnons," *Nat. Commun.*, vol. 5, p. 3452, 2014, doi: 10.1038/ncomms4452.

[415]   D. A. Bozhko, P. Clausen, A. V Chumak, Y. V Kobljanskyj, B. Hillebrands, and A. A. Serga, "Formation of Bose-Einstein magnon condensate via dipolar and exchange thermalization channels," *Low Temp. Phys.*, vol. 41, no. 10, p. 1024, 2015, doi: 10.1063/1.4932354.

[416]   Y. Tabuchi, S. Ishino, T. Ishikawa, R. Yamazaki, K. Usami, and Y. Nakamura, "Hybridizing Ferromagnetic Magnons and Microwave Photons in the Quantum Limit," *Phys. Rev. Lett*, vol. 113, no. 8, p. 83603, 2014.

[417]   L. Mihalceanu *et al.*, "Temperature-dependent relaxation of dipole-exchange magnons in yttrium iron garnet films," *Phys. Rev. B*, vol. 97, no. 21, p. 214405, 2018.

[418]   S. Kosen, A. F. Loo, D. A. Bozhko, L. Mihalceanu, and A. D. Karenowska, "Microwave magnon damping in YIG films at millikelvin temperatures," *APL Mater*, vol. 7, no. 10, p. 101120, 2019.

[419]   Y. Zhang, J. Xie, L. Deng, and L. Bi, "Growth of Phase Pure Yttrium Iron Garnet Thin Films on Silicon: The Effect of Substrate and Postdeposition Annealing Temperatures," *IEEE Trans. Magn*, vol. 51, no. 11, pp. 1–4, 2015.

[420]   I. Boventer, M. Pfirrmann, J. Krause, Y. Schön, M. Kläui, and M. Weides, "Complex temperature dependence of coupling and dissipation of cavity magnon polaritons from millikelvin to room temperature," *Phys. Rev. B*, vol. 97, no. 18, p. 184420, 2018.

[421]   I. Boventer, M. Kläui, R. Macêdo, and M. Weides, "Steering between level repulsion and attraction: broad tunability of two-port driven cavity magnon-polaritons," *New J. Phys*, vol. 21, p. 125001, 2019.

[422]   I. Boventer *et al.*, "Control of the coupling strength and linewidth of a cavity magnon-polariton," *Phys. Rev. Res.*, vol. 2, no. 1, p. 13154, 2020.

[423]   I. Proskurin, R. Macêdo, and R. L. Stamps, "Microscopic origin of level attraction for a coupled magnon-photon system in a microwave cavity," *New J. Phys*, vol. 21, no. 9, p. 95003, 2019.

[424]   T. Wolz *et al.*, "Introducing coherent time control to cavity magnon-polariton modes," *Commun. Phys*, vol. 3, p. 3, 2020.

[425]   M. Kostylev, A. B. Ustinov, A. V Drozdovskii, B. A. Kalinikos, and E. Ivanov, "Towards experimental observation of parametrically squeezed states of microwave magnons in yttrium iron garnet

films," *Phys. Rev. B*, vol. 100, no. 2, p. 20401, 2019.

[426] A. J. E. Kreil *et al.*, "From Kinetic Instability to Bose-Einstein Condensation and Magnon Supercurrents," *Phys. Rev. Lett.*, vol. 121, no. 7, p. 77203, 2018.

[427] L. Mihalceanu *et al.*, "Magnon Bose-Einstein condensate and supercurrents over a wide temperature range," *Ukr. J. Phys.*, vol. 64, no. 10, p. 927, 2019.

[428] R. Macêdo, K. L. Livesey, and R. E. Camley, "Using magnetic hyperbolic metamaterials as high frequency tunable filters," *Appl. Phys. Lett*, vol. 113, p. 121104, 2018.

[429] P. Clausen, D. A. Bozhko, V. I. Vasyuchka, B. Hillebrands, G. A. Melkov, and A. A. Serga, "Stimulated thermalization of a parametrically driven phonon magnon gas as a prerequisite for Bose-Einstein magnon condensation," *Phys. Rev. B*, vol. 91, no. 22, p. 220402, 2015.

[430] H. Maier-Flaig *et al.*, "Temperature-dependent magnetic damping of yttrium iron garnet spheres," *Phys. Rev. B*, vol. 95, p. 214423, 2017.

[431] X. Zhang, C.-L. Zou, L. Jiang, and H. X. Tang, "Cavity magnomechanics," *Sci. Adv.*, vol. 2, p. 1501286, 2016.

[432] K. An *et al.*, "Coherent long-range transfer of angular momentum between magnon Kittel modes by phonons," *Phys. Rev. B*, vol. 101, p. 60407, 2020.

[433] K. An *et al.*, "Bright and dark states of two distant macrospins strongly coupled by phonons," *arXiv: 2108.13272*, 2021.

[434] M. Eichenfield, J. Chan, R. M. Camacho, K. J. Vahala, and O. Painter, "Optomechanical crystals," *Nature*, vol. 78, pp. 72–78, 2009.

[435] J. Graf, S. Sharma, H. Huebl, and S. V. Kusminskiy, "Design of an optomagnonics crystal: Towards optimal magnon-phonon mode matching at the microscale," *Phys. Rev. Res.*, vol. 3, p. 13277, 2021.

[436] H. Huebl *et al.*, "High Cooperativity in Coupled Microwave Resonator Ferrimagnetic Insulator Hybrids," *Phys. Rev. Lett*, vol. 111, p. 127003, 2013.

[437] J. Li and S. Gröblacher, "Entangling the vibrational modes of two massive ferromagnetic spheres using cavity magnomechanics," *Quantum Sci. Technol*, vol. 6, p. 24005, 2021.

[438] T. Kikkawa *et al.*, "Magnon polarons in the spin Seebeck effect," *Phys. Rev. Lett*, vol. 117, pp. 207203 1----207203 6, 2016.

[439] B. Flebus *et al.*, "Magnon-polaron transport in magnetic insulators," *Phys. Rev. B*, vol. 95, pp. 144420 1----144420 11, 2017.

[440] B. Yang *et al.*, "Revealing thermally driven distortion of magnon dispersion by spin Seebeck effect in Gd3Fe5O12," *Phys. Rev. B*, vol. 103, pp. 054411 1----054411 6, 2021.

[441] R. Yahiro *et al.*, "Magnon polarons in the spin Peltier effect," *Phys. Rev. B*, vol. 101, pp. 024407 1----024407 7, 2020.

[442] T. Hioki, Y. Hashimoto, and E. Saitoh, "Bi-reflection of spin waves," *Commun. Phys*, vol. 3, p. 188, 2020.

[443] D. A. Bozhko *et al.*, "Bottleneck Accumulation of Hybrid Magnetoelastic Bosons," *Phys. Rev. Lett.*, vol. 118, no. 23, p. 237201, 2017, doi: 10.1103/PhysRevLett.118.237201.

[444] P. Frey, D. A. Bozhko, V. S. L'vov, B. Hillebrands, and A. A. Serga, "Double accumulation and anisotropic transport of magnetoelastic bosons in yttrium iron garnet films," *Phys. Rev. B*, vol. 104, no. 1, 2021, doi: 10.1103/PhysRevB.104.014420.

[445] T. Makiuchi *et al.*, "Parametron on magnetic dot: Stable and stochastic operation," *Appl. Phys. Lett*, vol. 118, p. 22402, 2021.

[446] B. Z. Rameshti *et al.*, "Cavity Magnonics," *arXiv:2106.09312*, 2021.

[447] S. V. Kusminskiy, H. X. Tang, and F. Marquardt, "Coupled spin-light dynamics in cavity optomagnonics'," *Phys. Rev. A*, vol. 94, p. 33821, 2016.

[448] V. A. S. V. Bittencourt, V. Feulner, and S. Viola Kusminskiy, "Magnon heralding in cavity optomagnonics," *Phys. Rev. A*, vol. 100, p. 13810, 2019.

[449] S. Scharma, V. A. S. V. Bittencourt, A. Karenowska, and S. V. Kusminskiy, "Spin cat states in ferromagnetic insulators'," *Phys. Rev. B*, vol. 103, p. 100403, 2021.

[450] C. A. Potts, V. A. S. V. Bittencourt, S. V. Kusminskiy, and J. P. Davis, "Magnon-Phonon Quantum Correlation Thermometry," *Phys. Rev. Appl.*, vol. 13, p. 64001, 2020.

[451] C. A. Potts, E. Varga, V. A. S. V. Bittencourt, S. V. Kusminskiy, and J. P. Davis, "Dynamical Backaction magnomechanics," *Phys. Rev. X*, vol. 11, p. 31053, 2021.

[452] V. Wachter *et al.*, "Optical signatures of the coupled spin-mechanics of a levitated magnetic microparticle," *arXiv2108.06214, Accept. J. Opt. Soc. Am. B https//doi.org/10.1364/JOSAB.440562*, 2021.

[453] T. S. Parvini, V. A. S. V. Bittencourt, and S. V. Kusminskiy, "Antiferromagnetic cavity optomagnonics," *Phys. Rev. Res.*, vol. 2, p. 22027, 2020.

[454] E. V. Boström, T. S. Parvini, J. W. McIver, A. Rubio, S. V. Kusminskiy, and M. A. Sentef, "All-optical generation of antiferromagnetic magnon currents via the magnon circular photogalvanic effect," *Phys. Rev. B*, vol. 104, no. 10, p. L100404, 2021.

[455] J. Graf, H. Pfeifer, F. Marquardt, and S. V. Kusminskiy, "Cavity optomagnonics with magnetic textures: Coupling a magnetic vortex to light," *Phys. Rev. B*, vol. 98, p. 241406, 2018.

[456] M. Harder *et al.*, "Level Attraction Due to Dissipative Magnon-Photon Coupling," *Phys. Rev. Lett*, vol. 121, p. 137203, 2018.

[457] B. Bhoi *et al.*, "Abnormal anticrossing effect in photon-magnon coupling," *Phys. Rev. B*, vol. 99, p. 134426, 2019.

[458] J. W. Rao *et al.*, "Level attraction and level repulsion of magnon coupled with a cavity anti-resonance," *New J. Phys*, vol. 21, p. 65001, 2019.

[459] Y. Yang, J. W. Rao, Y. S. Gui, B. M. Yao, W. Lu, and C. M. Hu, "Control of the Magnon-Photon Level Attraction in a Planar Cavity," *Phys. Rev. Appl.*, vol. 11, p. 54023, 2019.

[460] V. L. Grigoryan, K. Shen, and K. Xia, "Synchronized spin-photon coupling in a microwave cavity," *Phys. Rev. B*, vol. 98, p. 24406, 2018.

[461] M. Harder, B. M. Yao, Y. S. Gui, and C.-M. Hu, "Coherent and Dissipative Cavity Magnonics," *J. Appl. Phys*, vol. 129, p. 201101, 2021.

[462] Y. Wang *et al.*, "Nonreciprocity and Unidirectional Invisibility in Cavity Magnonics," *Phys. Rev. Lett*, vol. 123, p. 127202, 2019.

[463] Y. Yang, J. W. R. Yi-Pu Wang, Y. S. Gui, B. M. Yao, W. Lu, and C.-M. Hu, "Unconventional singularity in anti-parity-time symmetric cavity magnonics," *Phys. Rev. Lett*, vol. 125, p. 147202, 2020.

[464] Y. Li, W. Zhang, V. Tyberkevych, W.-K. Kwok, A. Hoffmann, and V. Novosad, "Hybrid magnonics: Physics, circuits, and applications for coherent information processing," *J. Appl. Phys.*, vol. 128, no. 13, 2020, doi: 10.1063/5.0020277.

[465] Y. Li *et al.*, "Strong Coupling between Magnons and Microwave Photons in On-Chip Ferromagnet-Superconductor Thin-Film Devices," *Phys. Rev. Lett.*, vol. 123, no. 10, p. 107701, 2019, doi: 10.1103/PhysRevLett.123.107701.

[466] J. T. Hou and L. Liu, "Strong Coupling between Microwave Photons and Nanomagnet Magnons," *Phys. Rev. Lett*, vol. 123, p. 107702, 2019.

[467] M. B. Jungfleisch *et al.*, "Insulating nanomagnets driven by spin torque," *Nano Lett*, vol. 17, p. 8, 2017.

[468] B. Bhoi, B. Kim, J. Kim, Y.-J. Cho, and S.-K. Kim, "Robust magnon-photon coupling in a planar-geometry hybrid of inverted split-ring resonator and YIG film," *Sci. Rep*, vol. 7, p. 11930, 2017.

[469] M. T. Kaffash, D. Wagle, and M. B. Jungfleisch, "Direct imaging of strong magnon-photon coupling in a planar geometry," *in preparation.* .

[470] Y. Li, "Coherent coupling of two remote magnonic resonators mediated by superconducting circuits," *Prep.*

[471] A. A. Gentile *et al.*, "Learning models of quantum systems from experiments," *Nat. Phys*, vol. 17, p. 1, 2020.

[472] V. I. Shnyrkov, A. A. S. Wu Yangcao, O. G. Turutanov, and V. Y. Lyakhno, "Frequency-tuned microwave photon counter based on a superconductive quantum interferometer," *Low Temp. Phys*, vol. 44, num. 3, 2018.

[473] M. A. Nielsen and I. L. Chuang, *Quantum Computation and Quantum Information*. Cambridge: Cambridge Press, 2000.

[474] X. Zhang, C. L. Zou, L. Jiang, and H. Tang, "Strongly coupled magnons and cavity microwave photons," *Phys. Rev. Lett*, vol. 113, p. 156401, 2014.

[475] Y. Wang and C. M. Hu, "Dissipative couplings in cavity magnonics," *J. Appl. Phys*, vol. 127, p. 130901, 2020.

[476] N. Crescini, C. Braggio, G. Carugno, R. D. Vora, A. Ortolan, and G. Ruoso, "Magnon-driven dynamics of a hybrid system excited with ultrafast optical pulses," *Commun. Phys*, vol. 3, p. 164, 2020.

[477] X. Zhang, A. Galda, X. Han, D. Jin, and V. M. Vinokur, "Broadband nonreciprocity enabled by strong coupling of magnons and microwave photons," *Phys. Rev. Appl*, vol. 13, p. 44039, 2020.

[478] J. Xu, C. Zhong, X. Han, D. Jin, L. Jinag, and X. Zhang, "Coherent

gate operations in hybrid magnonics," *Phys. Rev. Lett*, vol. 126, p. 207202, 2021.

[479] C. Potts and J. Davis, "Strong magnon–photon coupling within a tunable cryogenic microwave cavity," *Appl. Phys. Lett*, vol. 116, p. 263503, 2020.

[480] J. Xu, C. Zhong, X. Han, D. Jin, L. Jinag, and X. Zhang, "Floquet cavity electromagnonics," *Phys. Rev. Lett*, vol. 125, p. 237201, 2020.

[481] S. I. Y.Tabuchi, A. Noguchi, T. Ishikawa, R. Yamazaki, K. Usami, and Y. Nakamura, "Coherent coupling between a ferromagnetic magnon and a superconducting qubit," *Science (80-. ).*, vol. 349, pp. 405–408, 2015.

[482] A. A. Clerk, K. W. Lehnert, P. Bertet, J. R. Petta, and Y. Nakamura, "Hybrid quantum systems with circuit quantum electrodynamics," *Nat. Phys*, vol. 16, p. 257, 2020.

[483] X. Zhang, C.-L. Zou, N. Zhu, F. Marquardt, L. Jiang, and H. X. Tang, "Magnon dark modes and gradient memory"," *Nat. Commun*, vol. 6, p. 8914, 2015.

[484] R. Hisatomi *et al.*, "Bidirectional conversion between microwave and light via ferromagnetic magnons," *Phys. Rev. B*, vol. 93, p. 174427, 2016.

[485] I. A. Golovchanskiy *et al.*, "Ultrastrong photon-to-magnon coupling in multilayered heterostructures involving superconducting coherence via ferromagnetic layers," *Sci. Adv*, vol. 7, p. 8638, 2021.

[486] I. A. Golovchanskiy *et al.*, "Approaching Deep-Strong On-Chip Photon-To-Magnon Coupling," *Phys. Rev. Appl*, vol. 16, p. 34029, 2021.

[487] E. Lee-Wong *et al.*, "Nanoscale detection of magnon excitations with variable wavevectors through a quantum spin sensor," *Nano Lett*, vol. 20, p. 3284, 2020.

[488] L. Trifunovic, L. Pedrocchi, and D. Loss, "Long-distance entanglement of spin qubits via ferromagnet," *Phys. Rev*, vol. X, 3, p. 41023, 2013.

[489] C. Gonzalez-Ballestero, J. Gieseler, and O. Romero-Isart, "Quantum Acoustomechanics with a micromagnet"," *Phys. Rev. Lett*, vol. 124, p. 93602, 2020.

[490] C. Gonzalez-Ballestero, D. Hümmer, J. Gieseler, and O. Romero-Isart, "Theory of quantum acoustomagnonics and acoustomechanics with a micromagnet"," *Phys. Rev. B*, vol. 101, p. 125404, 2020.

[491] M. Elyasi, Y. M. Blanter, and G. E. W. Bauer, "Resources of nonlinear cavity magnonics for quantum information," *Phys. Rev. B*, vol. 101, p. 54402, 2020.

[492] M. Elyasi, E. Saitoh, and G. E. W. Bauer, "Theory of magnon parametrons"," *arXiv:2109.09117*, 2021.

[493] J. Li, S.-Y. Zhu, and G. S. Agarwal, "Magnon-Photon-Phonon Entanglement in Cavity Magnomechanics," *Phys. Rev. Lett*, vol. 121, p. 203601, 2018.

[494] G. Kirchmair *et al.*, "Observation of quantum state collapse and revival due to the single-photon Kerr effect," *Nature*, vol. 495, pp. 205–209, 2013.

[495] B. Vlastakis *et al.*, "Deterministically encoding quantum information using 100-photon Schrödinger cat states," *Science (80-. ).*, vol. 342, pp. 607–610, 2013.

[496] S. I. Y.Tabuchi, A. Noguchi, T. Ishikawa, R. Yamazaki, K. Usami, and Y. Nakamura, "Quantum magnonics: The magnon meets the superconducting qubit," *C. R. Phys*, vol. 17, pp. 729–739, 2015.

[497] D. Lachance-Quirion *et al.*, "Resolving quanta of collective spin excitations in a millimeter-sized ferromagnet," *Sci. Adv*, vol. 3, p. 1603150, 2017.

[498] C. L. Degen, F. Reinhard, and P. Cappellaro, "Quantum sensing," *Rev. Mod. Phys*, vol. 89, p. 35002, 2017.

[499] S. P. Wolski *et al.*, "Dissipation-based Quantum Sensing of Magnons with a Superconducting Qubit," *Phys. Rev. Lett*, vol. 125, p. 117701, 2020.

[500] A. Schneider *et al.*, "Transmon qubit in a magnetic field: Evolution of coherence and transition frequency," *Phys. Res. Res.*, vol. 1, no. 2, 2019, doi: 10.1103/PhysRevResearch.1.023003.